\documentclass[oneside, a4paper, 12pt]{book}
\usepackage[a-1b]{pdfx}
\usepackage[
  backend=biber,
  style=numeric,
  doi=true,
  url=false,
  isbn=false,
  eprint=false
]{biblatex}

\DeclareFieldFormat
  [article,inbook,incollection,inproceedings,patent,thesis,unpublished]
  {title}{#1}

\renewbibmacro{in:}{%
  \ifentrytype{article}{%
  }{%
    \printtext{\bibstring{in}\intitlepunct}%
  }%
}

\DeclareFieldFormat{pages}{#1}
\DeclareFieldFormat[article]{number}{\mkbibparens{#1}}

\renewbibmacro*{volume+number+eid}{%
  \setunit{\addcomma\space}%
  \printfield{volume}%
  \printfield{number}%
  \setunit{\addcomma\space}%
  \printfield{eid}%
  \setunit{\addcolon}%
  \printfield{pages}}

\renewbibmacro*{issue+date}{%
  \setunit{\addcomma\space}%
  \iffieldundef{issue}
    {\usebibmacro{date}}
    {\printfield{issue}%
     \setunit*{\addspace}%
     \usebibmacro{date}}%
\newunit}

\renewbibmacro*{note+pages}{%
  \printfield{note}%
  \newunit}

\usepackage[margin=30truemm]{geometry}

\usepackage{sty/thesis} 

\usepackage{graphicx}
\usepackage{latexsym}
\usepackage[figuresright]{rotating}
\usepackage{caption}

\usepackage{mathtools}
\usepackage{bussproofs}
\EnableBpAbbreviations{}
\newcommand{\shortstrut}{\vrule
  height \dimexpr\ht\strutbox-1.5pt\relax
  depth  \dimexpr\dp\strutbox-1.5pt\relax
  width  0pt}
\newcommand{\tallstrut}{\vrule
  height \dimexpr\ht\strutbox+2.5pt\relax
  depth  \dimexpr\dp\strutbox\relax
  width  0pt}

\usepackage{tikz}
\newcommand{\tikzexternalenable}{}
\newcommand{\tikzexternaldisable}{}
\usepackage{style-tikz}

\usepackage{amsmath,amssymb,amsfonts,amsthm}
\usepackage{mathrsfs}
\usepackage{ascmac}
\usepackage{cancel}
\usepackage{xcolor}
\usepackage{url}
\usepackage[percent]{overpic}
\usepackage{framed}

\let\newblock\relax
\usepackage[capitalise,nameinlink]{cleveref}

\usepackage{sty/algorithm} 
\usepackage{booktabs}     
\usepackage{xparse}

\usepackage{xspace}
\newcommand{\lgt}{\ensuremath{\lambda_\mathrm{GT}}\xspace}

\newcommand{\Xs}{\overrightarrow{X}}
\newcommand{\Ys}{\overrightarrow{Y}}
\newcommand{\Zs}{\overrightarrow{Z}}
\newcommand{\Ws}{\overrightarrow{W}}
\newcommand{\Vs}{\overrightarrow{V}}
\newcommand{\Us}{\overrightarrow{U}}

\newcommand{\Gs}{\overrightarrow{G}}

\newcommand{\thetas}{\overrightarrow{\theta}}

\newcommand{\Cons}{\mathrm{Cons}}

\newcommand{\Nil}{\mathrm{Nil}}
\newcommand{\Node}{\mathrm{Node}}
\newcommand{\Leaf}{\mathrm{Leaf}}
\newcommand{\LeafM}{\mathrm{LeafM}}
\newcommand{\one}{\mathtt{1}}
\newcommand{\two}{\mathtt{2}}

\newcommand{\dbllist}{\mathit{dbllist}}
\newcommand{\nat}{\mathit{nat}}

\newcommand{\li}{\mathit{list}}

\newcommand{\dom}{\mathrm{dom}}
\newcommand{\toto}{\longrightarrow}
\newcommand{\reduces}{\> \longrightarrow_{\mathrm{v}}\>}
\newcommand{\reducestwo}{\toto^2_{\mathrm{v}}}

\newcommand{\reducesp}[1]{\> \longrightarrow_{\mathrm{v}, {#1}}\>}

\newcommand{\reducespstar}[1]{\> \longrightarrow^\ast_{\mathrm{v}, {#1}}\>}

\newcommand{\lto}{\longrightarrow}
\newcommand{\fn}{\mathit{fn}}
\newcommand{\caseof}[4]{\mathbf{case}\;#1\;\mathbf{of}\;#2\to #3\;\texttt{|}\;\mathbf{otherwise} \to #4}
\newcommand{\letrec}[4]{\mathbf{let\ rec}\;#1\;#2\;\texttt{=}\;#3\;\mathbf{in}\;#4}
\newcommand{\letin}[3]{\mathbf{let}\;#1\;\texttt{=}\;#2\;\mathbf{in}\;#3}

\newcommand{\zero}{\textbf{0}}
\newcommand{\LI}[3]{(\{#1\} \multimap #2){\{#3\}}}
\newcommand{\LIempty}[2]{(\emptyset \multimap #1){\{#2\}}}

\newcommand{\Root}{\textit{pr}}
\newcommand{\FusedToRoot}{\textit{fused2PR}}
\newcommand{\areFused}[3]{#1 \bowtie_{#3} #2}
\newcommand{\CollectVars}{\mathit{vt}}

\newcommand{\seq}[1]{\overrightarrow{#1}}
\newcommand{\taus}{\overrightarrow{\tau}}
\newcommand{\ml}[1]{\begin{array}[b]{@{}l@{}l@{}l@{}l@{}}#1\end{array}}
\newcommand{\mlc}[1]{\begin{array}[b]{ccccc}#1\end{array}}
\newcommand{\narrowcolon}{\mathop{:}}

\newcommand{\paren}[1]{\bigl( #1 \bigr)}
\newcommand{\angled}[1]{\langle #1 \rangle}

\newcommand{\norm}[1]{\lvert #1 \rvert}
\newcommand{\rbracket}[1]{\left[ #1 \right]}

\newcommand{\TyVar}{\textsf{Ty-Var}}
\newcommand{\TyArrow}{\textsf{Ty-Arrow}}
\newcommand{\TyApp}{\textsf{Ty-App}}
\newcommand{\TyAlpha}{\textsf{Ty-Alpha}}
\newcommand{\TyCong}{\textsf{Ty-Cong}}
\newcommand{\TyProd}{\textsf{Ty-Prod}}
\newcommand{\TyCase}{\textsf{Ty-Case}}
\newcommand{\TyLIIntro}{\textsf{Ty-LI-Intro}}
\newcommand{\TyLITrans}{\textsf{Ty-LI-Trans}}
\newcommand{\TyLIElimZ}{\textsf{Ty-LI-Elim0}}
\newcommand{\TyLIIntroZ}{\textsf{Ty-LI-Intro0}}
\newcommand{\TyLIE}{\textsf{Ty-LI-E}}
\newcommand{\TyLIs}{\textsf{Ty-LI-}\ast}
\newcommand{\TyLetRec}{\textsf{Ty-LetRec}}
\newcommand{\TyVarA}{\textsf{V}}
\newcommand{\TyAlphaA}{\ensuremath{\alpha}}
\newcommand{\TyProdA}{\textsf{P}}
\newcommand{\TyLIIntroA}{\textsf{I}}
\newcommand{\TyLITransA}{\textsf{Tr}}
\newcommand{\TyLIElimZA}{\textsf{E0}}
\newcommand{\TyLIIntroZA}{\textsf{I0}}
\newcommand{\TyCongA}{\ensuremath{\equiv}}
\newcommand{\TyCaseA}{\textsf{C}}
\newcommand{\TyArrowA}{\textsf{Ar}}
\newcommand{\TyLIEA}{\textsf{E}}

\newcommand{\TyCaseC}{\textsf{Ty-Case}}

\newcommand{\RdBeta}{\textsf{Rd-$\beta$}}
\newcommand{\RdCaseMatch}{\textsf{Rd-Case1}}
\newcommand{\RdCaseOther}{\textsf{Rd-Case2}}
\newcommand{\RdLetRec}{\textsf{Rd-LetRec}}

\newcommand{\RdBetaD}{\textsf{Rd-$\beta$-D}}
\newcommand{\RdCaseMatchD}{\textsf{Rd-Case1-D}}
\newcommand{\RdCaseOtherD}{\textsf{Rd-Case2-D}}

\newcommand{\VarLinks}{\mathit{prs}}
\newcommand{\PatCond}{\mathit{patCond}}

\newcommand{\vinf}[1]{%
  \begin{array}[b]{c}
    \vdots \\
    #1
  \end{array}%
}

\newcommand{\Tx}{\mathcal{T}}
\newcommand{\Tc}{\widehat{\mathcal{T}}}

\newcommand{\eDLpop}{\ensuremath{e_{\mathsf{DLpop}}}}
\newcommand{\eLLT}{\ensuremath{e_{\mathsf{LLT}}}}
\newcommand{\eDLun}{\ensuremath{e_{\mathsf{DLfail}}}}
\newcommand{\eLappend}{\ensuremath{e_{\mathsf{append}}}}
\newcommand{\patDLun}{\ensuremath{\Tx_{\mathsf{DLfail}}}}

\newcommand{\GDL}{\ensuremath{G_\mathsf{DL3}}}
\newcommand{\Pdl}{\ensuremath{P_\textsf{DL-LLT}}}

\newtheorem{theorem}{Theorem}[section]
\newtheorem{prop}[theorem]{Proposition}
\newtheorem{corollary}[theorem]{Corollary}
\newtheorem{lemma}[theorem]{Lemma}
\newtheorem*{lemma*}{Lemma}
\newtheorem{example}[theorem]{Example}
\theoremstyle{definition}
\newtheorem{definition}[theorem]{Definition}

\crefname{theorem}{Theorem}{Theorems}
\Crefname{theorem}{Theorem}{Theorems}
\crefname{prop}{Proposition}{Propositions}
\Crefname{prop}{Proposition}{Propositions}
\crefname{corollary}{Corollary}{Corollaries}
\Crefname{corollary}{Corollary}{Corollaries}
\crefname{lemma}{Lemma}{Lemmas}
\Crefname{lemma}{Lemma}{Lemmas}
\crefname{example}{Example}{Examples}
\Crefname{example}{Example}{Examples}
\crefname{definition}{Definition}{Definitions}
\Crefname{definition}{Definition}{Definitions}

\newenvironment{mydef}[1][]
{\begin{definition}[#1]\pushQED{\qed}}
{\popQED\end{definition}}

\newcommand{\ctxx}[1]{%
  \tikz[baseline=1.5ex, scale=1]{%
    \begin{scope}[shift={(0,0)}]
      \node [ctxn] (C1)  at (0, 0)    {$#1$};
      \coordinate (p1AL) at ($(C1)+(-0.1,0.25)$) {};
      \coordinate (p1AR) at ($(C1)+(+0.1,0.25)$) {};
      \node [port] at (-0.400, -0.150) {1};
      \node [port] at (+0.250, +0.400) {2};
      \node [port] at (-0.250, +0.400) {3};
    \end{scope}
    \node [link] (LZ)  at (-1.0, 0.8) {$Z$};
    \node [link] (LW2) at (+1.0, 0.8) {$Y$};
    \node [link] (LX)  at (-1.0, 0.0) {$X$};
    \draw [out=0,   in=90, looseness=1.2] (LZ) to (p1AL);
    \draw [out=180, in=90, looseness=1.2] (LW2) to (p1AR);
    \draw []  (C1) to  (LX);
  }
}

\newcommand{\ctxu}[1]{%
  \tikz[font=\footnotesize, baseline=-0.5ex, scale=1]{%
    \node[ctxn,inner sep=3pt] (x) at (0, 0) {\(#1\)};
    \node[link] (X) at (0, 0.8) {$X$};
    \draw[] (x)--(X);
    \node[port] at (-0.15, 0.450) {1};
  }
}

\newcommand{\lltr}[1]{%
  \tikz[font=\normalsize, baseline=0.5ex, scale=1]{%
    \node[triangle,inner sep=1pt,font=\normalsize] (x) at (0, 0) {\(#1\)};
    \node[link] (X) at (+0.0, +0.8) {$X$};
    \node[link] (L) at (-0.8, -0.1) {$L$};
    \node[link] (R) at (+0.8, -0.1) {$R$};
    \node[port] at (-0.150, +0.500) {3};
    \node[port] at (-0.350, +0.150) {1};
    \node[port] at (+0.350, +0.150) {2};
    \draw[] ($(x.north)-(0,0.1)$)--(X);
    \draw[] (x)--(L);
    \draw[] (x)--(R);
  }
}

\NewDocumentCommand{\mkdbll}{m m g}{%
  \begin{scope}[shift={#2}]
    \node [tvar] (C#1) at (0, 0) {$\mathit{dbll}$};
    \coordinate (p#1AL) at ($(C#1)+(-0.15,0.27)$) {};
    \coordinate (p#1AR) at ($(C#1)+(+0.15,0.27)$) {};
    \coordinate (p#1L)  at ($(p#1AL)+(-0.35,0.53)$);
    \coordinate (p#1R)  at ($(p#1AR)+(+0.35,0.53)$);
    \node [port] at (-0.55, -0.150) {1};
    \node [port] at (+0.30, +0.400) {2};
    \node [port] at (-0.30, +0.400) {3};
    \IfNoValueF{#3}{\node [link,text=gray] at (-0.350, 1.000) {$(#3)$};}
  \end{scope}
}%

\newcommand{\dblltype}[3]{%
  \tikz[baseline=1.5ex, scale=1]{%
    \mkdbll{2}{(0,0)}
    \node [link] (C1) at (-1.0, 0.8) {$#3$}; 
    \node [link] (C3) at (+1.0, 0.8) {$#2$}; 
    \node [link] (LX) at (-1.0, 0.0) {$#1$}; 
    \draw [out=0,   in=90, looseness=1.5] (C1) to (p2AL);
    \draw [out=180, in=90, looseness=1.5] (C3) to (p2AR);
    \draw [] (C2) to (LX);
  }
}

\newcommand{\lltrtype}[3]{%
  \tikz[baseline=1ex, scale=1]{%
    \node[lltr] (x) at (0, 0) {\(\mathit{lltr}\)};
    \node[link] (X) at (+0, +1.0) {$#1$}; 
    \node[link] (L) at (-1, -0.1) {$#2$}; 
    \node[link] (R) at (+1, -0.1) {$#3$}; 
    \node[port] at (-0.150, 0.650) {3};
    \node[port] at (-0.450, 0.150) {1};
    \node[port] at (+0.450, 0.150) {2};
    \draw[] ($(x.north)-(0,0.1)$)--(X);
    \draw[] (x)--(L);
    \draw[] (x)--(R);
  }
}

\newcommand{\ctxyy}[2]{%
  \begin{scope}[shift={(#1,#2)}]
    \coordinate (t1) at (-2,  0);
    \coordinate (t2) at (0,   2);
    \coordinate (t3) at (2,   0);
    \coordinate (t4) at (1.7, 0);
    \coordinate (t5) at (0.6, 1);
    \coordinate (t6) at (0, 0.4);
    \coordinate (t7) at (-0.6, 1);
    \coordinate (t8) at (-1.7, 0);
    \draw[fill=ctxcolor,draw=linkcolor,dashed,rounded corners]
    (t1)--(t2)--(t3)--(t4)--(t5)--(t6)--(t7)--(t8)--cycle;
    \node[] at (0, 1.2) {\(y\)};
    \draw[] ($(t2)-(0,0.1)$) --++(0, 0.5);
    \draw[] (-1.7,0.3) --++(-0.6, -0);
    \draw[] ( 1.7,0.3) --++(+0.6, -0);
    \node[link] at (0,     2.6) {$X$};
    \node[link] at (-2.5, 0.35) {$L$};
    \node[link] at ( 2.5, 0.35) {$R$};
    \node[port] at (-0.25, 2.1) {6};
    \node[port] at (-1.9, 0.5)  {1};
    \node[port] at (-1.45, 0.1) {2};
    \node[port] at (-0.7, 0.8)  {3};
    \node[port] at (+0.7, 0.8)  {4};
    \node[port] at (+1.45, 0.1) {5};
    \node[port] at ( 1.9, 0.5)  {6};
  \end{scope}
}

\newcommand{\ctxyl}[2]{%
  \begin{scope}[shift={(#1,#2)}]
    \coordinate (t1) at (-2,  0)  ;
    \coordinate (t2) at (0,   2);
    \coordinate (t3) at (0.5, 1.5)  ;
    \coordinate (t6) at (-1,0)  ;
    \draw[fill=ctxcolor,draw=linkcolor,dashed, rounded corners]
    (t1)--(t2)--(t3)--(t6)--cycle;
    \node[] at (-0.5, 1) {\(y\)};
    \draw[] ($(t2)-(0,0.1)$) --++(0, 0.5);
    \draw[] (-1.7,0.3) --++(-0.6, -0);
    \node[link] (LX) at (0,     2.6) {$X$};
    \node[link] (LL) at (-2.5, 0.35) {$L$};
    \node[link] (LR) at ( 2.3, 0.35) {$R$};
    \node[port] at (-0.25, 2.1) {4};
    \node[port] at (-1.90, 0.5) {1};
    \node[port] at (-0.70, 0.0) {2};
    \node[port] at (+0.55, 1.5) {3};
  \end{scope}
}

\definecolor{ghostcolor}{RGB}{100,200,255}

\tikzset{
  ghost/.style={fill=white,draw=ghostcolor,text=ghostcolor},
  every node/.style={font=\small,thick},
}

\makeatletter
\newcommand\newblock{\hskip .11em\@plus.33em\@minus.07em}
\makeatother

\usepackage{listings}      
\usepackage[scaled]{inconsolata}

\lstdefinelanguage{lgt}
{
  morekeywords={let, rec, in, case, of, nu, otherwise,},
  comment=[l]{\%},
  morestring=[b]",
  alsodigit={-},
  alsoletter={&, _},
}

\definecolor{codeorange}{RGB}{220,100,0}
\definecolor{codecolor}{RGB}{0,0,200}
\definecolor{codeblue}{RGB}{0,50,200}
\definecolor{codelightblue}{RGB}{0,100,250}
\definecolor{codegreen}{RGB}{0,150,50}
\definecolor{codered}{RGB}{250,100,0}
\definecolor{codesky}{RGB}{0,150,250}

\lstdefinestyle{lgtinline}{
  language=lgt,
  basicstyle=\ttfamily,
  columns=fullflexible,
  keepspaces=true,
  breaklines=false
}

\lstdefinestyle{shellplain}{
  language=bash,
  basicstyle={\footnotesize\ttfamily},
  identifierstyle={},
  commentstyle={},
  keywordstyle={},
  stringstyle={},
  frame={tb},
  lineskip=0pt,
  breaklines=false,
  numbers=none,
  numberstyle={},
  xleftmargin=0pt,
  numbersep=5pt,
  stepnumber=1,
}

\newcommand{\nattonat}{\mathit{nat}(X) \mathop{\rightarrow} \mathit{nat}(X)}
\newcommand{\Tauu}{\mathcal{T}}

\newcommand{\subfigimage}[3][]{%
  \begin{overpic}[#1]{#2}
    \put(3,80){#3}
  \end{overpic}%
}

\newcommand{\linksize}{\scriptsize}

\newcommand{\colortt}[1]{\texttt{\textcolor{codecolor}{#1}}}

\NewDocumentCommand{\mkports}{m m m m m}{%
  \IfBlankF{#1}{\node [port] at (-0.375, -0.150)   {#1};}
  \IfBlankF{#2}{\node [port] at (+0.150, +0.375)   {#2};}
  \IfBlankF{#3}{\node [port] at (+0.375, -0.150)   {#3};}
  \IfBlankF{#4}{\node [port] at (-0.150, -0.375)   {#4};}
  \IfBlankF{#5}{\node [port] at (-0.150, +0.375-1) {#5};}
}

\ExplSyntaxOn

\NewDocumentCommand{\IfPosIntTF}{m m m}
{
  \regex_match:nnTF { \A [0-9]+ \Z } { #1 }
  { #2 }
  { #3 }
}

\ExplSyntaxOff

\NewDocumentCommand{\mkcons}{m m m m g}{%
  \begin{scope}[shift={#4}]
    \node [atom] (C#1) at (0, 0) {C};
    \coordinate (p#1AR) at ($(C#1)+(0,0.4)$) {};
    \coordinate (p#1AL) at ($(C#1)+(0,0.4)$) {};
    \coordinate (p#1R) at ($(p#1AR)+(+0.5,0.4)$);
    \coordinate (p#1L) at ($(p#1AL)+(-0.5,0.4)$);
    \IfNoValueTF{#5}{%
      \node [tvar] (N#1) at ($(C#1)+(0,-1)$) {\(\mathit{n}\)};
    }{%
      \IfPosIntTF{#5}{%
        \node [atom] (N#1) at ($(C#1)+(0,-1)$) {#5};
      }{%
        \node [ctxn] (N#1) at ($(C#1)+(0,-1)$) {$#5$};
      }%
    }%
    \mkports{2}{4}{3}{1}{1}
    \draw [] (C#1) to (N#1);
    \node [link,text=gray] at (0.50, -0.5) {$(#2)$};
    \IfBlankF{#3}{%
      \node [link,text=gray] at (0, 1) {$(#3)$};
    }%
  \end{scope}
}

\newcommand{\mknil}[2]{%
  \begin{scope}[shift={#2}]
    \node [atom] (C#1) at (0, 0) {N};
    \coordinate (p#1AR) at ($(C#1)+(0,0.4)$) {};
    \coordinate (p#1AL) at ($(C#1)+(0,0.4)$) {};
    \coordinate (p#1R) at ($(p#1AR)+(+0.5,0.4)$);
    \coordinate (p#1L) at ($(p#1AL)+(-0.5,0.4)$);
    \node [port] at (+0.150, +0.375) {2};
    \node [port] at (-0.375, -0.150) {1};
  \end{scope}
}

\newcommand{\mkconslinkleft}[2]{%
  \draw [in=180, out=0, looseness=1.50] (C#1) to (p#2L);
  \draw [in=90, out=0, rounded corners] (p#2L) to (p#2AL);
  \draw (p#2AL) to (C#2);
}

\newcommand{\mkconslinkright}[2]{%
  \draw [in=0, out=180, looseness=1.50, line width=4pt, white] (C#2) to (p#1R);
  \draw [in=0, out=180, looseness=1.50] (C#2) to (p#1R);
  \draw [in=90, out=180, rounded corners] (p#1R) to (p#1AR);
}

\newcommand{\mkconslink}[2]{%
  \mkconslinkleft{#1}{#2}
  \mkconslinkright{#1}{#2}
}

\NewDocumentCommand{\mkctxn}{m m m m m}{%
  \begin{scope}[shift={#3}]
    \node [ctxn] (C#1) at (0, 0) {$#2$};
    \coordinate (p#1AL) at ($(C#1)+(-0.1,0.3)$) {};
    \coordinate (p#1AR) at ($(C#1)+(+0.1,0.3)$) {};
    \coordinate (p#1L)  at ($(p#1AL)+(-0.5,0.5)$);
    \coordinate (p#1R)  at ($(p#1AR)+(+0.5,0.5)$);
    \node [port] at (+0.400, -0.150) {1};
    \node [port] at (+0.250, +0.450) {2};
    \node [port] at (-0.400, -0.150) {3};
    \node [port] at (-0.250, +0.450) {4};
    \IfBlankF{#4}{\node [link,text=gray] at (-0.4, 1) {$(#4)$};}%
    \IfBlankF{#5}{\node [link,text=gray] at (+0.4, 1) {$(#5)$};}%
  \end{scope}%
}

\NewDocumentCommand{\mkleafatom}{m}{%
  \tikz[baseline=-2pt, scale=1]{%
    \node [atom] (cons1) at (0, 0) {L};
    \IfPosIntTF{#1}{%
      \node [atom] (num1) at (0, -1) {#1};
    }{%
      \node [tvar] (num1) at (0, -1) {$\mathit{#1}$};
    }%
    \draw (cons1) -- (num1);
    \node [link] (p1) at (+0.85, 0) {$R$};
    \node [link] (p2) at (-0.85, 0) {$L$};
    \node [link] (p3) at (0, 0.95)  {$X$};
    \draw (cons1) to (p1);
    \draw [looseness=1.50,in=30,out=90] (cons1) to (p2);
    \draw (cons1) -- (p3);
    \mkports{}{3}{2}{1}{1}
    \node [link,text=gray] at (0.50, -0.5) {$(W_1)$};
  }%
}

\usepackage{fancyhdr}



\usetikzlibrary{calc,positioning}

\begin{document}

\begin{titlepage}
  \begin{center}
    \mbox{}
    \vspace{2truecm}

    {\Large
      \textbf{
        Functional Programming \\[0.2em]
        \quad with Hypergraph Transformation
      }
    }

    \vspace{3truecm}

    {\large
      \textit{Jin Sano}
    }

    \vspace{10truecm}

    {\normalsize
      Doctor of Engineering\\[0.1em]
      Department of Computer Science and Communications Engineering\\[0.1em]
      Graduate School of Fundamental Science and Engineering\\[0.1em]
      Waseda University\\[0.1em]
      July 2026
    }

  \end{center}
\end{titlepage}

\frontmatter
\section*{Abstract}

Declarative and safe manipulation of complex data structures remains a central challenge in programming language design.
Imperative languages provide a direct means of representing arbitrary structures through pointers and destructive updates, but this style of programming is often cumbersome and error-prone.
Moreover, the flexibility of such representations makes it difficult to ensure that the resulting structures satisfy programmer-intended invariants.
Functional languages, by contrast, support declarative manipulation through pattern matching and provide static guarantees for algebraic data types.
However, these mechanisms are most naturally suited to tree-shaped data and do not directly capture richer structures.
Graph transformation languages address this limitation by supporting graph manipulation through pattern matching and rewriting.
However, their state-transforming computational model does not integrate seamlessly with ordinary value- and function-oriented programming.

This dissertation addresses this gap by introducing \lgt{}, a programming language that integrates \emph{port-hypergraph transformation} into functional programming.
In \lgt{}, complex data structures can be represented as port hypergraphs and manipulated declaratively through pattern matching.
Its type system allows programmers to specify shape invariants and provides static guarantees that graph values conform to these invariants.

As a semantic foundation, we first formalise HyperLMNtal as a term-based port-hypergraph transformation formalism.
Building on this foundation, we define the syntax and operational semantics of the untyped \lgt{} language.
The language extends a call-by-value \(\lambda\)-calculus with port-hypergraph values and graph pattern matching, thereby enabling declarative operations over complex data structures.
We also implement a compact prototype interpreter, demonstrating that the core mechanisms of \lgt{} can be realised in an executable system.

We then develop a type system for \lgt{}.
Unlike type systems that specify only local node interfaces, this system can express global shape constraints on graphs.
We first formalise a core type system and prove its type soundness.
We also show that this type system can be viewed as a natural extension of existing graph-grammar-based typing frameworks for graph transformation languages.
We then extend the type system with \emph{linear implication types}, which provide a disciplined way to type incomplete data structures that naturally arise during pattern matching.
For the extended system, we establish type soundness in the presence of dynamic checks during pattern matching and prove conservativity over complete graph types.
We then identify a statically safe fragment of \lgt{} in which such dynamic checks can be safely eliminated.

Finally, we mechanise the definitions and meta-theoretic results in Isabelle/HOL.
This mechanisation strengthens confidence in the soundness arguments and provides a reliable foundation for future extensions of the language.

\section*{Acknowledgements}

First, I would like to express my deepest gratitude to my supervisor,
Professor Kazunori Ueda,
for his guidance and support throughout my doctoral studies.
His patient supervision, insightful advice, and continuous encouragement
were essential to the development of this research.

I am also deeply grateful to Professor Tachio Terauchi
for kindly serving as the chair of my dissertation committee
and for his valuable support in the final stage of my doctoral program.
I would also like to thank the other members of my dissertation committee,
Professor Kazue Sako and Professor Akimasa Morihata,
for their careful reading of my dissertation
and for their helpful comments and advice.

I would like to thank the anonymous reviewers of my papers,
especially those of IPSJ PRO (Information Processing Society of Japan, Special Interest Group on Programming),
for their careful reading and many useful comments.
Their feedback helped me clarify the ideas and contributions presented in this dissertation.

I am grateful to all the members of the Ueda Laboratory
for their support and for the research environment they created.
In particular,
I would like to thank Naoki Yamamoto for his advice and support.
As one of the pioneers in the research area of this dissertation,
he gave me helpful insight into my work.

There are many other colleagues and friends whom I would like to thank.
Although I cannot mention everyone individually,
I am grateful to all those who supported me throughout this period.

Finally,
I would like to thank my family for their support,
patience,
and understanding throughout this period.

\tableofcontents            
\newpage
\mainmatter

\chapter{Introduction}\label{ch:intro}

\begin{figure}[htbp]
  \setlength{\FrameSep}{0pt}
  \begin{framed}
    \centering
    \noindent
    \begin{tikzpicture}[scale=0.9]
  \path [use as bounding box] (0, -1) rectangle (16, 8.5);

  \begin{scope}[shift={(0, 6)}]
    \node [anchor=west] at (0, 1.75) {Difference list};
    \node [] (X) at (0,1) {};
    \node [atom] (c1) at (1,1) {C};
    \node [atom] (c2) at (2,1) {C};
    \node [atom] (c3) at (3,1) {C};
    \node [atom] (c4) at (4,1) {C};
    \node [] (Y) at (5,1) {};
    \draw (X)--(c1)--(c2)--(c3)--(c4)--(Y);

    \node [atom] (n1) at (1,0) {1};
    \draw (c1)--(n1);
    \node [atom] (n2) at (2,0) {2};
    \draw (c2)--(n2);
    \node [atom] (n3) at (3,0) {3};
    \draw (c3)--(n3);
    \node [atom] (n4) at (4,0) {4};
    \draw (c4)--(n4);
  \end{scope}

  \begin{scope}[shift={(0, 3)}]
    \node [anchor=west] at (0, 1.75) {Doubly linked list};
    \coordinate (x1) at (0.1, 1.15);
    \coordinate (x2) at (0.1, 0.85);
    \coordinate (y1) at (5, 1.15);
    \coordinate (y2) at (5, 0.85);
    \draw (x1)--(y1);
    \draw (x2)--(y2);

    \node [] (X) at (0,1) {};
    \node [atom] (c1) at (1,1) {C};
    \node [atom] (c2) at (2,1) {C};
    \node [atom] (c3) at (3,1) {C};
    \node [atom] (c4) at (4,1) {C};
    \node [atom] (c5) at (5,1) {C};

    \node [atom] (n1) at (1,0) {1};
    \draw (c1)--(n1);
    \node [atom] (n2) at (2,0) {2};
    \draw (c2)--(n2);
    \node [atom] (n3) at (3,0) {3};
    \draw (c3)--(n3);
    \node [atom] (n4) at (4,0) {4};
    \draw (c4)--(n4);
  \end{scope}

  \begin{scope}[shift={(0, 0)}]
    \node [anchor=west] at (0, 2) {Skip list};
    \node [] (X) at (0,1) {};
    \node [atom] (c1) at (1,1) {C};
    \node [atom] (c2) at (2,1) {C};
    \node [atom] (c3) at (3,1) {C};
    \node [atom] (c4) at (4,1) {C};
    \node [atom] (c5) at (5,1) {C};
    \draw (X)--(c1)--(c2)--(c3)--(c4)--(c5);

    \node [atom] (n1) at (1,0) {1};
    \draw (c1)--(n1);
    \node [atom] (n2) at (2,0) {2};
    \draw (c2)--(n2);
    \node [atom] (n3) at (3,0) {3};
    \draw (c3)--(n3);
    \node [atom] (n4) at (4,0) {4};
    \draw (c4)--(n4);

    \draw (0.1, 1.63) to[out=0,in=130] (c1);
    \draw (c1) to[out=50,in=130] (c3);
    \draw (c3) to[out=50,in=130] (c5);
  \end{scope}

  \begin{scope}[shift={(6.5, 4)}]
    \node [anchor=west] at (-0.5, 3.75) {Leaf-linked tree};

    \draw [blue!40!white, dashed, line width=0.75mm]
    (3.9, 1.2)--(5.25, -1.0);
    \draw [blue!40!white, dashed, line width=0.75mm]
    (4.85, 2.25)--(7.15, 1.15);

    \node [fill=blue!15!white, circle, minimum size=1.5cm]
    (circ1) at (4.5, 1.6) {};

    \node [] (X) at (3, 4) {};
    \node [atom] (c1) at (3, 3) {C};
    \node [atom] (c2) at (2, 2.5) {C};
    \node [atom] (c3) at (4, 2.5) {C};
    \node [atom] (c4) at (2.5, 1.5) {C};

    \draw (X)--(c1);
    \draw (c1)--(c2);
    \draw (c1)--(c3);
    \draw (c2)--(c4);

    \node [atom] (n1) at (1.2, 1.5) {1};
    \draw (c2)--(n1);
    \node [atom] (n2) at (2, 0.5) {2};
    \draw (c4)--(n2);
    \node [atom] (n3) at (3, 0.5) {3};
    \draw (c4)--(n3);
    \node [atom] (n4) at (3.5, 1.5) {4};
    \draw (c3)--(n4);
    \node [atom] (n5) at (4.6, 1.5) {5};
    \draw (c3)--(n5);

    \node [] (R) at (0.2, 1.5) {};
    \node [] (L) at (5.6, 1.5) {};
    \draw (R)--(n1) to[out=-20,in=160] (n2)--(n3) to[out=40,in=220] (n4)--(n5)--(L);

    \begin{scope}[shift={(0, -0.5)}]
      \node [fill=blue!15!white, circle, minimum size=2.7cm] (circ2) at (6.5, 0.3) {};
      \clip (6.5, 0.3) circle (1.5);

      \coordinate[] (x1) at (5, 0);
      \coordinate[] (x2) at (5.75, 1.65);
      \coordinate[] (x3) at (8, 0);

      \node [atom] (c6) at (4.85, 0) {C};
      \node [atom] (c7) at (5.75, 1.75) {C};

      \node [atom] (l5) at (7.2, 0) {5};

      \coordinate[] (h1) at (6.4, 0);
      \node [circle,fill=blue!40!white,inner sep=5pt] at (6.25, 0) {};
      \draw [very thick, blue] (c6) to[out=0,in=180] (h1);
      \draw [very thick, blue] (c7) to[out=-60,in=180] (h1);
      \draw [very thick, blue, ->] (h1)--(l5);
      \draw (l5)--(x3);
    \end{scope}

    \node [anchor=west] at (5, 3) {Hyperlink to};
    \node [anchor=west] at (5, 2.5) {connect $n > 0$ vertices.};
    \draw [blue!40!white, ->, line width=0.75mm]
    (7, 2) to[out=-90, in=90] (6.35, 0);

  \end{scope}

  \begin{scope}[shift={(6.5, -0.5)}]
    \node [anchor=west] at (-0.5, 3.75) {Threaded tree};
    \node [] (X) at (3, 4) {};
    \node [atom] (c1) at (3, 3) {C};
    \node [atom] (c2) at (2, 2.5) {C};
    \node [atom] (c3) at (4, 2.5) {C};
    \node [atom] (c4) at (2.5, 1.5) {C};

    \draw (X)--(c1);
    \draw (c1)--(c2);
    \draw (c1)--(c3);
    \draw (c2)--(c4);

    \node [atom] (n1) at (1.2, 1.5) {1};
    \draw (c2)--(n1);
    \node [atom] (n2) at (2, 0.5) {2};
    \draw (c4)--(n2);
    \node [atom] (n3) at (3, 0.5) {3};
    \draw (c4)--(n3);
    \node [atom] (n4) at (3.5, 1.5) {4};
    \draw (c3)--(n4);
    \node [atom] (n5) at (4.6, 1.5) {5};
    \draw (c3)--(n5);

    \node [] (R) at (0.2, 1.5) {};
    \node [] (L) at (5.6, 1.5) {};

    \draw (R)--(n1)  to[out=0, in=-90] (c2.south);
    \draw (c2.south) to[out=-90, in=160] (n2);

    \draw (n2) to[out=30, in=-90] (c4.south);
    \draw (c4.south) to[out=-90, in=150] (n3);

    \draw (n3) to[out=20, in=-60] (3, 1.45) to[out=120, in=-90] (c1.south);
    \draw (c1.south) to[out=-90, in=160] (n4);

    \draw (n4) to[out=30, in=-90] (c3.south);
    \draw (c3.south) to[out=-90, in=150] (n5)--(L);
  \end{scope}
\end{tikzpicture}
  \end{framed}
  \caption{Examples of complex data structures.}
  \label{fig:intro:hypergraphs}
\end{figure}
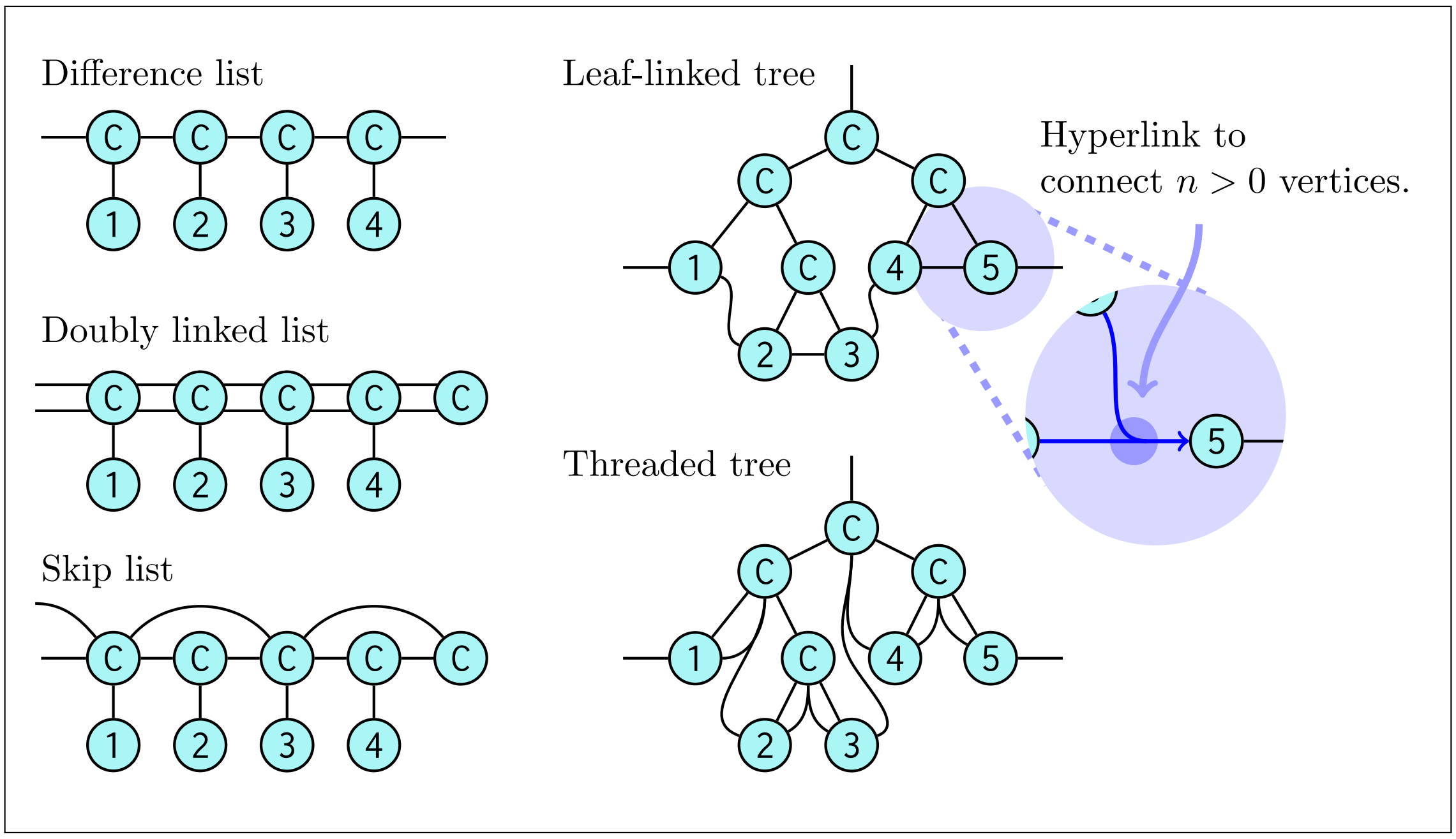

Programming is fundamentally concerned with transforming data into desired outcomes.
Although such data may take the form of simple atomic values, lists, or trees,
it may also exhibit more complex structures, including sharing, cycles, and multiple entry points (\cref{fig:intro:hypergraphs}).
Programs that manipulate such data structures are often difficult to write and understand.
Moreover, these programs are prone to errors, yet it remains difficult to
automatically check whether they satisfy the programmer's intent.
The central motivation of this dissertation is to support the manipulation of such complex data structures
in a manner that is declarative, statically safe, and compatible with a value- and function-oriented programming style.

In this dissertation, we propose a new language, \lgt,
which integrates hypergraph transformation into a call-by-value functional language.
\lgt treats hypergraphs as first-class data and enables declarative programming through pattern matching.
The type system of \lgt allows programmers to describe the shapes of data structures and
ensures that programs do not violate them.

In \cref{sec:intro:background}, we first review the strengths and limitations of existing programming paradigms.
Building on this discussion, \cref{sec:intro:objective} presents the research objectives of this dissertation.
\Cref{sec:intro:contributions} then summarises its main contributions
and describes the organisation of the remainder of the dissertation.

\section{Background}\label{sec:intro:background}

We first examine how existing programming paradigms address the manipulation of data structures.
Our focus is not limited to simple values, lists, and trees,
but extends to structures with richer connectivity, such as doubly linked lists,
skip lists, leaf-linked trees, threaded trees, and dataflow graphs.
Through this comparison, we clarify the strengths and limitations of each paradigm as a basis
for the research objectives presented in the next section.


\subsection{Imperative Programming}\label{sec:intro:imperative}

Imperative languages can represent and manipulate arbitrary data structures by means of pointers, references, and destructive updates.
This directness is useful for implementing complex and performance-critical data structures.
However, it also exposes programmers to low-level details of heap mutation,
which is often cumbersome and error-prone.

Existing memory-safety mechanisms alleviate some of these difficulties.
For example, garbage collection relieves programmers from explicitly deallocating heap memory.
By automatically reclaiming unreachable objects,
garbage collection prevents dangling pointers that would otherwise arise from explicit deallocation.
Nevertheless, a program may be memory-safe while still producing a heap that violates
the structural invariants intended by the programmer for complex data structures.
For example, under garbage collection, a doubly linked list may contain no dangling references,
yet fail to satisfy the mutual consistency between forward and backward links.

Besides garbage collection, linear and affine type systems provide another approach.
By restricting the sharing of references,
they make it possible to statically determine when a reference is no longer used,
thereby enabling memory deallocation to be inserted safely at compile time~\cite{wadler1990linear,walker2001regions}.
This idea has been implemented and widely adopted in practice,
most notably in Rust's ownership and borrowing system~\cite{rust}.
Rust prevents multiple mutable references to the same object,
thereby ruling out a common source of unsafe interference through aliasing.
However, this restriction also makes it difficult to express complex data structures that inherently involve sharing or cycles.
More fundamentally, mechanisms based on linearity, affinity, ownership, or borrowing do not by themselves guarantee
that the global structure of a complex data structure satisfies the programmer's intended invariants.

Verification techniques such as separation logic provide expressive frameworks
for specifying and proving properties of mutable heap structures~\cite{separation-logic}.
These techniques are powerful,
but they are primarily verification formalisms rather than lightweight typing frameworks
that ordinary programmers can use directly.

Taken together, these mechanisms address important aspects of heap manipulation.
However, they do not fully address the problem considered in this dissertation.
What remains is a language-level mechanism for expressing transformations of graph-shaped data structures directly,
while incorporating global shape invariants into a lightweight static type system.

\subsection{Functional Programming}\label{sec:intro:functional}

Functional programming languages provide a contrasting approach to data manipulation.
Rather than updating data structures in place,
they typically treat data as values and describe computation by constructing,
decomposing,
and composing such values.
This style supports declarative programming and makes programs easier to reason about,
especially when combined with static type systems and higher-order functions.

A central success of functional programming is its treatment of lists and trees.
Algebraic data types make the recursive structure of such data explicit at the type level,
and pattern matching allows programs to inspect values according to that structure.
The type system ensures that programs manipulate these data structures in ways consistent with their declared shapes,
so that \emph{well-typed programs won't go wrong} with respect to these structural assumptions.

The limitation is that the standard type mechanisms of functional languages are most naturally suited to trees.
Algebraic data types do not directly represent data structures that involve sharing, cycles, or multiple entry points.
Although such data structures can be encoded in functional languages,
these encodings do not usually make the intended global shape invariants part of the type.

For example,
one may define a node type with \texttt{prev} and \texttt{next} fields to represent a doubly linked list.
The type system can ensure that these fields refer to nodes of the appropriate type.
However,
the type of an individual node does not guarantee that forward and backward links are mutually consistent across the entire structure.
Ill-formed configurations may therefore remain well typed.

The goal of this dissertation is therefore not merely to show that complex data structures
can be encoded in a functional language.
Rather, it is to extend the advantages that functional programming derives from algebraic data types
to data structures whose shapes go beyond trees.

\subsection{Graph Transformation}\label{sec:intro:graph-rewriting}

For manipulating complex data structures, graph transformation is one of the most direct and expressive formalisms.
In these systems,
data structures are abstracted as graphs,
and computation is expressed as the rule-based transformation of graphs~\cite{algebraic-gt,handbook_graph_grammar}.
A rule typically consists of a pattern graph and a replacement graph.
Computation proceeds by finding an occurrence of the pattern in a host graph and
replacing it with the corresponding instance of the replacement.

This model provides a natural basis for modelling systems whose states are highly complex.
For example, LMNtal, a graph-transformation-based language pronounced ``elemental'',
is a representative language in this direction and is closely related to the language studied in this dissertation~\cite{lmntal2009}.
Its implementation, SLIM, provides a runtime system, state-space search,
and LTL model checking for LMNtal programs~\cite{slim-github}.
In particular, its state-space search has been reported to scale to more than \(10^9\) states
on machines with tens of cores.

However,
the programming model of graph rewriting languages differs from the one pursued in this dissertation.
Graph rewriting languages are typically based on a state-transformation model,
in which a graph is repeatedly modified by applying rewrite rules.
This view is natural for modelling and verification,
but it does not fit directly with ordinary value- and function-oriented programming.

\section{Research Objectives}\label{sec:intro:objective}

The objective of this dissertation is to develop a foundation
for manipulating complex data structures
in a manner that is declarative, statically safe,
and compatible with a value- and function-oriented programming style.
To this end,
we propose \lgt~\cite{sano2023,sano-icgt2023},
a purely functional programming language that integrates port hypergraph transformation
into a call-by-value functional calculus.
The design of \lgt is guided by the requirements summarised in \cref{tab:intro:requirements},
where port hypergraphs are referred to simply as graphs.

\begin{table}[htb]
  \centering
  \caption{Design requirements of \lgt.}
  \label{tab:intro:requirements}
  \begin{tabular}{@{}p{0.12\linewidth}p{0.78\linewidth}@{}}
    \toprule
    \textbf{ID} & \textbf{Requirement}                                                                                                                                                                         \\
    \midrule
    L1          & Graphs should be composable in a direct and natural manner.                                                                                                                                  \\
    L2          & Graphs should be first-class values that can be bound to variables and passed through programs.                                                                                              \\
    L3          & Graphs should be decomposable by pattern matching.                                                                                                                                           \\
    L4          & Functions should remain first-class entities, as in conventional functional languages, while interacting smoothly with graph values.                                                         \\
    \midrule
    T1          & Users should be able to define types that describe structural properties of graphs, and graph values should be checked against these types.                                                  \\
    T2          & Intermediate graph structures arising during pattern matching and transformation should be typable.                                                                                          \\
    T3          & The type system should provide static guarantees to the greatest extent possible, while clearly identifying any remaining dynamic checks and the conditions under which they can be reduced. \\
    \bottomrule
  \end{tabular}
\end{table}

In \lgt,
data structures involving sharing and cycles are represented as \textit{port hypergraphs}.
A port hypergraph is a generalisation of a graph
in which a hyperedge can connect an arbitrary number of nodes through designated attachment points,
called \textit{ports}.
By using port hypergraphs as values,
\lgt treats complex data structures not as implicit heap configurations manipulated by pointers,
but as explicit data values in a functional language.

For manipulating such data structures,
\lgt provides two basic forms of graph manipulation.
Graph construction is expressed by composition,
and graph decomposition is expressed by pattern matching.
Thus,
programs can be written according to the structure of the data,
without relying on explicit pointer-based updates.
At the same time,
functions remain first-class,
so graph manipulation can be combined with the compositional programming style of functional languages.

For safety,
\lgt is equipped with a type system.
The type system allows programmers to define graph types that describe structural properties of port hypergraphs
and checks that graph values conform to these specifications.
Its purpose is not merely to check local node interfaces,
such as the number and arrangement of ports,
but also to ensure global structural constraints of data structures.

\section{Contributions and Thesis Structure}\label{sec:intro:contributions}

This dissertation makes the following contributions.

\begin{description}
  \item[Core Language]
        We first establish the core language foundation of \lgt.
        \Cref{ch:hyperlmntal} formalises HyperLMNtal,
        a term-based formalism for port hypergraph transformation.
        Using this formalisation as a basis,
        \cref{ch:untyped} presents the untyped \lgt language,
        which integrates port hypergraph transformation into a call-by-value \(\lambda\)-calculus.
        After defining the syntax and operational semantics of \lgt,
        \cref{ch:impl} demonstrates its practical feasibility
        through the development of a compact interpreter.
  \item[Type System]
        We then develop type systems for \lgt.
        \Cref{ch:core-type} introduces a core type system for \lgt
        and demonstrates its expressiveness through a range of examples.
        In particular,
        we establish its type soundness,
        and prove that the core type system extends
        existing grammar-based graph typing.
        Building on this core type system,
        \cref{ch:limp-type} analyses its limitations,
        in particular its inability to type incomplete graphs,
        which often arise in pattern matching.
        It then extends the type system with linear implication types,
        thereby enabling the typing of incomplete graph structures.
        Finally,
        \Cref{ch:fully-static} investigates how dynamic checks in pattern matching
        can be reduced by introducing static constraints.
        It identifies a statically safe fragment of \lgt
        and analyses its soundness under the refined typing discipline.
  \item[Mechanised Verification]
        \Cref{ch:formal} mechanises core parts of the meta-theory of \lgt in Isabelle/HOL,
        providing a mechanised development of the meta-theory
        and strengthening the reliability of the formal results.

\end{description}

\Cref{ch:related} further clarifies the contributions of this dissertation
by situating this work within the broader landscape of existing research,
with particular attention to approaches that treat graphs as objects of computation,
pattern matching beyond ordinary tree patterns,
and static reasoning techniques for data structures.
Finally,
\cref{ch:conclusion} concludes the dissertation
and outlines directions for future research.

\paragraph{Relationship to prior publications.}
Parts of this dissertation are based on the author's previously published and accepted papers.
The port-hypergraph formalism and the syntax-driven treatment of HyperLMNtal are based in part on~\cite{sano2021,sano-ba}.
The core language design of \lgt and its original type system are based on~\cite{sano2023}.
The reference interpreter presented in \cref{ch:impl} is based on~\cite{sano-icgt2023}.
The extension with linear implication types and the treatment of incomplete graph structures in \cref{ch:limp-type} and \cref{ch:fully-static} are based on~\cite{sano2025linear}.
The presentation in this dissertation revises, reorganises, and extends these results
to provide a unified account of \lgt.

\subsection*{Preliminaries}\label{sec:intro:preliminaries}

We introduce some syntactic conventions as follows:
\(\seq{E}\) denotes
\(E_1,
\dots,
E_n\)
or
\(E_1 \dots E_n\)
for some \(n\ (\geq 0)\) and
some syntactic entity
\(E_i\ (1\le i\le n)\).
The length $n$ of $\seq{E}$ is denoted as $\norm{\seq{E}}$.
\(S\{s\}\) denotes a set \(S\)
such that \(s \in S\).
Hypergraphs and hyperlinks (hyperedges) may be simply denoted as graphs and links,
respectively.


\chapter{Formalising Hypergraph Transformation}\label{ch:hyperlmntal}

\newcommand{\equivby}[1]{\equiv_{\mbox{\scriptsize #1}}}

In order to integrate hypergraph transformation into a functional programming
language, we need a formalism whose syntax and semantics are compatible with
the usual term-based presentation of programming languages.
For this reason,
not every graph transformation formalism is suitable for our purpose.

Many standard approaches to graph transformation are formulated in an algebraic
style.
Graphs are typically defined in terms of sets of vertices and edges, graph
equivalence is described through suitable bijections or isomorphisms, and graph
rewriting is often formalised by categorical constructions such as double-pushout
rewriting~\cite{handbook_graph_grammar}.
These frameworks provide a powerful and well-established mathematical foundation
for graph transformation.
However, they do not directly provide a term-based syntax of graphs or a
syntax-directed operational semantics.
Their definitions are not primarily organised so that the meaning of a
whole graph is obtained compositionally from the meanings of its syntactic parts.
Although it may be possible to embed such algebraic formalisms into a functional
language, doing so would make the subsequent design of substitution, pattern
matching, and especially a compositional type system substantially less direct.

We therefore take LMNtal~\cite{lmntal2009} (pronounced as ``elemental'') as our starting point.
LMNtal is a computational model and programming language based on hierarchical
graph rewriting, and it provides a term-based presentation of graph structures and
rewriting.
This makes it closer in spirit to the syntax-directed style used in the
\(\lambda\)-calculus and its variants~\cite{sos}.

At the same time, data structures in general may involve not only binary links
between objects, but also shared components and multi-way relationships.
It is therefore natural for our purposes to use a representation in which a
single connection may relate an arbitrary number of endpoints.
However, ordinary LMNtal is not itself a hypergraph transformation formalism:
its links are restricted to connect at most two endpoints.
In contrast,
the SLIM implementation of LMNtal includes an extension that supports
hyperlinks, namely links that may connect an arbitrary number of endpoints~\cite{slim-github,hyperlmntal}.
We refer to this hypergraph transformation extension to LMNtal as HyperLMNtal.
While HyperLMNtal provides the kind of hypergraph structure needed for \lgt{},
it has not previously been given a formal semantics suitable for use as the
foundation of a typed functional language.

This chapter addresses this gap by formalising HyperLMNtal as a term-based and
syntax-directed hypergraph transformation formalism.
HyperLMNtal serves as the underlying model of graph structure and transformation
in \lgt{}, and provides the basis for integrating graph-based computation into a
functional setting.
In this dissertation, we focus on a restricted fragment, called
\emph{Flat HyperLMNtal}, in which hierarchical structure is disallowed and graphs
are formed solely by nodes and hyperlinks.
For the purpose of formalisation, we separate hypergraphs and rewrite rules in
the presentation, although they are unified in the original formulation.
In the following, we use the term \emph{HyperLMNtal} to refer to this formal
system and, for simplicity, refer to hyperlinks as \emph{links} and hypergraphs as
\emph{graphs}.

The remainder of this chapter is organised as follows.
\Cref{sec:hyperlmntal:syntax} defines the syntax of HyperLMNtal graphs and
rewrite rules.
\Cref{sec:hyperlmntal:congruence} introduces link substitution and structural
congruence, which specify when two graph terms are regarded as representing the
same graph.
\Cref{sec:flat-hyperlmntal-op-sem} defines the reduction relation of HyperLMNtal.
Finally, \cref{sec:hyperlmntal-proof} proves basic properties of structural
congruence, including the elimination of unused link creation, \(\alpha\)-conversion
of link names, and fundamental properties of fusions.

\section{Syntax of Graphs and Rewrite Rules}\label{sec:hyperlmntal:syntax}

HyperLMNtal is composed of two syntactic categories.
\begin{itemize}
  \item
        \(X\)
        denotes a \emph{Link Name}.

  \item
        \(p\)
        denotes an \emph{Atom Name}.

\end{itemize}
The only preserved atom name is
\(\bowtie\), where an atom \(X \bowtie Y\), called a \emph{fusion},
fuses the link \(X\) and the link \(Y\) into a single link.

\begin{figure}[htbp]
  \begin{framed}
    \centering
    \begin{tabular}{rccll}%
      Graph        & \(G\) & $::=$   & \(\zero\)               & Null               \\
                   &       & $|$     & \(p (\seq{X})\)         & Atom               \\
                   &       & \(|\)   & \((G, G)\)              & Molecule           \\
                   &       & \(|\)   & \(\nu X.G\)             & Hyperlink creation \\
      \\
      Rewrite Rule & \(r\) & \(::=\) & \(G \longrightarrow G\) & Rule
    \end{tabular}
  \end{framed}
  \caption{Syntax of HyperLMNtal.}\label{table:hyperlmntal:syntax}
\end{figure}

The syntax of HyperLMNtal is given in \cref{table:hyperlmntal:syntax}.
We abbreviate \(\nu X_1. \dots \nu X_n. G\) to \(\nu X_1 \dots X_n. G\),
which can be denoted as \(\nu \seq{X}. G\).
The pair of the name \(p\) and the arity \(n = \norm{\seq{X}}\) of an atom \(p(\seq{X})\)
is referred to as the \emph{functor}\footnote{%
  Synonym of function symbol and function object; not to be confused with functors in category theory.
}
of the atom and is written as \(p/n\).

\begin{figure}[htbp]
  \begin{framed}
    \[
      \begin{aligned}
        \mathit{fn}(\zero)      & \triangleq \emptyset                              \\
        \mathit{fn}(p(\seq{X})) & \triangleq \{\seq{X}\}                            \\
        \mathit{fn}((G_1, G_2)) & \triangleq \mathit{fn}(G_1) \cup \mathit{fn}(G_2) \\
        \mathit{fn}(\nu X.G)    & \triangleq \mathit{fn}(G) \setminus \{X\}         \\
      \end{aligned}
    \]
  \end{framed}
  \caption{The set of free link names.}\label{table:hyperlmntal:free-names}
\end{figure}

The set of free link names in hypergraph \(G\) is denoted as \(\mathit{fn}{(G)}\),
which is defined inductively in \cref{table:hyperlmntal:free-names}.

\begin{mydef}[Syntactic Convention]\label{def:abbreviation}
  For convenience, we introduce some abbreviation rules.
  \begin{enumerate}
    \item
          An atom \(p()\) (with no arguments) is written also as \(p\).
    \item
          A sequence of nested
          binders
          \( \nu X_1.\,\dots\,. \nu X_n. G \)
          can be abbreviated to
          \( \nu X_1 \dots X_n. G \)
          or \(\nu \Xs. G \).
          When $n=0$, $\nu . G$ is written also as $G$.
    \item
          The \textit{term notation} allows
          \[\nu Y.(p_1 (\Xs,Y,\Zs), p_2 (\Ws,Y)),\]
          where $Y \notin \{\Xs,\Zs,\Ws\}$,
          to be written as
          \[ p_1(\Xs, p_2(\Ws),\Zs). \]
          That is, an atom $p_2 (\Ws,Y)$ may be embedded into the
          position of the other occurrence of $Y$ as $p_2 (\Ws)$.
          Here, we allow $\Xs,\Zs,\Ws$ to contain such embedded
          atoms (rather than links) introduced by the term notation.
  \end{enumerate}
\end{mydef}

An example of the term notation will be given as soon as
Structural Congruence is introduced in \cref{sec:hyperlmntal:congruence}.

Rules have the form \(G \longrightarrow G\).
The two \(G\)s are called the left-hand side (LHS) and
the right-hand side (RHS), respectively.

\begin{definition}[Syntactic condition on rules]
  A rule \(G_1 \longrightarrow G_2\) should satisfy
  \(\mathit{fn}(G_1) \supseteq \mathit{fn}(G_2)\).
\end{definition}

The condition indicates that we must denote a new hyperlink
in the scope of a \(\nu\) (new) on the RHS of a rule.

A doubly linked list is a fundamental data structure frequently used in programming
due to its efficiency in supporting bidirectional traversal.
In HyperLMNtal, a doubly linked list is naturally represented as a graph,
where each \(\Cons\) node maintains hyperlinks (i.e., pointers) to both its predecessor and successor nodes,
and a \(\Nil\) node terminates the list.
The formal encoding of a doubly linked list in this language is given as follows:
Here,
the first argument of each \(\Cons\) atom is connected to its value;
for example, \(1, 2\), and \(3\).
The second and the third argument point to the previous and next elements,
respectively.
The last argument is the access point to the current node,
which also applies to the \(\Nil\) node.
The first argument of the \(\Nil\) node points to the previous node.
\begin{align}\label{eq:dbllist-eg}
  \GDL \triangleq \begin{array}{lll}
                    \nu W_1 W_2 W_3 W_4 W_5.( \\
                    \quad \Cons(W_1,X,W_2,Z),
                    1(W_1),                   \\
                    \quad \Cons(W_3,Z,W_4,W_2),
                    2(W_3),                   \\
                    \quad \Cons(W_5,W_2,Y,W_4),
                    3(W_5),                   \\
                    \quad \Nil(W_4, Y)        \\
                    )
                  \end{array}
\end{align}

The structure of the doubly linked list $\GDL$ 
can be visualised as follows:
\begin{align}
  \begin{tikzpicture}
  \mkcons{1}{W_1}{}{(0,0)}{1}
  \mkcons{2}{W_3}{W_2}{(1.5,0)}{2}
  \mkcons{3}{W_5}{W_4}{(3.0,0)}{3}
  \mknil{4}{(4.5,0)}
  %
  \node [link] (C0) at (-1.0, 0.8) {$Z$};
  \node [link] (LX) at (-1.0, 0.0) {$X$};
  \node [link] (C5) at (+5.5, 0.8) {$Y$};
  \draw (C1) to (LX);
  \mkconslinkleft{0}{1}
  \mkconslink{1}{2}
  \mkconslink{2}{3}
  \mkconslink{3}{4}
  \mkconslinkright{4}{5}
\end{tikzpicture}
\end{align}

In this figure, \(\mathrm{C}\) denotes \(\mathrm{Cons}\),
and \(\mathrm{N}\) represents \(\mathrm{Nil}\).
The numbers around each atom indicate the atom's ports to which links
are attached.
Arguments numbers are attached to the arguments of an atom.
Although the link \(X\) does not play a significant role in this example,
it is essential for defining a well-structured inductive representation
and for enabling a
systematic implementation of programs that manipulate this
structure.

\section{Structural Congruence}\label{sec:hyperlmntal:congruence}

The semantics of HyperLMNtal comes with two major ingredients,
structural congruence \(\equiv\)
and reduction relation \(\rightsquigarrow\) on graphs.
Structural congruence defines what graphs (represented in the
syntax of \cref{table:hyperlmntal:syntax}) are essentially the same.
This subsection defines structural congruence.

\begin{definition}[Link Substitution]
  \(G\angled{ Y_1, \dots, Y_n / X_1, \dots, X_n }\)
  is a \emph{link substitution} that replaces all free occurrences of
  \(X_i\) with \(Y_i\)
  as defined in \cref{table:hyperlmntal:hyperlink-substitution}.
  Here, the \(X_1, \dots, X_n\) should be mutually distinct.
  Note that, if a free occurrence of \(X_i\) occurs at a location where \(Y_i\) would not be free,
  \(\alpha\)-conversion may be required.

  \begin{figure}[htbp]
    \begin{framed}
      \centering
      \begin{tabular}{@{}r@{\hspace{0.5em}}c@{\hspace{0.5em}}l@{}}
        \(\zero\sigma\)                     & \(=\) & \(\zero\)
        \vspace{0.8em}                                                                     \\

        \(p(\seq{X})\sigma\)                & \(=\) & \(p(X_1\sigma, \ldots, X_n\sigma) \) \\
                                            &
        \multicolumn{2}{l}{%
          where \(
          X\angled{\seq{Z}/\seq{Y}} =
          \left\{
          \begin{array}{ll}
            Z_i & \mbox{if } X = Y_i              \\
            X   & \mbox{if } X \notin \{\seq{Y}\}
          \end{array}
          \right.
          \)
          \vspace{0.8em} }
        \\

        \((G_1, G_2)\sigma\)                & \(=\) & \((G_1 \sigma, G_2 \sigma)\)
        \vspace{0.8em}                                                                     \\
        \(
        (\nu X.G)\angled{\seq{Z}/\seq{Y}}\) & \(=\) &                                      \\
        \multicolumn{3}{l}{%
          \(\left\{
          \begin{array}{ll}
            \nu X.G\angled{\seq{Z'}/\seq{Y'}} & \mbox{if } X = Y_i\ \land                                  \\
                                              & \seq{Z'} = Z_1, \dots, Z_{i - 1}, Z_{i + 1}, \dots, Z_n    \\
                                              & \seq{Y'} = Y_1, \dots, Y_{i - 1}, Y_{i + 1}, \dots, Y_n
            \\[2mm]
            \nu X.G\angled{\seq{Z}/\seq{Y}}   & \mbox{if } X \notin \{\seq{Y}\} \land X \notin \{\seq{Z}\}
            \\[2mm]
            \nu W.(G\angled{W/X})\angled{\seq{Z}/\seq{Y}}
                                              & \mbox{if } X \notin \{\seq{Y}\} \land X \in \{\seq{Z}\}    \\
                                              & \land W \notin \mathit{fn}(G) \land W \notin \{\seq{Z}\}
          \end{array}
          \right.
          \)
        }
      \end{tabular}
    \end{framed}
    \caption{Hyperlink substitution.}\label{table:hyperlmntal:hyperlink-substitution}
  \end{figure}
\end{definition}

\begin{definition}[Structural Congruence]
  We define the relation \(\equiv\) on graphs as the minimal equivalence relation
  satisfying the rules shown in \cref{table:hyperlmntal:equiv}.
  \begin{figure}[htbp]
    \begin{framed}
      \centering
      \begin{tabular}{c}
        (E1) \((\mathbf{0}, G) \equiv G\)
        \quad
        (E2)   \((G_1, G_2) \equiv (G_2, G_1)\)
        \\[2.0mm]
        (E3) \((G_1, (G_2, G_3)) \equiv ((G_1, G_2), G_3)\)
        \\[2.0mm]
        (E4) \(\dfrac{G_1 \equiv G_2}{(G_1, G_3) \equiv (G_2, G_3)}\)
        \qquad (E5) \(\dfrac{G_1 \equiv G_2}{\nu X.G_1 \equiv \nu X.G_2}\)
        \\[4.0mm]
        (E6) \(\nu X.(X \bowtie Y, G)  \equiv \nu X.G\langle Y / X \rangle\)
        \\
        where \(X \in \fn(G) \lor Y \in \fn(G)\)
        \\[2.0mm]
        (E7) \(\nu X.\nu Y.X \bowtie Y \equiv \mathbf{0}\)
        \qquad
        (E8) \(\nu X.\mathbf{0} \equiv \mathbf{0}\)
        \\[2.0mm]
        (E9) \(\nu X.\nu Y.G \equiv \nu Y.\nu X.G\)
        \\[2.0mm]
        (E10) \(\nu X.(G_1, G_2) \equiv (\nu X.G_1, G_2)\)
        where \(X \notin \fn(G_2)\)
      \end{tabular}
    \end{framed}
    \caption{Structural congruence on HyperLMNtal graphs.}\label{table:hyperlmntal:equiv}
  \end{figure}
\end{definition}

Two graphs related by \(\equiv\) are essentially the same and are convertible
to each other in zero steps.
(E1), (E2) and (E3) are the characterisation of molecules as multisets.
(E4) and (E5) are structural rules that make \(\equiv\) a congruence.
(E6) and (E7) are concerned with fusions.
(E7) says that a closed fusion is equivalent to \(\zero\).
(E6) is an absorption law of \(\bowtie\),
which says that a fusion can be absorbed by connecting hyperlinks.
(E8), (E9) and (E10) are concerned with hyperlink creations.

We give two important theorems showing that the symmetry of
\(\bowtie\) and \(\alpha\)-conversion can be derived from the rules of
\cref{table:hyperlmntal:equiv}.

\begin{theorem}[Symmetry of \(\bowtie\)]\label{th:symmetry-of-bowtie}
  \[X \bowtie Y \equiv Y \bowtie X\]
\end{theorem}
\begin{proof}
  We give this soon in \cref{sec:hyperlmntal-proof}.
\end{proof}

\begin{theorem}[\(\alpha\)-conversion of hyperlinks]\label{th:alpha-equiv}
  Bound link names are \(\alpha\)-convertible in HyperLMNtal, i.e.,
  \[\nu X.G \equiv \nu Y.G\angled{Y/X} \text{ where } Y \notin \mathit{fn} (G)\]
\end{theorem}
\begin{proof}
  We give this soon in \cref{sec:hyperlmntal-proof}.
\end{proof}

Because of the symmetry of \(\bowtie\),
(E6) says that an atom can emit a fusion as well.

\begin{example}[Term Notation]
  Consider
  \[\begin{array}{l@{\qquad}l}
      \nu Z_1.\nu Z_2.\nu Z_3.\nu Z_4.( &       \\
      \quad \Node(Z_1, Z_2, X),         & (a_1) \\
      \quad \quad \Node(Z_3, Z_4, Z_1), & (a_2) \\
      \quad \qquad \Leaf(Z_3),          & (a_3) \\
      \quad \qquad \Leaf(Z_4),          & (a_4) \\
      \quad \quad \Leaf(Z_2),           & (a_5) \\
      ).
    \end{array}\]
  Since structural congruence allows the reordering of atoms
  (E2)(E3) and the reordering and movement of binders (E9)(E10),
  the term notation (\cref{def:abbreviation}~(3))
  allows us to write the above graph as
  \[
    \Node(\Node(\Leaf, \Leaf), \Leaf, X)
  \]
  by embedding $a_3$ and $a_4$ into $a_2$ and then embedding $a_2$
  and $a_5$ into $a_1$.
\end{example}

It is a subject for future work to
elucidate the relationship between the structural congruence rules and
graph isomorphism, including
the soundness and completeness of the structural congruence rules.
However, these properties are irrelevant to the validity of the semantics of HyperLMNtal and of \lgt{}
introduced in the next chapter.
The graphs handled in HyperLMNtal and \lgt{} are
HyperLMNtal graphs, which are defined inductively from the beginning,
rather than graphs
in ordinary algebraic graph transformation formalisms \cite{algebraic-gt}.
Therefore, the structural congruence rules need not correspond to
graph isomorphism.
We have run examples to confirm that equivalence based on
structural congruence is practical for HyperLMNtal and \lgt{}.

\section{Reduction Relation}\label{sec:flat-hyperlmntal-op-sem}

We give the reduction relation of HyperLMNtal that defined
the small-step semantics of the language.  Note, however, that
\lgt described in the next section has its own operational
semantics without incorporating the reduction relation described here.
We nevertheless introduce the reduction relation of HyperLMNtal here
because it serves as the basis of the graph types of
\lgt described in \cref{ch:core-type}.

\begin{figure}[htbp]
  \begin{framed}
    \centering
    \begin{tabular}{lc}
      (R1) & \(\dfrac{G_1 \rightsquigarrow_{P} G_2}{(G_1, G_3) \rightsquigarrow_{P}  (G_2, G_3)} \)
      \vspace{1em}                                                                                                                                                     \\
      (R2) & \(\dfrac{G_1 \rightsquigarrow_{P} G_2}{\nu X.G_1 \rightsquigarrow_{P}  \nu X.G_2} \)
      \vspace{1em}                                                                                                                                                     \\
      (R3) & \(\dfrac{\hspace{1em} G_1 \equiv G_2 \hspace{2em} G_2 \rightsquigarrow_{P} G_3 \hspace{2em} G_3 \equiv G_4 \hspace{1em}}{G_1 \rightsquigarrow_{P} G_4} \)
      \vspace{1em}                                                                                                                                                     \\
      (R4) & \(\dfrac{(G_1 \longrightarrow G_2) \in P}{G_1 \rightsquigarrow_{P} G_2} \)                                                                                \\
    \end{tabular}
  \end{framed}
  \caption{Reduction relation on HyperLMNtal graphs.}\label{table:hyperlmntal-trans}
\end{figure}

\begin{definition}[Reduction relation]
  For a set \{P\} of rewrite rules,
  the reduction relation \(\rightsquigarrow_P\) on graphs is defined
  as the minimal relation
  satisfying the rules in \cref{table:hyperlmntal-trans}.
\end{definition}

We illustrate the use of rewrite rules with a simple example.
The following rules generate doubly linked lists.

\begin{example}[Rewrite rules to generate a doubly linked list]\label{ex:prod-dbllist-01}
  The rewrite rules to generate a doubly linked list can be defined as follows.
  \[\begin{array}{rcl}
      \mathit{dbllist}(X,Y,Z) & \longrightarrow & \Nil(X, Z), Y \bowtie Z          \\
      \mathit{dbllist}(X,Y,Z) & \longrightarrow & \nu W_1 W_2.(\Cons(W_1,X,W_2,Z),
      \mathit{nat}(W_1),\mathit{dbllist}(Z,Y,W_2))                                 \\
      \mathit{nat}(X)         & \longrightarrow & \mathrm{Zero} (X)                \\
      \mathit{nat}(X)         & \longrightarrow & \mathrm{Succ} (\mathit{nat}, X)
    \end{array}\]
\end{example}

Intuitively, the first rule terminates the list by producing a \(\Nil\) cell,
whereas the second rule extends the list by producing a \(\Cons\) cell
and recursively generating the remaining part of the list.
The auxiliary predicate \(\mathit{nat}\) generates natural-number labels for list elements.

The rewrite rules for producing \(\Nil\) and \(\Cons\) cells are illustrated as follows.
\begin{align}\label{eq:dbllist-prod-rules-eg-vis}
  \begin{array}{ccc}
    \dblltype{X}{Y}{Z}
     &
    \longrightarrow
     &
    \tikz[baseline=1.5ex, scale=1]{%
      \mknil{2}{(0,0)}
      \node [link] (C1) at (-1, 0.8) {$Z$};
      \node [link] (LX) at (-1, 0.0) {$X$};
      \node [link] (C3) at (+1, 0.8) {$Y$};
      \mkconslinkleft{1}{2}
      \draw (C2) to (LX);
      \mkconslinkright{2}{3}
    }
    \mid
    \tikz[baseline=1.5ex, scale=1]{%
      \mkcons{2}{W_1}{}{(0,0)}
      \mkdbll{3}{(1.5,0)}{W_2}
      \node [link] (C1) at (-1.0, 0.8) {$Z$};
      \node [link] (LX) at (-1.0, 0.0) {$X$};
      \node [link] (C4) at (+2.5, 0.8) {$Y$};
      \mkconslinkleft{1}{2}
      \draw (C2) to (LX);
      \mkconslink{2}{3}
      \mkconslinkright{3}{4}
    }
  \end{array}
\end{align}

\begin{figure}[htbp]
  \begin{framed}
    \input{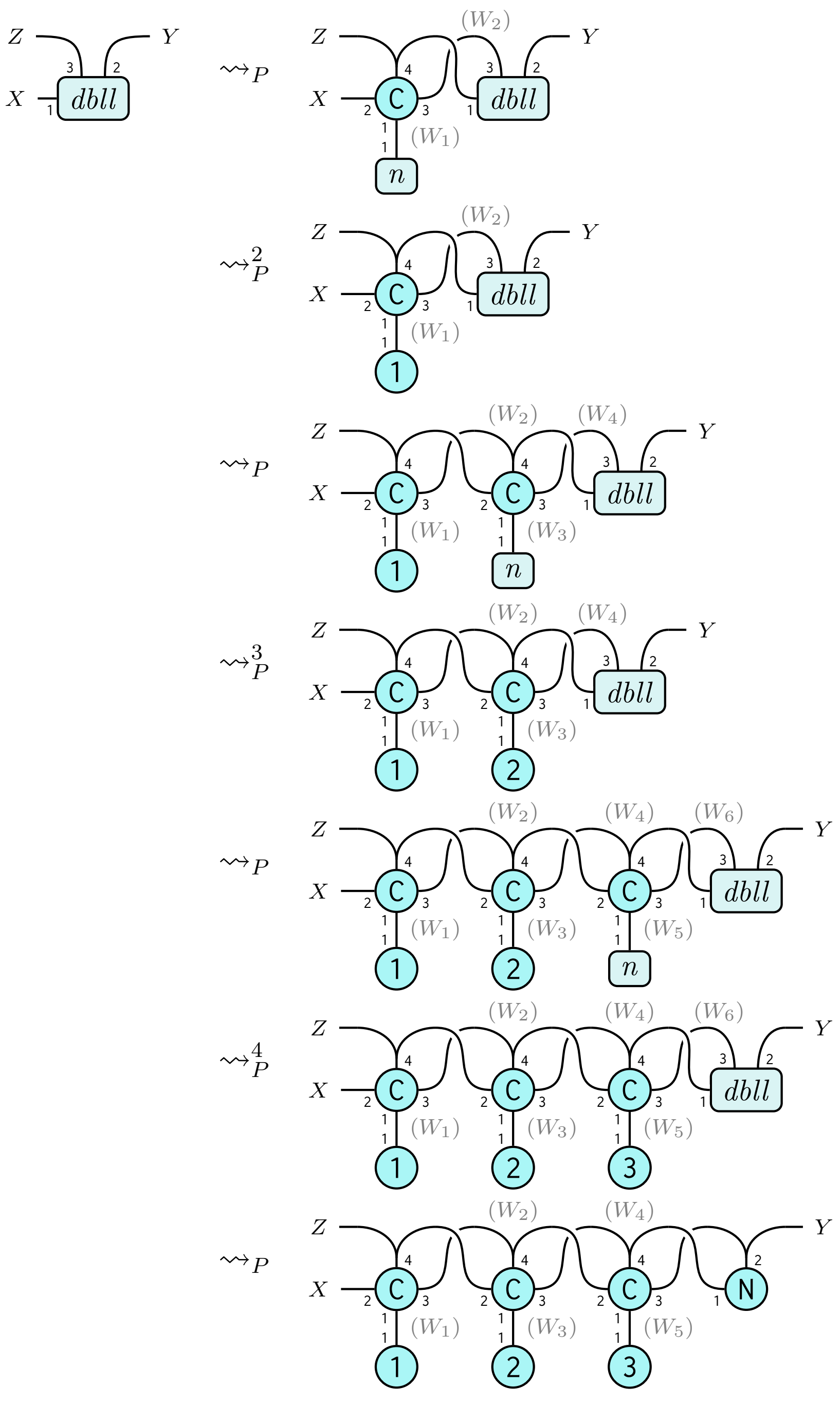}
  \end{framed}
  \caption{Generation of a doubly linked list.}\label{fig:hyperlmntal:dbllist-gen}
\end{figure}

Applying these rules to \(\mathit{dbllist}(X,Y,Z)\) generates doubly linked lists.
One possible result is the graph shown in \cref{fig:hyperlmntal:dbllist-gen},
which corresponds to the doubly linked list in \eqref{eq:dbllist-eg}.
In this example, natural numbers are abstracted as unary nodes for simplicity.

This example will be used again in \cref{ch:limp-type},
where such rewrite rules are interpreted as specifications of the shapes of data structures.

\section{Properties of HyperLMNtal}\label{sec:hyperlmntal-proof}

This section establishes several elementary properties of the structural congruence
of HyperLMNtal.
These properties are used to justify the informal interpretation of link creation,
link substitution, and fusion.
In particular, we first show that a link creation whose bound link name does not
occur in the graph is semantically irrelevant.
We then prove that substituting a link name for itself leaves a graph syntactically
unchanged.
Although these facts appear elementary, they are needed in the proofs of
\(\alpha\)-conversion for link names and the symmetry of fusion.
Finally, we show that fusions behave as identifications of link names.
This gives a formal basis for treating fusions as inducing an equivalence-like
structure over free link names.

\begin{lemma}[Elimination of unused link creation]\label{lem:absorb-link-creation}
  \[
    \nu X.G \equiv G
    \quad \mbox{where } X \notin \mathit{fn}(G).
  \]
\end{lemma}

\begin{proof}{}{}
  The lemma states that binding a link name that does not occur free in the graph
  has no observable effect under structural congruence.
  The proof moves the unused binder to the empty graph and then removes it.

  \[
    \begin{array}{ll@{\qquad}l}
       &
      \nu X.G
       &
      \\

      \equiv_{\mbox{\scriptsize E5}}
       &
      \nu X.(\zero, G)
       &
      \because G \equiv_{\scriptsize \mbox{E1}} (\zero, G)
      \\

      \equiv_{\mbox{\scriptsize E10}}
       &
      (\nu X.\zero, G)
       &
      \because X \notin \mathit{fn}(G)
      \\

      \equiv_{\mbox{\scriptsize E4}}
       &
      (\zero, G)
       &
      \because \nu X.\zero \equiv_{\mbox{\scriptsize E8}} \zero
      \\

      \equiv_{\mbox{\scriptsize E1}}
       &
      G.
       &
    \end{array}
  \]
\end{proof}

\begin{lemma}[Elimination of a futile link substitution]\label{lem:futile-link-substitution}
  \[
    G\angled{X/X} = G.
  \]
\end{lemma}

This lemma states that replacing every free occurrence of \(X\) by \(X\) leaves
the graph unchanged as a syntactic object.
This is slightly subtler than it may first appear.
Since link substitution is defined capture-avoidantly, one must rule out the
possibility that the substitution triggers an \(\alpha\)-conversion of a bound link
name, which would yield a structurally congruent but syntactically different graph.

\begin{proof}{}{}
  We prove the statement by induction on the structure of \(G\).
  The cases for \(\zero\), \(p(X_1,\ldots,X_m)\), \((G,Q)\), and rewrite rules are
  immediate from the definition of link substitution and the induction hypotheses.

  It remains to consider the binder case.
  Suppose that the graph is of the form \(\nu Y.G\).
  By the definition of link substitution, we have
  \[
    (\nu Y.G)\angled{X/X}
    \triangleq
    \begin{cases}
      \nu Y.G,
       & \mbox{if } Y = X,
      \\[0.5em]
      \nu Y.G\angled{X/X},
       & \mbox{if } Y \neq X.
    \end{cases}
  \]
  In the second case, the induction hypothesis gives \(G\angled{X/X}=G\), and
  hence the result is again \(\nu Y.G\).
  Moreover, the capture-avoiding branch of the substitution definition cannot be
  selected in this situation, since it would require both \(Y \neq X\) and
  \(Y = X\).
  Therefore, no \(\alpha\)-conversion is performed, and the result is syntactically
  identical to the original graph.
\end{proof}

\begin{proof}[Proof of \cref{th:alpha-equiv}]
  We prove that bound link names are \(\alpha\)-convertible.
  The proof relies on \cref{lem:absorb-link-creation} and
  \cref{lem:futile-link-substitution}.
  We distinguish two cases, depending on whether the link name \(X\) actually
  occurs free in \(G\).
  When \(X\) occurs free, the proof introduces a pair of fusions that mediate the
  renaming from \(X\) to \(Y\).
  When \(X\) does not occur free, the proof first adds a vacuous occurrence of
  \(X\) by using a closed fusion, reduces the problem to the previous case, and
  then eliminates the redundant binding again.

  First, suppose that \(X \in \mathit{fn}(G)\).
  We derive both \(\nu X.G\) and \(\nu Y.G\angled{Y/X}\) from the same graph
  \[
    \nu X.\nu Y.(Y \bowtie X, (X \bowtie Y, G)).
  \]
  On the one hand, this graph is congruent to \(\nu X.G\):
  \[
    \begin{array}{ll}
       &
      \nu X. \nu Y. (Y \bowtie X, (X \bowtie Y, G))
      \\

      \equivby{E5, E3}
       &
      \nu X. \nu Y. ((Y \bowtie X, X \bowtie Y), G)
      \\

      \equivby{E5, E10}
       &
      \nu X. (\nu Y. (Y \bowtie X, X \bowtie Y), G)
      \\
       &
      \hspace{2em}\because
      Y \notin \mathit{fn}(G)
      \\

      \equivby{E5, E6}
       &
      \nu X. (\nu Y. X \bowtie X, G)
      \\
       &
      \hspace{2em}\because
      (X \bowtie Y)\angled{X/Y} = X \bowtie X
      \\

      \equivby{E5, E10}
       &
      \nu X. \nu Y. (X \bowtie X, G)
      \\
       &
      \hspace{2em}\because
      Y \notin \mathit{fn}(G)
      \\

      \equivby{E9}
       &
      \nu Y. \nu X. (X \bowtie X, G)
      \\

      \equivby{E6, \cref{lem:futile-link-substitution}}
       &
      \nu Y. \nu X. G
      \\
       &
      \hspace{2em}\because
      G\angled{X/X} = G
      \\

      \equivby{\cref{lem:absorb-link-creation}}
       &
      \nu X. G
      \\
       &
      \hspace{2em}\because
      Y \notin \mathit{fn}(G).
    \end{array}
  \]
  On the other hand, the same graph is congruent to \(\nu Y.G\angled{Y/X}\):
  \[
    \begin{array}{ll}
       &
      \nu X. \nu Y. (Y \bowtie X, (X \bowtie Y, G))
      \\

      \equivby{E2, E3, E5, E9}
       &
      \nu Y. \nu X. (X \bowtie Y, (Y \bowtie X, G))
      \\

      \equivby{E5, E6}
       &
      \nu Y. \nu X. (Y \bowtie Y, G\angled{Y/X})
      \\

      \equivby{E5, \cref{lem:absorb-link-creation}}
       &
      \nu Y. (Y \bowtie Y, G\angled{Y/X})
      \\
       &
      \hspace{2em}\because
      X \notin \mathit{fn}((Y \bowtie Y, G\angled{Y/X}))
      \\

      \equivby{E6}
       &
      \nu Y.G\angled{Y/X}
      \\
       &
      \hspace{2em}\because
      (G\angled{Y/X})\angled{Y/Y} = G\angled{Y/X}
      \mbox{ by \cref{lem:futile-link-substitution}}.
    \end{array}
  \]
  Therefore, \(\nu X.G \equiv \nu Y.G\angled{Y/X}\).

  Now suppose that \(X \notin \mathit{fn}(G)\).
  In this case, the binder \(\nu X\) is unused.
  Nevertheless, we cannot simply ignore this case in the proof of
  \(\alpha\)-conversion, because link substitution may in general involve
  \(\alpha\)-conversion.
  We therefore reduce this case to the previous one by adding a closed fusion that
  contains \(X\).
  This closed fusion is congruent to the empty graph, but it makes \(X\) occur in
  the intermediate graph.

  \[
    \begin{array}{ll}
       &
      \nu X. G
      \\

      \equivby{\cref{lem:absorb-link-creation}}
       &
      G
      \\

      \equivby{E1}
       &
      (\zero, G)
      \\

      \equivby{E4, E7}
       &
      (\nu X. \nu X. X \bowtie X, G)
      \\

      \equivby{E4, \cref{lem:absorb-link-creation}}
       &
      (\nu X. X \bowtie X, G)
      \\
       &
      \hspace{2em}\because X \notin \mathit{fn}(\nu X. X \bowtie X)
      \\

      \equivby{E10}
       &
      \nu X. (X \bowtie X, G)
      \\
       &
      \hspace{2em}\because X \notin \mathit{fn}(G)
      \\

      \equivby{the previous case}
       &
      \nu Y. (X \bowtie X, G)\angled{Y/X}
      \\
       &
      \hspace{2em}\because X \in \mathit{fn}((X \bowtie X, G))
      \\

      =
       &
      \nu Y. (Y \bowtie Y, G\angled{Y/X})
      \\

      \equivby{E10}
       &
      (\nu Y. Y \bowtie Y, G\angled{Y/X})
      \\
       &
      \hspace{2em}\because
      X \notin \mathit{fn}(G), \mbox{ and hence }
      Y \notin \mathit{fn}(G\angled{Y/X})
      \\

      \equivby{E4, \cref{lem:absorb-link-creation}}
       &
      (\nu Y. \nu Y. Y \bowtie Y, G\angled{Y/X})
      \\

      \equivby{E7}
       &
      (\zero, G\angled{Y/X})
      \\

      \equivby{E1}
       &
      G\angled{Y/X}
      \\

      \equivby{\cref{lem:absorb-link-creation}}
       &
      \nu Y. G\angled{Y/X}.
    \end{array}
  \]
  This completes the proof.
\end{proof}

\begin{proof}[Proof of \cref{th:symmetry-of-bowtie}]
  We show that fusion is symmetric under structural congruence.
  The idea is to construct a graph with a fresh local link \(Z\) connected by
  fusions to both \(X\) and \(Y\).
  Eliminating \(Z\) in one order yields \(X \bowtie Y\), while eliminating it in
  the other order yields \(Y \bowtie X\).

  First, by absorbing the fusion \(Z \bowtie X\), we obtain
  \[
    \begin{array}{ll}
       &
      \nu Z. (Z \bowtie X, Z \bowtie Y)
      \\

      \equivby{E6}
       &
      \nu Z. (X \bowtie Y)
      \\
       &
      \hspace{2em}\because (Z \bowtie Y)\angled{X/Z} = X \bowtie Y
      \\

      \equivby{\cref{lem:absorb-link-creation}}
       &
      X \bowtie Y.
    \end{array}
  \]
  Conversely, after exchanging the two fusions, absorbing \(Z \bowtie Y\) yields
  \[
    \begin{array}{ll}
       &
      \nu Z. (Z \bowtie X, Z \bowtie Y)
      \\

      \equivby{E2, E5}
       &
      \nu Z. (Z \bowtie Y, Z \bowtie X)
      \\

      \equivby{E6}
       &
      \nu Z. (Y \bowtie X)
      \\
       &
      \hspace{2em}\because (Z \bowtie X)\angled{Y/Z} = Y \bowtie X
      \\

      \equivby{\cref{lem:absorb-link-creation}}
       &
      Y \bowtie X.
    \end{array}
  \]
  Therefore, \(X \bowtie Y \equiv Y \bowtie X\).
\end{proof}

We next record several consequences of the preceding results.
These properties clarify how fusions behave under structural congruence.
They are useful, for example, when proving confluence of graph normalisation up
to isomorphism of local link names.
Intuitively, a fusion \(X \bowtie Y\) identifies the two free link names \(X\) and
\(Y\).
The following theorems make this intuition precise by showing that fused link
names may be renamed, reflexive fusions may be inserted or removed, duplicate
fusions may be contracted, and fusions may be composed transitively.

\begin{theorem}[Renaming along a fusion]\label{th:free-fusion}
  \[
    (X \bowtie Y, p(Y))
    \equiv
    (X \bowtie Y, p(X)).
  \]
\end{theorem}

\begin{proof}
  This theorem states that, in the presence of a fusion \(X \bowtie Y\), an
  occurrence of \(X\) may be replaced by \(Y\), and conversely, inside the same
  graph.
  The proof introduces a fresh local link \(Z\), uses two fusions to transfer the
  occurrence of \(X\) to \(Y\), and then eliminates the unused binder.

  \[
    \begin{array}{lll}
       &
      (X \bowtie Y, p(X))
      \\

      \equivby{(E6)}
       &
      \nu Z. (Z \bowtie X, (Z \bowtie Y, p(Z)))
      \\

      \equivby{(E2), (E3), (E4)}
       &
      \nu Z. (Z \bowtie Y, (Z \bowtie X, p(Z)))
      \\

      \equivby{(E6)}
       &
      \nu Z. (Y \bowtie X, p(Y))
      \\

      \equivby{\cref{th:symmetry-of-bowtie}}
       &
      \nu Z. (X \bowtie Y, p(Y))
      \\

      \equivby{\cref{lem:absorb-link-creation}}
       &
      (X \bowtie Y, p(Y)).
    \end{array}
  \]
\end{proof}

Here, \(p(X)\) may be any atom or fusion containing \(X\) as a free link name.

\begin{theorem}[Reflexivity of fusion]\label{th:fusion-reflexivity}
  \[
    (X \bowtie X, p(X))
    \equiv
    p(X).
  \]
\end{theorem}

\begin{proof}
  This theorem shows that a fusion of a link name with itself carries no additional
  information.
  Equivalently, a reflexive fusion may be inserted or removed whenever the link
  name occurs elsewhere in the graph.

  \[
    \begin{array}{lll}
       &
      p(X)
      \\

      \equivby{\cref{lem:absorb-link-creation}}
       &
      \nu Z. p(X)
      \\

      \equivby{(E6)}
       &
      \nu Z. (Z \bowtie X, p(X))
      \\

      \equivby{(E6)}
       &
      \nu Z. (Z \bowtie Z, (Z \bowtie X, p(X)))
      \\

      \equivby{(E2), (E3), (E4), (E5)}
       &
      \nu Z. (Z \bowtie X, (Z \bowtie Z, p(X)))
      \\

      \equivby{(E6)}
       &
      \nu Z. (X \bowtie X, p(X))
      \\

      \equivby{\cref{lem:absorb-link-creation}}
       &
      (X \bowtie X, p(X)).
    \end{array}
  \]
\end{proof}

\begin{theorem}[Contraction of fusions]\label{th:fusion-contraction}
  \[
    X \bowtie Y
    \equiv
    (X \bowtie Y, X \bowtie Y).
  \]
\end{theorem}

\begin{proof}
  This theorem states that duplicate copies of the same fusion do not change the
  graph up to structural congruence.
  Thus, once two link names are fused, recording the same identification more than
  once is redundant.

  \[
    \begin{array}{lll}
       &
      X \bowtie Y
      \\

      \equivby{\cref{th:fusion-reflexivity}}
       &
      (X \bowtie X, X \bowtie Y)
      \\

      \equivby{(E2), \cref{th:free-fusion}}
       &
      (X \bowtie Y, X \bowtie Y).
    \end{array}
  \]
\end{proof}

\begin{theorem}[Transitivity of fusion]\label{th:transitive-fusion}
  \[
    (X \bowtie Y, Y \bowtie Z)
    \equiv
    (X \bowtie Z, (X \bowtie Y, Y \bowtie Z)).
  \]
\end{theorem}

\begin{proof}
  This theorem expresses the transitive behaviour of fusions.
  If \(X\) is fused with \(Y\) and \(Y\) is fused with \(Z\), then the fusion
  between \(X\) and \(Z\) may be added without changing the graph up to structural
  congruence.
  The proof first inserts a reflexive fusion on \(X\), and then repeatedly applies
  renaming along existing fusions.

  \[
    \begin{array}{lll}
       &
      (X \bowtie Y, Y \bowtie Z)
      \\

      \equivby{\cref{th:fusion-reflexivity}}
       &
      (X \bowtie X, (X \bowtie Y, Y \bowtie Z))
      \\

      \equivby{(E2), (E3), (E4), \cref{th:free-fusion}}
       &
      (X \bowtie Y, (X \bowtie Y, Y \bowtie Z))
      \\

      \equivby{(E2), (E3), (E4), \cref{th:free-fusion}}
       &
      (X \bowtie Z, (X \bowtie Y, Y \bowtie Z)).
    \end{array}
  \]
\end{proof}

The graphs of HyperLMNtal and their equivalence relation,
which is given by structural congruence,
differ from the usual graph-theoretic treatment,
where graphs are defined in terms of vertices,
edges,
and incidence relations,
and equivalence is given by graph isomorphism.
It is therefore not immediate how HyperLMNtal graphs correspond to graphs in the usual graph-theoretic sense.
Nevertheless,
the results established in this section suggest a natural way to relate the two views.
Nakamura studies spatial existential positive logics for hyperedge replacement grammars,
where formulas are interpreted as graphs or graph languages modulo graph isomorphism~\cite{nakamura2022spatial}.
Following a similar line of argument,
HyperLMNtal graphs are expected to be interpretable as ordinary hypergraphs whose free links are connected modulo the equivalence relation generated by fusions.
A formal proof of this correspondence is left for future work.

\chapter{The Untyped \lgt Language}\label{ch:untyped}

In the previous chapter, we formalised HyperLMNtal as a term-based foundation
for hypergraph transformation.
This chapter builds on that foundation and introduces \lgt,
the programming language proposed in this dissertation.
Whereas HyperLMNtal provides a formal account of graphs and graph
transformation, \lgt integrates graphs into a call-by-value
\(\lambda\)-calculus.

The language is developed to satisfy the following design requirements
in \cref{tab:untyped:requirements}.
The central idea of \lgt is to treat graphs as immutable, first-class values.
This extends the familiar treatment of algebraic data in functional programming
to graph-structured data, while retaining graph-specific operations such as
composition and pattern matching.
A distinctive feature of the language is that functions and graphs are not
separate layers: functions may occur as node labels inside graphs, and graph
templates may occur inside expressions.

\begin{table}[htb]
  \centering
  \caption{Design requirements of the core language of \lgt .}
  \label{tab:untyped:requirements}
  \begin{tabular}{@{}p{0.12\linewidth}p{0.78\linewidth}@{}}
    \toprule
    \textbf{ID} & \textbf{Requirement}                                                                                                                 \\
    \midrule
    L1          & Graphs should be composable in a direct and natural manner.                                                                          \\
    L2          & Graphs should be first-class values that can be bound to variables and passed through programs.                                      \\
    L3          & Graphs should be decomposable by pattern matching.                                                                                   \\
    L4          & Functions should remain first-class entities, as in conventional functional languages, while interacting smoothly with graph values. \\
    \bottomrule
  \end{tabular}
\end{table}

The remainder of this chapter formalises the untyped core of \lgt.
\Cref{sec:untyped:syntax} presents the syntax of the language.
\Cref{sec:untyped:semantics} defines its operational semantics.
Together, these definitions show how the language realises the design requirements
outlined above within a call-by-value functional framework.
\Cref{sec:untyped:examples} presents a series of motivating examples,
illustrating how graph-based programs can be naturally expressed and manipulated in \lgt.

\section{Syntax}\label{sec:untyped:syntax}

The syntax of \lgt consists of
two mutually dependent ingredients:
(i) term-based hypergraph representations as values
and (ii) expressions that manipulate these structures.
To realise Requirement L1,
we adopt the graph definition style of HyperLMNtal~\cite{sano-ba,sano2021},
which offers both a term-based syntax and compositional semantics.
\Cref{table:lgt-syntax-concept} illustrates how the graph
definitions of HyperLMNtal
have been extended and adapted into a functional language framework
in a mutually dependent manner.

\newcommand{\tikznd}[2]{
  \tikz[remember picture, baseline=(#1.base)]{%
    \node [rounded corners, fill=gray!10!white] (#1) {#2};
  }
}

\begin{figure}[htbp]
  \tikzexternaldisable
  \setlength{\fboxrule}{0.8pt}
  \small
  \begin{minipage}[t]{0.4\textwidth}
    \centering
    \small
    \textbf{Extending graphs}
  \end{minipage}
  \hfill
  \begin{minipage}[t]{0.4\textwidth}
    \centering
    \small
    \textbf{\kern-5pt Extending \(\lambda\)-expressions}
  \end{minipage}
  \\[1mm]
  \begin{minipage}[t]{0.4\textwidth}
    \small
    \framebox[\textwidth]{\parbox{\textwidth}{%
        \centering
        \tikznd{n11}{\small HyperLMNtal Graphs} \\[4pt]
        \textit{augmented with}                       \\[4pt]
        \tikznd{n12}{\small \(\lambda\)-abstractions} \\
        (embedded as node labels)
      }}%
  \end{minipage}
  \hfill
  \begin{minipage}[t]{0.4\textwidth}
    \small
    \hfill
    \framebox[\textwidth]{\parbox{\textwidth}{%
        \centering
        \tikznd{n21}{\small \(\lambda\)-expressions} \\[4pt]
        \textit{augmented with}                      \\[4pt]
        \tikznd{n22}{\small Graph Templates}\\
        (graphs with variables)
      }}%
  \end{minipage}
  \begin{tikzpicture}[remember picture,overlay]
    \path let
    \p1 = (n11),
    \p2 = (n22),
    in
    coordinate (c11) at (-0.5\textwidth, \y1)
    coordinate (c22) at (-0.5\textwidth, \y2);

    \node [fill=white, text=white, inner sep=0]
    (N1) at (c11) {$\langle$ \quad includes \quad $\rangle$\kern-20pt};
    \node [fill=white, text=white, inner sep=0]
    (N2) at (c22) {$\langle$ \quad has \quad $\rangle$\kern-5pt};

    \draw[<-] (n11) to (N1.east);
    \draw[] (n12) to (N2.east);

    \draw[->] (N2.east) to[out=0,in=180] (n21.west);
    \draw[] (N1.east) to[out=0,in=180] (n22.west);

    \node [fill=white, below=0.1mm of c11, inner xsep=0, inner ysep=0.4mm]
    {\small $\langle$includes$\rangle$};
    \node [fill=white, below=0.1mm of c22, inner xsep=0, inner ysep=0.4mm]
    {\small $\langle$has$\rangle$};
  \end{tikzpicture}
  \tikzexternalenable
  \caption{Conceptual overview of the \lgt syntax.}
  \label{table:lgt-syntax-concept}
\end{figure}

We first introduce the following syntactic categories:

\begin{itemize}
  \item \textbf{Link Names} \(X, Y, \dots\)
        represent individual edges in the hypergraph.
  \item \textbf{Constructor Name} \(C\)
        represents
        node labels of hypergraphs.
  \item \textbf{Graph Variable Names}%
        \footnote{These variables correspond to \textit{process contexts}
          in HyperLMNtal~\cite{hyperlmntal}.  The term \textit{graph
            variable names} is used in this dissertation to avoid confusion
          with other uses of the term ``context'' in programming language
          theory.}
        \(x, y, \dots\)
        denote placeholders for subgraphs.
\end{itemize}

\begin{mydef}[Syntax of \lgt]\label{syntax}
  The syntax of \emph{values} and \emph{expressions} in
  \lgt is defined by \cref{table:lgt-syntax}.
\end{mydef}

\begin{figure}[htbp]
  \begin{framed}
    \centering
    \begin{tabular}{rccl}
      Value      & \(G\)
                 & \(::=\)
                 &
      \(\mathbf{0}
      \mid p(\Xs)
      \mid X \bowtie Y
      \mid (G, G)
      \mid {\nu X.G}
      \)
      \\[1mm]
      Expression & \(e\)
                 & \(::=\)
                 &
      \(T
      \mid (\caseof{e}{T}{e}{e})
      \mid (e\; e)\)
      \\[1mm]
      Template   & \(T\)
                 & \(::=\)
                 &
      \(\mathbf{0}
      \mid p(\Xs)
      \mid X \bowtie Y
      \mid (T, T)
      \mid \nu X.T
      \mid x[\Xs]
      \)
      \\[1mm]
      Atom Name  & \(p\)
                 & \(::=\)
                 &
      \(C
      \mid \lambda\,x[\Xs].e\)
    \end{tabular}
  \end{framed}
  \caption{Syntax of the \lgt Language.}
  \label{table:lgt-syntax}
\end{figure}

In \lgt,
values are hypergraphs,
which are written in a term-based syntax
to ensure
high affinity with programming languages~\cite{sano2021}.

\begin{itemize}
  \item
        The term \(\mathbf{0}\) denotes the empty graph.

  \item
        An \emph{atom}
        \(p(\seq{X})\)
        stands for a \emph{node} of a data structure with the label \(p\) and
        totally ordered links \(\seq{X}\).
        The individual links are regarded as attached to
        \textit{ports} ($1^\textrm{st}$, $2^\textrm{nd}$, $\dots$)
        of an atom rather than the atom itself, and such
        (hyper)graphs are often called \textit{port (hyper)graphs}.

        An atom name
        (i.e., a node label)
        is either a constructor
        (as in \(\mathrm{Nil} (X)\) and
        \(\mathrm{Cons} (Y, Z, X)\))
        or a \(\lambda\)-abstraction.

        A \(\lambda\)-abstraction atom has the form
        \((\lambda\, x [\seq{X}]. e) (\seq{Y})\).
        Here, \(x [\seq{X}]\),
        where \(\seq{X}\) is a sequence of
        \textit{pairwise distinct} links,
        denotes a \emph{graph variable},
        which corresponds to and extends
        a variable in functional languages.
        We need this extension
        (with the \([\seq{X}]\))
        because, unlike standard tree structures
        with single roots, a graph
        may have an arbitrary number of roots
        that are ``access points'' to
        the graph from outside.

        The \(\lambda\)-abstraction atom
        \((\lambda\, x [\seq{X}]. e) (\seq{Y})\)
        takes a graph with free links
        (i.e., links not hidden by $\nu X$ described soon below)
        \(\seq{X}\),
        binds it to the graph variable \(x [\seq{X}]\), and
        returns the value
        obtained by evaluating the expression \(e\)
        with the bound graph variable.
        The $\Ys$ attached to the abstraction represent the links
        of the atom whose name (node label) is the abstraction.
        Those links are used for forming graphs containing
        functions but becomes unnecessary and disappears upon
        $\beta$-reduction (as will be defined in \cref{table:lgt-reduction}).

  \item
        A \emph{fusion},
        \(X \bowtie Y\),
        fuses the link \(X\) and the link \(Y\) into a single link.

  \item
        A \emph{molecule},
        \((G, G)\),
        stands for
        the composition or gluing of graphs.

  \item
        A \emph{hyperlink creation},
        \(\nu X.G\), hides the link name \(X\) in \(G\).
        In \lgt, all links are
        \emph{free links} as long as they are not hidden.
        The set of free links in \(T\) is denoted as \(\fn (T)\)
        defined inductively as in \cref{table:untyped:free-names}.
        \begin{figure}[htbp]
          \begin{framed}
            \[
              \begin{aligned}
                \fn(\zero)      & \triangleq \emptyset              \\
                \fn(p(\seq{X})) & \triangleq \{\seq{X}\}            \\
                \fn(X\bowtie Y) & \triangleq \{X,Y\}                \\
                \fn((T_1, T_2)) & \triangleq \fn(T_1) \cup \fn(T_2) \\
                \fn(\nu X.T)    & \triangleq \fn(T) \setminus \{X\} \\
                \fn(x[\seq{X}]) & \triangleq \{\seq{X}\}            \\[-.3em]
              \end{aligned}
            \]
          \end{framed}
          \caption{The set of free link names over \lgt graph templates.}\label{table:untyped:free-names}
        \end{figure}
        Links that do not occur free are called \emph{local links}.
\end{itemize}

Since values are represented as graphs, and graphs may include
$\lambda$-abstractions as node labels,
the language
naturally supports higher-order programming.
Furthermore, values can be graphs containing one or more $\lambda$-abstractions.
This allows the representation of not only lists or trees of functions
(as in standard functional languages)
but also
dataflow graphs with functions assigned to each node,
as soon demonstrated in the next chapter.

A \lgt
program is called an \emph{expression}
which is either of the following:
\begin{itemize}
  \item
        \(T\) is a graph with zero or more graph variables.

  \item
        \((\caseof{e_1}{T}{e_2}{e_3})\)
        evaluates \(e_1\),
        checks whether it matches the graph template \(T\), and
        reduces to \(e_2\) or \(e_3\).
        We require that
        graph variables names occurring in pattern $T$
        are pairwise distinct
        and that
        the link names in each
        graph variables are pairwise distinct.
  \item
        \((e_1\; e_2)\) is an application.
\end{itemize}

\begin{mydef}[Syntactic Convention]\label{SyntacticConvention}
  For convenience, we introduce
  an additional abbreviation rule to
  the HyperLMNtal abbreviation rules in \cref{def:abbreviation}.
  \begin{itemize}
    \item
          As in other functional languages,
          we provide the form
          \(\letin{x[\Xs]}{e'}{e}\)
          as an abbreviation for
          \(((\lambda\,x[\Xs].e)()\;e')\),
          which can be written also (by (1)) as
          \(((\lambda\,x[\Xs].e)\;e')\). \qedhere
  \end{itemize}
\end{mydef}

\section{Operational Semantics}\label{sec:untyped:semantics}

We introduce two forms of
substitution in \lgt.
\textit{Link substitution}
\(\sigma\) replaces occurrences of free link names within a graph with
other link names.
For example, \(T\angled{\Zs/\Ys}\) denotes the substitution of all
free occurrences of \(\Ys\) in \(T\) with \(\Zs\).
\textit{Graph substitution}
\(\theta\) replaces occurrences of free graph variables within an
expression with specified graph templates (or graphs).
For example, \(e[T/x[\Xs]]\) denotes the replacement of all
occurrences of \(x[\Xs]\) in the expression \(e\) with the graph template
\(T\).
We require that the link names in $\Xs$ are pairwise distinct and
also that $\fn(T) = \{\Xs\}$, that is, the substitution does not
change the set of free links.
Their formal definitions, which avoid capture
(of local links or local graph variables) in a standard manner.

\begin{mydef}[Link Substitution]
  A \emph{link substitution},
  \(T\angled{ Z_1, \dots, Z_n / Y_1, \dots, Y_n }\),
  that replaces all free occurrences of
  \(Y_i\) with \(Z_i\)
  is defined as
  in \cref{table:hyperlink-substitution-lgt}.
  Here, the \(Y_1, \dots, Y_n\) should be pairwise distinct.
  Note that, if a free occurrence of \(Y_i\) occurs at a
  location where \(Z_i\) would not be free,
  \(\alpha\)-conversion may be required.

  \begin{figure}[htbp]
    \begin{framed}
      \centering
      \begin{tabular}{@{}r@{\hspace{0.5em}}c@{\hspace{0.5em}}l@{}}
        \(\zero\angled{\Zs/\Ys}\)
                                       & \(\triangleq \) & \(\zero\)                                                \\[1pt]
        \(p(\Xs)\angled{\Zs/\Ys}\)     & \(\triangleq \) & \(p(X_1\angled{\Zs/\Ys}, \ldots, X_n\angled{\Zs/\Ys}) \) \\
        \multicolumn{3}{r}{%
          where \(
          X\angled{\Zs/\Ys} =
          \left\{
          \begin{array}{@{}ll}
            Z_i & \mbox{if } X = Y_i          \\
            X   & \mbox{if } X \notin \{\Ys\}
          \end{array}
          \right.
          \)
        }                                                                                                           \\[3pt]
        \(x[\Xs]\angled{\Zs/\Ys}\)     & \(\triangleq \) & \(x[X_1\angled{\Zs/\Ys}, \ldots, X_n\angled{\Zs/\Ys}] \) \\
        \multicolumn{3}{r}{%
          where \(
          X\angled{\Zs/\Ys} =
          \left\{
          \begin{array}{@{}ll}
            Z_i & \mbox{if } X = Y_i          \\
            X   & \mbox{if } X \notin \{\Ys\}
          \end{array}
          \right.
          \)
        }                                                                                                           \\[3pt]

        \((T_1, T_2)\angled{\Zs/\Ys}\) & \(\triangleq \) & \((T_1 \angled{\Zs/\Ys}, T_2 \angled{\Zs/\Ys})\)         \\[3pt]
        \(
        (\nu X.T)\angled{\Zs/\Ys}\)    & \(\triangleq \) &                                                          \\
        \multicolumn{3}{l}{%
          \(\left\{
          \begin{array}{@{}l@{~~}l@{}}
            \nu X.T\angled{\seq{Z'}/\seq{Y'}} & \mbox{if } X = Y_i\ \land                               \\
                                              & \seq{Z'} = Z_1, \dots, Z_{i - 1}, Z_{i + 1}, \dots, Z_n \\
                                              & \seq{Y'} = Y_1, \dots, Y_{i - 1}, Y_{i + 1}, \dots, Y_n \\[1pt]
            \nu X.T\angled{\Zs/\Ys}           & \mbox{if } X \notin \{\Ys\} \land X \notin \{\Zs\}      \\[3pt]
            \nu W.(T\angled{W/X})\angled{\Zs/\Ys}
                                              & \mbox{if } X \notin \{\Ys\} \land X \in \{\Zs\}         \\
                                              & \land W \notin \mathit{fn}(T) \land W \notin \{\Zs\}
          \end{array}
          \right.
          \)
        }
      \end{tabular}
    \end{framed}
    \caption{(Hyper)link substitution over \lgt graph templates.}\label{table:hyperlink-substitution-lgt}
  \end{figure}
\end{mydef}

Before defining graph substitution, we define the notion of \emph{free
  functors}, where a functor, written $x/n$,
stands for the name $x$ of a graph variable $x[\Xs]$
with its arity $n$ which is equal to $|\Xs|$.
We regard two graph variables with the same name and the
same arity and within the same scope as referring to the same variable
with possibly renamed free links.

\begin{mydef}[Free functors of an expression]\label{def:free-functor}

  We define free functors of an expression \(e\), \(\mathit{ff}(e)\),
  in \cref{table:free-functor}.

  \begin{figure}[htbp]
    \begin{framed}
      \centering
      \begin{tabular}{r@{\hspace{0.5em}}c@{\hspace{0.5em}}l}
        \multicolumn{3}{l}{\(\mathit{ff} (\caseof{e_1}{T}{e_2}{e_3}) \triangleq \)}                                                                                  \\
                                                     &                 & \(\mathit{ff}(e_1) \cup (\mathit{ff}(e_2) \setminus \mathit{ff}(T)) \cup \mathit{ff}(e_3)\) \\[3pt]
        \(\mathit{ff}((e_1\; e_2))\)                 & \(\triangleq \) & \(\mathit{ff}(e_1) \cup \mathit{ff}(e_2)\)                                                  \\[3pt]
        \(\mathit{ff} (x [\Xs])\)                    & \(\triangleq \) & \(\{x/\norm{\Xs}\}\)                                                                        \\[1pt]
        \(\mathit{ff} (v\,(\Xs))\)                   & \(\triangleq \) & \(\emptyset\)                                                                               \\[1pt]
        \(\mathit{ff} ((\lambda\, x[\Xs].e) (\Ys))\) & \(\triangleq \) & \(\mathit{ff}(e) \setminus \{x/\norm{\Xs}\}\)                                               \\[1pt]
        \(\mathit{ff} ((T_1, T_2))\)                 & \(\triangleq \) & \(\mathit{ff}(T_1) \cup \mathit{ff}(T_2)\)                                                  \\[1pt]
        \(\mathit{ff} (\nu X. T)\)                   & \(\triangleq \) & \(\mathit{ff} (T)\)                                                                         \\
      \end{tabular}
    \end{framed}
    \caption{Free functors of an expression.}\label{table:free-functor}
  \end{figure}
\end{mydef}

Free functors are not to be confused with free link names.



\begin{mydef}[Graph Substitution]
  We define capture-avoiding substitution
  \(\theta\)
  of a graph variable
  \(x [\Xs]\)
  with a template \(T\)
  in \(e\),
  written \(e [T / x [\Xs]]\),
  as in \cref{table:graph-substitution}.
  The definition is standard except that it handles the substitution of
  the free links of graph variables in the third rule.
\end{mydef}

\begin{figure}[htbp]
  \begin{framed}
    \centering
    \begin{tabular}{@{}r@{~}c@{~}l}
      \((T_1, T_2)\theta\)       & \(\triangleq \)
                                 & \((T_1 \theta, T_2 \theta)\)                                                           \\[3pt]
      \((\nu X. T)\theta\)       & \(\triangleq \)
                                 & \(\nu X. T\theta\)                                                                     \\[3pt]
      \((x [\Xs])[T / y [\Ys]]\) & \(\triangleq \)                                 &
      if \(x/\norm{\Xs} = y/\norm{\Ys}\)
      then \(T{\angled{\Xs/\Ys}}\) else \(x [\Xs]\)                                                                       \\[1pt]
      \((C (\Xs))\theta\)        & \(\triangleq \)                                 & \(C (\Xs)\)                          \\[1pt]
      \multicolumn{3}{@{}l}{%
      \(((\lambda\, x [\Xs].e) (\Zs))[T / y [\Ys]]~\triangleq \)}                                                         \\
                                 & \multicolumn{2}{@{}l}{                                                                 %
        \hspace*{-20pt}if \(x/\norm{\Xs} = y/\norm{\Ys}\) then
      \((\lambda\, x [\Xs].e) (\Zs)\)}                                                                                    \\
                                 & \multicolumn{2}{@{}l}{                                                                 %
      \hspace*{-20pt}else if \(x/\norm{\Xs} \notin \mathit{ff}(T)\) then
      \((\lambda\, x [\Xs].e[T / y [\Ys]]) (\Zs)\)}                                                                       \\
                                 & \multicolumn{2}{@{}l}{                                                                 %
      \hspace*{-20pt}else
      \((\lambda\, z[\Xs].e [z [\Xs] / x [\Xs]] [T / y [\Ys]]) (\Zs)\)}                                                   \\
                                 &                                                 & \hfill{} where \(z/\norm{\Xs} \notin
      \mathit{ff}(e)\cup\mathit{ff}(T)\)                                                                                  \\[1pt]
      \multicolumn{3}{@{}l}{%
      \((\caseof{e_1}{T}{e_2}{e_3})\theta\)}                                                                              \\
                                 & \(\triangleq \)
                                 & \(\caseof{e_1\theta}{T}{e_2\theta}{e_3\theta}\)                                        \\[3pt]
      \((T_1\; T_2)\theta\)      & \(\triangleq \)
                                 & \((T_1 \theta\;\; T_2 \theta)\)                                                        \\
    \end{tabular}
  \end{framed}
  \caption{Graph substitution.}\label{table:graph-substitution}
\end{figure}

To establish a formal foundation for pattern matching,
we introduce the equivalence relation \(\equiv\) on graphs
and graph templates.

\begin{mydef}[Structural Congruence]\label{def:untyped:congruence}
  The equivalence relation \(\equiv\) on graphs
  and graph templates is defined as the minimal equivalence relation
  that satisfies the rules given in \cref{table:untyped:congruence}.
\end{mydef}

\begin{figure}[htbp]
  \begin{framed}
    \centering
    \begin{tabular}{c}
      (E1) \((\mathbf{0}, T) \equiv T\)
      \quad
      (E2)   \((T_1, T_2) \equiv (T_2, T_1)\)
      \\[2.0mm]
      (E3) \((T_1, (T_2, T_3)) \equiv ((T_1, T_2), T_3)\)
      \\[2.0mm]
      (E4) \(\dfrac{T_1 \equiv T_2}{(T_1, T_3) \equiv (T_2, T_3)}\)
      \qquad (E5) \(\dfrac{T_1 \equiv T_2}{\nu X.T_1 \equiv \nu X.T_2}\)
      \\[4.0mm]
      (E6) \(\nu X.(X \bowtie Y, T)  \equiv \nu X.T\langle Y / X \rangle\)
      \\
      where \(X \in \fn(T) \lor Y \in \fn(T)\)
      \\[2.0mm]
      (E7) \(\nu X.\nu Y.X \bowtie Y \equiv \mathbf{0}\)
      \qquad
      (E8) \(\nu X.\mathbf{0} \equiv \mathbf{0}\)
      \\[2.0mm]
      (E9) \(\nu X.\nu Y.T \equiv \nu Y.\nu X.T\)
      \\[2.0mm]
      (E10) \(\nu X.(T_1, T_2) \equiv (\nu X.T_1, T_2)\)
      where \(X \notin \fn(T_2)\)
    \end{tabular}
  \end{framed}
  \caption{Congruence rules over \lgt graph templates.}
  \label{table:untyped:congruence}
\end{figure}



\begin{mydef}[Small Step Semantics of \lgt]\label{SmallStepSemantics}
  The \lgt language is given
  a small-step, contextual semantics using the reduction relation
  defined in \cref{table:lgt-reduction},
  where the $\thetas$ here stands for a sequence of graph substitutions
  $[T_1/x_1[\overrightarrow{X_1}]]\dots
    [T_n/x_n[\overrightarrow{X_n}]]$ such that each
  $T_i$ is a template not containing graph variables (i.e., a \emph{value}),
  making the substitutions order-independent.
\end{mydef}

\begin{figure}[htbp]
  \begin{framed}
    \centering
    \raggedright Evaluation context $E$:
    \[
      E ::= [\,] \mid (\caseof{E}{T}{e}{e}) \mid (E\; e) \mid (G\; E)
    \]

    \medskip

    \raggedright Reduction relation $\reduces$:
    \begin{prooftree}
      \AXC{\(G\equiv T\thetas\)}
      \RightLabel{\RdCaseMatch{}}
      \UIC{\(
        E[\caseof{G}{T}{e_1}{e_2}]
        \reduces
        E[e_1\thetas]
        \)}
    \end{prooftree}

    \begin{prooftree}
      \AXC{\(\nexists \thetas.(G \equiv T\thetas)\)}
      \RightLabel{\RdCaseOther{}}
      \UIC{\(
        E[\caseof{G}{T}{e_1}{e_2}]
        \reduces
        E[e_2]
        \)}
    \end{prooftree}

    \begin{prooftree}
      \AXC{$G_1 \equiv (\lambda\, x [\Xs].e) (\Ys)$}
      \AXC{\(\fn(G_2) = \{\Xs\}\)}
      \RightLabel{\RdBeta{}}
      \BIC{\(
        E[G_1\ G_2]
        \reduces
        E[e [G_2 / x[\Xs]]]
        \)}
    \end{prooftree}
  \end{framed}
  \caption{Reduction relation of \lgt.}
  \label{table:lgt-reduction}
\end{figure}

\RdCaseMatch{} and \RdCaseOther{} are the reduction rules for graph pattern
matching.
If matching succeeds (\RdCaseMatch{}), we apply to $\eDLpop$ the graph
substitution obtained by the matching.

Note that the substitution $\theta$ in \RdCaseMatch{} is not necessarily unique.
Issues such as the decidability of determining whether \RdCaseMatch{} or
\RdCaseOther{} applies, as well as the design of a practical language
implementation, are beyond the scope of this paper.
Nevertheless, we conjecture that this decision problem can be reduced to
the subgraph isomorphism problem, which is known to be decidable.
A proof that graphs in \lgt, their equivalence, and the corresponding
notion of matching correspond to graphs in graph theory, graph isomorphism,
and subgraph isomorphism, respectively, is left for future work.

\RdBeta{} applies a function to a value.
The definition is standard except that
(i) we need to check the correspondence of free links and
(ii) we use graph substitution instead of the standard substitution in
the \(\lambda\)-calculus.
Note that, while the free links of $G$ must be exactly $\Xs$,
the graph variable $x[\Xs]$ representing $G$
may be used in $e$ in arbitrarily many times, each
with its free links renamed as $x[\overrightarrow{X'}]$,
$x[\overrightarrow{X''}]$ etc.

It is worth noting that the above small-step semantics does not
guarantee the uniqueness of the evaluation result due to the
possibly nondeterministic graph pattern matching in \RdCaseMatch{}.


The untyped \lgt can easily encode
single-step graph rewriting using a
HyperLMNtal rewrite rule.
The rule \(G_1 \rightarrow G_2\) matches a graph \(G\), which has a subgraph \(G_1\)
and rewrites the subgraph to \(G_2\).
One-step rewriting using
this rule can be encoded into a function
\((\lambda\,g[\seq{Y}].\ \caseof{g[\seq{Y}]}{\allowbreak (G_1, x[\seq{X}])}{(G_2, x[\seq{X}])}{g[\seq{Y}]})(Z)\),
where \(\{\seq{Y}\} \triangleq \mathit{fn}(G)\)
and \(\{\seq{X}\} \triangleq \mathit{fn}(G_1)\).
This function takes \(G\) and returns the graph obtained by applying the rule
(or the original graph if the rule is not applicable).
Repeated rule application can be encoded in a standard
manner using a fixed-point combinator.

\section{Examples}\label{sec:untyped:examples}

In the \lgt language,
it is possible to express transformations on data structures
such as removing the last element from a doubly linked list or applying a function to
the leaves of a leaf-linked tree in a purely declarative manner.
These transformations rely on pattern matching and avoid the need for explicit
destructive updates to the heap or pointer manipulations.

Recall we have represented doubly linked lists in a HyperLMNtal graph in \eqref{eq:dbllist-eg}.
The program for removing the last element from a doubly linked list
can be expressed straightforwardly using pattern matching in \lgt.

The function can be visually represented as
\begin{align}\label{eq:dbllist-removal-eg-vis}
  \eDLpop \triangleq
  \begin{array}{l@{}}
    (\lambda\,\ctxx{x}.                            \\[1.5mm]
    \quad \mathbf{case}\; \ctxx{x}\; \mathbf{of}\; \\[1.5mm]
    \qquad
    \tikz[baseline=-0.5ex]{%
      \mkctxn{1}{y}{(0,0)}{}{W_2}
      \mkcons{2}{W_3}{W_1}{(1.5,0)}{w}
      \mknil{3}{(3,0)}
      \node [link] (C0) at (-1, 0.8) {$Z$};
      \node [link] (LX) at (-1, 0.0) {$X$};
      \node [link] (C4) at (+4, 0.8) {$Y$};
      \draw (C1) to (LX);
      \mkconslinkleft{0}{1}
      \mkconslink{1}{2}
      \mkconslink{2}{3}
      \mkconslinkright{3}{4}
    }
    \to
    \tikz[baseline=-0.5ex]{%
      \mkctxn{1}{y}{(0,0)}{}{W_2}
      \mknil{2}{(1.5,0)}
      \node [link] (C0) at (-1.0, 0.8) {$Z$};
      \node [link] (LX) at (-1.0, 0.0) {$X$};
      \node [link] (C3) at (+2.5, 0.8) {$Y$};
      \draw (C1) to (LX);
      \mkconslinkleft{0}{1}
      \mkconslink{1}{2}
      \mkconslinkright{2}{3}
    }
    \\[12mm]
    \quad \texttt{|}\; \ \mathbf{otherwise} \qquad
    \to \ctxx{x}                                   \\
    )(W)
  \end{array}%
\end{align}

\noindent
(when it is given as a unary atom with the link $W$).
Alternatively, the function can be expressed in our formal notation
as
\begin{align}\label{eq:dbllist-removal-eg}
  \eDLpop \triangleq
  \begin{array}{l}
    (\lambda\,x[X,Y,Z].                                 \\
    \quad \mathbf{case}\; x[X,Y,Z]\; \mathbf{of}\;      \\
    \qquad \nu W_1 W_2 W_3. (                           \\
    \qquad\quad y[W_1,W_2,X,Z],                         \\
    \qquad\quad \Cons(W_3,W_2,Y,W_1),                   \\
    \qquad\quad z[W_3],                                 \\
    \qquad\quad \Nil(W_1,Y)                             \\
    \qquad) \to \nu W.(y[Y,W,X,Z], \Nil(W,Y))           \\
    \quad\: \texttt{|}\;\mathbf{otherwise} \to x[X,Y,Z] \\
    )(W).
  \end{array}
\end{align}

As a function, only the node label ($\lambda$-abstraction) of the
atom matters, and the role of the $W$ is to accommodate the function
inside a linked structure.  If a function need not be put in a linked
structure, it can be represented as the label of a nullary atom, in
which case the $(W)$ above is omitted
(as in $\eLLT$ defined later in \eqref{eq:lltree-traverse-eg-vis}).

To illustrate the evaluation of this function, consider the doubly linked list
$\GDL$ defined in \eqref{eq:dbllist-eg}.
Applying \(\eDLpop\) to $\GDL$ results in the following
two-step reduction (\RdBeta{} followed by \RdCaseMatch{}):
\begin{align*}
   &
  \Biggl(
  \eDLpop \qquad
  \tikz[baseline=-0.5ex]{%
    \mkcons{1}{W_1}{}{(0,0)}{1}
    \mkcons{2}{W_3}{W_2}{(1.5,0)}{2}
    \mkcons{3}{W_5}{W_4}{(3.0,0)}{3}
    \mknil{4}{(4.5,0)}
    \node [link] (C0) at (-1.0, 0.8) {$Z$};
    \node [link] (LX) at (-1.0, 0.0) {$X$};
    \node [link] (C5) at (+5.5, 0.8) {$Y$};
    \draw (C1) to (LX);
    \mkconslinkleft{0}{1}
    \mkconslink{1}{2}
    \mkconslink{2}{3}
    \mkconslink{3}{4}
    \mkconslinkright{4}{5}
  }
  \Biggr)
  \notag
  \\
   & \reducestwo
  \quad
  \tikz[baseline=-0.5ex]{%
    \mkcons{1}{W_1}{}{(0,0)}{1}
    \mkcons{2}{W_3}{W_2}{(1.5,0)}{2}
    \mknil{3}{(3,0)}
    \node [link] (C0) at (-1.0, 0.8) {$Z$};
    \node [link] (LX) at (-1.0, 0.0) {$X$};
    \node [link] (C4) at (+4.0, 0.8) {$Y$};
    \draw (C1) to (LX);
    \mkconslinkleft{0}{1}
    \mkconslink{1}{2}
    \mkconslink{2}{3}
    \mkconslinkright{3}{4}
  }
  \notag
\end{align*}

In this example,
firstly, the application reduces to a case expression
using
\RdBeta{}.
Then, the case expression is evaluated with \RdCaseMatch{}.
In this evaluation step,
the graph variable\
\tikz[baseline=1.5ex, scale=1]{%
  \begin{scope}[shift={(0,0)}]
    \node [ctxn] (C1)  at (0, 0)    {$y$};
    \coordinate (p1AL) at ($(C1)+(-0.1,0.3)$) {};
    \coordinate (p1AR) at ($(C1)+(+0.1,0.3)$) {};
    \coordinate [] (p1L) at ($(p1AL)+(-0.5,0.2)$);
    \coordinate [] (p1R) at ($(p1AR)+(+0.5,0.2)$);
    \node [port] at (+0.400, -0.150) {1};
    \node [port] at (+0.250, +0.450) {2};
    \node [port] at (-0.400, -0.150) {3};
    \node [port] at (-0.250, +0.450) {4};
  \end{scope}
  %
  \node [link] (LZ)  at (-1.0, 0.8)  {$Z$};
  \node [link] (LX)  at (-1.0, 0.0)  {$X$};
  \node [link] (LW2) at (+1.0, 0.8) {$W_2$};
  \node [link] (LW1) at (+1.0, 0.0) {$W_1$};
  \draw [out=0, in=90, looseness=1.50] (LZ) to (p1AL);
  \draw []  (C1) to  (LX);
  %
  \draw [out=180, in=90, looseness=1.50] (LW2) to (p1AR);
  \draw []  (C1) to  (LW1);
}
matches the first two \(\mathrm{Cons}\) nodes containing the numbers 1 and 2,
and it is directly terminated with \(\mathrm{Nil}\).

Another example is a program that applies a function to the
values of leaves in a leaf-linked tree,
a tree in which the leaves are explicitly linked,
allowing efficient traversal.
Such a tree can be illustrated as follows.
\begin{align}\label{eq:lltree-eg-vis}
   & \lltr{t} \triangleq
  \quad
  \notag
  \\[-6pt]
   &
  \quad
  \tikz[baseline=-0.5ex, scale=1]{%
    \begin{scope}[shift={(0, 2)}]
      \node [atom] (node1) at (0,0) {N};
      \node [port] at (0.100, 0.395)   {3};
      \node [port] at (-0.395, -0.100) {1};
      \node [port] at (0.395, -0.100)  {2};
    \end{scope}
    \begin{scope}[shift={(-1.4, 1.2)}]
      \node [atom] (node2) at (0,0) {N};
      \node [port] at (0.100, 0.395)   {3};
      \node [port] at (-0.395, -0.100) {1};
      \node [port] at (0.395, -0.100)  {2};
      \node [link,text=gray] at (0.500, 0.800) {$(W_1)$};
    \end{scope}
    \begin{scope}[shift={(1.4, 1.2)}]
      \node [atom] (node3) at (0,0) {N};
      \node [port] at (0.100, 0.395)   {3};
      \node [port] at (-0.395, -0.100) {1};
      \node [port] at (0.395, -0.100)  {2};
      \node [link,text=gray] at (-0.500, 0.800) {$(W_2)$};
    \end{scope}
    \begin{scope}[shift={(-0.8, 0)}]
      \node [atom] (node4) at (0,0) {N};
      \node [port] at (0.100, 0.395)   {3};
      \node [port] at (-0.395, -0.100) {1};
      \node [port] at (0.395, -0.100)  {2};
      \node [link,text=gray] at (0.100, 0.700)   {$(W_3)$};
    \end{scope}
    %
    \begin{scope}[shift={(-2.6, 0)   }]
      \node [atom,fill=black!60,text=white] (leaf1) at (0, 0) {L};
      \node [port] at (0.100, 0.395)   {3};
      \node [port] at (0.395, 0.100)   {2};
      \node [port] at (-0.100, -0.395) {1};
      \node [port] at (-0.100, -0.7)   {1};
      \node [link,text=gray] at (-0.510, -0.5)   {$(W_8)$};
    \end{scope}
    \begin{scope}[shift={(-1.6, -1.3)}]
      \node [atom] (leaf2) at (0, 0) {L};
      \node [port] at (0.100, 0.395)   {3};
      \node [port] at (0.395, 0.100)   {2};
      \node [port] at (-0.100, -0.395) {1};
      \node [port] at (-0.100, -0.7)   {1};
      \node [link,text=gray] at (-0.45, 0.9)   {$(W_4)$};
      \node [link,text=gray] at (0.410, -0.5) {$(W_9)$};
    \end{scope}
    \begin{scope}[shift={(-0.395, -1.3)}]
      \node [atom] (leaf3) at (0, 0) {L};
      \node [port] at (0.100, 0.395)   {3};
      \node [port] at (0.395, 0.100)   {2};
      \node [port] at (-0.100, -0.395) {1};
      \node [port] at (-0.100, -0.7)   {1};
      \node [link,text=gray] at (0.40, 0.820) {$(W_5)$};
      \node [link,text=gray] at (0.495, -0.5)  {$(W_{10})$};
    \end{scope}
    \begin{scope}[shift={(1.1, -0.7)}]
      \node [atom] (leaf4) at (0,0) {L};
      \node [port] at (0.100, 0.395)   {3};
      \node [port] at (0.395, 0.100)   {2};
      \node [port] at (-0.100, -0.395) {1};
      \node [port] at (-0.100, -0.7)   {1};
      \node [link,text=gray] at (0.300, 1)    {$(W_6)$};
      \node [link,text=gray] at (0.495, -0.5) {$(W_{11})$};
    \end{scope}
    \begin{scope}[shift={(2.6, 0) }]
      \node [atom] (leaf5) at (0, 0) {L};
      \node [port] at (0.100, 0.395)   {3};
      \node [port] at (0.395, -0.100)  {2};
      \node [port] at (-0.100, -0.395) {1};
      \node [port] at (-0.100, -0.7)   {1};
      \node [link,text=gray] at (-0.100, 0.9) {$(W_7)$};
      \node [link,text=gray] at (0.495, -0.5) {$(W_{12})$};
    \end{scope}
    \node [atom,below=0.8cm] (num1) at (leaf1) {1};
    \node [atom,below=0.8cm] (num2) at (leaf2) {2};
    \node [atom,below=0.8cm] (num3) at (leaf3) {3};
    \node [atom,below=0.8cm] (num4) at (leaf4) {4};
    \node [atom,below=0.8cm] (num5) at (leaf5) {5};
    \node [link] (LL) at (-4, 0)  {$L$};
    \node [link] (LX) at (0, 3)   {$X$};
    \node [link] (LR) at (4, 0)   {$R$};
    \draw (node1) -- (LX);
    \draw (node1) -- (node2);
    \draw (node1) -- (node3);
    \draw (node2) -- (node4);
    \draw [looseness=1.50,in=90,out=-130] (node2) to (leaf1);
    \draw [looseness=1.50,in=90,out=-130] (node4) to (leaf2);
    \draw [looseness=1.50,in=90,out=-50 ] (node4) to (leaf3);
    \draw [looseness=1.50,in=90,out=-130] (node3) to (leaf4);
    \draw [looseness=1.50,in=90,out=-50 ] (node3) to (leaf5);
    \draw [in=90,out=30,looseness=1.50] (LL) to (leaf1.north);
    \draw [in=90,out=30,looseness=1.50] (leaf1) to (leaf2);
    \draw [in=90,out=30,looseness=1.50] (leaf2) to (leaf3);
    \draw [in=90,out=30,looseness=1.50] (leaf3) to (leaf4);
    \draw [in=90,out=30,looseness=1.50] (leaf4) to (leaf5);
    \draw (leaf5) to (LR);
    \draw  (leaf1) -- (num1);
    \draw  (leaf2) -- (num2);
    \draw  (leaf3) -- (num3);
    \draw  (leaf4) -- (num4);
    \draw  (leaf5) -- (num5);
  }
\end{align}

In this context,
\(\mathrm{N}\) represents a node,
while \(\mathrm{L}\) represents a leaf.
To facilitate traversal, we introduce a distinguished leaf,
which is visually marked as
\tikz[baseline=-0.5ex, scale=0.5]{
  \node [atom,inner sep=1pt,fill=black!60,text=white]
  (leaf1) at (0, 0)    {\(\mathrm{L}\)};
}.
This marked leaf serves as an indicator for traversal operations.

The program that applies a given function to the
value of a marked leaf of a leaf-linked tree
can be represented as follows:

\begin{align}\label{eq:lltree-traverse-eg-vis}
   & \eLLT \triangleq
  (\lambda\,\ctxu{f}.\:
  (\lambda\,\lltr{x}.\:
  \mathbf{case}\; \lltr{x}\; \mathbf{of}
  \notag
  \\
   &
  \begin{array}{@{}l@{}}
    \tikz[baseline=6ex, scale=1]{%
      \ctxyy{0}{0}
      \begin{scope}[shift={(-0.6, -0.5)}]
        \node [atom,fill=black!60,text=white] (leaf1) at (0, 0) {L};
        \node [port] at (+0.150, +0.395) {3};
        \node [port] at (+0.395, +0.150) {2};
        \node [port] at (-0.150, -0.395) {1};
        \node [port] at (-0.150, -0.700) {1};
        \node [link,text=gray] at (0.410, -0.5)   {$(W_1)$};
        \node [ctxn,below=0.8cm] (num1) at (leaf1)  {$z$};
        \draw (leaf1) -- (num1);
        \node [link,text=gray] at (0.435, 0.85)   {$(W_2)$};
      \end{scope}
      \begin{scope}[shift={(+0.6, -0.5)}]
        \node [atom] (leaf2) at (0, 0) {L};
        \node [port] at (+0.150, +0.395) {3};
        \node [port] at (+0.395, +0.150) {2};
        \node [port] at (-0.150, -0.395) {1};
        \node [port] at (-0.150, -0.700) {1};
        \node [link,text=gray] at (0.410, -0.5)   {$(W_3)$};
        \node [ctxn,below=0.8cm] (num2) at (leaf2)  {$w$};
        \draw (leaf2) -- (num2);
        \node [link,text=gray] at (-0.420, 0.50)   {$(W_4)$};
        \node [link,text=gray] at (1.010, 0.25)   {$(W_5)$};
      \end{scope}
      \coordinate (p1) at (-1.4, 0.2);
      \coordinate (p2) at (+1.4, 0.2);
      \coordinate (p3) at (-0.6, 0.9);
      \coordinate (p4) at (+0.6, 0.9);
      \draw [looseness=1.50,out=0,in=90]    (p1) to (leaf1);
      \draw [looseness=1.50,out=-90,in=90]  (p3) to (leaf1);
      \draw [looseness=1.50,out=-90 ,in=90] (p4) to (leaf2);
      \draw [looseness=1.50,out=0,in=90] (leaf1) to (leaf2);
      \draw [looseness=1.50,out=0,in=180] (leaf2) to (p2);
    }
    \hspace{-2mm}
    \to
    \begin{array}[t]{l}
      \mathbf{let}\;\ctxu{z'}\;\texttt{=}\;{\ctxu{f} \: \ctxu{z}}
      \\[2mm]
      \mathbf{in}
      \hspace{-3mm}
      \tikz[baseline=10ex, scale=1]{%
        \ctxyy{0}{0}
        \begin{scope}[shift={(-0.6, -0.5)}]
          \node [atom] (leaf1) at (0, 0) {L};
          \node [port] at (+0.150, +0.395) {3};
          \node [port] at (+0.395, +0.150) {2};
          \node [port] at (-0.150, -0.395) {1};
          \node [port] at (-0.150, -0.700) {1};
          \node [link,text=gray] at (0.410, -0.5)   {$(W_1)$};
          \node [ctxn,below=0.8cm] (num1) at (leaf1)  {$z'$};
          \draw (leaf1) -- (num1);
          \node [link,text=gray] at (0.435, 0.85)   {$(W_2)$};
        \end{scope}
        \begin{scope}[shift={(+0.6, -0.5)}]
          \node [atom,fill=black!60,text=white] (leaf2) at (0, 0) {L};
          \node [port] at (+0.150, +0.395) {3};
          \node [port] at (+0.395, +0.150) {2};
          \node [port] at (-0.150, -0.395) {1};
          \node [port] at (-0.150, -0.700) {1};
          \node [link,text=gray] at (0.410, -0.5)   {$(W_3)$};
          \node [ctxn,below=0.8cm] (num2) at (leaf2)  {$w$};
          \draw (leaf2) -- (num2);
          \node [link,text=gray] at (-0.420, 0.50)   {$(W_4)$};
          \node [link,text=gray] at (1.010, 0.25)   {$(W_5)$};
        \end{scope}
        \coordinate (p1) at (-1.4, 0.2);
        \coordinate (p2) at (+1.4, 0.2);
        \coordinate (p3) at (-0.6, 0.9);
        \coordinate (p4) at (+0.6, 0.9);
        \draw [looseness=1.50,out=0,in=90]    (p1) to (leaf1);
        \draw [looseness=1.50,out=-90,in=90]  (p3) to (leaf1);
        \draw [looseness=1.50,out=-90 ,in=90] (p4) to (leaf2);
        \draw [looseness=1.50,out=0,in=90] (leaf1) to (leaf2);
        \draw [looseness=1.50,out=0,in=180] (leaf2) to (p2);
      }
    \end{array}
    \\[-0mm]
    \:
    \texttt{|}\;\mathbf{otherwise}
    \to \mathbf{case}\; \lltr{x}\; \mathbf{of}\; \\[0.5mm]
    \tikz[baseline=5ex, scale=1]{%
      \ctxyl{0}{0}
      \begin{scope}[shift={(0.8, 0)}]
        \node [atom,fill=black!60,text=white] (leaf1) at (0, 0) {L};
        \node [port] at (+0.150, +0.395) {3};
        \node [port] at (+0.395, +0.000) {2};
        \node [port] at (-0.150, -0.395) {1};
        \node [port] at (-0.150, -0.700) {1};
        \node [link,text=gray] at (0.50, -0.5)    {$(W_1)$};
        \node [link,text=gray] at (0.40, 0.8)     {$(W_2)$};
        \node [ctxn,below=0.8cm] (num1) at (leaf1)  {$z$};
        \draw  (leaf1) -- (num1);
      \end{scope}
      \coordinate (p1) at (-0.70, 0.2);
      \coordinate (p2) at (0.3, 1.35);
      \draw [looseness=1.50,out=0,in=90] (p1) to (leaf1);
      \draw [looseness=1.50,out=-45,in=90] (p2) to (leaf1);
      \draw [looseness=1.50,out=30,in=180] (leaf1) to (LR);
    }
    \hspace{-2mm}
    \to
    \begin{array}[t]{l}
      \mathbf{let}\;\ctxu{z'}\;\texttt{=}\;{\ctxu{f} \: \ctxu{z}}
      \\[2mm]
      \mathbf{in}
      \hspace{-2mm}
      \tikz[baseline=10ex, scale=1]{%
        \ctxyl{0}{0}
        \begin{scope}[shift={(0.8, 0)}]
          \node [atom] (leaf1) at (0, 0) {L};
          \node [port] at (+0.150, +0.395) {3};
          \node [port] at (+0.395, +0.000) {2};
          \node [port] at (-0.150, -0.395) {1};
          \node [port] at (-0.150, -0.700) {1};
          \node [link,text=gray] at (0.50, -0.5)    {$(W_1)$};
          \node [link,text=gray] at (0.40, 0.8)     {$(W_2)$};
          \node [ctxn,below=0.8cm] (num1) at (leaf1)  {$z'$};
          \draw  (leaf1) -- (num1);
        \end{scope}
        \coordinate (p1) at (-0.70, 0.2);
        \coordinate (p2) at (0.3, 1.35);
        \draw [looseness=1.50,out=0,in=90] (p1) to (leaf1);
        \draw [looseness=1.50,out=-45,in=90] (p2) to (leaf1);
        \draw [looseness=1.50,out=30,in=180] (leaf1) to (LR);
      }
    \end{array}
    \\[-5mm]
    \: \texttt{|}\;\mathbf{otherwise}
    \to \lltr{x}                                 \\[2mm]
    ))
  \end{array}
\end{align}

Applying the function \(\eLLT\) to a function
that returns a successor of a given number,
we obtain a function that produces a tree
in which the marked leaf contains the successor value,
and the marker is moved one step forward.
This is illustrated
as follows using the leaf-linked tree
$t[L,R,X]$ defined in \eqref{eq:lltree-eg-vis}.

\begin{align}
   &
  \Biggl(\biggl(\eLLT
    \quad
    \tikz[baseline=-0.5ex, scale=0.5]{%
      \node [atom,font=\footnotesize] (node1) at (0, 0) {$+1$};
      \draw (node1) --++(0, 1.2);
      \node [link] at (0,     1.5) {$X$};
      \node [port] at (-0.25,  0.9) {1};
    }\biggr)
  \;\;
  \lltr{t}
  \Biggr)
  \notag
  \\[-5mm]
   & \reducestwo
  \qquad
  \tikz[baseline=-0.5ex, scale=1]{%
    \begin{scope}[shift={(0, 2)}]
      \node [atom] (node1) at (0,0) {N};
      \node [port] at (+0.100, +0.395) {3};
      \node [port] at (-0.395, -0.100) {1};
      \node [port] at (+0.395, -0.100) {2};
    \end{scope}
    \begin{scope}[shift={(-1.4, 1.2)}]
      \node [atom] (node2) at (0,0) {N};
      \node [port] at (+0.100, +0.395) {3};
      \node [port] at (-0.395, -0.100) {1};
      \node [port] at (+0.395, -0.100) {2};
      \node [link,text=gray] at (0.500, 0.800) {$(W_1)$};
    \end{scope}
    \begin{scope}[shift={(1.4, 1.2)}]
      \node [atom] (node3) at (0,0) {N};
      \node [port] at (-0.100, +0.395) {3};
      \node [port] at (-0.395, -0.100) {1};
      \node [port] at (+0.395, -0.100) {2};
      \node [link,text=gray] at (-0.500, 0.800) {$(W_2)$};
    \end{scope}
    \begin{scope}[shift={(-0.8, 0)}]
      \node [atom] (node4) at (0,0) {N};
      \node [port] at (+0.100, +0.395) {3};
      \node [port] at (-0.395, -0.100) {1};
      \node [port] at (+0.395, -0.100) {2};
      \node [link,text=gray] at (0.100, 0.700)   {$(W_3)$};
    \end{scope}
    %
    \begin{scope}[shift={(-2.6, 0)   }]
      \node [atom] (leaf1) at (0, 0) {L};
      \node [port] at (+0.100, +0.395) {3};
      \node [port] at (+0.395, +0.100) {2};
      \node [port] at (-0.100, -0.395) {1};
      \node [port] at (-0.100, -0.700) {1};
      \node [link,text=gray] at (-0.510, -0.5)   {$(W_8)$};
    \end{scope}
    \begin{scope}[shift={(-1.6, -1.3)}]
      \node [atom,fill=black!60,text=white] (leaf2) at (0, 0) {L};
      \node [port] at (+0.100, +0.395) {3};
      \node [port] at (+0.395, +0.100) {2};
      \node [port] at (-0.100, -0.395) {1};
      \node [port] at (-0.100, -0.700) {1};
      \node [link,text=gray] at (-0.45, 0.9)   {$(W_4)$};
      \node [link,text=gray] at (0.410, -0.5) {$(W_9)$};
    \end{scope}
    \begin{scope}[shift={(-0.395, -1.3)}]
      \node [atom] (leaf3) at (0, 0) {L};
      \node [port] at (+0.100, +0.395) {3};
      \node [port] at (+0.395, +0.100) {2};
      \node [port] at (-0.100, -0.395) {1};
      \node [port] at (-0.100, -0.700) {1};
      \node [link,text=gray] at (0.40, 0.820) {$(W_5)$};
      \node [link,text=gray] at (0.495, -0.5)  {$(W_{10})$};
    \end{scope}
    \begin{scope}[shift={(1.1, -0.7)}]
      \node [atom] (leaf4) at (0,0) {L};
      \node [port] at (+0.100, +0.395) {3};
      \node [port] at (+0.395, +0.100) {2};
      \node [port] at (-0.100, -0.395) {1};
      \node [port] at (-0.100, -0.700) {1};
      \node [link,text=gray] at (0.300, 1)    {$(W_6)$};
      \node [link,text=gray] at (0.495, -0.5) {$(W_{11})$};
    \end{scope}
    \begin{scope}[shift={(2.6, 0) }]
      \node [atom] (leaf5) at (0, 0) {L};
      \node [port] at (+0.100, +0.395) {3};
      \node [port] at (+0.395, -0.100) {2};
      \node [port] at (-0.100, -0.395) {1};
      \node [port] at (-0.100, -0.700) {1};
      \node [link,text=gray] at (-0.100, 0.9)    {$(W_7)$};
      \node [link,text=gray] at (0.495, -0.5) {$(W_{12})$};
    \end{scope}
    \node [atom,below=0.8cm] (num1) at (leaf1) {2};
    \node [atom,below=0.8cm] (num2) at (leaf2) {2};
    \node [atom,below=0.8cm] (num3) at (leaf3) {3};
    \node [atom,below=0.8cm] (num4) at (leaf4) {4};
    \node [atom,below=0.8cm] (num5) at (leaf5) {5};
    \node [link] (LL) at (-4, 0)   {$L$};
    \node [link] (LX) at (+0, 3)   {$X$};
    \node [link] (LR) at (+4, 0)   {$R$};
    \draw (node1) -- (LX);
    \draw (node1) -- (node2);
    \draw (node1) -- (node3);
    \draw (node2) -- (node4);
    \draw [looseness=1.50,in=90,out=-130] (node2) to (leaf1);
    \draw [looseness=1.50,in=90,out=-130] (node4) to (leaf2);
    \draw [looseness=1.50,in=90,out=-50 ] (node4) to (leaf3);
    \draw [looseness=1.50,in=90,out=-130] (node3) to (leaf4);
    \draw [looseness=1.50,in=90,out=-50 ] (node3) to (leaf5);
    \draw [in=90,out=30,looseness=1.50] (LL) to (leaf1.north);
    \draw [in=90,out=30,looseness=1.50] (leaf1) to (leaf2);
    \draw [in=90,out=30,looseness=1.50] (leaf2) to (leaf3);
    \draw [in=90,out=30,looseness=1.50] (leaf3) to (leaf4);
    \draw [in=90,out=30,looseness=1.50] (leaf4) to (leaf5);
    \draw (leaf5) to (LR);
    \draw (leaf1) -- (num1);
    \draw (leaf2) -- (num2);
    \draw (leaf3) -- (num3);
    \draw (leaf4) -- (num4);
    \draw (leaf5) -- (num5);
  }\notag
\end{align}

Using a fixed-point operator (i.e., \texttt{let rec}),
we can define the \texttt{map} function, which applies a given function
to every leaf in a leaf-linked tree.
Since \texttt{let rec} has already been implemented \cite{sano-icgt2023}
and will be introduced in the next chapter (\cref{ch:impl}),
we focus here on the more subtle issues involved in handling graphs in \lgt{}
and leave the formal definition of \texttt{let rec} to
Appendix~\ref{chapter:letrec}.

\chapter{Implementing the Untyped \lgt Language}\label{ch:impl}

The semantics of the untyped \lgt language differs substantially
from that of the ordinary $\lambda$-calculus; therefore,
it is non-trivial to realise an executable implementation.
To address this issue, we develop a compact reference interpreter for untyped \lgt\footnote{%
  The interpreter is available at
  \url{https://github.com/sano-jin/lambda-gt-alpha/tree/icgt2023}.
  The visualiser is available at
  \url{https://sano-jin.github.io/lambda-gt-playground}.
}.

Although this dissertation does not prove the correctness of the execution algorithm
with respect to the reduction semantics,
the implementation is kept small and explicit so that this correspondence can be studied in future work.
Our implementation consists of
about 500 lines of OCaml~\cite{ocaml} code as shown in \cref{table:loc}.
This is about half the size
of the reference interpreter of
the graph transformation-based language GP 2~\cite{GP2},
which is about 1,000 lines of Haskell code~\cite{gp2-ref-impl}.

\begin{table}[tb]
  \caption{LOC of the interpreter.}\label{table:loc}
  \centering
  \small
  \begin{tabular}{lr@{\qquad}lr@{\qquad}lr}
    \toprule
    parser/parser.mly    & 84 & eval/preprocess.ml  & 34 & parser/parse.ml       & 10  \\
    parser/lexer.mll     & 57 & eval/pushout.ml     & 34 & eval/match.ml         & 9   \\
    eval/eval.ml         & 55 & util/list\_extra.ml & 29 & util/option\_extra.ml & 9   \\
    eval/syntax.ml       & 48 & eval/postprocess.ml & 24 & util/combinator.ml    & 7   \\
    eval/match\_atoms.ml & 38 & parser/syntax.ml    & 20 & bin/main.ml           & 3   \\
    eval/match\_ctxs.ml  & 35 & util/util.ml        & 19 & Total                 & 515 \\
    \bottomrule
  \end{tabular}
\end{table}

The remainder of this chapter is organised as follows.
\Cref{sec:impl:intro} discusses the implementation challenges of the \lgt language.
\Cref{sec:impl-programming} presents representative programming examples and
highlights the role of graph pattern matching, including cases that require
fusion.
\Cref{sec:impl:algorithm} then describes the design and implementation of the
reference interpreter.
Finally, \Cref{sec:impl-related} discusses the position of the implementation and
outlines future directions.

\section{Implementation Challenges}\label{sec:impl:intro}

The previous chapter defined \lgt as a call-by-value functional language in which
graphs are treated as immutable, first-class values.
This chapter focuses on the implementation issues that arise from that design.
The main difficulty is that graph values cannot be treated in the same way as
ordinary tree-shaped algebraic data.
Graphs may contain local links, free links, and fusions, and these structures must
be handled correctly when graphs are copied, substituted, matched, and recomposed.

The first issue is graph copying.
During evaluation, a graph value may be duplicated, for example when a graph
variable occurs more than once.
Such duplication cannot be implemented as a naive copy of a syntactic tree.
Local links must be renamed freshly so that the copied graph is independent of the
original graph, whereas free links must be preserved because they form the
external interface of the graph.
Thus, copying a graph requires a distinction between internal connectivity and
external connectivity.

The second issue is graph pattern matching.
In \lgt, patterns may contain graph variables that bind subgraphs of the host
graph.
A successful match therefore has to decompose the host graph into the explicit
part described by the pattern and the remaining subgraphs assigned to graph
variables.
Moreover, matching is performed modulo structural congruence inherited from
HyperLMNtal.
Consequently, a matcher cannot merely compare graph terms syntactically.

The third issue is the treatment of fusions.
Some matches require fusions that are not explicitly present in the syntactic
representation of the host graph.
In such cases, the implementation must determine whether the missing connections
required by the pattern can be supplied by appropriate fusions.
This operation, which we call fusion completion, is one of the main differences
between pattern matching in \lgt and ordinary pattern matching for algebraic data.

To investigate these issues,
we implemented a reference interpreter of approximately 500 lines of OCaml
code.
Graph values are represented and manipulated using persistent data structures.
The implementation includes preprocessing of graph values, atom matching with link
environments, fusion completion, matching of graph variables, graph substitution,
and evaluation.
When a graph variable occurs non-linearly, the interpreter copies the corresponding
graph value rather than sharing it by mutation.
The implementation is not intended to be efficient;
instead, it is intended to make the operational content of the untyped calculus
explicit.

\section{Execution Example}\label{sec:impl-programming}

\lgt naturally extends existing functional languages to support graphs as
first-class data.
For example, \cref{fig:dlist-reduction} shows the reduction sequence for
a program that concatenates two singleton difference lists
(a.k.a. list segments).
The \(\lambda\)-abstraction atom is a function that takes
two difference lists, both having \(X\) (head) and \(Y\) (tail) as free
links, and returns their concatenation also having \(X\) and \(Y\)
as its free links.

\begin{figure}[htbp]
  \small
  \begin{framed}
    \[\begin{array}{llll}
        (\lambda\: x[Y, X].\
        (\lambda\: y[Y, X].~\nu Z. (x[Z, X], y[Y, Z]) )(Z) )(Z)
        \quad
        \mathrm{Cons} (1, Y, X)
        \quad
        \mathrm{Cons} (2, Y, X)
        \\[5mm]
        \reduces
        \quad
        (\lambda\: y[Y, X].~\nu Z. (\mathrm{Cons} (1, Z, X), y[Y, Z]) )(Z)
        \quad
        \mathrm{Cons} (2, Y, X)
        \\[5mm]
        \reduces
        \quad
        \nu Z. (\mathrm{Cons} (1, Z, X), \mathrm{Cons} (2, Y, Z))
      \end{array}\]
  \end{framed}
  \caption{An example of a reduction: concatenation of difference lists.}\label{fig:dlist-reduction}
\end{figure}

In addition to graph construction, \lgt supports graph deconstruction by means of pattern matching.
\Cref{table:dlist-rotate1} shows the function that rotates a difference list
by taking the last node of the difference list and reconnecting it to the head.
If \(\mathrm{Cons} (1, \mathrm{Cons} (2, Y), X)\) is applied to the function,
it will return
\(\mathrm{Cons} (2, \mathrm{Cons} (1, Y), X)\).

\begin{figure}[htbp]
  \begin{framed}
    \[\begin{array}{llll}
        (\lambda\, x[Y, X].\
        \mathbf{case}\ x[Y, X]\ \mathbf{of} \\
        \qquad y[\mathrm{Cons} (z, Y), X] \rightarrow \mathrm{Cons} (z, y[Y], X)
        \quad|\quad \mathbf{otherwise} \rightarrow x[Y, X]
        ) (Z)
      \end{array}\]
  \end{framed}
  \caption{Rotate function on a difference list.}\label{table:dlist-rotate1}
\end{figure}

Without the term notation,
the graph template used in the matching
(on the second line of \cref{table:dlist-rotate1})
can be written as
\(\nu W. (y[W, X],  \mathrm{Cons} (Z, Y, W), z[Z])\).
The matching against this template proceeds as follows:
\[\begin{array}{cl}
     & \nu W Z. (\mathrm{Cons} (1, W, X), \mathrm{Cons} (Z, Y, W), 2(Z))
    \\
    \equiv
     & \nu W Z. (y[W, X],  \mathrm{Cons} (Z, Y, W), z[Z])
    \left[
      \mathrm{Cons} (1, W, X) / y[W, X],\:
      2(Z) / z[Z]
      \right]
  \end{array}\]

In order to implement the language
correctly,
we need to handle corner cases,
most notably
matching that needs to exploit fusion.
Consider the case where a singleton list 
\(\nu Z. (\mathrm{Cons} (Z, Y, X), 1(Z))\)
is given to the function.
We need a subgraph with free links
\(W\) and \(X\),
the free links of the graph variable \(y[W, X]\)
in the graph template,
which do not exist in the list.
Thus the matching would not proceed without
supplying subgraphs.

In this case,
we need to first supply a fusion using (E6) of \cref{table:hyperlmntal:equiv}
and match \(y[W, X]\) against the supplied fusion
as follows:
\[\begin{array}{cl}
     & \mathrm{Cons} (1, Y, X)                               \\
    \equiv
     & \nu W Z. (W \bowtie X, \mathrm{Cons} (Z, Y, W), 1(Z)) \\
    =
     & \nu W Z. (y[W, X],  \mathrm{Cons} (Z, Y, W), z[Z])
    \left[
      W \bowtie X / y[W, X],\:
      1(Z) / z[Z]
      \right]
  \end{array}\]
Therefore, the program will result in
\(\nu W. (\mathrm{Cons} (1, W, X), Y \bowtie W)\),
which is congruent to \(\mathrm{Cons} (1, Y, X)\).
Since pattern matching involving fusions
does not appear in ordinary ADT
matching in functional languages,
and
formalisation and implementation of
fusions is not common in ordinary graph transformation,
it has been a major challenge to deal with fusions properly.

\section{Algorithm}\label{sec:impl:algorithm}


In our implementation,
values, i.e., graphs, are represented with
lists of atoms without link creations inside;
i.e., in prenex normal form.
We call them \emph{host graphs}.
Links are classified into free links with \texttt{string} names and
local links \(\alpha\)-converted to fresh \texttt{int}eger ids.
From now on,
\(L_i\) denotes a local link with the unique id \(i\) and
\(F_X\) denotes a free link with the name \(X\).
Fusions with local link\@(s) are absorbed beforehand.
The interpreter transforms a graph template
to a pair of a list of the atoms and a list of the graph variables.
For example, the graph template
\(y[\mathrm{Cons} (z, Y), X]\),
the pattern in the \cref{table:dlist-rotate1},
and the host graph \(\mathrm{Cons} (1, Y, X)\) are transformed into
the following,
respectively:
\begin{align}
  \langle
  \left[ \mathrm{Cons} (L_{1001}, F_Y, L_{1000}) \right],
  \left[ y \left[ L_{1000}, F_X \right], z \left[ L_{1001} \right] \right]
  \rangle
  \label{eq:eq1}
  \\
  [ \mathrm{Cons} (L_{0}, F_Y, F_X), 1 (L_{0}) ]
  \label{eq:eq2}
\end{align}


The interpreter first tries to match all the atoms
in the template to the host graph
and then match the graph variables in the remaining host graph.
The interpreter backtracks if the consequent matching fails.
As a reference interpreter, optimisation of matching
strategies is left to future implementation.


To match an atom, we need to check that they have the same name
and the correspondence of links.
Since local link names are \(\alpha\)-convertible,
it is not sufficient to check that they have the same link names.
We maintain a \emph{link environment} \(\sigma\),
a mapping from the local link names in a template to the link names in a host graph.
\Cref{table:check-link-name} shows
the matching of a link name \(X\) in a template with a link name \(Y\) in a host graph
using a link environment \(\sigma\).
Free links in the template match links with the same names in the host graph (line 3).
On the other hand, the matching of local links in the template (line 5--7) is more flexible
since we can \(\alpha\)-convert them.

\begin{figure}[htbp]
  \begin{framed}
    \begin{algorithm}[H]
      \DontPrintSemicolon{}
      \Let{%
        \(\mathit{check\_link}\)
        \(\sigma\)
        \(X\)
        \(Y\)
      }{%
        \Switch{\((X, Y)\)}{%
          \lCase{\((F_X, F_Y)\)}{\leIf*{\(X = Y\)}{\(\mathrm{Some}\ \sigma\)}{\(\mathrm{None}\)}}%
          \lCase{\((F_X, L_i)\)}{\(\mathrm{None}\)}
          \Case{\((L_i, Y)\)}{%
            \leIf{\(L_i \mapsto Z \in \sigma\)}{%
              \leIf{\(Y = Z\)}{\(\mathrm{Some}\ \sigma\)}
              {\(\mathrm{None}\)}
            }{\(\mathrm{Some}\ (\sigma \cup \{L_i \mapsto Y\})\)}
          }
        }
      }
    \end{algorithm}
  \end{framed}
  \caption{Link name matching.}\label{table:check-link-name}
\end{figure}

For example,
the atom \(\mathrm{Cons} (L_{1001}, F_Y, L_{1000})\)
in the template~(\ref{eq:eq1}) matches the atom
\(\mathrm{Cons} (L_{0}, F_Y, F_X)\)
in the host graph~(\ref{eq:eq1})
with the environment \(\sigma_1 \triangleq
\{L_{1000} \mapsto F_X, L_{1001} \mapsto L_{0} \}\),
and the remaining host graph will be
\(G_1 \triangleq [ 1 (L_{0}) ]\).


After all the atoms in the template have matched,
we substitute link names in the host graph
with inverse of the obtained link environment.
For example,
the remaining host graph \(G_1\)
becomes
\(G_2 \triangleq [ 1 (L_{1001}) ]\).

If the mapping is not injective, we supply fusions to the host graph.
For example, if we obtain the link environment
\(\{ L_{1000} \mapsto L_{0}, L_{1001} \mapsto L_{0} \}\),
we should supply
\(L_{1000} \bowtie L_{1001}\).
If a local link in the template is mapped to a free link in the host graph,
we also need to supply a fusion
since a local link does not match a free link without such a treatment.
For example,
in the link environment \(\sigma_1\), we have
\( L_{1000} \mapsto F_X \).
Thus, we should supply
\(L_{1000} \bowtie F_X\)
and obtain
\(
G_3 \triangleq [ 1 (L_{1001}), L_{1000} \bowtie F_X ]
\)
as the remaining subgraph.


To match graph variables,
the interpreter collects connected graphs (atoms) that have free links of the graph variables.
For example,
\(y [ L_{1000}, F_X ]\)
and
\(z [ L_{1001} ]\),
the graph variables in (\Ref{eq:eq1}),
can match host graphs
\(L_{1000} \bowtie F_X\)
and
\(1 (L_{1001})\),
the subgraphs in \(G_3\),
respectively.
After the matching, we obtain the graph substitution
\begin{equation}\label{eq:graph-subst1}
  \begin{array}{l}
    [
      L_{1000} \bowtie F_X / y \left[ L_{1000}, F_X \right],
      1 (L_{1001}) / z \left[ L_{1001} \right]
    ]\ .
  \end{array}
\end{equation}


Graph substitution can be done by adding the matched atom\@(s) \(G\)
to the host graph,
updating link names.
If we have matched \(G\) with \(x[\overrightarrow{X}]\)
and the template has \(x[\overrightarrow{Y}]\),
we update \(G\) with the substitution
\(\angled{\seq{Y}/\seq{X}}\).
We also reassign ids for the local links
because composing graphs may result in a conflict of ids.

With the obtained graph substitution~(\ref{eq:graph-subst1}),
we can instantiate the template on the right-hand side of \(\rightarrow\)
in \cref{table:dlist-rotate1},
which will result in
\([	L_{1000} \bowtie F_X, \mathrm{Cons} (L_{1001}, \allowbreak
    F_Y, L_{1000}), 1 (L_{1001}) ]\).
Our implementation absorbs fusions as much as possible after graph substitutions.
Thus, it will be normalised, and the ids are reassigned, to obtain
\([ \mathrm{Cons} (L_{0}, F_Y, F_X), 1 (L_{0}) ]\).


The rest of the evaluator is implemented just like those for functional languages.
It takes an environment, a mapping from graph variables to the matched subgraphs,
and evaluates an expression with the environment.

To use it as the backend of the visualiser with breakpoints,
we have slightly changed the evaluator
and implemented it in a continuation-passing style.
This allows us to stop the evaluation at each breakpoint
and proceed from the point after getting a signal from a user.

\section{Demonstrations}\label{sec:appendix}

We first describe the concrete syntax supported by our interpreter and then present several examples,
some of which are visualised using the visualiser implemented on top of the interpreter.

\Cref{table:lgt-concrete-syntax} gives the concrete syntax of the language we have implemented.
In addition to the syntax of \cref{table:lgt-syntax},
we implemented standard constructs including
\lstinline[style=lgtinline]|let| and \lstinline[style=lgtinline]|let rec|
expressions and arithmetic operations.
Comments start at \lstinline[style=lgtinline]|
As usual,
\lstinline[style=lgtinline]|let|
defines a variable that has a value on the right-hand side of
\lstinline[style=lgtinline]|=|, if \(n = 0\).
If \(n \geq 1\), this defines a curried \(n\)-ary function.
Recursive \(n\)-ary functions can be defined with \lstinline[style=lgtinline]|let rec|s.
We designed a syntax that is easy to parse and
did not yet implement the term-notation abbreviation.
We do not intend to provide a full-fledged language
but to study the feasibility of the language
with a concise implementation.

\newcommand{\pc}{\colortt{,}}
\newcommand{\lmem}{\colortt{\char '173}}
\newcommand{\rmem}{\colortt{\char '175}}
\newcommand{\lpar}{\colortt{(}}
\newcommand{\rpar}{\colortt{)}}
\newcommand{\lbra}{\colortt{[}}
\newcommand{\rbra}{\colortt{]}}
\newcommand{\lang}{\colortt{<}}
\newcommand{\rang}{\colortt{>}}
\newcommand{\dottt}{\colortt{.}}
\newcommand{\nutt}{\colortt{nu}}
\newcommand{\mem}[1]{\lmem{#1}\rmem}

\newcommand{\caseoftt}[4]{%
  \colortt{case}\ #1\ \colortt{of}\ #2 \>\colortt{->}\> #3\ \colortt{|}\ \colortt{otherwise} \>\colortt{->}\> #4%
}%

\newcommand{\letintt}[3]{%
  \colortt{let}\ #1\ \colortt{=}\ #2\ \colortt{in}\ #3%
}%
\newcommand{\letrecintt}[3]{%
  \colortt{let\ rec}\ #1\ \colortt{=}\ #2\ \colortt{in}\ #3%
}%

\begin{figure}[htbp]
  \begin{framed}
    \centering
    \begin{tabular}{rclll}
      \(e\)           & \(::=\) & \(\lmem T \rmem
      \mathrel{|} \caseoftt{e}{\lmem T\rmem}{e}{e}
      \mathrel{|} e\; e
      \mathrel{|} e \mathrel{\mathit{op}} e
      \)
      \\
                      & \(|\)   & \(
      \letintt{x \lbra\seq{X}\rbra\
        x_1 \lbra\seq{X_1}\rbra  \dots\ x_n \lbra\seq{X_n}\rbra }{e}{e}
      \)              \hfill \((n \geq 0)\)
      \\
                      & \(|\)   & \(
      \letrecintt{x \lbra\seq{X}\rbra\
        x_1 \lbra\seq{X_1}\rbra  \dots\ x_n \lbra\seq{X_n}\rbra }{e}{e}
      \)              \hfill \((n \geq 1)\)
      \\[0.5mm]
      \(\mathit{op}\) & \(::=\) & \(\colortt{+}
      \mathrel{|} \colortt{*}
      \mathrel{|} \colortt{-}
      \mathrel{|} \colortt{<}
      \mathrel{|} \colortt{=}
      \)
      \\[0.5mm]
      \(T\)           & \(::=\) & \(C\lpar{}\seq{X}\rpar{}
      \mathrel{|} \lpar\colortt{\char`\\} x \lbra\seq{X}\rbra\dottt e \rpar \lpar\seq{Y}\rpar
      \mathrel{|} X \>\colortt{><}\> Y
      \mathrel{|} T\pc T
      \mathrel{|} \nutt \>X \dottt T
      \mathrel{|} x \lbra \seq{X} \rbra
      \)
      \\[0.5mm]
      \(X\)           & \(::=\) & {\color{codecolor}\verb|_[A-Z][a-zA-Z0-9_']*|}
      \\
      \(C\)           & \(::=\) & {\color{codecolor}\verb|[A-Z][a-zA-Z0-9_']*|}
      \(|\) {\color{codecolor}\verb|[1-9][0-9]*|}
      \(|\) {\color{codecolor}\verb|0|}
      \\
      \(x\)           & \(::=\) & {\color{codecolor}\verb|[a-z][a-zA-Z0-9_']*|}
    \end{tabular}
  \end{framed}
  \caption{Concrete syntax of \lgt.}\label{table:lgt-concrete-syntax}
\end{figure}

\begin{figure}[htbp]
\begin{lstlisting}[language=lgt]
%  Append two difference lists.

let dlist1[_Y, _X] = {nu _Z1 _Z2. (
    nu _X1. (Cons (_X1, _Z1, _X),  1 (_X1)),
    nu _X1. (Cons (_X1, _Z2, _Z1), 2 (_X1)),
    nu _X1. (Cons (_X1, _Y, _Z2),  3 (_X1))
    )}
in
let dlist2[_Y, _X] = {nu _Z1. (
    nu _X1. (Cons (_X1, _Z1, _X), 4 (_X1)),
    nu _X1. (Cons (_X1, _Y, _Z1), 5 (_X1))
    )}
in
let append[_X] =
  {(\ x[_Y, _X]. {(\ y[_Y, _X]. 
    {nu _Z. (x[_Z, _X], y[_Y, _Z])})(_X)})(_X)}
in
  {append[_X]} {dlist1[_Y, _X]} {dlist2[_Y, _X]}

% {nu _L0 _L1 _L2 _L3 _L4 _L5 _L6 _L7 _L8. (
%   Cons (_L1, _L2, _X),  1 (_L1), 
%   Cons (_L3, _L4, _L2), 2 (_L3), 
%   Cons (_L5, _L0, _L4), 3 (_L5), 
%   Cons (_L6, _L7, _L0), 4 (_L6), 
%   Cons (_L8, _Y, _L7),  5 (_L8)
% )}
\end{lstlisting}
  \caption{Append function on a difference list (in the concrete syntax).}\label{fig:dlist-append}
\end{figure}

\begin{figure}[htbp]
\begin{lstlisting}[language=lgt]
%  Rotate a difference list (push an element to front from back).

let rotate[_X] = {(\dlist[_Y, _X]. 
  case {dlist[_Y, _X]} of
    {nu _W1 _W2. (dlist[_W2, _X], Cons(_W1, _Y, _W2), h[_W1])} 
    -> {nu _W1 _W2. (Cons(_W1, _W2, _X), h[_W1], dlist[_Y, _W2])} 
    | otherwise -> { dlist[_Y, _X] })(_X)}
in
let dlist[_Y, _X] = {nu _Z1 _Z2 _Z3 _Z4. (
    nu _X1. (Cons(_X1, _Z1, _X),  1(_X1)),
    nu _X1. (Cons(_X1, _Z2, _Z1), 2(_X1)),
    nu _X1. (Cons(_X1, _Z3, _Z2), 3(_X1)),
    nu _X1. (Cons(_X1, _Z4, _Z3), 4(_X1)),
    nu _X1. (Cons(_X1, _Y,  _Z4), 5(_X1))
    )}
in
  {rotate[_X]} {dlist[_Y, _X]})

% {nu _Z1 _Z2 _Z3 _Z4. (
%   nu _X1. (Cons(_X1, _Z1, _X),  5(_X1)),
%   nu _X1. (Cons(_X1, _Z2, _Z1), 1(_X1)),
%   nu _X1. (Cons(_X1, _Z3, _Z2), 2(_X1)),
%   nu _X1. (Cons(_X1, _Z4, _Z3), 3(_X1)),
%   nu _X1. (Cons(_X1, _Y,  _Z4), 4(_X1))
% )}
\end{lstlisting}
  \caption{Rotate function on a difference list (in the concrete syntax).}\label{fig:dlist-rotate}
\end{figure}

We now present some concrete programs that were used above
to illustrate the reduction examples in \cref{sec:impl-programming}.
\Cref{fig:dlist-append} shows the \lstinline[style=lgtinline]|append| function for difference lists,
and \cref{fig:dlist-rotate} shows the \lstinline[style=lgtinline]|rotate| function for difference lists.
Evaluating these programs yields the outputs shown in the commented lines.
These examples illustrate how \lgt{} supports both the composition and decomposition of graphs.

Beyond simple graph composition and decomposition,
some operations require specific nodes to be traced during matching.
For such operations, we can use some atoms as markers.
For example, a \lstinline[style=lgtinline]|map| function that applies a function to the elements stored
in the leaves of a leaf-linked tree can be written as shown in \cref{table:lltree-traversal}.

\begin{figure}[htbp]
\begin{lstlisting}[language=lgt]
% Map a function to the leaves of a leaf-linked tree.

let succ[_X] x[_X] = {x[_X]} + {1(_X)} in

let map[_X] f[_X] x[_L, _R, _X] =
  let rec helper[_X] x2[_L, _R, _X] =
    case {Log} {x2[_L, _R, _X]} of
      {nu _L2 _R2 _X2 _X3. (
        y [_L, _R, _X, _L2, _R2, _X2],
        M (_L2)), Leaf (_X3, _L2, _R2, _X2), z [_X3]}
      -> let z2[_X] = {f[_X]} {z[_X]} in
        {helper[_X]}
        {nu _L2 _R2 _X2 _X3 _X4. (
          y [_L, _R, _X, _L2, _R2, _X2],
          Leaf (_X3, _L2, _R2, _X2), z2 [_X3], M (_R2))}
    | otherwise -> case {x2[_L, _R, _X]} of
      { y[_L, _R, _X], M (_R) } -> { y[_L, _R, _X] }
    | otherwise -> {Error, x2[_L, _R, _X]}
  in {helper [_X]} {x[_L, _R, _X], M (_L)}
in

{map[_X]} {succ[_X]} ({Log} {
  nu _X1 _X2 _X10. (Node (_X1, _X2, _X),
    nu _X3 _X4 _X7. (Node (_X3, _X4, _X1),
      nu _X8. (Leaf (_X8 ,_L, _X7, _X3), 1 (_X8)),
      nu _X9. (Leaf (_X9, _X7, _X10, _X4), 2 (_X9))
   ),
   nu _X5 _X6 _X11. (Node (_X5, _X6, _X2),
      nu _X12. (Leaf (_X12 ,_X10, _X11, _X5), 3 (_X12)),
      nu _X13. (Leaf (_X13, _X11, _R, _X6), 4 (_X13))
))})
\end{lstlisting}
  \vspace{-3mm}
  \caption{Mapping a function to the leaves of a leaf-linked tree in \lgt.}\label{table:lltree-traversal}
\end{figure}

\begin{figure}[htbp]
  \begin{framed}
    \centering
    \begin{tikzpicture}[font=\small,scale=1.2]
  \node [] (In) at (0, 4) {$\mathit{In}$};

  \node [rounded corners, draw, fill=atomcolor] (node1)
  at (0, 3) {$\lambda n. (n, 1)$};

  \node [rounded corners, draw, fill=atomcolor] (node2)
  at (0, 2) {$\lambda (i, \_). (i = 0)$};

  \node [rounded corners, draw, fill=atomcolor] (node3)
  at (0, 1) {$\lambda (i, k). (i - 1, k \times i)$};

  \node [rounded corners, draw, fill=atomcolor] (node4)
  at (3, 2) {$\lambda (\_, k). k$};

  \node [] (Out) at (3, 1) {$\mathit{Out}$};

  \draw [->] (In)--(node1);
  \draw [->] (node1)--(node2);
  \draw [->] (node2) edge node[anchor=west, name=la] {False} (node3);
  \draw [->] (node2) edge node[anchor=south, name=la] {True} (node4);
  \draw [->, rounded corners] (node3)--++(0, -0.5)--++(-1.9, 0)--++(0, 1.5)--(node2);
  \draw [->] (node4)--(Out);
\end{tikzpicture}
  \end{framed}
  \caption{Dataflow graph (factorial of \(n\)).}\label{images:dataflow-graph}
\end{figure}
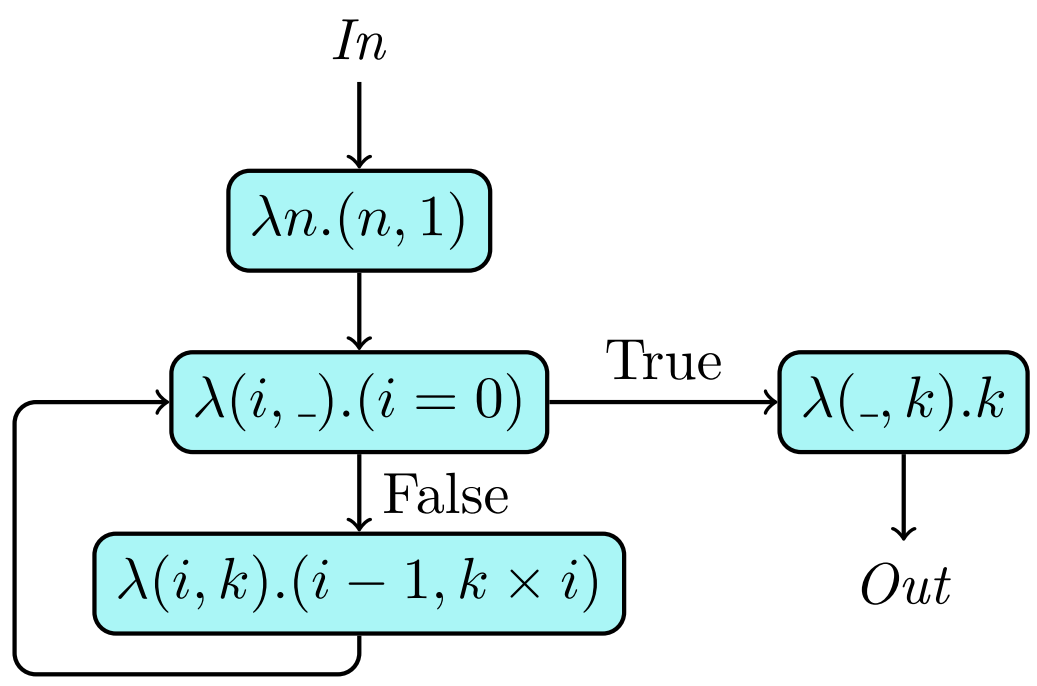

\begin{figure}[htbp]
  \input{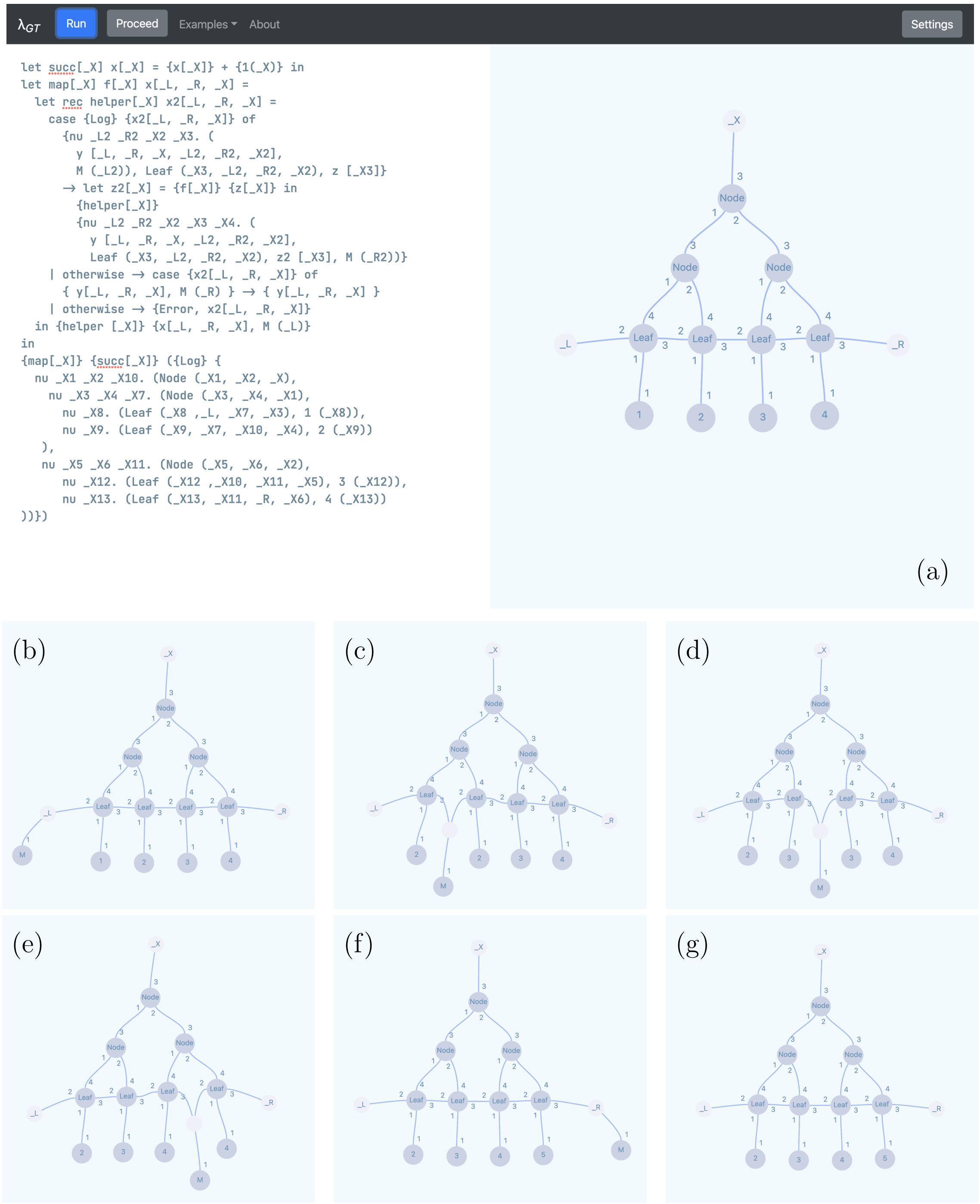}
  \caption{Demo of mapping a function to the leaves of a leaf-linked tree.}\label{demo:lltree}
\end{figure}

The visualiser can be used for tracing such programs.
We allow \lstinline[style=lgtinline]|Log| atoms, which are identity functions serving as breakpoints.
The tool visualises graphs applied to the atom at each breakpoint.
The demo in \cref{demo:lltree} shows visualising each step of
function application to the leaves of a leaf-linked tree.
To run a code,
users press the \texttt{Run} button and then
press \texttt{Proceed} to go to the next breakpoint.
\Cref{demo:lltree} (a) is the view obtained by pressing the \texttt{Run} button
after filling in the code.
\Cref{demo:lltree} (b)--(g) shows each graph at the breakpoints,
from which computation can be resumed with the \texttt{Proceed} button.

\lgt is capable of handling graphs that include functions.
\Cref{images:dataflow-graph} illustrates a dataflow graph that calculates the factorial of an input number.
The graph has input and output links from and to the outside and functions at the nodes.
There are two kinds of nodes;
nodes with an input and an output,
and a node with an input and two outputs (a branch).
This can be written as \cref{table:dataflow-dsl}, lines 2--28, in \lgt.

The evaluator of such a dataflow graph is shown in
\cref{table:dataflow-dsl}, lines 31--52.
We use a marker \lstinline[style=lgtinline]|M| to trace the value on the flow
(lines 53--54).
Lines 33--37 deal with a node with an input and an output
(\lstinline[style=lgtinline]|N2|).
They apply the embedded function to the value.
Lines 39--45 deal with a node with an input and two outputs
(\lstinline[style=lgtinline]|N3|).
They decide which way to go by applying the embedded function to the value.
Line 47 terminates the evaluator when the marker is pointing at the output link
\lstinline[style=lgtinline]|_Out|.
We have run the code on our interpreter
and checked that the interpreter returns
\lstinline[style=lgtinline]|{120(_Z)}| (\(= 5!\))
after printing 14 intermediate graphs.

We can even build a larger dataflow from components.
For example, we can compose the graph connecting the output and the input of them:\\
\lstinline[style=lgtinline]|{nu _X. (dataflow[_In,_X], dataflow[_X,_Out])}|.
Applying \lstinline[style=lgtinline]|{run[_Z]}| to \lstinline[style=lgtinline]|{3(_Z)}| and the graph returns
\lstinline[style=lgtinline]|{720(_Z)}| (\(= (3!)!\)).

\begin{figure}[tb!]
\begin{lstlisting}[language=lgt]
% let f1 n = (n,1)
let f1[_Z] n[_Z] = {nu _N1 _N2.(T2(_N1,_N2,_Z), n[_N1], 1(_N2))} in

% let pred1 (i,_) = i = 0
let pred1[_Z] v[_Z] = case {v[_Z]} of
  {nu _I _K.(T2(_I,_K,_Z), i[_I], k[_K])} -> {i[_Z]} = {0(_Z)}
  | otherwise -> {Error1} in

% let f2 (i,k) = (i - 1,k * i)
let f2[_Z] v[_Z] = case {v[_Z]} of
  {nu _I _K.(T2(_I,_K,_Z), i[_I], k[_K])} ->
    let i'[_Z] = {i[_Z]} - {1(_Z)} in
    let k'[_Z] = {k[_Z]} * {i[_Z]} in
    {nu _I _K.(T2(_I,_K,_Z), i'[_I], k'[_K])}
  | otherwise -> {Error2} in

% let f3 (_,k) = k
let f3[_Z] v[_Z] = case {v[_Z]} of
  {nu _I _K.(T2(_I,_K,_Z), i[_I], k[_K])} -> {k[_Z]}
  | otherwise -> {Error3} in

% The dataflow to caliculate the factorial of an input number.
let dataflow[_In,_Out] = {nu _X1 _X2 _X3 _F1 _F2 _P.(
  N2(_F1,_In,_X1), f1[_F1],
  N3(_P,_X1,_X3,_X2), pred1[_P],
  N2(_F2,_X2,_X1), f2[_F2],
  N2(_F3,_X3,_Out), f3[_F3]
)} in

% The evaluator
let rec proceed[_Z] g[_In,_Out] =
  case {Log} {g[_In,_Out]} of
    {nu _X _Y _F _V.(N2(_F,_X,_Y), f[_F],
       M(_V,_X), v[_V], rest[_X,_Y,_In,_Out])} ->
     let v'[_Z] = {f[_Z]} {v[_Z]} in
     {proceed[_Z]} {nu _X _Y _F _V.(N2(_F,_X,_Y), f[_F],
                      M(_V,_Y), v'[_V], rest[_X,_Y,_In,_Out])}
  | otherwise -> case {g[_In,_Out]} of
    {nu _X _Y1 _Y2 _P _V.(N3(_P,_X,_Y1,_Y2), pred[_P],
       M(_V,_X), v[_V], rest[_X,_Y1,_Y2,_In,_Out])} ->
    {proceed[_Z]} (case {pred[_Z]} {v[_Z]} of
       {True(_Z)} -> {nu _X _Y1 _Y2 _P _V.(N3(_P,_X,_Y1,_Y2), 
        pred[_P], M(_V,_Y1), v[_V], rest[_X,_Y1,_Y2,_In,_Out])}
     | otherwise  -> {nu _X _Y1 _Y2 _P _V.(N3(_P,_X,_Y1,_Y2), 
        pred[_P], M(_V,_Y2), v[_V], rest[_X,_Y1,_Y2,_In,_Out])})
  | otherwise -> case {g[_In,_Out]} of
    {nu _V.(M(_V,_Out), v[_V], rest[_In,_Out])} -> {v[_Z]}
  | otherwise -> {Error4} in

% Initialise with a marker
let run[_Z] v[_Z] g[_In,_Out] =
  {proceed[_Z]} {nu _V.(g[_In,_Out], M(_V,_In), v[_V])} in

% The main code
{run[_Z]} {5(_Z)} {dataflow[_In,_Out]} % 5!
\end{lstlisting}
  \caption{Embedding a dataflow language in \lgt.}\label{table:dataflow-dsl}
\end{figure}

\section{Future Directions}\label{sec:impl-related}

This chapter has presented a compact prototype interpreter for the untyped
\lgt{} language.
The implementation is not intended to serve as an efficient runtime system.
Rather, its purpose is to examine whether the core mechanisms of the untyped
calculus can be realised in an implementation.

A further direction for future work is to formally verify the prototype interpreter.
The interpreter has deliberately been kept small and explicit so that its
relationship to the reduction semantics can be studied in a precise manner.
Although this dissertation does not prove the correctness of the interpreter,
its implementation separates the main semantic operations, including graph
copying, freshening of local links, preservation of free links, graph substitution,
fusion completion, and graph pattern matching, into small functional
components.
This structure is intended to make it possible to relate each implementation
component to the corresponding rule or auxiliary operation in the formal
semantics.
A natural next step is therefore to mechanise this correspondence in a proof
assistant and establish the correctness of the interpreter with respect to the
untyped reduction semantics.

Another future work is to develop efficient graph pattern
matching algorithms for \lgt{}.
Such algorithms must handle not only explicit graph patterns but also multiple
graph variables and fusion completion.
A practical implementation will require strategies for controlling the search
space while preserving the declarative meaning of matching.
Clarifying the complexity of this matching problem and identifying useful
sub-classes of patterns are also important topics for future work.

A further direction is to develop an efficient runtime system while preserving
the immutable-value semantics of \lgt{}.
The prototype interpreter explicitly copies graph values when necessary, for
example when a graph variable occurs non-linearly.
This design is simple and faithful to the intended semantics, but it is not
efficient.
A practical implementation should avoid unnecessary copying and may internally
use destructive updates when doing so is safe and does not change the observable
functional semantics.

To support such optimisations, additional static analyses will be needed.
One possible direction is to analyse the direction, or polarity, of links
\cite{Ueda2014}.
Such information may be combined with ownership checking \cite{ownership2005} to
determine when a graph can be updated in place without introducing unwanted
aliasing or violating structural invariants.
Based on these analyses, a compiler could translate \lgt{} programs into
lower-level code that internally uses references or pointers, while preserving
the source-level semantics in which graphs are immutable values.

\chapter{Core Type System for \lgt}\label{ch:core-type}

The preceding chapters introduced the untyped core of \lgt{} and showed that its
main operational mechanisms can be implemented as an executable prototype.
The next step is to equip this language with a type system that can describe and
check structural properties of graph values.
A value in \lgt{} is a hypergraph with explicit free links, and these links form
the interface through which the value may be composed with other graphs.
Thus, a type must describe not only a data constructor at a root, but also the
connectivity of the graph and the free links through which it is exposed.
Designing such a type system is substantially different from designing a type
system for ordinary tree-shaped algebraic data.

Recall from \cref{tab:intro:requirements} that the type system of \lgt{} is
intended to satisfy the following requirements
as restated in \cref{tab:type:requirements}.

\begin{table}[htb]
  \centering
  \caption{Design requirements of the type system of \lgt.}
  \label{tab:type:requirements}
  \begin{tabular}{@{}p{0.12\linewidth}p{0.78\linewidth}@{}}
    \toprule
    \textbf{ID} & \textbf{Requirement}                                                                                                                                                                         \\
    \midrule
    T1          & Users should be able to define types that describe structural properties of graphs, and graph values should be checked against these types.                                                  \\
    T2          & Intermediate graph structures arising during pattern matching and transformation should be typable.                                                                                          \\
    T3          & The type system should provide static guarantees to the greatest extent possible, while clearly identifying any remaining dynamic checks and the conditions under which they can be reduced. \\
    \bottomrule
  \end{tabular}
\end{table}

The core type system introduced in this chapter satisfies the first requirement,
Requirement~T1.
It allows graphs to be specified by production rules
and assigns types to graph values, graph variables, and \(\lambda\)-abstraction atoms.
It does not yet fully satisfy the remaining two requirements,
Requirements~T2 and T3.
In particular, case expressions still rely on dynamic checks to ensure that
the subgraphs bound by graph pattern matching satisfy the type annotations in the pattern.
A more refined account of intermediate graph structures requires
the extension with linear implication types introduced in the next chapter.
The role of the present chapter is to establish the foundational graph type system
on which these later refinements are built.

The remainder of this chapter is organised as follows.
\Cref{sec:core-type:syn-rules} defines the syntax and the typing rules of the type system.
\Cref{sec:core-type:examples} presents examples of graph types definable by
production rules, including difference lists, doubly linked lists, and
leaf-linked trees.
\Cref{sec:type-soundness} proves type soundness by establishing
preservation and progress.
Finally,
\cref{sub:fgt-reduction} relates the core type system to reduction-based graph
typing frameworks such as Structured Gamma and LMNtal ShapeType.


\section{Formal Definition}\label{sec:core-type:syn-rules}


Our type system is formulated with four ingredients, (i) type names,
(ii) types (or type atoms), (iii) production rules, and (iv) typing
rules.  We describe them step by step.

Firstly, we introduce \emph{type names} \(\alpha, \beta, \dots\).
In concrete syntax, they are distinguished by using identifiers
starting with lowercase letters (as opposed to constructor names
starting with uppercase letters).

Secondly, we define the syntax of \emph{type atoms} (or simply \emph{types)}
in \lgt as follows:

\begin{mydef}[Types of \lgt]\label{OriginalType}
  \[
    \tau\>\> ::= \>\>(\tau \to \tau)(\Xs) \mid \alpha (\Xs)  \qedhere\]
\end{mydef}

Thirdly, we introduce \emph{production rules of hypergraphs}
based on hypergraph grammar, which acts as a
counterpart of type definitions in standard functional languages.

\begin{mydef}[Production Rule]\label{ProductionRule}
  $$
    \alpha (\Xs)
    \lto
    \nu \Zs.(C(\Ys), \seq{W \bowtie U}, \taus)
  $$
  For each production rule,
  the set of the free links on the left- and the right-hand sides
  must be equal, that is, we require that
  $$
    \{\Xs\} = \paren{\{\Ys\} \cup \fn(\seq{W \bowtie U}) \cup \fn(\taus)}
    \setminus \{\Zs\}.
  $$
  Without loss of generality,
  we also require that links in $\fn(\seq{W \bowtie U})$ are free
  links of the right-hand side, i.e.,
  $\fn(\seq{W \bowtie U}) \cap \{\Zs\} = \emptyset$.
  This can be achieved by absorbing the fusions of local links
  by preprocessing with structural congruence.
\end{mydef}

Compared to the original definition \cite{sano2023},
we have restricted the form of production rules to
what could be called a ``standard form''to ensure
that each production rule has exactly one constructor atom on the
right-hand side.
This does not reduce the expressive power of the types,
as rules with multiple constructor atoms can be represented by
introducing auxiliary type names.

The types \((\tau \to \tau)(\Xs)\) and \(\alpha (\Xs)\)
correspond to arrow types and type
names in existing functional languages.
However, since the values, which are graphs, have free links in
\lgt,
the types also have free links.

Finally, we define \emph{typing relation} and \emph{typing rules}.

\begin{mydef}[Typing Environment]
  A \emph{typing environment} \(\Gamma\) is
  a set of bindings of the form \(x[\Xs]: \tau\), where
  we require that (i) $\{\Xs\}= \fn(\tau)$
  to ensure the consistency of free links, and that (ii)
  the links in $\Xs$ are pairwise distinct.
  We also require that the graph variables names occurring in a typing
  environment are pairwise distinct.
\end{mydef}

\begin{mydef}[Typing Relation]\label{def:TypingRelation}
  A \emph{typing relation} is represented as
  \(
  \Gamma \vdash_P e: \tau,
  \)
  where \(\Gamma\) is the typing environment,
  \(P\) is a set of production rules,
  \(e\) is an expression, and \(\tau\) is its type,
  and
  $\fn(e) = \fn(\tau)$.
\end{mydef}

A fundamental characteristic of the type system of \lgt
is that each type \(\tau\) contains zero or more free links,
which directly correspond to the free links present in the hypergraph values.

We introduce explicit type annotations to \(\lambda\)-abstractions
\(
\lambda\, x[\Xs]: \tau\,.\,e
\).
In addition, we allow type annotations for graph variables occurring
within expressions.
In a case expression $(\caseof{e_1}{T}{e_2}{e_3})$,
each graph variable occurring in \(T\) must occur with
a type annotation.
In the representation of the production rules,
rules with identical left-hand sides can be enumerated using \(\mid\),
following the convention of formal grammars.

\begin{mydef}[Collecting Graph Variables with Type Annotations]
  \begin{align*}
    \CollectVars(\zero)        & \triangleq \emptyset                                \\[-2pt]
    \CollectVars(p(\Xs))       & \triangleq \emptyset                                \\[-2pt]
    \CollectVars(x[\Xs]: \tau) & \triangleq \{x[\Xs]: \tau\}                         \\[-2pt]
    \CollectVars((T_1, T_2))   & \triangleq \CollectVars(T_1) \cup \CollectVars(T_2) \\[-2pt]
    \CollectVars(\nu X.T)      & \triangleq \CollectVars(T) \qedhere
  \end{align*}
\end{mydef}



From this point on, we define typing rules for deriving typing relations.

\begin{mydef}[Typing Rules]
  The typing relation \(\Gamma \vdash_P e: \tau\)
  is deduced by the inference rules in \cref{table:typing-rules}.
\end{mydef}

\begin{figure}[htbp]
  \begin{framed}
    \begin{prooftree}
      \AXC{$x[\Xs]:\tau \in \Gamma$}
      \RightLabel{\TyVar{}}
      \UIC{$\Gamma \vdash_P x[\Zs] : \tau\angled{\Zs/\Xs}$}
    \end{prooftree}

    \begin{prooftree}
      \AXC{$\Gamma, x[\Xs]: \tau_1 \vdash_P e : \tau_2$}
      \RightLabel{\TyArrow{}}
      \UIC{$\Gamma \vdash_P (\lambda\, x[\Xs] : \tau_1.e)(\Zs) : (\tau_1 \to \tau_2) (\Zs)$}
    \end{prooftree}

    \begin{prooftree}
      \AXC{$\Gamma \vdash_P e_1 : (\tau_1 \rightarrow \tau_2) (\Zs)$}
      \AXC{$\Gamma \vdash_P e_2 : \tau_1$}
      \RightLabel{\TyApp{}}
      \BIC{$\Gamma \vdash_P (e_1\; e_2) : \tau_2$}
    \end{prooftree}

    \begin{prooftree}
      \AXC{$\Gamma \vdash_P T : \tau$}
      \AXC{$Y \notin \fn(T)$}
      \RightLabel{\TyAlpha{}}
      \BIC{$\Gamma \vdash_P T\angled{Y/X} : \tau \angled{Y/X}$}
    \end{prooftree}

    \begin{prooftree}
      \AXC{$\Gamma \vdash_P T : \tau$}
      \AXC{$T \equiv T'$}
      \RightLabel{\TyCong{}}
      \BIC{$\Gamma \vdash_P T' : \tau$}
    \end{prooftree}

    \begin{prooftree}
      \AXC{$\Gamma \vdash_P T_1 : \tau_1$}
      \AXC{$\dots$}
      \AXC{$\Gamma \vdash_P T_n : \tau_n$}
      \RightLabel{\TyProd{}}
      \TIC{$\Gamma
          \mathrel{\vdash_P} \nu
          \Zs.(C(\Ys), \seq{U \bowtie V}, T_1, \dots, T_n) : \alpha (\Xs)$}
    \end{prooftree}
    \vspace{-1mm}
    \raggedleft where
    $(\alpha (\Xs) \lto \nu
      \Zs.(C(\Ys), \seq{U \bowtie V}, \tau_1, \dots, \tau_n)) \in P$

    \begin{prooftree}
      \AXC{$\Gamma \vdash_P e_1 : \tau_1$}
      \AXC{$\Gamma, \CollectVars(T) \vdash_P e_2 : \tau_2$}
      \AXC{$\Gamma \vdash_P e_3 : \tau_2$}
      \RightLabel{\TyCase{}}
      \TIC{$\Gamma \vdash_P (\caseof{e_1}{T}{e_2}{e_3}) : \tau_2$}
    \end{prooftree}
  \end{framed}
  \caption{Typing rules of \lgt.}
  \label{table:typing-rules}
\end{figure}

\TyVar{}, \TyArrow{}, and \TyApp{},
are analogous to those of conventional functional type systems
except that the free links of variables must be properly handled.
In contrast,
\TyAlpha{}, \TyCong{}, and \TyProd{} are specific
to the hypergraph incorporation of \lgt.
\TyCong{} incorporates structural congruence
into the reduction relation.
\TyAlpha{} \(\alpha\)-converts the free link names of both the graph and
its type.
\TyProd{} incorporates production rules to the type system.
Each $\tau_i$ denotes a type atom occurring in the right-hand
side
of a production rule, and each \(T_i\) denotes an expression
of type $\tau_i$.
By simultaneously replacing each $\tau_i$
by the corresponding $T_i$,
the right-hand side of the production rule
is transformed into an expression of type $\alpha(\Xs)$.
\TyCase{} defines the typing rule for the case expression.



The type system proposed here is not fully static;
instead, it relied on dynamic checking for case expressions.
\begin{mydef}[Reduction with Dynamic Type Checking in Case Expressions]
  The reduction relation $\reducesp{P}$ denotes a reduction relation
  parameterised by the production rules $P$.
  The reduction rules for case expressions with dynamic checking are defined in
  \cref{fig:refined-rd-case}.
  The remaining reduction rule, \RdBeta{}, is defined as in
  \cref{table:lgt-reduction}, except that it uses $\reducesp{P}$.
\end{mydef}
In dynamic type checking for case expressions,
after pattern matching has been performed,
the runtime system checks whether there exists a graph substitution \( \thetas \)
that satisfies \( G \equiv T\thetas \),
and checks that all type annotations
for the graph variables within $T$ are satisfied.
Here, $P$ denotes the set of production rules
used to type the case expression.
If the matching with the type constraints succeeds,
evaluation proceeds along
\RdCaseMatchD{};
otherwise,
evaluation follows \RdCaseOtherD{}.

The fundamental reason why static type checking is difficult in pattern matching
stems from the structural complexity of the underlying computational model.
Unlike conventional functional programming languages that operate on tree-like data structures,
the language under consideration employs graph-based structures.
Graphs inherently exhibit a higher degree of complexity compared to trees,
introducing significant challenges in enforcing static type constraints.
Furthermore, as demonstrated earlier,
the expressiveness of the pattern matching mechanism in this language surpasses
that of conventional algebraic data type (ADT) pattern matching.
For instance, it permits direct matching on the tail of a list and the leaves of a leaf-linked tree,
which extends beyond the capabilities typically supported in ADT-based pattern matching.
This enhanced expressiveness introduces additional difficulties in establishing a
fully-static type system.

\begin{figure}[htbp]
  \begin{framed}
    \begin{prooftree}
      \AXC{$\begin{array}{c}
            G \equiv T\thetas
            \land
            \bigwedge_{x[\Xs]: \tau \in \CollectVars(T)}
            \paren{\emptyset \vdash_P x[\Xs]\thetas: \tau}
          \end{array}$}
      \RightLabel{\RdCaseMatchD{}}
      \UIC{$\mathmakebox[\dimexpr\width-6pt\relax][l]{%
            E[\caseof{G}{T}{e_1}{e_2}]\!\reducesp{P}\! E[e_1\thetas]}$}
    \end{prooftree}

    \medskip\smallskip

    \begin{prooftree}
      \AXC{$
          \not\exists \thetas.
          \paren{
            G \equiv T\thetas
            \land
            \bigwedge_{x[\Xs]: \tau \in \CollectVars(T)}
            \paren{\emptyset \vdash_P x[\Xs]\thetas: \tau}
          }$}
      \RightLabel{\RdCaseOtherD{}}
      \UIC{$\mathmakebox[\dimexpr\width-2pt\relax][l]{%
            E[\caseof{G}{T}{e_1}{e_2}] \reducesp{P} E[e_2]}$}
    \end{prooftree}
  \end{framed}
  \caption{Reduction relation for case expressions in dynamic type checking.}
  \label{fig:refined-rd-case}
\end{figure}

Again, the substitution $\theta$ in \RdCaseMatchD{} is not necessarily unique
even we have dynamic type checking.

In this work, we mainly focus on the theoretical design of the language.
Under the current reduction semantics, however, the language does not report
errors when type-inconsistent matching occurs.
For an actual implementation of the language, an alternative design choice
would be possible.
In particular, when pattern matching succeeds but type checking fails,
the evaluator could report a type error and terminate execution.
Even if this modification is adopted, the subsequent soundness discussion
can be carried out in essentially the same way, provided that
\texttt{Type-error} is allowed as a possible state.

\section{Examples}\label{sec:core-type:examples}

We now illustrate the expressive power of the core type system of \lgt
through a series of concrete examples.
We begin with data structures that are familiar from ordinary functional
programming, and then move on to complex data structures that cannot be
directly captured by standard algebraic data types.

Algebraic data types (ADTs) can be expressed in a straightforward way.
In this sense, our language and type system can be regarded as
a natural extension of functional languages and their type systems.

\begin{example}[Type of a natural number]
  The type of natural numbers connected to a free link \(X\)
  is denoted by \(\mathit{nat}(X)\),
  with the following production rules.
  \[\begin{array}{lcl}
      \mathit{nat}(X) & \longrightarrow & \mathrm{Zero} (X)               \\
      \mathit{nat}(X) & \longrightarrow & \mathrm{Succ} (\mathit{nat}, X)
    \end{array}\]
  (Recall that the RHS of the latter rule is a shorthand of
  \(\nu N.\mathrm{Succ} (N, X),\mathit{nat}\,(N) \) (Def.~\ref{def:abbreviation}).)
\end{example}

The interest of \lgt, however, is not limited to ordinary ADTs.
Since types are defined over graphs rather than over trees, the type system
can describe data structures with sharing, back links, and other non-tree
connections.
A doubly linked list is a simple example of such data structure.

\begin{example}[Type of a doubly linked list (Restatement of \cref{ex:prod-dbllist-01})]\label{ex:prod-dbllist}
  The production rules for a doubly linked list can be defined as follows.
  \[\begin{array}{lcl}
      \mathit{dbllist}(X,Y,Z) & \longrightarrow & \Nil(X, Z), Y \bowtie Z          \\
      \mathit{dbllist}(X,Y,Z) & \longrightarrow & \nu W_1 W_2.(\Cons(W_1,X,W_2,Z),
      \mathit{nat}(W_1),\mathit{dbllist}(Z,Y,W_2))                                 \\
    \end{array}\]
  Here, we omit the production rules for \(\mathit{nat}\).
\end{example}

The production rules for doubly linked lists are illustrated as follows.
\begin{align}\label{eq:prod-rules-eg-vis-dbllist}
  \begin{array}{ccc}
    \dblltype{X}{Y}{Z}
     &
    \longrightarrow
     &
    \tikz[baseline=1.5ex, scale=1]{%
      \mknil{2}{(0,0)}
      \node [link] (C1) at (-1, 0.8) {$Z$};
      \node [link] (LX) at (-1, 0.0) {$X$};
      \node [link] (C3) at (+1, 0.8) {$Y$};
      \mkconslinkleft{1}{2}
      \draw (C2) to (LX);
      \mkconslinkright{2}{3}
    }
    \mid
    \tikz[baseline=1.5ex, scale=1]{%
      \mkcons{2}{W_1}{}{(0,0)}
      \mkdbll{3}{(1.5,0)}{W_2}
      \node [link] (C1) at (-1.0, 0.8) {$Z$};
      \node [link] (LX) at (-1.0, 0.0) {$X$};
      \node [link] (C4) at (+2.5, 0.8) {$Y$};
      \mkconslinkleft{1}{2}
      \draw (C2) to (LX);
      \mkconslink{2}{3}
      \mkconslinkright{3}{4}
    }
  \end{array}
\end{align}

Using the typing rules \TyAlpha{}, \TyCong{}, and \TyProd{},
we can derive the following typing relation for an example graph.
\begin{align}
  \emptyset \vdash_{\Pdl}
  \tikz[baseline=-0.5ex, scale=1]{%
    \mkcons{1}{W_1}{}{(0,0)}{1}
    \mkcons{2}{W_3}{W_2}{(1.5,0)}{2}
    \mknil{3}{(3,0)}
    %
    \node [link] (C0) at (-1.0, 0.8) {$Z$};
    \node [link] (LX) at (-1.0, 0.0) {$X$};
    \node [link] (C4) at (+4.0, 0.8) {$Y$};
    \draw (C1) to (LX);
    \mkconslinkleft{0}{1}
    \mkconslink{1}{2}
    \mkconslink{2}{3}
    \mkconslinkright{3}{4}
  }
  :
  \dblltype{X}{Y}{Z}\notag
\end{align}

A leaf-linked tree is a graph with three free links, say \(X,L,R\).
It is a tree whose root is represented by \(X\), and whose leaves form
a difference list represented by \(L\) and \(R\).

\begin{example}[Type of a leaf-linked tree]\label{ex:prod-lltree}
  \[\begin{array}{lcl}
      \mathit{lltree}(L, R, X) & \longrightarrow & L \bowtie X, \mathrm{Leaf} (\mathit{nat}, R, X)                        \\
      \mathit{lltree}(L, R, X) & \longrightarrow & \nu Y. \mathrm{Node} (\mathit{lltree}(L, Y), \mathit{lltree}(Y, R), X) \\
    \end{array}\]
\end{example}

The production rules for leaf-linked trees are illustrated as follows.
\begin{align}\label{eq:prod-rules-eg-vis}
  \begin{array}{ccc}
    \lltrtype{X}{L}{R}
     &
    \longrightarrow
     &
    \mkleafatom{n}
    \mid
    \tikz[baseline=5.4ex, scale=1]{%
      \begin{scope}[shift={(0,1.3)}]
        \node [atom] (cons1) at (0, 0) {N};
        \node [link] (LX) at (0, 1) {$X$};
        \node [port] at (+0.1, +0.4) {3};
        \node [port] at (-0.2, -0.4) {1};
        \node [port] at ( 0.2, -0.4) {2};
      \end{scope}
      \begin{scope}[shift={(-0.75, 0)}]
        \node [lltr] (x1) at (0, 0) {$\mathit{lltr}$};
        \node [link,text=gray] at (-0.1, 1.0) {$(W_1)$};
        \node[port] at (+0.200, 0.550) {3};
        \node[port] at (-0.450, 0.150) {1};
        \node[port] at (+0.450, 0.150) {2};
      \end{scope}
      \begin{scope}[shift={(+0.75, 0)}]
        \node [lltr] (x2) at (0.0, 0) {$\mathit{lltr}$};
        \node [link,text=gray] at (0.1, 1.0) {$(W_2)$};
        \node[port] at (-0.200, 0.550) {3};
        \node[port] at (-0.450, 0.150) {1};
        \node[port] at (+0.450, 0.150) {2};
      \end{scope}
      \node [link] (LL) at (-1.75, -0.1) {$L$};
      \node [link] (LR) at (+1.75, -0.1) {$R$};
      \node [link,text=gray] at (0, -0.3) {$(Y)$};
      \draw (cons1)--(LX);
      \draw (cons1)--($(x1.north)-(0,0.1)$);
      \draw (cons1)--($(x2.north)-(0,0.1)$);
      \draw (x1)--(x2);
      \draw (x1)--(LL);
      \draw (x2)--(LR);
    }
  \end{array}
\end{align}\relax

As in the case of doubly linked lists, the typing of a concrete graph can be
derived using \TyAlpha{}, \TyCong{}, and \TyProd{}.
The following figure shows an example typing relation.
\begin{align}
  \emptyset \vdash_{\Pdl}
  \lltr{t}
  :
  \lltrtype{X}{L}{R}
\end{align}

A threaded tree is similar to a leaf-linked tree, but each non-terminal node
has access to the rightmost leaf of its left subtree and the leftmost leaf
of its right subtree.

\begin{example}[Type of a threaded tree]\label{ex:prod-thtree}
  \[\begin{array}{lcl}
      \mathit{thtree}(L, R, X) & \longrightarrow & L \bowtie X, \mathrm{Leaf} (\mathit{nat}, R, X) \\
      \mathit{thtree}(L, R, X) & \longrightarrow & \mathrm{Node}
      (\mathit{thtree}(L, X), \mathit{thtree}(X, R), X)                                            \\
    \end{array}\]
\end{example}

The examples above demonstrate that the type system can describe
complex data structures whose intended shapes cannot be represented by
ordinary algebraic data types alone.
We next focus on difference-list-like structures, which are particularly useful
for explaining how links can represent open-ended data structures.

A difference list can be understood as a list equipped with an
additional link to the last element.
This data structure has been popular since the early days of logic
programming, where links are represented as logical variables.
It allows two lists to be appended in constant time.
In functional programming, a difference list can be implemented
using a higher-order function that takes a subsequent list
and returns the entire list; however, we wish to represent such data
structures in a first-order setting.

\begin{example}[Type of a difference list]\label{ex:prod-dlist}
  The production rules for difference lists can be defined as follows.
  \[\begin{array}{lcl}
      \mathit{nodes}(Y, X) & \longrightarrow & X \bowtie Y                                \\
      \mathit{nodes}(Y, X) & \longrightarrow & \Cons (\mathit{nat}, \mathit{nodes}(Y), X) \\
    \end{array}\]
\end{example}

The production rules for difference lists are illustrated as follows.
\begin{align}\label{eq:prod-rules-eg-vis-difflist}
  \begin{array}{ccc}
    \tikz[font=\footnotesize, baseline=-0.5ex, scale=1]{%
      \node [tvar] (cons1) at (0, 0) {$\mathit{nodes}$};
      \node [link] (X) at (-1.3, 0) {$X$}; 
      \node [link] (Y) at ( 1.3, 0) {$Y$}; 
      \node [port] at (-0.75, +0.150) {2};
      \node [port] at (+0.75, +0.150) {1};
      \draw (cons1)--(X);
      \draw (cons1)--(Y);
    }
     &
    \longrightarrow
     &
    \tikz[baseline=-0.5ex, scale=1]{%
      \begin{scope}[shift={(0,0)}]
        \fill (0,0) circle (1.5pt);
      \end{scope}
      %
      \node [link] (LZ) at (-0.5, 0) {$X$};
      \node [link] (LY) at (+0.5, 0) {$Y$};
      \draw (LZ) -- (LY);
    }
    \mid
    \tikz[baseline=-0.4ex, scale=1]{%
      \begin{scope}[shift={(0,0)}]
        \node [atom] (C1)  at (0, 0) {C};
        \node [tvar] (N1)  at (0,-1) {$\mathit{n}$};
        \draw (C1) to (N1);
        \mkports{3}{}{2}{1}{1}
        \node [link,text=gray] at (0.50, -0.5) {$(W_1)$};
      \end{scope}
      \begin{scope}[shift={(1.5,0)}]
        \node [tvar] (C2) at (0, 0) {$\mathit{nodes}$};
        \node [port] at (-0.75, +0.150) {1};
        \node [port] at (+0.75, +0.150) {2};
        \node [link,text=gray] at (-1, 0.5)  {$(W_2)$};
      \end{scope}
      %
      \node [link] (LX) at (-1, 0) {$X$};
      \node [link] (LY) at (+3, 0) {$Y$};
      %
      \draw (LX) to (C1);
      \draw (C1) to (C2);
      \draw (C2) to (LY);
    }
  \end{array}
\end{align}

\begin{figure}[htbp]
  \hrulefill
  \vspace{-5mm}
  \begin{prooftree}
    \AXC{}
    \RightLabel{\TyVarA{}}
    \UIC{$\ml{\Gamma\{succ[Z_1]: (\nattonat) (Z_1)\} \vdash_P \\
          \quad succ[Z_1] : (\nattonat) (Z_1)}$}
    \RightLabel{\TyVarA{}}

    \AXC{}
    \RightLabel{\TyProdA{}}
    \UIC{\(\Gamma \vdash_{P\{P_1\}} X \bowtie Y : \mathit{nodes}(Y, X)\)}
    \RightLabel{\TyAlphaA{}}
    \UIC{\(\Gamma \vdash_P Z_2 \bowtie Y : \mathit{nodes}(Z_2, X)\)}

    \RightLabel{\TyProdA{}}
    \BIC{
      \tikz[remember picture,baseline=(upper-root1.base)]
      \node [font=\normalsize] (upper-root1)
      {$\nu Z_1 Z_2. (\Cons (Z_1, Z_2, X), succ [Z_1], Z_2 \bowtie Y)$};
    }
  \end{prooftree}

  \bigskip

  \begin{prooftree}
    \AXC{
      \tikz[remember picture,baseline=(lower-hole.base)]
      \node[draw, dashed, rectangle, inner sep=3pt, font=\normalsize] (lower-hole)
      {$\color{white}Cons$};
    }
    \AXC{$\nu Z_1 Z_2. (\Cons (Z_1, Z_2, X), succ [Z_1], Z_2 \bowtie Y) \equiv
        \Cons (succ, Y, X)$}
    \RightLabel{\TyCongA{}}
    \BIC{$\Gamma \vdash_P \Cons (succ, Y, X) : \mathit{nodes}(Y, X)$}
  \end{prooftree}

  \begin{tikzpicture}[remember picture,overlay]
    \draw[->,thick] (upper-root1.south) to[looseness=1,out=-90,in=90] (lower-hole.north);
  \end{tikzpicture}

  Here,
  \TyVar{},  \TyProd{},  \TyAlpha{},  and \TyCong{}  andare abbreviated as
  \TyVarA{}, \TyProdA{}, \TyAlphaA{}, and \TyCongA{} respectively.

  \hrulefill
  \caption{Type checking a difference list.}\label{fig:dlist-typing-example}
\end{figure}

\begin{example}[Typing a difference list with functions]\label{ex:typing-dlist}
  Since \lgt and its type system treat functions as first-class citizens,
  it is even possible to define a difference list whose elements are functions.
  \Cref{fig:dlist-typing-example} shows that the graph
  \(G = \Cons (succ, Y, X)\)
  has type
  \(\mathit{nodes}(Y, X)\)
  under the type environment
  \(\Gamma = succ [Z_1]: (\nattonat) (Z_1)\)
  and production rules \(P=\{P_1,P_2\}\) where
  \[ \begin{array}{lclcl}
      \mathit{nodes}(Y, X) & \longrightarrow & X \bowtie Y                             & \cdots & P_1 \\
      \mathit{nodes}(Y, X) & \longrightarrow & \Cons (\nattonat, \mathit{nodes}(Y), X) & \cdots & P_2 \\
    \end{array}\]

\end{example}

A doubly-linked difference list is a list with four free
links, two different links for each end.
Although the (hyper)links of \lgt and HyperLMNtal
are undirected, we are interested in using them to model
directed hyperlinks (roughly corresponding to pointers in
imperative languages) that are to be `followed' in one direction.
As with difference lists, the addition of elements to the tail of
the list can be done in constant time, as desired in
representing dequeues.
\begin{example}[Type of a doubly-linked difference list]\label{ex:prod-dbl}
  \[\begin{array}{lcl}
      \mathit{nodes}(F', B, B', F) & \longrightarrow & F \bowtie B,B' \bowtie F'                                       \\
      \mathit{nodes}(F', B, B', F) & \longrightarrow & \nu X. \Cons (\mathit{nat}, F', \mathit{nodes}(X, B, B'), X, F) \\
    \end{array}\]
\end{example}

By extending the type definition of difference lists,
we can define a type for unbounded-level skip lists.
This also implies that we can define types for skip lists
with a nil node at the end and/or with a fixed level.
\begin{example}[Type of difference skip lists]\label{ex:prod-skip}
  \[\begin{array}{lcl}
      \mathit{nodes}(Y, X)   & \longrightarrow & X \bowtie Y                                                           \\
      \mathit{nodes}(Y, X)   & \longrightarrow & \Cons (\mathit{nat}, \mathit{forks}\,(Y), X)                          \\
      \mathit{forks}\,(Y, X) & \longrightarrow & \mathrm{Next} (\mathit{nodes}(Y), X)                                  \\
      \mathit{forks}\,(Y, X) & \longrightarrow & \nu Z.\mathrm{Fork} (Z, \mathit{forks}\,(Z), X), \mathit{nodes}(Y, Z) \\
    \end{array}\]
\end{example}

We also show a visualisation of the production rules as follows.
\begin{align}\label{eq:prod-rules-eg-vis-diffskiplist}
  \begin{array}{ccl}
    \tikz[baseline=-0.5ex, scale=1]{%
      \node [tvar] (cons1) at (0, 0) {$\mathit{nodes}$};
      \node [link] (X) at (-1.3, 0) {$X$}; 
      \node [link] (Y) at ( 1.3, 0) {$Y$}; 
      \node [port] at (-0.75, +0.150) {2};
      \node [port] at (+0.75, +0.150) {1};
      \draw (cons1)--(X);
      \draw (cons1)--(Y);
    }
     &
    \longrightarrow
     &
    \tikz[baseline=-0.5ex, scale=1]{%
      \begin{scope}[shift={(0,0)}]
        \fill (0,0) circle (1.5pt);
      \end{scope}
      \node [link] (LZ) at (-0.5, 0) {$X$};
      \node [link] (LY) at (+0.5, 0) {$Y$};
      \draw (LZ) -- (LY);
    }
    \quad
    \mid
    \quad
    \tikz[baseline=-0.4ex, scale=1]{%
      \begin{scope}[shift={(0,0)}]
        \node [atom] (C1)  at (0, 0) {C};
        \node [tvar] (N1)  at (0,-1) {$\mathit{n}$};
        \draw (C1) to (N1);
        \mkports{3}{}{2}{1}{1}
        \coordinate (p1) at ($(C1)+(0.5,0)$) {};
        \coordinate (p2) at ($(p1)+(0,0.7)$);
        \coordinate (p3) at ($(p2)+(1,0)$);
      \end{scope}
      \begin{scope}[shift={(1.5,0)}]
        \node [tvar] (C2) at (0, 0) {$\mathit{nodes}$};
        \node [port] at (-0.150, 0.450) {1};
        \node [port] at (+0.750, 0.150) {2};
      \end{scope}
      %
      \node [link] (LX) at (-1.0, 0) {$X$};
      \node [link] (LY) at (+2.8, 0) {$Y$};
      %
      \draw (LX) to (C1);
      \draw (C2) to (LY);
      \draw (C1) to (p1);
      \draw (p1) to (p2);
      \draw (p2) to (p3);
      \draw (p3) to (C2);
    }
    \\
    \tikz[font=\footnotesize, baseline=-0.5ex, scale=1]{%
      \node [tvar] (cons1) at (0, 0) {$\mathit{forks}$};
      \node [link] (X) at (-1.2, 0) {$X$}; 
      \node [link] (Y) at (+1.2, 0) {$Y$}; 
      \node [port] at (-0.700, +0.150) {2};
      \node [port] at (+0.700, +0.150) {1};
      \draw (cons1)--(X);
      \draw (cons1)--(Y);
    }
     &
    \longrightarrow
     &
    \tikz[baseline=-0.4ex, scale=1]{%
      \begin{scope}[shift={(0,0)}]
        \node [atom] (N1)  at (0, 0)    {N};
        \node [port] at (-0.150, +0.350) {1};
        \node [port] at (+0.350, -0.150) {2};
      \end{scope}
      \node [link] (LX) at (0, 1) {$X$};
      \node [link] (LY) at (1, 0) {$Y$};
      \draw (LX) to (N1);
      \draw (N1) to (LY);
    }
    \quad
    \mid
    \quad
    \tikz[baseline=-0.4ex, scale=1]{%
      \begin{scope}[shift={(0,0)}]
        \node [tvar] (C1) at (0, 0) {$\mathit{forks}$};
        \node [port] at (-0.150, +0.450) {2};
        \node [port] at (+0.700, +0.150) {1};
      \end{scope}
      \begin{scope}[shift={(0,1)}]
        \node [atom] (N1)  at (0, 0) {F};
        \draw (C1) to (N1);
        \mkports{}{3}{2}{1}{}
        \coordinate (p1) at (1,0) {};
        \coordinate (P1) at (1,-1);
      \end{scope}
      \fill (P1) circle (1.5pt);
      \begin{scope}[shift={(2,0)}]
        \node [tvar] (C2) at (0, 0) {$\mathit{nodes}$};
        \node [port] at (-0.75, 0.150) {1};
        \node [port] at (+0.75, 0.150) {2};
        \node [link,text=gray] at (-0.6, 1)  {$(Z)$};
      \end{scope}
      %
      \node [link] (LX) at (0.0, 2) {$X$};
      \node [link] (LY) at (3.3, 0) {$Y$};
      %
      \draw (LX) to (N1);
      \draw (C1) to (C2);
      \draw (C2) to (LY);
      \draw (N1) to (p1);
      \draw (p1) to (P1);
    }
  \end{array}
\end{align}

The visualised example of a difference skip list is as follows.
\begin{align}\label{eq:prod-rules-eg-vis-diffskiplist-eg}
  \tikz[baseline=-0.4ex, scale=1]{%
    \begin{scope}[shift={(0,0)}]
      \node [atom] (C1) at (0, 0) {C};
      \node [atom] (n1) at (0,-1) {1};
      \draw (C1) to (n1);
      \mkports{2}{}{3}{1}{1}
    \end{scope} %
    \coordinate (p11) at ($(C1)+(0.5,0)$);
    \coordinate (p12) at ($(p11)+(0, 3)$);
    \coordinate (p13) at ($(p12)+(0.5,0)$);
    \draw (C1) to (p11);
    \draw (p11) to (p12);
    \draw (p12) to (p13);
    \node [link] (LX) at (-1, 0) {$X$};
    \draw (LX) to (C1);
    \begin{scope}[shift={(1,0)}]
      \node [atom] (N1) at (0, 0) {N};
      \mkports{}{1}{2}{}{}
    \end{scope}
    \begin{scope}[shift={(1,1)}]
      \node [atom] (F11) at (0, 0) {F};
      \mkports{}{3}{1}{2}{}
      \coordinate (pF11) at (2.5,0);
      \coordinate (PF11) at (2.5,-1);
      \draw (F11) to (pF11);
      \draw (pF11) to (PF11);
      \fill (PF11) circle (1.5pt);
    \end{scope}
    \begin{scope}[shift={(1,2)}]
      \node [atom] (F12) at (0, 0) {F};
      \mkports{}{3}{1}{2}{}
      \coordinate (pF12) at (4.5,0);
      \coordinate (PF12) at (4.5,-1);
      \draw (F12) to (pF12);
      \draw (pF12) to (PF12);
      \fill (PF12) circle (1.5pt);
    \end{scope}
    \draw (p13) to (F12);
    \draw (F12) to (F11);
    \draw (F11) to (N1);
    \begin{scope}[shift={(2,0)}]
      \node [atom] (C2) at (0, 0) {C};
      \node [atom] (n2) at (0,-1) {2};
      \draw (C2) to (n2);
      \mkports{2}{}{3}{1}{1}
    \end{scope}
    \draw (N1) to (C2);
    \coordinate (p21) at ($(C2)+(0.5,0)$);
    \coordinate (p22) at ($(p21)+(0, 0.5)$);
    \coordinate (p23) at ($(p22)+(0.5,0)$);
    \draw (C2) to (p21);
    \draw (p21) to (p22);
    \draw (p22) to (p23);
    \begin{scope}[shift={(3,0)}]
      \node [atom] (N2) at (0, 0) {N};
      \mkports{}{1}{2}{}{}
    \end{scope}
    \begin{scope}[shift={(4,0)}]
      \node [atom] (C3) at (0, 0) {C};
      \node [atom] (n3) at (0,-1) {3};
      \draw (C3) to (n3);
      \mkports{2}{}{3}{1}{1}
    \end{scope}
    \draw (p23) to (N2);
    \draw (N2) to (C3);
    \coordinate (p31) at ($(C3)+(0.5,0)$);
    \coordinate (p32) at ($(p31)+(0, 1.5)$);
    \coordinate (p33) at ($(p32)+(0.5,0)$);
    \draw (C3) to (p31);
    \draw (p31) to (p32);
    \draw (p32) to (p33);
    \begin{scope}[shift={(5,0)}]
      \node [atom] (N3) at (0, 0) {N};
      \mkports{}{1}{2}{}{}
    \end{scope}
    \begin{scope}[shift={(5,1)}]
      \node [atom] (F31) at (0, 0) {F};
      \mkports{}{3}{1}{2}{}
      \coordinate (pF31) at (0.5,0);
      \coordinate (PF31) at (0.5,-1);
      \draw (F31) to (pF31);
      \draw (pF31) to (PF31);
      \fill (PF31) circle (1.5pt);
    \end{scope}
    \draw (p33) to (F31);
    \draw (F31) to (N3);
    \node [link] (LY) at (6, 0) {$Y$};
    \draw (N3) to (LY);
  }
\end{align}

The examples in this section will also be used in the later development.
For this reason, we collect the production rules for doubly linked lists and
leaf-linked trees in a single definition below.
Here, \(\mathrm{LeafM}\) denotes a marked leaf atom.
It has the same link structure as \(\mathrm{Leaf}\), but it is distinguished
as a separate atom symbol so that marked leaves can be referred to in later
examples.

\begin{align}\label{eq:prod-rules-eg}
  \Pdl \triangleq
  \left\{
  \begin{array}{@{}l@{}}
    \mathit{dbllist}(X,Y,Z)  \lto                                                 \\
    \quad \;\; \Nil(X, Z), Y \bowtie Z                                            \\[1mm]
    \quad \mid \nu W_1 W_2.(\Cons(W_1,X,W_2,Z),                                   \\
    \qquad \mathit{nat}(W_1),\mathit{dbllist}(Z,Y,W_2))                           \\[3mm]
    \mathit{nat}(X) \lto 0(X) \mid 1(X) \mid 2(X) \mid \dots                      \\[3mm]
    \mathit{lltree}(L,R,X) \lto                                                   \\
    \quad \;\; \nu W_1. (\Leaf(W_1,R,X), L \bowtie X, \mathit{nat}(W_1))          \\[1mm]
    \quad \mid \nu W_1. (\mathrm{LeafM}(W_1,R,X), L \bowtie X, \mathit{nat}(W_1)) \\[1mm]
    \quad \mid \nu Y W_1 W_2.(\Node(W_1, W_2,X),                                  \\
    \qquad \mathit{lltree}(L,Y, W_1), \mathit{lltree}(Y,R, W_2))
  \end{array}\right.
\end{align}

\section{Type Soundness}\label{sec:type-soundness}

This section proves the type soundness of \lgt's type system.
More precisely, we establish the standard type preservation and progress properties.
Type preservation states that evaluation preserves typing: if a well-typed expression takes an evaluation step, then the resulting expression is also well typed.
Progress states that a well-typed closed expression is either a value or can take an evaluation step.
Together, these properties ensure that well-typed programs do not lose type consistency during evaluation and do not get stuck in a semantically invalid state, that is, a non-value state for which the operational semantics defines no transition.

It is important, however, to clarify what is guaranteed by type soundness in this dissertation.
The meaning of type soundness depends on what the type system is capable of expressing.
In an ordinary functional language such as OCaml, for example, one may implement a doubly linked list using references.
In such an implementation, one can define a type of nodes with two reference fields and ensure that each node has this type.
However, this type alone does not guarantee the intended shape of a doubly linked list.
In particular, it does not ensure that the two references of each node point respectively to the previous and next nodes in the list.
A structure in which the \texttt{next} reference of a node points to the node itself may still be well typed in the ordinary sense.
Such a program may therefore satisfy the usual type soundness theorem of the host language: it does not reach a stuck state caused by a violation of the language's operational semantics.
Nevertheless, the structure is not a valid doubly linked list as intended by the programmer.

The distinguishing feature of \lgt's type system is that it is designed to express and enforce structural properties of graph data.
Thus, the type soundness result proved in this section has a stronger significance for the intended applications of the language.
It does not merely show that evaluation preserves conventional type consistency.
Rather, it shows that graph values whose shapes are specified by types remain compatible with those specifications throughout evaluation, and that well-typed programs do not become stuck during graph transformation.
In this sense, the type system supports programming that is closer to the programmer's structural intention.

For simplicity, we assume that every graph variable name has
its own arity;
i.e., we disallow any graph variable name to occur with different
arities (and stand for different variables).

\begin{lemma}[Weakening]\label{lemma-weakening}
  If \(\Gamma \vdash_P e : \tau\) and
  \(x[\Xs] \not\in \mathrm{dom}(\Gamma)\), then
  \(\Gamma, x[\Xs]: \tau' \vdash_P e : \tau\).
\end{lemma}

\begin{proof} By induction on the derivation tree of the typing relation
  \(\Gamma \vdash_P e : \tau\).
  For a derivation tree consisting of a single \TyVar{}, augmenting
  the environment with a fresh binding does not invalidate the
  derivation.  For all other derivation trees, the final step of the
  derivation (with a rule other than \TyVar{}) does not depend on the
  content of the environment $\Gamma$, which can therefore be augmented.
\end{proof}

The following Substitution Lemma plays the key role in establishing
the preservation property.

\begin{lemma}[Substitution Lemma]\label{lemma-substitution-lemma}
  If \(\Gamma, x[\Xs]:\tau' \vdash_P e : \tau\) and
  \(\Gamma \vdash_P G : \tau'\), then
  \(\Gamma \vdash_P e[G / x[\Xs]] : \tau\).
\end{lemma}

The proof technique is standard, i.e.,
by induction on the derivation tree of the typing relation and case
analysis by the typing rule used in the final step, but
the handling of two kinds of substitution (graph substitution and
link substitution), structural congruence, and so on, needs special care.

\begin{proof}
  By induction on the derivation tree of the typing relation
  \(\Gamma, x[\Xs]:\tau' \vdash_P e : \tau\).
  We split the cases by the rule used in the final step.

  \noindent
  \textbf{Case \TyVar{}:} 
  By assumption, we have a proof of the form

  \begin{prooftree}
    \AXC{$y[\Zs] \in \dom(\Gamma, x[\Xs]:\tau')$}
    \RightLabel{\TyVar{}}
    \UIC{$\Gamma, x[\Xs]:\tau' \vdash_P y[\Ys] : \tau'\angled{\Ys/\Zs}$}
  \end{prooftree}

  and a proof ending with
  \(\Gamma \vdash_P G : \tau'\).  The goal is to derive
  \[
    \Gamma \vdash_P y[\Ys][G / x[\Xs]]  : \tau'\angled{\Ys/\Zs}.
  \]

  \begin{enumerate}
    \def\labelenumi{\arabic{enumi}.}
    \item If \(y = x\),
          since we allow only one graph variable with the same name
          in a single type
          environment, we have \(y[\Zs] = x[\Xs]\) and \(\Zs = \Xs\).
          From the assumption \(\Gamma \vdash_P G : \tau'\), we
          have \(\Gamma \vdash_P G\angled{\Ys/\Zs} : \tau'\angled{\Ys/\Zs}\)
          using \TyAlpha{}.
          Since \(y = x\), \(y[\Ys][G/x[\Xs]] =
          G\angled{\Ys/\Xs} = G\angled{\Ys/\Zs}\), which allows us to
          rewrite the last typing relation to
          \(\Gamma \vdash_P y[\Ys][G/x[\Xs]] : \tau'\angled{\Ys/\Zs}\).

    \item If \(y \neq x\), we have
          (i) \(y[\Ys][G/x[\Xs]] = y[\Ys]\) and (ii)
          \(y[\Zs] \in \dom(\Gamma)\) from the premise of the above \TyVar{}.
          From (ii) we have
          \(\Gamma \vdash_P y[\Ys] : \tau'\angled{\Ys/\Zs}\) by \TyVar{},
          which is the same as
          \(\Gamma \vdash_P y[\Ys][G/x[\Xs]] : \tau'\angled{\Ys/\Zs}\) by (i).
  \end{enumerate}

  \noindent
  \textbf{Case \TyArrow{}:} 
  By assumption, we have a proof ending with the following form.

  \begin{prooftree}
    \def\ScoreOverhang{0pt}
    \def\defaultHypSeparation{\hskip .1in}
    \def\labelSpacing{2pt}
    \AXC{$\Gamma, x[\Xs]:\tau', y[\Ys]: \tau_1 \vdash_P e_1 : \tau_2$}
    \RightLabel{\TyArrow{}}
    \UIC{$\Gamma, x[\Xs]\narrowcolon \tau' \vdash_P
        (\lambda\, y[\Ys]\narrowcolon \tau_1.e_1)(\Zs)
        \narrowcolon\, (\tau_1\mathord{\to}\tau_2) (\Zs)$}
  \end{prooftree}

  Since we do not allow two or more graph variables with the same name
  (though the name conflict could be circumvented by renaming),
  we can assume \(y \neq x\).

  From the induction hypothesis on the premise of the above
  derivation step
  \[ \Gamma, y[\Ys]: \tau_1, x[\Xs]:\tau' \vdash_P e_1 : \tau_2 \]
  and the assumption \(\Gamma \vdash_P G : \tau'\), we have
  \[ \Gamma, y[\Ys]:\tau_1 \vdash_P e_1[G/x[\Xs]] : \tau_2. \]
  By applying \TyArrow{}, we have
  \[ \Gamma \vdash_P
    (\lambda y[\Ys]:\tau_1.\, e_1[G/x[\Xs]])(\Zs) : (\tau_1 \to \tau_2)(\Zs). \]
  Because \(y \neq x\), we can move the substitution outwards and get
  \[ \Gamma \vdash_P
    (\lambda y[\Ys]:\tau_1.\, e_1)(\Zs)[G/x[\Xs]] : (\tau_1 \to \tau_2)(\Zs). \]

  \noindent
  \textbf{Case \TyApp{}:} 
  By assumption, we have a proof ending with the following form.

  \begin{prooftree}
    \def\labelSpacing{2pt}
    \def\defaultHypSeparation{\hskip .05in}
    \AXC{$\begin{array}{@{}r@{~}l@{}}
          \Gamma, x[\Xs]\narrowcolon\tau' & \vdash_P
          e_1\narrowcolon\,(\tau_1 \rightarrow \tau) (\Zs) \\
          \Gamma, x[\Xs]\narrowcolon\tau' & \vdash_P
          e_2\narrowcolon\tau_1
        \end{array}$}
    \RightLabel{\TyApp{}}
    \UIC{$\Gamma, x[\Xs]\narrowcolon\tau' \vdash_P (e_1\; e_2)\narrowcolon\tau$}
  \end{prooftree}

  For the two premises of the above \TyApp{},
  the induction hypothesis gives us
  \(\Gamma \vdash_P e_1[G/x[\Xs]] : (\tau_1 \to \tau)(\Zs)\)
  and
  \(\Gamma \vdash_P e_2[G/x[\Xs]] : \tau_1\).
  From these, \TyApp{}
  gives us
  \[
    \Gamma \vdash_P e_1[G/x[\Xs]]\; e_2[G/x[\Xs]] : \tau.
  \]
  Because
  \(e[G/x[\Xs]] = (e_1[G/x[\Xs]]\; e_2[G/x[\Xs]])\), we get
  \[
    \Gamma \vdash_P (e_1\; e_2)[G/x[\Xs]] : \tau.
  \]
  and thus the lemma holds.

  \noindent
  \textbf{Case \TyAlpha{}:} 
  By assumption, we have a proof ending with the following form.

  \begin{prooftree}
    \AXC{$\Gamma, x[\Xs]:\tau' \vdash_P T : \tau_1$}
    \AXC{$Y \notin \mathit{fn}(T)$}
    \RightLabel{\TyAlpha{}}
    \BIC{$\Gamma, x[\Xs]:\tau' \vdash_P T\angled{Y/Z} : \tau_1\angled{Y/Z}$}
  \end{prooftree}

  By applying the induction hypothesis to the first premise
  \(\Gamma, x[\Xs]:\tau' \vdash_P T : \tau_1\), we obtain
  \(\Gamma \vdash_P T[G/x[\Xs]] : \tau_1\).
  Applying \TyAlpha{} to this typing relation, we obtain
  \(\Gamma \vdash_P T[G/x[\Xs]]\angled{Y/Z} : \tau_1\angled{Y/Z}\).

  The remaining step to reach the final goal
  \(\Gamma \vdash_P T\angled{Y/Z}[G/x[\Xs]] : \tau_1\angled{Y/Z}\)
  is to prove
  \[
    T[G/x[\Xs]]\angled{Y/Z} = T\angled{Y/Z}[G/x[\Xs]].
  \]
  This is proved by induction on the form of $T$.  We focus on the
  base case of $T$ being a graph variable $y[\Ws]$ because it is the
  most non-obvious case that actually performs substitution,
  whereas the structural induction steps are straightforward.

  If $y\ne x$, the both sides are just $y[\Ws]\angled{Y/Z}$ and the
  claim holds trivially.

  If $y=x$, the left-hand side is $G\angled{\Ws/\Xs}\angled{Y/Z}$,
  while the right-hand side is
  $y[\Ws\angled{Y/Z}][G/x[\Xs]]$ $=$
  $G\angled{\Ws\angled{Y/Z}/\Xs}$.

  By the definition of graph substitution, $\fn(G)=\{\Xs\}$; therefore
  it suffices to consider how each free link $X_i$ in $G$
  is affected by the
  $\angled{Y/Z}$, for $1\le i \le |\Xs|$.
  If $W_i=Z$, the link substitution applied to
  $G$ behaves as $\angled{Y/X_i}$ for both sides.
  If $W_i\ne Z$, the link substitution applied to
  $G$ behaves as $\angled{W_i/X_i}$ for both sides.  Therefore
  $G\angled{\Ws/\Xs}\angled{Y/Z} = G\angled{\Ws\angled{Y/Z}/\Xs}$.

  \noindent
  \textbf{Case \TyCong{}:} 
  By assumption, we have a proof ending with the following form.

  \begin{prooftree}
    \AXC{$\Gamma, x[\Xs]:\tau'  \vdash_P T : \tau$}
    \AXC{$T \equiv T'$}
    \RightLabel{\TyCong{}}
    \BIC{$\Gamma, x[\Xs]:\tau'  \vdash_P T' : \tau$}
  \end{prooftree}

  By applying the induction hypothesis to the first premise
  of \TyCong{}, we have \(\Gamma \vdash_P T[G/x[\Xs]] : \tau\).

  Structural congruence rules (\cref{table:untyped:congruence})
  enjoy the following substitution property
  \[
    T \equiv T'
    \implies
    T[G/x[\Xs]] \equiv T'[G/x[\Xs]],
  \]
  which is straightforward by noticing that the requirement $\fn(G)=\{\Xs\}$
  of graph substitution ensures that the side conditions of (E6) and
  (E10) on free links are not affected by graph substitution.

  Now using $T[G/x[\Xs]] \equiv T'[G/x[\Xs]]$ as the second premise of
  \TyCong{}, we get
  \[
    \Gamma \vdash_P T'[G/x[\Xs]] : \tau.
  \]

  \noindent
  \textbf{Case \TyProd{}:} 
  By assumption, we have a proof ending with the form

  \begin{prooftree}
    \AXC{$\Gamma, x[\Xs]:\tau' \vdash_P T_1 : \tau_1$}
    \AXC{$\dots$}
    \AXC{$\Gamma, x[\Xs]:\tau' \vdash_P T_n : \tau_n$}
    \RightLabel{\TyProd{}}
    \TIC{$\Gamma, x[\Xs]:\tau' \vdash_P
        \nu \Zs.(C(\Ys), T_1, \dots, T_n, \seq{U \bowtie V}) : \alpha (\Ws)$}
  \end{prooftree}
  where
  \(\alpha (\Ws) \lto \nu \Zs.(C(\Ys), \tau_1, \dots, \tau_n, \seq{U
    \bowtie V}) \in P\).

  By applying the induction hypothesis to the premise
  \(\Gamma, x[\Xs]:\tau' \vdash_P T_i : \tau_i\), we have
  \(\Gamma \vdash_P T_i[G/x[\Xs]] : \tau_i\),
  from which another application of \TyProd{} gives us
  \[\Gamma \vdash_P \\
    \nu \Zs.(C(\Ys), T_1[G/x[\Xs]], \dots, T_n[G/x[\Xs]], \seq{U \bowtie V}) : \alpha(\Ws).
  \]
  From the definition of graph substitution, we have
  \begin{align}
     & (\nu \Zs.(C(\Ys), T_1, \dots, T_n, \seq{U \bowtie V}))[G/x[\Xs]] \nonumber  \\[-3pt]
     & = \nu \Zs.(C(\Ys), T_1[G/x[\Xs]], \dots, T_n[G/x[\Xs]], \seq{U \bowtie V}),
  \end{align}
  which allows us to factor out the substitution and get
  \[
    \Gamma
    \vdash_P \nu \Zs.(C(\Ys), T_1, \dots, T_n, \seq{U \bowtie
      V})[G/x[\Xs]] : \alpha(\Ws).
  \]

  \noindent
  \textbf{Case \TyCase{}:} 
  By assumption, we have a proof ending with the form

  \begin{prooftree}
    \AXC{$\begin{array}{r@{~}l}
          \Gamma                  & \vdash_P e_1 : \tau_1
          \\
          \Gamma, \CollectVars(T) & \vdash_P e_2 : \tau_2
        \end{array}$}
    \AXC{$\begin{array}{r@{~}l}
          \CollectVars(T) & \vdash_P T : \tau_1
          \\
          \Gamma          & \vdash_P e_3 : \tau_2
        \end{array}$}
    \RightLabel{\TyCase{}}
    \BIC{$\ml{\Gamma \vdash_P\shortstrut (\caseof{e_1}{T}{e_2}{e_3}) : \tau_2}$}
  \end{prooftree}%

  where each graph variable occurring
  in \(T\) must occur with
  a type annotation as
  $x_i[\overrightarrow{X_i}]:\sigma_i$,
  and \(\Gamma'\) = \{$\overrightarrow{x[\overrightarrow{X}]:\sigma}$\}.

  By applying the induction hypothesis to the premises,
  we have the following:
  \[\begin{array}{r@{~}l}
      \Gamma                  & \vdash_P e_1[G / x[\Xs]] : \tau_1, \\
      \CollectVars(T)         & \vdash_P T : \tau_1,               \\
      \Gamma, \CollectVars(T) & \vdash_P e_2[G / x[\Xs]] : \tau_2, \\
      \Gamma                  & \vdash_P e_3[G / x[\Xs]] : \tau_2.
    \end{array}\]
  Using them as premises of \TyCase{}, we have
  \begin{align}
    \Gamma \vdash_P
    (\caseof{e_1[G/x[\Xs]]}{T}{e_2[G/x[\Xs]]}{e_3[G/x[\Xs]]}) : \tau_2
  \end{align}
  By factoring out the substitution, we obtain
  \[
    \Gamma
    \vdash_P (\caseof{e_1}{T}{e_2}{e_3})[G/x[\Xs]] : \tau_2.
  \]
\end{proof}

Before proving the preservation properties, we
we show type preservation of evaluation contexts.
\begin{lemma}[Evaluation Context Preservation]\label{ECP}
  If
  \(\Gamma \vdash_P E[e] : \tau\), there exists \(\tau'\) such that
  \(\Gamma \vdash_P e : \tau'\), and for all \(e'\) such that
  \(\Gamma \vdash_P e' : \tau'\), \(\Gamma \vdash_P E[e'] : \tau\) holds.
\end{lemma}

\begin{proof}
  Based on straightforward induction on the structure of evaluation contexts $E$.
  If \(E = [\,]\), the claim holds trivially.
  Otherwise, i.e., if
  $E[e]$ is $\textbf{case}\ E'[e] \ \textbf{of}\ \dots$ or
  \((E'[e]\,e_2)\) or \((G\,E'[e])\),
  $E'[e]$ and $E'[e']$ have the same type by the induction hypothesis, and
  the final step of the derivation tree with \TyCase{} or \TyApp{}
  establishes \(\Gamma \vdash_P E[e'] : \tau\).
\end{proof}

\begin{theorem}[Preservation / Subject Reduction]\label{theorem-preservation-subject-reduction}
  If \(\Gamma \vdash_P e : \tau\) and \(e \reducesp{P} e'\), then
  \(\Gamma \vdash_P e' : \tau\).
\end{theorem}

\begin{proof}
  By \cref{table:lgt-reduction}, without loss of generality,
  we can assume that $e$ is of the form
  \(E[e_0]\), where (i) \(E\) is an evaluation context and
  (ii) \(e_0\) is
  a redex such that \(e_0 \reducesp{P} e_0'\) for some \(e_0'\),
  which implies \(E[e_0] \reducesp{P} E[e_0']\).
  We prove
  \[
    \Gamma \vdash_P e_0 : \tau_0 \implies \Gamma \vdash_P e_0' : \tau_0
  \]
  by case analysis on the rule used for reduction:
  \RdBetaD{}, \RdCaseMatchD{}, or \RdCaseOtherD{}.

  \noindent
  \textbf{Case \textsf{Rd-}\(\boldsymbol{\beta}\)\textsf{-D}}:
  The reduction is of the following form.
  \begin{prooftree}
    \AXC{$G_1 \equiv (\lambda\, x [\Xs].e) (\Ys)$}
    \AXC{\(\fn(G_2) = \{\Xs\}\)}
    \RightLabel{\RdBetaD{}}
    \BIC{\(E[G_1\ G_2] \reducesp{P} E[e [G_2 / x[\Xs]]]\)}
  \end{prooftree}
  Since an application can be typed only with \TyApp{},
  the typing of the subject \(e_0\) must use \TyApp{}, whose premises are
  \begin{itemize}
    \item
          \(\Gamma \vdash_P G_1 : (\tau_1 \to \tau_0)(\Ys)\)
          and
    \item
          \(\Gamma \vdash_P G_2 : \tau_1\).
  \end{itemize}
  In the derivation of
  \(\Gamma \vdash_P G_1 : (\tau_1 \to \tau)(\Ys)\),
  \TyArrow{} must be used, whose premise is
  \(\Gamma, x[\Xs]:\tau_1 \vdash_P e_1 : \tau_0\).
  With this and the substitution lemma, we have
  \[
    \Gamma \vdash_P e_1[G_2/x[\Xs]] : \tau_0.
  \]%

  \noindent
  \textbf{Case \RdCaseMatchD{}:}
  The reduction is of the form
  \begin{prooftree}
    \def\ScoreOverhang{5pt}
    \AXC{$G \equiv T\thetas
      \land
      \bigwedge_{i} \paren{\emptyset \vdash_P x_i[\Xs_i]\thetas: \sigma_i}$}
    \RightLabel{\RdCaseMatchD{}}
    \UIC{$(\caseof{G}{T}{e_1}{e_2}) \reducesp{P} e_1\thetas$}
  \end{prooftree}
  where
  \(\thetas\mathord{=} [G_1/x_1[\Xs_1]] \dots [G_n/x_n[\Xs_n]]\)
  such that \(\{x_1[\Xs_1],\)\\
  \(\dots, x_n[\Xs_n]\}\) is the set of graph
  variables in $T$.
  The typing of this case expression can be done only with \TyCase{}:
  %
  \begin{prooftree}
    \def\defaultHypSeparation{\hskip .2in}
    \def\ScoreOverhang{0pt}
    \AXC{$\Gamma \vdash_P G : \tau'$}
    \AXC{$(\Gamma, \Gamma') \vdash_P e_1 : \tau_0$}
    \AXC{$\Gamma \vdash_P e_2 : \tau_0$}
    \RightLabel{\TyCase{}}
    \TIC{$\Gamma \vdash_P (\caseof{G}{T}{e_1}{e_2}) : \tau_0$}
  \end{prooftree}
  where each graph variable occurring
  in \(T\) must occur with a type annotation
  $x_i[\overrightarrow{X_i}]:\sigma_i$,
  and
  \(\Gamma' = x_1[\Xs_1]:\sigma_1, \dots, x_n[\Xs_n]:\sigma_n\).
  By applying the substitution lemma $n$ times, from the second premise
  \[\Gamma, x_1[\Xs_1]:\sigma_1, \dots, x_n[\Xs_n]:\sigma_n \vdash_P e_1 : \tau_0,\]
  we can get
  \(\Gamma \vdash_P e_1[G_1/x_1[\Xs_1]] \dots [G_n/x_n[\Xs_n]] : \tau_0\)
  provided that we have $n$ typing relations
  \[\Gamma \vdash_P G_1:\sigma_1, \dots, \Gamma \vdash_P G_n:\sigma_n.\]
  These $n$ typing relations can be obtained
  by applying the weakening lemma to the family of premises of \RdCaseMatchD{}
  because $G_i=x_i[\Xs_i]\thetas$.
  Thus we have derived
  \(\Gamma \vdash_P e_1\thetas : \tau_0\).

  \noindent
  \textbf{Case \RdCaseOtherD{}:}
  The typing of the case expression must use \TyCase{} above, which contains
  \(\Gamma \vdash_P e_2 : \tau_0\); therefore we have
  \(\Gamma \vdash_P e_0' : \tau_0\).

  To summarise, let \(e = E[e_0]\).
  If \(e_0 \reducesp{P} e_0'\) and \(\Gamma \vdash_P e_0 : \tau_0\),
  then we have  \(\Gamma \vdash_P e_0' : \tau_0\) and
  \(\Gamma \vdash_P E[e_0'] : \tau\) from \cref{ECP}.
  Therefore,
  \[
    \Gamma \vdash_P e : \tau \ \land \ e \reducesp{P} e'
    \implies
    \Gamma \vdash_P e' : \tau
  \]
  and the claim is established.
\end{proof}

\begin{theorem}[Progress]\label{theorem-progress}
  If \(\emptyset \vdash_P e : \tau\), then either \(e\) is a value or there
  exists \(e'\) such that \(e \reducesp{P} e'\).
\end{theorem}

\begin{proof}
  By induction on the structure of the expression \(e\).

  If \(e\) is a template $T$, the assumption that the typing
  environment is empty implies that $T$ contains no type variable
  and hence is a value.

  If $e$ is a case expression, by the induction hypothesis,
  either the evaluation of the condition
  makes progress,
  or the condition is a value, and
  the case expression is reduced to one of the two branches (logically trivial case analysis).

  If $e$ is an application $(e_1\, e_2)$ and
  \(e_1\) is a value, \(e_1\) must be a $\lambda$-atom because $e$ is
  well-typed.
  If \(e_2\) is a value as well,
  then $(e_1\, e_2)$ is a \(\beta\)-redex;
  otherwise, by the induction hypothesis
  \(e_2 \reducesp{P} e_2'\) and hence
  \((e_1\, e_2) \reducesp{P}\ (e_1\, e_2')\).
  If \(e_1\) is not a value, by the induction hypothesis
  \(e_1 \reducesp{P} e_1'\) and hence \(e \reducesp{P} (e_1'\,e_2)\).
\end{proof}

\begin{corollary}[Type Soundness]\label{corollary-type-soundness}
  If \(\emptyset \vdash_P e : \tau\) and
  \(e \reducespstar{P} e'\) with
  \(e'\) normal (no further steps), then \(e'\) is a value of type
  \(\tau\).
\end{corollary}

\begin{proof}
  Follows from repeated application of
  \cref{theorem-preservation-subject-reduction} and
  \cref{theorem-progress}.
\end{proof}

\section{Connection to Grammar-Based Graph Typing}\label{sub:fgt-reduction}

This section relates the core type system of \lgt
to grammar-based typing frameworks for graph structures,
such as Structured Gamma~\cite{structuredgamma}
and LMNtal ShapeType~\cite{yamamoto2024}.
Structured Gamma is a first-order graph rewriting system
developed for representing and reasoning about
the shapes of pointer data structures.
The ideas of Structured Gamma were later adapted to LMNtal,
whose graph representation is, roughly speaking,
dual to that of Structured Gamma,
leading to the design and implementation of LMNtal ShapeType.

Despite several syntactic differences,
such as the duality of nodes and links
and the presence or absence of hyperlinks,
Structured Gamma and LMNtal ShapeType can essentially handle
the graph structures of \lgt
as long as graph variables and \(\lambda\)-abstraction atoms are not involved.
In those systems,
typing can be understood as reduction from a type atom
to a graph generated by production rules.
The following definition abstracts this reduction-based view.

\begin{definition}[Typing relation in Structured Gamma/LMNtal ShapeType]\label{def:structured-gamma}
  A graph $G$ has type $\alpha (\seq{X})$ under production rules $P$
  if and only if
  $\alpha (\seq{X}) \rightsquigarrow^{\ast}_{P} G$
  and
  \(G\) does not contain type variables (i.e., non-terminal symbols).
\end{definition}

The core type system of \lgt extends this reduction-based view.
The main difference is that a graph term in \lgt may contain
graph variables and \(\lambda\)-abstraction atoms.
Such components cannot be generated directly by production rules
in the sense of Structured Gamma.
Therefore,
to compare the two typing relations,
we replace each graph variable and each \(\lambda\)-abstraction atom
by the type atom assigned to it.
Under this replacement,
typing in the core type system corresponds to reduction
from the target type.


\begin{theorem}[The Core Type System of \lgt and HyperLMNtal Reduction]\label{th:fgt-stgamma}
  \[\begin{array}{l}
      \Gamma
      \vdash_P
      T : \tau\, (\seq{X}) \\
      \qquad\Leftrightarrow
      \tau\, (\seq{X})
      \rightsquigarrow^{\ast}_{P}
      T
      {\overrightarrow{[\tau_i / x_i [\seq{X_i}]]}}^{i}
      {[\overrightarrow{\tau_j / {(\lambda \dots)}_j (\seq{W_j})}]}^{j}
    \end{array}\]

  where
  \begin{itemize}
    \item
          \(\Gamma =
          {\overrightarrow{x_i [\seq{X_i}]: \tau_i}}^{i}
          \),

    \item
          \({(\lambda \dots)}_{j} (\seq{W_j})\)
          are all the \(\lambda\)-abstraction atoms in \(T\), and
    \item
          \(\Gamma \vdash_P {(\lambda \dots)}_j (\seq{W_j}) : \tau_j\).

  \end{itemize}

\end{theorem}

\begin{proof}
  For \(\Rightarrow\),
  we can prove by induction on the last applied typing rules.
  For \(\Leftarrow\),
  We prove by induction on the length of the reduction
  \(\rightsquigarrow^{\ast}_{P}\).
\end{proof}


If \(T\) contains neither graph variables nor \(\lambda\)-abstraction atoms,
the replacement in \cref{th:fgt-stgamma} is empty.
In this case,
the theorem states that the typing relation of the core type system of \lgt
coincides with the reduction-based typing relation
used in Structured Gamma and LMNtal ShapeType.
Thus,
the core type system of \lgt can be seen as an extension of
reduction-based graph typing to graph variables and
\(\lambda\)-abstraction atoms.
This connection allows us to reuse intuitions and techniques
from Structured Gamma,
LMNtal ShapeType,
and graph parsing based on graph grammars.

\begin{example}[
    Theorem~\ref{th:fgt-stgamma} on the difference list example]

  Here, we see that \cref{th:fgt-stgamma} holds on
  \cref{ex:typing-dlist}.
  Recall that
  \(succ [Z_1]: (\nattonat) (Z_1) \vdash_P \Cons\allowbreak (succ,\allowbreak Y,\allowbreak X) : \mathit{nodes}(Y, X)\)
  holds in the type system,
  which can also be shown using HyperLMNtal reduction as follows.
  \[\begin{array}{l}
      \mathit{nodes}(Y, X)                                                          \\
      \rightsquigarrow_{P_2}
      \nu Z_1 Z_2. (\Cons (Z_1, Z_2, X), (\nattonat) (Z_1), \mathit{nodes}(Y, Z_2)) \\
      \rightsquigarrow_{P_1}
      \nu Z_1 Z_2. (\Cons (Z_1, Z_2, X), (\nattonat) (Z_1), Z_2 \bowtie Y)          \\
      \equiv
      \nu Z_1. (\Cons (Z_1, Y, X), (\nattonat) (Z_1))                               \\
      =
      \Cons (succ , Y, X) [(\nattonat) (Z_1) / succ [Z_1]]
    \end{array}\]

\end{example}

We now give the proof of \cref{th:fgt-stgamma}.
We first record a lemma stating that reductions are stable under renaming a free link.

\begin{lemma}\label{lem:redux-alpha}
  If
  \(G_1
  \rightsquigarrow^{\ast}_{P}
  G_2
  \)
  then
  \(G_1\angled{Y/X}
  \rightsquigarrow^{\ast}_{P}
  G_2\angled{Y/X}
  \)
\end{lemma}

\begin{proof}
  By (R1), (R2), (R3) and
  \(G_1
  \rightsquigarrow_{P}
  G_2
  \),
  we can show
  \(\nu X.(X \bowtie Y, G_1)
  \rightsquigarrow_{P}
  \nu X.(X \bowtie Y, G_2)
  \).
  Thus
  \(G_1\angled{Y/X}
  \rightsquigarrow_{P}
  G_2\angled{Y/X}
  \) by (R3).
  Then
  we can obtain
  \(G_1\angled{Y/X}
  \rightsquigarrow^{\ast}_{P}
  G_2\angled{Y/X}
  \)
  by induction on the length of the reduction
  \(\rightsquigarrow^{\ast}_{P}\)
\end{proof}

\begin{proof}[Proof of \cref{th:fgt-stgamma}]

  We write
  \({\overrightarrow{[\tau_i / x_i [\seq{X_i}]]}}^{i}\)
  as
  \(\theta_x\)
  and
  \({[\overrightarrow{\tau_j / {(\lambda \dots)}_j (\seq{W_j})}]}^{j}\)
  as
  \(\theta_\lambda\).

  We firstly prove \(\Rightarrow\).
  We split the cases by the last applied typing rules.

  \vspace{0.5em}\noindent%
  Case \TyVar{}:
  \begin{quote}
    \(T = x[\seq{X}]\)
    and
    \(x[\seq{X}]: \tau \in \Gamma\).
    Thus
    \(T[\tau / x[\seq{X}], \dots]\theta_\lambda
    = \tau
    \rightsquigarrow^{\ast}_{P}
    \tau
    \).
  \end{quote}

  \vspace{0.5em}\noindent%
  Case \TyArrow{}:
  \begin{quote}
    \(T = (\lambda \dots)(\seq{X})\)
    where
    \(
    \Gamma \vdash_P
    (\lambda \dots)(\seq{X}):
    \tau
    \).
    Thus
    \(T \theta_x [\tau / (\lambda \dots)(\seq{X})]
    = \tau
    \rightsquigarrow^{\ast}_{P}
    \tau
    \).
  \end{quote}

  \vspace{0.5em}\noindent%
  Case \TyCong{}:
  \begin{quote}
    Suppose the antecedent of \TyCong{} was
    \(\Gamma \vdash_P T' : \tau\)
    where
    \(T \equiv T'\).
    By induction hypothesis,
    \(\tau
    \rightsquigarrow^{\ast}_{P}
    T' \theta_x \theta_\lambda
    \).
    Since
    \(
    T\theta_x \theta_\lambda
    \equiv
    T'\theta_x \theta_\lambda
    \),
    we can show
    \(\tau
    \rightsquigarrow^{\ast}_{P}
    T \theta_x \theta_\lambda
    \)
    using (R3).
  \end{quote}

  \vspace{0.5em}\noindent%
  Case \TyAlpha{}:
  \begin{quote}
    Suppose the antecedent of \TyAlpha{} was
    \(\Gamma \vdash_P T' : \tau'\)
    where
    \(T = T'\angled{Y/X}\)
    and
    \(\tau = \tau'\angled{Y/X}\).
    By induction hypothesis,
    \(\tau'
    \rightsquigarrow^{\ast}_{P}
    T' \theta_x \theta_\lambda
    \).
    Here,
    we can show that
    \(T\theta_x \theta_\lambda
    =
    T'\theta_x \theta_\lambda\angled{Y/X}
    \).
    Therefore, by \cref{lem:redux-alpha},
    \(\tau'\angled{Y/X}
    \rightsquigarrow^{\ast}_{P}
    T \theta_x \theta_\lambda\angled{Y/X}
    \).
  \end{quote}

  \vspace{0.5em}\noindent%
  Case \TyProd{}:
  \begin{quote}
    Suppose the antecedents of \TyProd{} was
    \({\overrightarrow{\Gamma \vdash_P T_i : \tau_i}}^{i}\)
    where
    \(T = \Tauu[{\overrightarrow{T_i / \tau_i}}^{i}]\).
    By induction hypothesis,
    \(\tau_i
    \rightsquigarrow^{\ast}_{P}
    T_i \theta_x \theta_{\lambda i}
    \).
    Therefore, using (R1), (R2), and (R3),
    we can show
    \(\Tauu'
    \rightsquigarrow^{\ast}_{P}
    \Tauu'[T_i \theta_x \theta_{\lambda i} / \tau_i]
    \) for any \(\Tauu'\).
    Thus,
    we can have
    \(\tau_i
    \rightsquigarrow^{\ast}_{P}
    \Tauu_0
    \rightsquigarrow^{\ast}_{P}
    \dots
    \rightsquigarrow^{\ast}_{P}
    \Tauu_n
    \)
    where
    \(\Tauu_i\) is inductively defined as
    \(\Tauu_0 = \Tauu\)
    and
    \(\Tauu_{i + 1} = \Tauu_i[T_i \theta_x \theta_{\lambda i} / \tau_i]\),
    in which
    \(\Tauu_n = T\theta_x \theta_\lambda\).
  \end{quote}
  \vspace{1em}

  Then, we prove \(\Leftarrow\).
  by induction on the length of the reduction
  \(\rightsquigarrow^{\ast}_{P}\).
  We write
  \({\overrightarrow{[x_i [\seq{X_i}] / \tau_i]}}^{i}\)
  as
  \(\theta_x^{-1}\)
  and
  \({[\overrightarrow{{(\lambda \dots)}_j (\seq{W_j})} / \tau_j]}^{j}\)
  as
  \(\theta_\lambda^{-1}\).
  Then, the proposition can be rewritten as
  \[
    \tau
    \rightsquigarrow^{\ast}_{P}
    \Tauu
    \Rightarrow
    \Gamma \vdash_P
    \Tauu \theta_x^{-1} \theta_\lambda^{-1}:
    \tau.
  \].

  \vspace{0.5em}\noindent%
  Case \(\tau = \Tauu\)
  (The length of
  \(\rightsquigarrow^{\ast}_{P}\)
  is zero):
  \begin{quote}
    Follows by \TyVar{} or \TyArrow{}
    depending on whether the
    \(\tau\)
    is replaced with the graph context in
    \(\theta_x^{-1}\)
    or the
    \(\lambda\)-abstraction atom in
    \(\theta_\lambda^{-1}\).
  \end{quote}

  \vspace{0.5em}\noindent%
  Case \(
  \tau
  \rightsquigarrow^{\ast}_{P}
  \Tauu'
  \rightsquigarrow_{P}
  \Tauu
  \)
  (The length of
  \(\rightsquigarrow^{\ast}_{P}\)
  is \(n > 0\)):
  \begin{quote}
    Suppose the production rule applied to reduce from
    \(\Tauu'\)
    to
    \(\Tauu\)
    was
    \(\alpha(\seq{Y}) \longrightarrow \Tauu''\).
    Using (R1), (R2), and (R3),
    we can obtain (new)
    \(
    \tau
    \rightsquigarrow^{\ast}_{P}
    \Tauu'
    \rightsquigarrow_{P}
    \Tauu
    \)
    which satisfies
    \(
    \Tauu
    =
    \Tauu'[\Tauu'' / \alpha(\seq{Y})]
    \).
    By induction hypothesis,
    we can obtain the derivation tree of
    \begin{align}\label{math:all}
      \Gamma \vdash_P \Tauu' \theta_x^{-1} \theta_\lambda^{-1} : \tau.
    \end{align}
    Since \(\Tauu'\) contains \(\alpha(\seq{X})\),
    there exists a derivation of
    \begin{align}\label{math:prev}
      \Gamma' \vdash_P \alpha \theta_x^{-1} \theta_\lambda^{-1} : \alpha(\seq{X}).
    \end{align}
    in the tree.
    Since
    \begin{align}\label{math:after}
      \Gamma' \vdash_P \Tauu'' \theta_x^{-1} \theta_\lambda^{-1} : \alpha(\seq{X}).
    \end{align}
    holds immediately by \TyProd{},
    we can replace the derivation tree of
    (\Ref{math:prev})
    with
    that of
    (\Ref{math:after})
    in
    that of
    (\Ref{math:all}),
    which will result in the derivation tree of the desired typing relation.
  \end{quote}
\end{proof}

\chapter{Introducing Linear Implication Types}\label{ch:limp-type}

In the previous chapter,
we introduced a core type system for \lgt{}.
The core system allows programmers to define graph types
that describe the intended structure of graph values,
and it can type many examples involving structures such as
doubly linked lists and leaf-linked trees.
However,
the system still has a fundamental limitation:
it has no uniform mechanism for typing graph fragments
that arise as intermediate results of graph decomposition.

This limitation becomes apparent when pattern matching separates
a graph value into a visible part and a remaining context.
The remaining part is often not itself a complete value of the original graph type.
Rather,
it is an incomplete graph structure:
it becomes a complete graph only when supplied with a missing part of an appropriate type.

For example,
consider the graph variables \(y[\dots]\)
that occur in the definitions of
\(\eDLpop\) in \eqref{eq:dbllist-removal-eg-vis}
and \(\eLLT\) in \eqref{eq:lltree-traverse-eg-vis}.
These variables denote graph fragments that are meaningful parts of
doubly linked lists or leaf-linked trees,
but they are not themselves complete values of the corresponding graph types.
The following figures illustrate such incomplete graph structures.

\begin{align}\label{eq:incomplete-gctx-eg-vis}
  \tikz[baseline=-0.5ex, scale=1.0]{%
    \mkctxn{1}{y}{(0,0)}{}{W_2}
    \begin{scope}[shift={(1.5,0)}]
      \node [atom, ghost] (C2)  at (0, 0)    {C};
      \coordinate (p2AL) at ($(C2)+(0,0.4)$);
      \coordinate (p2AR) at ($(C2)+(0,0.4)$);
      \coordinate (p2R) at ($(p2AR)+( 0.5,0.4)$);
      \coordinate (p2L) at ($(p2AL)+(-0.5,0.4)$);
      \node [ctxn, ghost] (N2)  at ($(C2)+(0,-1)$) {$w$};
      \draw [ghost] (C2) to (N2);
      \node [text=ghostcolor, port] at (-0.375, -0.150) {4};
      \node [text=ghostcolor, port] at (+0.150, +0.375) {2};
      \node [text=ghostcolor, port] at (+0.375, -0.150) {1};
      \node [text=ghostcolor, port] at (-0.150, -0.375) {3};
      \node [text=ghostcolor, port] at (-0.150, +0.375-1) {1};
      \node [link,text=gray] at (0, 1) {$(W_1)$};
      \node [text=ghostcolor,link] at (0.450, -0.5) {$(W_3)$};
    \end{scope}
    \begin{scope}[shift={(3,0)}]
      \node [atom, ghost] (C3)  at (0, 0) {N};
      \coordinate (p3A) at ($(C3)+(0,0.4)$);
      \coordinate (p3R) at ($(p3A)+( 0.5,0.4)$);
      \coordinate (p3L) at ($(p3A)+(-0.5,0.4)$);
    \end{scope}
    %
    \node [link] (C0) at (-1, 0.8)  {$Z$};
    \node [link] (LX) at (-1, 0.0)  {$X$};
    \node [text=ghostcolor,link] (LY) at (4, 0.8)   {$Y$};
    \draw (C1) to  (LX);
    \mkconslinkleft{0}{1}
    \mkconslink{1}{2}
    \draw [ghost] (p2AL) to (C2);
    \draw [in=180, out=0, looseness=1.50,ghost] (C2) to (p3L);
    \draw [in=90, out=0, rounded corners,ghost] (p3L) to (p3A);
    \draw [ghost] (p3A) to (C3);
    \draw [in=0, out=180, looseness=1.50, line width=4pt,white] (C3) to (p2R);
    \draw [in=0, out=180, looseness=1.50, ghost] (C3) to (p2R);
    \draw [in=90, out=180, rounded corners, ghost]  (p2R) to (p2AL);
    \draw [in=0, out=180, looseness=1.50, ghost] (LY) to (p3R);
    \draw [in=90, out=180, rounded corners, ghost]  (p3R) to (p3A);
  }
\end{align}

\begin{align}\label{eq:incomplete-gctx-eg-vis2}
  \tikz[baseline=6ex, scale=1]{%
    \ctxyy{0}{0}
    \begin{scope}[shift={(-0.6, -0.5)}]
      \node [atom,ghost,fill=ghostcolor,text=white] (leaf1) at (0, 0) {L};
      \node [port,text=ghostcolor] at (0.150, 0.395)   {3};
      \node [port,text=ghostcolor] at (0.395, 0.150)   {2};
      \node [port,text=ghostcolor] at (-0.150, -0.395) {1};
      \node [port,text=ghostcolor] at (-0.150, -0.7)   {1};
      \node [text=ghostcolor,link] at (0.410, -0.5)   {$(W_1)$};
      \node [ctxn,below=0.8cm,ghost] (num1) at (leaf1)  {$z$};
      \draw [ghost]      (leaf1) -- (num1);
      \node [text=gray,port] at (0.435, 0.85)   {$(W_2)$};
    \end{scope}
    \begin{scope}[shift={(+0.6, -0.5)}]
      \node [atom,ghost] (leaf2) at (0, 0) {L};
      \node [port,text=ghostcolor] at (0.150, 0.395)   {3};
      \node [port,text=ghostcolor] at (0.395, 0.150)   {2};
      \node [port,text=ghostcolor] at (-0.150, -0.395) {1};
      \node [port,text=ghostcolor] at (-0.150, -0.7)   {1};
      \node [text=ghostcolor,link] at (0.410, -0.5)   {$(W_3)$};
      \node [ctxn,below=0.8cm,ghost] (num2) at (leaf2)  {$w$};
      \draw [ghost]      (leaf2) -- (num2);
      \node [link,text=gray] at (-0.420, 0.50)   {$(W_4)$};
      \node [link,text=gray] at (1.010, 0.25)   {$(W_5)$};
    \end{scope}
    \coordinate (p1) at (-1.4, 0.2);
    \coordinate (p2) at (+1.4, 0.2);
    \coordinate (p3) at (-0.6, 0.9);
    \coordinate (p4) at (+0.6, 0.9);
    \draw [looseness=1.50,out=0,in=90] (leaf1) to (leaf2);
    \draw [looseness=1.50,out=0,in=90]    (p1) to (leaf1);
    \draw [looseness=1.50,out=-90,in=90]  (p3) to (leaf1);
    \draw [looseness=1.50,out=-90 ,in=90] (p4) to (leaf2);
    \draw [looseness=1.50,out=0,in=180] (leaf2) to (p2);
  }
\end{align}

\begin{align}\label{eq:incomplete-gctx-eg-vis3}
  \tikz[baseline=5ex, scale=1]{%
    \ctxyl{0}{0}
    \begin{scope}[shift={(0.8, 0)}]
      \node [atom,ghost,fill=ghostcolor,text=white] (leaf1) at (0, 0) {L};
      \node [port,text=ghostcolor] at (0.150, 0.395)   {3};
      \node [port,text=ghostcolor] at (0.395, 0.0)   {2};
      \node [port,text=ghostcolor] at (-0.150, -0.395) {1};
      \node [port,text=ghostcolor] at (-0.150, -0.7)   {1};
      \node [text=ghostcolor,link] at (0.460, -0.5)   {$(W_1)$};
      \node [link] at (0.400, 0.8)   {$W_2$};
      \node [ctxn,below=0.8cm,ghost] (num1) at (leaf1)  {$z$};
      \draw [ghost]      (leaf1) -- (num1);
    \end{scope}
    \coordinate (p1) at (-0.70, 0.2);
    \coordinate (p2) at (0.4, 1.4);
    \draw [looseness=1.50,out=0,in=90] (p1) to (leaf1);
    \draw [looseness=1.50,out=-45,in=90] (p2) to (leaf1);
    \draw [looseness=1.50,out=30,in=180,ghostcolor] (leaf1) to (LR);
    \node [link,text=gray] at (LR)  {$R$};
  }
\end{align}

As a consequence,
even if the arguments of the functions are given the intended type annotations,
the core type system cannot derive the following typings:
\begin{align}
  \emptyset \vdash_{\Pdl} & \eDLpop : (dbllist(X,Y,Z) \to dbllist(X,Y,Z))(W),
  \label{eq:dbllist-typing-relation-eg}
  \\
  \emptyset \vdash_{\Pdl} & \eLLT :
  \begin{array}[t]{lll}
    (((\mathit{nat}(X) \to \mathit{nat}(X))(X) \to \\
    \quad lltree(X,Y,Z))(X) \to lltree(X,Y,Z))(X).
  \end{array}
  \label{eq:lltree-typing-relation-eg}
\end{align}

One possible workaround is to require users to define separate graph types
for every incomplete structure that may arise during graph transformation.
For instance,
difference lists can be introduced as a separate type from ordinary lists.
This approach is sometimes useful,
but it does not scale well.
As graph structures become more complex,
the number and variety of intermediate graph fragments also increase,
and requiring separate user-defined types for all of them
places an undesirable burden on the programmer.
Moreover,
such auxiliary type definitions obscure the relationship between
an incomplete graph fragment and the complete graph type from which it is derived.

This chapter addresses this problem by introducing
\emph{linear implication types}, written \(\multimap\).
Intuitively,
a linear implication type describes an incomplete graph structure:
a value of type \(A \multimap B\) can be understood as a graph fragment
that becomes a value of type \(B\) when supplied with a missing graph of type \(A\).
Thus,
linear implication types provide a uniform way to describe graph contexts
or partially constructed graphs without requiring the user to define
a new graph type for each such context.

The introduction of linear implication types significantly increases
the expressive power of the type system.
It enables the typing of intermediate graph structures
that arise naturally during graph pattern matching and graph transformation,
thereby addressing Requirement~T2 stated in \cref{ch:untyped}.
At the same time,
the extension is not merely a matter of adding a new type constructor.
Because graph fragments interact through free links and fusions,
linear implication types must be designed carefully so that
graph composition,
pattern matching,
and substitution remain compatible with type safety.

It is also important to clarify the scope of this additional expressive power.
Linear implication types are intended to make incomplete graph structures explicit
in types;
they should not act as a hidden mechanism for typing ordinary complete graph
values.
For this reason,
we also prove a conservativity result:
if a judgement whose assumptions and conclusion do not mention linear
implication types is derivable in the extended system,
then it can be derived without using the linear implication rules.
This shows that linear implication types extend the system precisely for
incomplete graph structures,
while leaving the typing power for ordinary complete graph values unchanged.

The remainder of this chapter is organised as follows.
\Cref{sec:ext-type} introduces linear implication types
and presents the extended typing rules.
\Cref{sec:li-free} discusses the design of free links
in the presence of linear implication types,
highlighting subtle interactions with graph composition.
\Cref{sec:sec3-limp-example} presents representative typing examples,
including detailed proof trees that illustrate the expressiveness
of the extended type system.
\Cref{sec:sec3-soundness} proves type soundness for the extended type system,
showing that preservation and progress continue to hold in the presence of
linear implication types and dynamic checks in case expressions.
Finally,
\cref{sec:elim-limprules} proves an eliminability result for linear implication
rules.
It shows that, for judgements whose assumptions and conclusions do not mention
linear implication types, any derivation in the extended system can be transformed
into one that does not use the linear implication rules.
This establishes that the extension is conservative with respect to complete
graph types.

\section{The Extended Type System}\label{sec:ext-type}

We extend the type system of \lgt
by incorporating the \emph{linear implication type}
\[(\{\taus\} \multimap \tau)\{\Xs\}\]
into the type system.
The $\{\taus\}$ on the left operand of $\multimap$ is a multiset
that allows us to reorder its elements whenever necessary.
Let $\emptyset$ stand for an empty $\{\taus\}$.
The necessity of explicitly specifying the set of free links \(\{\Xs\}\)
will be discussed later in \cref{sec:li-free}.
Unlike atoms and arrow types, whose free links are represented as ordered
sequences, the free links of a linear implication type form an unordered set.
Accordingly, atoms and arrow types annotate free links using parentheses
``('' and ``)'', reflecting their ordered nature, whereas linear implication
types use braces ``\{'' and ``\}'' to emphasise that free links are treated
as a set.

The incorporation of this type is motivated by the \emph{Magic Wand}
operator in Separation Logic~\cite{separation-logic}
and the \emph{Difference Type} in LMNtalGG~\cite{yamamoto2024}.
This extension provides a formal mechanism for explicitly managing
the type of incomplete structures.

As will be explained shortly, we adopt this form of linear implication
(as opposed to a standard form and currying
$\tau_1 \multimap \tau_2 \multimap \dots \multimap \tau_n$)
to smoothly relate it to production rules.

We refine the type in \cref{OriginalType} as follows.
\begin{mydef}[Syntax of the Type, Revised]
  \[\tau ::=
    \LI{\taus}{\tau}{\Xs}
    \mid (\tau \to \tau)(\Xs)
    \mid \alpha (\Xs) \qedhere
  \]
\end{mydef}

In order to keep the theory not too complicated, henceforth we assume that
the above extension does not apply to types used in the following contexts
(though we believe that this restriction is not essential):
\begin{enumerate}
  \item
        types occurring in the left- and right-hand sides of $\multimap$ and $\to$
        are $\multimap$-free,
  \item types occurring in the right-hand sides of production rules
        (\cref{ProductionRule}) are $\multimap$-free, and
  \item the first subexpression of a \textbf{case} expression cannot be
        of a linear implication type.
\end{enumerate}
The above three restrictions are essentially
to avoid the generation and handling
of curried forms of $\multimap$ which we are
going to dispense with (as explained above) in proofs etc.

The linear implication type \(\LI{\seq{\tau_1}}{\tau_2}{\Xs}\)
represents a structure with free links \(\Xs\),
which attains the type \(\tau_2\) when any (sub)graphs of types
\(\seq{\tau_1}\) is supplied.
This differs fundamentally from arrow types \((\tau_1 \to \tau_2)(\Xs)\).
Unlike the function type \((\taus \to \tau)(\Xs)\),
which represent the type of functions that contain expressions,
the linear implication type \(\LI{\taus}{\tau}{\Xs}\)
is for operations on data structures based on
the decomposition and composition of subgraphs.

We give an intuitive explanation of how linear implication
enables the typing of data structures with missing elements.

A doubly linked list with a missing tail list, namely \(\Cons 2\) and \(\Nil\),
is depicted as follows.
\begin{align}\label{eq:dbllist-diff-eg-vis}
   & \emptyset \vdash_{\Pdl}
  \tikz[baseline=-0.5ex, scale=1]{%
    \mkcons{1}{W_1}{}{(0,0)}{1}
    \begin{scope}[shift={(1.5,0)}]
      \node [atom,ghost] (C2)  at (0, 0)    {C};
      \coordinate (p2AR) at ($(C2)+(0,0.4)$) {};
      \coordinate (p2AL) at ($(C2)+(0,0.4)$) {};
      \coordinate (p2R) at ($(p2AR)+( 0.5,0.4)$);
      \coordinate (p2L) at ($(p2AL)+(-0.5,0.4)$);
      \node [atom,ghost] (N2)  at ($(C2)+(0,-1)$) {2};
      \draw [ghost] (C2) to (N2);
      \node [port,text=ghostcolor] at (-0.375, -0.150)   {4};
      \node [port,text=ghostcolor] at (+0.150, +0.375)   {2};
      \node [port,text=ghostcolor] at (+0.375, -0.150)   {1};
      \node [port,text=ghostcolor] at (-0.150, -0.375)   {3};
      \node [port,text=ghostcolor] at (-0.150, +0.375-1) {1};
      \node [link] at (0, 1) {$W_2$};
      \node [text=ghostcolor,link] at (0.500, -0.5) {$(W_3)$};
    \end{scope}
    \begin{scope}[shift={(3,0)}]
      \node [atom,ghost] (C4)  at (0, 0)    {N};
      \coordinate (p4AL) at ($(C4)+(0,0.4)$) {};
      \coordinate (p4R) at ($(p4AL)+( 0.5,0.4)$);
      \coordinate (p4L) at ($(p4AL)+(-0.5,0.4)$);
      \node [port,text=ghostcolor] at (+0.150, +0.375) {2};
      \node [port,text=ghostcolor] at (-0.375, -0.150) {1};
    \end{scope}
    %
    \node [link] (C0) at (-1.0, 0.8)  {$Z$};
    \node [link] (LX) at (-1.0, 0.0)  {$X$};
    \node [link,text=ghostcolor] (LY) at (4, 0.8)   {$Y$};
    %
    \draw (C1) to  (LX);
    \mkconslinkleft{0}{1}
    \mkconslink{1}{2}
    \draw [ghost] (p2AL) to (C2);
    %
    \draw [in=180, out=0, looseness=1.50,ghost] (C2) to (p4L);
    \draw [in=90, out=0, rounded corners,ghost] (p4L) to (p4AL);
    \draw [ghost] (p4AL) to (C4);
    \draw [line width=4pt,white,in=0, out=180, looseness=1.50] (C4) to (p2R);
    \draw [in=0, out=180, looseness=1.50,ghost] (C4) to (p2R);
    \draw [in=90, out=180, rounded corners,ghost]  (p2R) to (p2AL);
    %
    \draw [in=0, out=180, looseness=1.50,ghost] (LY) to (p4R);
    \draw [in=90, out=180, rounded corners,ghost]  (p4R) to (p4AL);
  }
  :
  \dblltype{Z}{Y}{W_2}
  \multimap
  \dblltype{X}{Y}{Z}
\end{align}

A leaf-linked tree with two missing leaves, namely \(\LeafM 1\) and \(\Leaf 2\),
can be visualised as follows.
\begin{align}\label{eq:lltree-diff-eg-vis}
   & \emptyset \vdash_{\Pdl}
  \tikz[baseline=-0.5ex, scale=1]{%
    \begin{scope}[shift={(0, 2)}]
      \node [atom] (node1) at (0,0) {N};
      \node [port] at (0.100, 0.395)   {3};
      \node [port] at (-0.395, -0.100) {1};
      \node [port] at (0.395, -0.100)  {2};
    \end{scope}
    \begin{scope}[shift={(-1.4, 1.2)}]
      \node [atom] (node2) at (0,0) {N};
      \node [port] at (0.100, 0.395)   {3};
      \node [port] at (-0.395, -0.100) {1};
      \node [port] at (0.395, -0.100)  {2};
      \node [link,text=gray] at (0.500, 0.800) {$(W_1)$};
    \end{scope}
    \begin{scope}[shift={(1.4, 1.2)}]
      \node [atom] (node3) at (0,0) {N};
      \node [port] at (0.100, 0.395)   {3};
      \node [port] at (-0.395, -0.100) {1};
      \node [port] at (0.395, -0.100)  {2};
      \node [link,text=gray] at (-0.500, 0.800) {$(W_2)$};
    \end{scope}
    \begin{scope}[shift={(-0.8, 0)}]
      \node [atom] (node4) at (0,0) {N};
      \node [port] at (0.100, 0.395)   {3};
      \node [port] at (-0.395, -0.100) {1};
      \node [port] at (0.395, -0.100)  {2};
      \node [link,text=gray] at (0.100, 0.700)   {$(W_3)$};
    \end{scope}
    %
    \begin{scope}[shift={(-2.6, 0)   }]
      \node [atom,ghost,fill=ghostcolor,text=white] (leaf1) at (0, 0) {L};
      \node [port,text=ghostcolor] at (0.100, 0.395)   {3};
      \node [port,text=ghostcolor] at (0.395, 0.100)   {2};
      \node [port,text=ghostcolor] at (-0.100, -0.395) {1};
      \node [port,text=ghostcolor] at (-0.100, -0.7)   {1};
      \node [text=ghostcolor,link] at (-0.510, -0.5)   {$(W_8)$};
      \node [] at (0, 0.9)   {$L$};
    \end{scope}
    \begin{scope}[shift={(-1.6, -1.3)}]
      \node [atom,ghost] (leaf2) at (0, 0) {L};
      \node [port,text=ghostcolor] at (0.100, 0.395)   {3};
      \node [port,text=ghostcolor] at (0.395, 0.100)   {2};
      \node [port,text=ghostcolor] at (-0.100, -0.395) {1};
      \node [port,text=ghostcolor] at (-0.100, -0.7)   {1};
      \node [link] at (-0.35, 0.9)   {$W_4$};
      \node [text=ghostcolor,link] at (0.410, -0.5) {$(W_9)$};
    \end{scope}
    \begin{scope}[shift={(-0.395, -1.3)}]
      \node [atom] (leaf3) at (0, 0) {L};
      \node [port] at (0.100, 0.395)   {3};
      \node [port] at (0.395, 0.100)   {2};
      \node [port] at (-0.100, -0.395) {1};
      \node [port] at (-0.100, -0.7)   {1};
      \node [link] at (0.30, 0.820) {$W_5$};
      \node [link,text=gray] at (0.495, -0.5)  {$(W_{10})$};
    \end{scope}
    \begin{scope}[shift={(1.1, -0.7)}]
      \node [atom] (leaf4) at (0,0) {L};
      \node [port] at (0.100, 0.395)   {3};
      \node [port] at (0.395, 0.100)   {2};
      \node [port] at (-0.100, -0.395) {1};
      \node [port] at (-0.100, -0.7)   {1};
      \node [link,text=gray] at (0.300, 1)    {$(W_6)$};
      \node [link,text=gray] at (0.495, -0.5) {$(W_{11})$};
    \end{scope}
    \begin{scope}[shift={(2.6, 0) }]
      \node [atom] (leaf5) at (0, 0) {L};
      \node [port] at (0.100, 0.395)   {3};
      \node [port] at (0.395, -0.100)  {2};
      \node [port] at (-0.100, -0.395) {1};
      \node [port] at (-0.100, -0.7)   {1};
      \node [link,text=gray] at (-0.100, 0.9)    {$(W_7)$};
      \node [link,text=gray] at (0.495, -0.5) {$(W_{12})$};
    \end{scope}
    \node [atom,below=0.8cm,ghost] (num1) at (leaf1) {1};
    \node [atom,below=0.8cm,ghost] (num2) at (leaf2) {2};
    \node [atom,below=0.8cm]       (num3) at (leaf3) {3};
    \node [atom,below=0.8cm]       (num4) at (leaf4) {4};
    \node [atom,below=0.8cm]       (num5) at (leaf5) {5};
    \node [link,text=white] (LL) at (-4, 0)   {$L$};
    \node [link] (LX) at (0, 3)   {$X$};
    \node [link] (LR) at (4, 0)   {$R$};
    \draw (node1) -- (LX);
    \draw (node1) -- (node2);
    \draw (node1) -- (node3);
    \draw (node2) -- (node4);
    \draw [looseness=1.50,in=90,out=-130] (node2) to (leaf1);
    \draw [looseness=1.50,in=90,out=-130] (node4) to (leaf2);
    \draw [looseness=1.50,in=90,out=-50 ] (node4) to (leaf3);
    \draw [looseness=1.50,in=90,out=-130] (node3) to (leaf4);
    \draw [looseness=1.50,in=90,out=-50 ] (node3) to (leaf5);
    \draw [in=90,out=30,looseness=1.50] (LL) to (leaf1.north);
    \draw [in=90,out=30,looseness=1.50] (leaf1) to (leaf2);
    \draw [in=90,out=30,looseness=1.50] (leaf2) to (leaf3);
    \draw [in=90,out=30,looseness=1.50] (leaf3) to (leaf4);
    \draw [in=90,out=30,looseness=1.50] (leaf4) to (leaf5);
    \draw (leaf5) to (LR);
    \draw [ghost] (leaf1) -- (num1);
    \draw [ghost] (leaf2) -- (num2);
    \draw (leaf3) -- (num3);
    \draw (leaf4) -- (num4);
    \draw (leaf5) -- (num5);
  }
  \notag                     \\
   & :
  \left\{
  \lltrtype{L}{L}{W_4}
  \;\raisebox{-3.8ex}{,}\;
  \lltrtype{W_4}{W_4}{W_5}
  \right\}
  \multimap
  \lltrtype{X}{L}{R}
\end{align}

Using the linear implication type,
we can successfully annotate the types of the graph variables in
$\eDLpop$ and $\eLLT$ defined in \eqref{eq:dbllist-removal-eg-vis} and \eqref{eq:lltree-traverse-eg-vis}.
For example, the graph variable
\(y[W_1,W_2,X,Z]\) in $\eDLpop$
can be type-annotated as follows.
\begin{align}
  \tikz[baseline=-0.5ex, scale=1]{%
    \mkctxn{1}{y}{(0,0)}{}{W_2}
    \begin{scope}[shift={(1.5,0)}]
      \node [atom,ghost] (C2)  at (0, 0)    {C};
      \coordinate (p2A) at ($(C2)+(0,0.4)$) {};
      \coordinate (p2R) at ($(p2A)+( 0.5,0.4)$);
      \coordinate (p2L) at ($(p2A)+(-0.5,0.4)$);
      \node [ctxn,ghost] (N2)  at ($(C2)+(0,-1)$) {$w$};
      \draw [ghost] (C2) to (N2);
      \node [port,text=ghostcolor] at (-0.375, -0.150)   {4};
      \node [port,text=ghostcolor] at (+0.150, +0.375)   {2};
      \node [port,text=ghostcolor] at (+0.375, -0.150)   {1};
      \node [port,text=ghostcolor] at (-0.150, -0.375)   {3};
      \node [port,text=ghostcolor] at (-0.150, +0.375-1) {1};
      \node [link] at (0, 1) {$W_1$};
      \node [text=ghostcolor,link] at (0.500, -0.5) {$(W_3)$};
    \end{scope}
    \begin{scope}[shift={(3,0)}]
      \node [atom,ghost] (C3) at (0, 0) {N};
      \coordinate (p3A) at ($(C3)+(0,0.4)$) {};
      \coordinate (p3R) at ($(p3A)+( 0.5,0.4)$);
      \coordinate (p3L) at ($(p3A)+(-0.5,0.4)$);
      \node [port,text=ghostcolor] at (+0.150, +0.375) {4};
      \node [port,text=ghostcolor] at (-0.375, -0.150) {2};
    \end{scope}
    %
    \node [link] (C0) at (-1, 0.8)  {$Z$};
    \node [link] (LX) at (-1, 0)  {$X$};
    \node [text=ghostcolor,link] (LY) at (4, 0.8)   {$Y$};
    \mkconslinkleft{0}{1}
    \draw (C1) to  (LX);
    \draw [in=180, out=0, looseness=1.50] (C1) to (p2L);
    \draw [in=90, out=0, rounded corners] (p2L) to (p2A);
    \draw [ghost] (p2A) to (C2);
    \draw [line width=4pt,white,in=0, out=180, looseness=1.50] (C2) to (p1R);
    \draw [in=0, out=180, looseness=1.50] (C2) to (p1R);
    \draw [in=90, out=180, rounded corners]  (p1R) to (p1AR);
    \draw [in=180, out=0, looseness=1.50,ghost] (C2) to (p3L);
    \draw [in=90, out=0, rounded corners,ghost] (p3L) to (p3A);
    \draw [ghost] (p3A) to (C3);
    \draw [line width=4pt,white,in=0, out=180, looseness=1.50] (C3) to (p2R);
    \draw [in=0, out=180, looseness=1.50,ghost] (C3) to (p2R);
    \draw [in=90, out=180, rounded corners,ghost]  (p2R) to (p2A);
    \draw [in=0, out=180, looseness=1.50,ghost] (LY) to (p3R);
    \draw [in=90, out=180, rounded corners,ghost]  (p3R) to (p3A);
  }
  \!\!:
  \dblltype{W_2}{Y}{W_1}
  \multimap
  \dblltype{X}{Y}{Z}
  \label{eq:dbllist-diff-gctx-eg-vis}
\end{align}

This represents that the graph variable \(y[W_1,W_2,X,Z]\)
matches a subgraph that is an incomplete graph
that will become a doubly linked list
if we supply a doubly linked list at the end.

Formally, the graph variable $y$ 
can be type-annotated as
\begin{align}\label{eq:dbllist-diff-gctx-eg}
  y[W_1,W_2,X,Z]:
  \LI{\mathit{dbllist}(W_2,Y,W_1)}{\mathit{dbllist}(X,Y,Z)}{W_1,W_2,X,Z}.
\end{align}

We define the free links of the linear implication type
$\LI{\seq{\tau_1}}{\tau_2}{\Xs}$ as
\[\fn(\LI{\seq{\tau_1}}{\tau_2}{\Xs}) \triangleq \{\Xs\}.\]
We discuss in \cref{sec:li-free}
why linear implication types require the explicit notation \(\Xs\) of
free links.

For defining the typing rules for the linear implication type later,
we define link substitution for linear implication types as follows.

\begin{mydef}[Link Substitution, Extended]
  Given a substitution \(\sigma\),
  the substitution of a linear implication type is defined as
  \[\begin{array}[b]{l}
      (\LI{\tau_1, \dots, \tau_n}{\tau}{X_1, \dots, X_n})
      \sigma \triangleq \\
      \quad \LI{\tau_1 \sigma, \dots, \tau_n \sigma}{\tau\sigma}{
        X_1\sigma, \dots, X_n\sigma}.
    \end{array} \qedhere \]
\end{mydef}

We then present the newly added typing rules.
\begin{mydef}[Newly Added Typing Rules]
  The newly added typing rules that incorporate linear implication types
  are defined in \cref{table:revised-typing-rules}.
\end{mydef}

\begin{figure}[htbp]
  \begin{framed}
    \bigskip
    \bigskip

    \begin{prooftree}
      \AXC{}
      \RightLabel{\TyLIIntro{}}
      \UIC{$\Gamma
          \vdash_P
          (C(\Ys), \seq{W \bowtie U})
          : \LI{\taus}{\alpha(\Xs)}{\Zs}$}
    \end{prooftree}
    \raggedright where
    $(\alpha (\Xs) \lto \nu \Zs.(C(\Ys), \seq{W \bowtie U}, \taus))\in P$
    and
    $\{\Zs\} = \{\Ys\} \cup \mathit{fn}(\seq{W \bowtie U})$.

    \bigskip
    \bigskip
    \bigskip

    \begin{prooftree}
      \AXC{$\Gamma \vdash_P T_1 : \LI{\tau_2, \seq{\tau_1}}{\tau_1}{\Xs}$}
      \AXC{$\Gamma \vdash_P T_2 : \LI{\seq{\tau_2}}{\tau_2}{\Ys}$}
      \RightLabel{$\TyLITrans{}$}
      \BIC{$\Gamma \vdash_P \nu \Zs.(T_1, T_2) :
          \LI{\seq{\tau_1},\seq{\tau_2}}{\tau_1}{\Ws}$}
    \end{prooftree}%

    \raggedright where
    \[\begin{array}[t]{cl@{\qquad}clll}
        \text{(i)}   & \{\Xs, \Ys\} \setminus \{\Zs\} = \{\Ws\},
                     &
        \text{(ii)}  & \{\Xs\} \cap \{\Ys\} \subseteq \fn(\tau_2),                                          \\
        \text{(iii)} & \{\Xs\} \cap \fn(\seq{\tau_2}) \subseteq \fn(\tau_2),
                     &
        \text{(iv)}  & \paren{\{\Ys\} \cup \fn(\seq{\tau_2})} \cap \fn(\seq{\tau_1}) \subseteq \fn(\tau_2), \\
        \text{(v)}   & \fn(\tau_1) \subseteq \{\Ws\} \cup \fn(\seq{\tau_1},\seq{\tau_2}),                   \\[2pt]
      \end{array}\]
    \raggedright
    and $\seq{\tau_i}$ denotes $\tau_{i1}, \dots, \tau_{in}$
    and is different from $\tau_i$.

    \bigskip
    \bigskip
    \bigskip

    \begin{prooftree}
      \AXC{$\Gamma \vdash_P
          T : \LIempty{\tau}{\Ys}$}
      \RightLabel{\TyLIElimZ{}}
      \UIC{$\Gamma \vdash_P \nu\Zs.T : \tau$}
    \end{prooftree}
    \raggedright
    where $\fn(T) = \{\Ys\}$
    and $\fn(T) \setminus \{\Zs\} = \fn(\tau)$.

    \bigskip
    \bigskip
    \bigskip

    \begin{prooftree}
      \AXC{$\Gamma \vdash_P T : \tau$}
      \RightLabel{\TyLIIntroZ{}}
      \UIC{$\Gamma \vdash_P
          T : \LIempty{\tau}{\Ys}$}
    \end{prooftree}
    \raggedright
    where $\fn(T) = \fn(\tau) = \{\Ys\}$.

    \bigskip
    \bigskip
  \end{framed}
  \caption{Newly added typing rules.}
  \label{table:revised-typing-rules}
\end{figure}

Let us explain the individual rules.

\TyLIIntro{} is intuitively read as: ``a graph (atom) $C(\Ys)$ has
the type which states that $C(\Ys)$ is obtained from a graph of type
$\alpha(\Xs)$ by removing graphs of types $\tau_1, \dots, \tau_n$.''
\TyLIIntro{} is designed in this form to smoothly relate production
rules and the linear implication type.  Indeed, the form of the linear
implication type that allows a multiset of types as its antecedent is
exactly motivated by this relation.

\TyLITrans{} states the transitivity of linear implication;
it states that two linear implications can be fused by cancelling $\tau_2$.
The free link condition (i) comes from the constraint that free links of
the subject and its type must have the same set of free links.
The remaining conditions (ii)–(v) are technical side constraints
whose detailed roles are explained in \cref{sec:elim-limprules}:
Conditions (ii), (iv), and (v) are applied in Hyperlink Creation Localisation (Step~2),
and Condition (iii) is used in Proof Tree Normalisation (Step~3)
to ensure that condition (ii) holds.
where we also clarify how and where each condition is employed.
\TyLITrans{} is thought of a generalisation of the standard elimination
rule of linear implication.

These introduction and elimination rules are accompanied by two `base
cases', namely \TyLIElimZ{} and \TyLIIntroZ{}, for handling linear
implication with empty antecedents.

The standard elimination rule of linear implication
could be represented as follows.
\begin{mydef}[Elimination of Linear Implication]
  ~ 
  \begin{prooftree}
    \AXC{$\begin{array}{@{}c@{}}
          \Gamma \vdash_P
          y[\Ys] : \LI{\tau_1, \dots, \tau_n}{\tau}{\Ys}
          \\
          \Gamma \vdash_P T_1 : \tau_1
          \quad \dots \quad
          \Gamma \vdash_P T_n : \tau_n
        \end{array}$}
    \RightLabel{\TyLIE{}}
    \UIC{$\Gamma
        \vdash_P \nu \Xs.(y[\Ys],
        T_1, \dots, T_n) : \tau$}
  \end{prooftree}
  where $\fn(\nu \Xs.(y[\Ys],T_1, \dots, T_n)) = \fn(\tau)$.
\end{mydef}

We note that this rule
is admissible and can be derived easily using \TyLITrans{},
\TyLIElimZ{}, and \TyLIIntroZ{}.
The constraint on free links,
$\fn(\nu \Xs.(y[\Ys],T_1, \dots, T_n)) = \fn(\tau)$,
is derived from the
free link conditions of \TyLITrans{}, \TyLIElimZ{}, and \TyLIIntroZ{}.

The key typing rule of \lgt is \TyProd{} in \cref{table:typing-rules}.
This rule becomes admissible in the linear-implication extended system.
Later, in \cref{sec:elim-limprules},
we use this result in the reverse direction to establish that the extension with
linear implication types is conservative for complete graph types.
\begin{theorem}[Admissibility of \TyProd{}]\label{thm:typrod-admissibility}
  For any typing derivation in \lgt,
  the same typing judgement can be derived without using \TyProd{}.
\end{theorem}
\begin{proof}
  \Cref{fig:ty-prod-admissible-eg} shows how \TyProd{} can be derived
  from \TyLIIntro{}, \TyLIIntroZ{}, \TyLITrans{}, \TyLIElimZ{} for any fixed $n$.
\end{proof}

\begin{sidewaysfigure}[htbp]
  \hrulefill

  \raggedright
  \TyProd{} (Restatement of \cref{table:typing-rules}):
  \begin{prooftree}
    \AXC{$\paren{\alpha (\Xs) \lto \nu
          \Zs.(C(\Ys), \seq{U \bowtie V}, \tau_1, \dots, \tau_n)} \in P$}
    \AXC{$\Gamma \vdash_P T_1 : \tau_1$}
    \AXC{$\dots$}
    \AXC{$\Gamma \vdash_P T_n : \tau_n$}
    \RightLabel{$\TyProdA{}$}
    \QuaternaryInfC{$\Gamma \vdash_P
        \nu \Zs.(C(\Ys), \seq{U \bowtie V}, T_1, \dots, T_n) : \alpha (\Xs)$}
  \end{prooftree}

  \bigskip

  \(\TyProd{}\) derived from other rules:
  \begin{prooftree}
    \AXC{$\paren{\alpha (\Xs) \lto \nu
          \Zs.(C(\Ys), \seq{U \bowtie V}, \tau_1, \dots, \tau_n)} \in P$}
    \RightLabel{$\TyLIIntroA{}$}
    \UIC{$\ml{\Gamma \vdash_P (C(\Ys), \seq{U \bowtie V})
          : \LI{\tau_1,\tau_2,\dots,\tau_n}{\alpha(\Xs)}{\Us,\Vs,\Ys}}$}
    \AXC{$\Gamma \vdash_P T_1 : \tau_1$}
    \RightLabel{$\TyLIIntroZA{}$}
    \UIC{$\Gamma \vdash_P T_1
        : \LIempty{\tau_1}{\Zs_1}$}
    \RightLabel{$\TyLITransA{}$}
    \BIC{$\mlc{\Gamma \vdash_P (C(\Ys), \seq{U \bowtie V},T_1)
          : \LI{\tau_2,\dots,\tau_n}{\alpha(\Xs)}{\Zs_1,\Us,\Vs,\Ys}
          \\
          \ddots
          \\
          \Gamma \vdash_P (C(\Ys), \seq{U \bowtie V},T_1,\dots,T_{n-1})
          : \LI{\tau_n}{\alpha(\Xs)}{\Zs_{n-1},\dots,\Zs_1,\Us,\Vs,\Ys}
        }$}
    \AXC{$\Gamma \vdash_P T_n : \tau_n$}
    \RightLabel{$\TyLITransA{}$}
    \UIC{$\Gamma \vdash_P T_n : \LIempty{\tau_n}{\Zs_n}$}
    \RightLabel{$\TyLITransA{}$}
    \BIC{$\Gamma \vdash_P (C(\Ys), \seq{U \bowtie V},T_1, \dots,T_{n-1},T_n)
        : \LIempty{\alpha(\Xs)}{\Zs_n,\Zs_{n-1},\dots,\Zs_1,\Us,\Vs,\Ys}$}
    \RightLabel{$\TyLIElimZA{}$}
    \UIC{$\Gamma \vdash_P \nu \Zs.
        (C(\Ys), \seq{U \bowtie V},T_1,\dots, T_{n-1},T_n)
        : \alpha(\Xs)$}
  \end{prooftree}

  Here,
  \TyProd{},  \TyLIIntro{},  \TyLITrans{},  \TyVar{},  \TyLIElimZ{},  \TyLIIntroZ{}, andare abbreviated as
  \TyProdA{}, \TyLIIntroA{}, \TyLITransA{}, \TyVarA{}, \TyLIElimZA{}, and \TyLIIntroZA{} respectively.

  \hrulefill
  \caption{Admissibility of \TyProd{}.}
  \label{fig:ty-prod-admissible-eg}
\end{sidewaysfigure}

We also modify the \TyCase{} rule
so that the consistency between the type expected by the pattern
and the type of the matching expression in a \textbf{case} expression
can be checked.
We introduce the revised typing rule for \textbf{case} expressions,
shown in \cref{def:revised-ty-case}.

\begin{mydef}[Revised \TyCase{}]\label{def:revised-ty-case}
  ~ 
  \begin{prooftree}
    \AXC{$\begin{array}{r@{~}l}
          \Gamma                  & \vdash_P e_1 : \tau_1
          \\
          \Gamma, \CollectVars(T) & \vdash_P e_2 : \tau_2
        \end{array}$}
    \AXC{$\begin{array}{r@{~}l}
          \CollectVars(T) & \vdash_P T : \tau_1
          \\
          \Gamma          & \vdash_P e_3 : \tau_2
        \end{array}$}
    \RightLabel{\TyCase{}}
    \BIC{$\Gamma \vdash_P (\caseof{e_1}{T}{e_2}{e_3}) : \tau_2$}
  \end{prooftree}%
\end{mydef}

\TyCase{} is the revised typing rule for case expressions.
This rule defines the type assignment for conditional expressions
that perform structural analysis on graph-based terms.
Given an expression \(e_1\) of type \(\tau_1\),
a pattern \(T\) that also has type \(\tau_1\),
and two branches \(e_2\) and \(e_3\),
the expression \(\caseof{e_1}{T}{e_2}{e_3}\) is
given
type \(\tau_2\).

\section{Free Links of Linear Implication Types}\label{sec:li-free}

In this section, we examine the need of the explicit notation of free
links \(\{\Xs\}\) in linear
implication types \(\LI{\tau_1}{\tau_2}{\Xs}\).

The free links of $\lambda$-abstraction atoms are distinct from the free
links of the input--output graph of a function.
For linear implication types, however, one might expect that the free
links can be inferred from the free links of their constituent types,
since some of them originate from the argument graph.
For example, consider the type annotation \eqref{eq:dbllist-diff-gctx-eg}.
The free links of the expression \(y[W_1,W_2,X,Z]\) is
\(\{W_1,W_2,X,Z\}\).
One may think that this is obtained as the symmetric difference of the
sets of free links of the two operands, i.e.,
$(\{W_2,Y,W_1\}\setminus \{X,Y,Z\}) \cup
  (\{X,Y,Z\}\setminus \{W_2,Y,W_1\}$,
which gives \(\{W_1,W_2,X,Z\}\) in this particular example.
However, it is not obvious whether this idea generalises to all
cases.


To illustrate this point, let \(T_0\) be a graph (or
a graph variable) such that
\(\Gamma \vdash_P T_0:\LI{\tau_1}{\tau_2}{\Xs}\) for some
\(\Gamma\) and \(P\).
That is, \(T_0\) represents a graph that will acquire type
\(\tau_2\) upon supplying a graph of type \(\tau_1\).
Let \(T_1\) be a graph of type \(\tau_1\),
and \(T_2\) be a graph \(\nu \Xs.(T_0,T_1)\)
satisfying \(\fn(\nu \Xs.(T_0,T_1))=\fn(T_2)\),
which, by the definition of \(T_0\), is given the type \(\tau_2\).

\begin{figure}[htb]
  \centering
  \begin{tikzpicture}[scale=0.5]
    \draw[dashed] (-0.5,-0.5) circle [radius=3.5];

    \coordinate (X1) at (-1, 4);
    \coordinate (X2) at (-1, -3);
    \coordinate (X3) at (-5, 1);
    \coordinate (X4) at (-3, -1);
    \coordinate (X5) at (3, -3);
    \coordinate (X6) at (1, -3);

    \filldraw[black] (X1) circle (4pt) node[anchor=east] {\(\Xs_1\)};
    \filldraw[black] (X2) circle (4pt) node[anchor=east] {\(\Xs_2\)};
    \filldraw[black] (X3) circle (4pt) node[anchor=east] {\(\Xs_3\)};
    \filldraw[black] (X4) circle (4pt) node[anchor=north] {\(\Xs_4\)};
    \filldraw[black] (X5) circle (4pt) node[anchor=north] {\(\Xs_5\)};
    \filldraw[black] (X6) circle (4pt) node[anchor=north] {\(\Xs_6\)};

    \draw[] (0,0) -- (X3);
    \draw[] (0,0) -- (X4);
    \draw[] (0,0) -- (X5);
    \draw[] (0,0) -- (X6);
    \fill [white] (0,0) circle [radius=2.25];

    \draw[] (0, 2.25) -- (X1);
    \draw[] (-1, 1.83) -- (X1);

    \draw[] (0, -2.25) -- (X2);
    \draw[] (-1, -1.83) -- (X2);

    \coordinate (p1) at (-0.9,1.8);
    \coordinate (p2) at (-0.9,-1.8);

    \draw[] (p1)
    to [out=160,in=45] (-1.59099025767,1.59099025767)
    arc (135:225:2.25)
    to [out=-45,in=-160] (p2)
    to [out=180-160,in=180+160] cycle;

    \draw[fill=black!10]
    (0,2.25)
    to [out=180,in=180-60] (-0.25,1.8)
    to [out=360-60,in=90] (0.35,0)
    to [out=180+90,in=60] (-0.25,-1.8)
    to [out=180+60,in=180] (0,-2.25)
    arc (-90:90:2.25)
    --cycle;

    \node at (-1,0) {\(T_0\)};
    \node at (1.5,0) {\(T_1\)};
    \node at (-2.5,1.5) {\(T_2\)};
  \end{tikzpicture}
  \caption{Free links of graphs visualised.}
  \label{fig:li-freelinks-vis}
\end{figure}
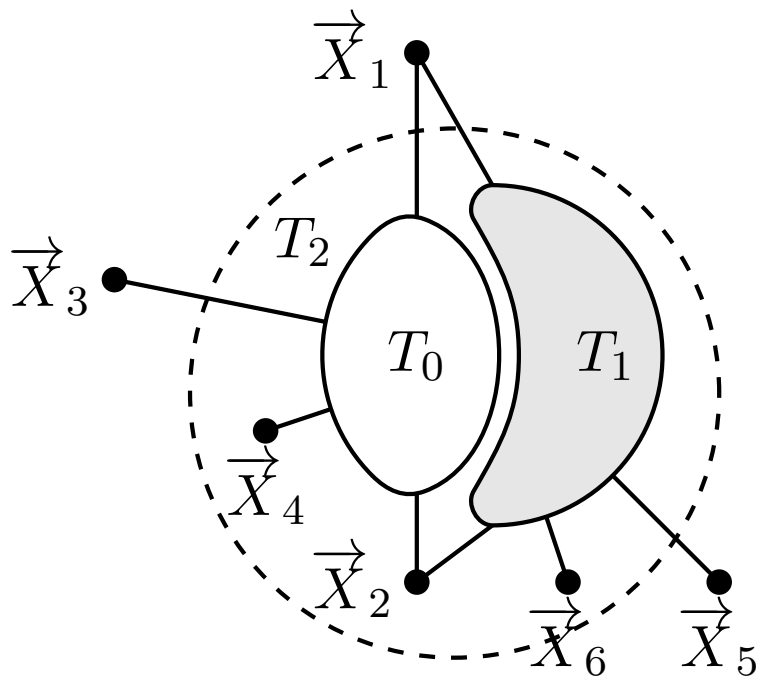

The free link sets of the expressions \(T_0\), \(T_1\), and
\(T_2 \triangleq \nu \Xs.(T_0,T_1)\)
an be classified as in
\cref{table:revision-comparison},
with each category labelled from 1 to 8.
Note that the free links \(\Xs_8\) do not occur anywhere and that
the free links \(\Xs_7\) are not possible by definition.
The remaining six categories are visualised in
\cref{fig:li-freelinks-vis}.

\begin{table}[htb]
  \caption{Free links of graphs enumerated.}
  \centering
  \begin{tabular}{ccccl}
    \toprule
              &
    \(\mathit{fn}(T_0)\)
              &
    \(\mathit{fn}(T_1)\)
              &
    \(\mathit{fn}(T_2)\)
    \\
    \(\Xs_1\) & \(\in\)    & \(\in\)    & \(\in\)    \\
    \(\Xs_2\) & \(\in\)    & \(\in\)    & \(\notin\) \\
    \(\Xs_3\) & \(\in\)    & \(\notin\) & \(\in\)    \\
    \(\Xs_4\) & \(\in\)    & \(\notin\) & \(\notin\) \\
    \(\Xs_5\) & \(\notin\) & \(\in\)    & \(\in\)    \\
    \(\Xs_6\) & \(\notin\) & \(\in\)    & \(\notin\) \\
    \(\Xs_7\) & \(\notin\) & \(\notin\) & \(\in\)    \\
    \(\Xs_8\) & \(\notin\) & \(\notin\) & \(\notin\) \\
    \bottomrule
  \end{tabular}
  \label{table:revision-comparison}
\end{table}

Among the free links, \(\Xs_2\),
\(\Xs_4\) and \(\Xs_6\) may be ignored as they
are absorbed by $\nu \Xs$.
It remains crucial to distinguish between \(\Xs_1\) and \(\Xs_5\), where
\(\Xs_1\) belongs to \(\fn(T_0)\), whereas \(\Xs_5\) does not.
However, using only the free link sets of \(T_1\) and \(T_2\),
it is not possible to differentiate between \(\Xs_1\) and \(\Xs_5\).
Consequently, even when \(\fn(\tau_1) = \fn(T_1)\) and \(\fn(\tau_2) =
\fn(T_2)\) hold,
the free link set of \(T_0: \tau_1 \multimap \tau_2\)
cannot be determined solely from $\tau_1$ and $\tau_2$.
To resolve this issue,
an explicit annotation of the free link set  \(\Xs \triangleq
\fn(T_0)\)
is incorporated into the type, denoted as \(\LI{\tau_1}{\tau_2}{\Xs}\).


The above observation is in contrast with
a typing framework of LMNtal (without hyperlinks)
that incorporates the notion of \emph{difference types}
\cite{yamamoto2024}
that can be thought of as a counterpart of to our linear implication
types for incomplete hypergraphs.
In \cite{yamamoto2024}, such explicit annotation is unnecessary
because the language enforces a structural constraint
whereby a link occurring exactly twice within a term is classified as
a local link, which means that any link that is a free link in both
\(T_0\) and \(T_1\) must necessarily be a local link in \(T_2\).
This property allows the set of free links in \(T_0\)
to be derived solely from the free link sets of \(T_1\) and \(T_2\)
as their symmetric difference.

\section{Examples}\label{sec:sec3-limp-example}

In this section, we present an application of the linear implication type.
We first show the typing of a segment of a doubly linked list,
and then a program that removes the last element of a doubly linked list.

\subsubsection{Doubly Linked List Segments}

First we revisit the production rules for doubly linked lists in
$\Pdl$ \eqref{eq:prod-rules-eg} (\cref{sec:core-type:examples}):
\begin{align}
  \begin{array}{@{}lll}
    \mathit{dbllist}(X,Y,Z)  \lto
    \Nil(X, Z), Y \bowtie Z \\[1mm]
    \quad \mid \nu W_1 W_2.(\Cons(W_1,X,W_2,Z), \mathit{nat}(W_1),\mathit{dbllist}(Z,Y,W_2)).
  \end{array}
\end{align}

Now we consider the typing of a fragment of a doubly linked list that consists of two adjacent $\Cons{}$ nodes with $\mathit{nat}$ elements:
\begin{equation}
  \begin{array}{l}
    \nu W_1. (\Cons(W_1, X, W_2, Z), 3(W_1)), \\
    \nu W_3. (\Cons(W_3, Z, W_4, W_2), 5(W_3)).
  \end{array}
\end{equation}

This graph has the following type under $\Pdl$, as shown in \cref{fig:diff-dbllist-typing-derivation-eg}:
\[\LI{\mathit{dbllist}(W_2,Y,W_4)}{\mathit{dbllist}(X,Y,Z)}{W_4,W_2,X,Z}.\]

\begin{figure}[htbp]
  \hrulefill

  \bigskip
  \bigskip

  \raggedright
  {
    \AXC{}
    \RightLabel{\TyLIIntroA{}}
    \UIC{$\ml{\vdash_{\Pdl} \Cons(W_1,X,W_2,Z) \\
          : \LI{\mathit{nat}(W_1), \mathit{dbllist}(Z,Y,W_2)}{\\
            \qquad \mathit{dbllist}(X,Y,Z)}{W_1,X,W_2,Z}}$}
    \AXC{}
    \RightLabel{\TyLIIntroA{}}
    \UIC{$\ml{\vdash_{\Pdl} 3(W_1) \\
          : \LIempty{\mathit{nat}(W_1)}{W_1}}$}
    \RightLabel{\TyLITransA{}}
    \BIC{
      \tikz[remember picture,baseline=(upper-root1.base)]
      \node [font=\normalsize] (upper-root1)
      {$\ml{\vdash_{\Pdl} \nu W_1. (\Cons(W_1,X,W_2,Z),3(W_1)) \\
            : \LI{\mathit{dbllist}(Z,Y,W_2)}{\mathit{dbllist}(X,Y,Z)}{X,W_2,Z}}$};
    }
    \DisplayProof
  }

  \bigskip
  \bigskip

  \raggedleft
  {
    \AXC{}
    \RightLabel{\TyLIIntroA{}}
    \UIC{$\ml{\vdash_{\Pdl} \Cons(W_3,Z,W_4,W_2) \\
          : \LI{\mathit{nat}(W_3), \mathit{dbllist}(W_2,Y,W_4)}{\\
            \qquad \mathit{dbllist}(Z,Y,W_2)}{W_3,Z,W_4,W_2}}$}

    \AXC{}
    \RightLabel{\TyLIIntroA{}}
    \UIC{$\ml{\vdash_{\Pdl} 5(W_3) \\
          : \LIempty{\mathit{nat}(W_3)}{W_3}}$}

    \RightLabel{\TyLITransA{}}
    \BIC{
      \tikz[remember picture,baseline=(upper-root2.base)]
      \node[inner sep=0pt] (upper-root2)
      {$\ml{\vdash_{\Pdl} \nu W_3. (\Cons(W_3,Z,W_4,W_2),5(W_3)) \\
            : \LI{\mathit{dbllist}(W_2,Y,W_4)}{\mathit{dbllist}(Z,Y,W_2)}{Z,W_4,W_2}}$};
    }
    \DisplayProof
  }

  \bigskip

  \centering
  \begin{prooftree}
    \AXC{
      \tikz[remember picture,baseline=(lower-hole1.base)]
      \node[draw, dashed, rectangle, inner sep=3pt, font=\normalsize] (lower-hole1)
      {$\color{white}Cons$};
    }
    \AXC{\hspace{8cm}}
    \AXC{
      \tikz[remember picture,baseline=(lower-hole2.base)]
      \node[draw, dashed, rectangle, inner sep=3pt, font=\normalsize] (lower-hole2)
      {$\color{white}Cons$};
    }
    \RightLabel{\TyLITransA{}}
    \TIC{$\ml{\vdash_{\Pdl}
          \nu W_1. (\Cons(W_1, X, W_2, Z), 3(W_1)),
          \nu W_3. (\Cons(W_3, Z, W_4, W_2), 5(W_3)) \\
          : \LI{\mathit{dbllist}(W_2,Y,W_4)}{\mathit{dbllist}(X,Y,Z)}{W_4,W_2,X,Z}}$}
  \end{prooftree}
  \begin{tikzpicture}[remember picture,overlay]

    \begin{scope}[even odd rule]
      \clip (current page.south west) rectangle (current page.north east)
      (-0.3\linewidth,3.4) rectangle (0.3\linewidth,7.2);
      \draw[->,thick,dashed] (upper-root1.south) to[looseness=0.5,out=-90,in=90] (lower-hole1.north);
      \draw[->,thick] (upper-root2.south) to[looseness=1,out=-90,in=90] (lower-hole2.north);
    \end{scope}
  \end{tikzpicture}

  Here,
  \TyLIIntro{}  and \TyLITrans{}  andare abbreviated as
  \TyLIIntroA{} and \TyLITransA{} respectively.

  \hrulefill
  \caption{Typing derivation of a fragment of a doubly linked list.}
  \label{fig:diff-dbllist-typing-derivation-eg}
\end{figure}

Next, we consider the type of a complete doubly linked list adding a
$\Nil$ node to the above fragment:
\begin{equation}
  \begin{array}{l}
    \nu W_2 W_4. (                                    \\
    \quad \nu W_1. (\Cons(W_1, X, W_2, Z), 3(W_1)),   \\
    \quad \nu W_3. (\Cons(W_3, Z, W_4, W_2), 5(W_3)), \\
    \quad \Nil(W_2, W_4), Y \bowtie W_4               \\
    ).
  \end{array}
\end{equation}

This graph has type $\mathit{dbllist}(X,Y,Z)$.
We give the typing derivation of the graph in \cref{fig:dbllist-typing-derivation-eg}.
Note that we can type a complete graph without missing parts in this way without using \TyProd{},
which is consistent with the admissibility of \TyProd{} established in \cref{thm:typrod-admissibility}.

\begin{figure}[htbp]
  \hrulefill

  \begin{prooftree}
    \AXC{(\cref{fig:diff-dbllist-typing-derivation-eg})}
    \AXC{}
    \RightLabel{\TyLIIntroA{}}
    \UIC{$\ml{\vdash_{\Pdl} \Nil(W_2, W_4), Y \bowtie W_4 \\
          : \LIempty{\mathit{dbllist}(W_2,Y,W_4)}{W_2,Y,W_4}}$}
    \RightLabel{\TyLITransA{}}
    \BIC{$\vdash_{\Pdl} \nu W_2 W_4. (G, \Nil(W_2, W_4), Y \bowtie W_4)
        : \LIempty{\mathit{dbllist}(X,Y,Z)}{X,Y,Z}$}
    \RightLabel{\TyLIElimZA{}}
    \UIC{$\vdash_{\Pdl} \nu W_2 W_4. (G, \Nil(W_2, W_4), Y \bowtie W_4) : \mathit{dbllist}(X,Y,Z)$}
  \end{prooftree}
  where $G = \nu W_1. (\Cons(W_1, X, W_2, Z), 3(W_1)), \nu W_3. (\Cons(W_3, Z, W_4, W_2), 5(W_3))$.

  Here,
  \TyLIIntro{}, \TyLITrans{}, and \TyLIElimZ{} andare abbreviated as
  \TyLIIntroA{}, \TyLITransA{}, and \TyLIElimZA{} respectively.

  \hrulefill
  \caption{Typing derivation of complete doubly linked lists.}
  \label{fig:dbllist-typing-derivation-eg}
\end{figure}

\subsubsection{Pattern Matching at the List Tail}

In this section, we present an application of the linear implication type
by assigning types to a program that removes the last element of a doubly linked list.

First, consider the expression \(\eDLpop\)
\eqref{eq:dbllist-removal-eg} (\cref{sec:untyped:examples}).
In this case, we refine the case expression template
by explicitly adding type annotations to the graph variable \(y\) and \(z\),
illustrated as follows: 

\begin{align}\label{eq:dbllist-removal-annot-eg}
  \eDLpop' \triangleq
  \begin{array}{l}
    (\lambda\,x[X,Y,Z]: \mathit{dbllist}(X,Y,Z).        \\
    \quad \mathbf{case}\; x[X,Y,Z]\; \mathbf{of}\;      \\
    \qquad \nu W_1 W_2 W_3. (                           \\
    \quad\qquad y[W_1,W_2,X,Z]:                         \\
    \qquad\qquad \LI{\mathit{dbllist}(W_2,Y,W_1)}{      \\
      \qquad\qquad\quad  \mathit{dbllist}(X,Y,Z)}{W_1,W_2,X,Z}
    ,                                                   \\
    \quad\qquad \Cons(W_3,W_2,Y,W_1),                   \\
    \quad\qquad z[W_3]: \mathit{nat}(W_3),              \\
    \quad\qquad \Nil(W_1,Y)                             \\
    \qquad) \to \nu W.(y[Y,W,X,Z], \Nil(W,Y))           \\
    \quad\: \texttt{|}\;\mathbf{otherwise} \to x[X,Y,Z] \\
    )(W).
  \end{array}
\end{align}

We now demonstrate that, given \( \Pdl \) \eqref{eq:prod-rules-eg}
and \( \eDLpop' \) \eqref{eq:dbllist-removal-annot-eg},
we can derive the typing judgement:

\begin{align}\label{eq:dbllist-removal-type-judge-eg}
  \emptyset \vdash_{\Pdl} \eDLpop' : (\mathit{dbllist}(X,Y,Z) \to \mathit{dbllist}(X,Y,Z))(W).
\end{align}


For $\eDLpop'$ \eqref{eq:dbllist-removal-annot-eg},
the type environment obtained by collecting all graph variables and
their corresponding type annotations is given as follows:
\begin{align}
  \Gamma' \triangleq
  \begin{array}{lll}
    y[W_1,W_2,X,Z]:\LI{\mathit{dbllist}(W_2,Y,W_1)}{\mathit{dbllist}(X,Y,Z)}{W_1,W_2,X,Z}, \\
    z[W_3]: \mathit{nat}(W_3).
  \end{array}
\end{align}

We employ the newly introduced \TyLIE{} rule
to eliminate linear implication types,
thereby ensuring consistency in the pattern-matching process within the case expression
as shown in
\cref{fig:dbllist-removal-typing-derivation-eg-1}.
Furthermore, we verify the type of the reconstructed graph after pattern matching
using the graph variable \(y\) as in \cref{fig:dbllist-removal-typing-derivation-eg-2}
Finally,
in the application of the \TyCase{} rule,
we employ the type environment
\(\Gamma \triangleq x[X,Y,Z]: dbllist(X,Y,Z)\),
together with \(\Gamma'\), to establish the typing judgement
as shown in \cref{fig:dbllist-removal-typing-derivation-eg}.

\begin{figure*}[tb]
  \hrulefill

  \begin{prooftree}
    \AXC{}
    \RightLabel{\TyVarA{}}
    \UIC{$\Gamma' \vdash_{\Pdl} z[W_1] : \mathit{nat}(W_1)$}
    \AXC{}
    \RightLabel{\TyProdA{}}
    \UIC{$\ml{\Gamma' \vdash_{\Pdl} \Nil(X,Z), Y \bowtie Z \\
          : \mathit{dbllist}(X,Y,Z)}$}
    \RightLabel{\TyAlphaA{}}
    \UIC{$\ml{\Gamma' \vdash_{\Pdl} \Nil(Z,W_2), Y \bowtie W_2 \\
          : \mathit{dbllist}(Z,Y,W_2)}$}
    \RightLabel{\TyProdA{}}
    \BIC{$\ml{\Gamma' \vdash_{\Pdl} \nu W_1 W_2.(\Cons(W_1,X,W_2,Z),z[W_1],\Nil(Z,W_2),Y \bowtie W_2) \\
      : \mathit{dbllist}(X,Y,Z)}$}
    \RightLabel{\TyAlphaA{}}
    \UIC{
    \tikz[remember picture,baseline=(upper-root1.base),inner sep=0]
    \node [font=\normalsize] (upper-root1)
    {$\ml{\Gamma' \vdash_{\Pdl} \nu W_3W_4.(\Cons(W_3,W_2,W_4,W_1),z[W_3],
      \Nil(W_1,W_4), Y \bowtie W_4) \\
      : \mathit{dbllist}(W_2,Y,W_1)}$};
    }
  \end{prooftree}

  \bigskip
  \bigskip

  \begin{prooftree}
    \AXC{}
    \RightLabel{\TyVarA{}}
    \UIC{$\ml{\Gamma'\{y[W_1, W_2, X, Z]\\
          \quad : \LI{\mathit{dbllist}(W_2,Y,W_1)}{%
            \mathit{dbllist}(X,Y,Z)}{W_1,W_2,X,Z)} \\
          \vdash_{\Pdl} y[W_1, W_2, X, Z] \\
          : \LI{\mathit{dbllist}(W_2,Y,W_1)}{%
            \mathit{dbllist}(X,Y,Z)}{W_1,W_2,X,Z}
        }$}
    \AXC{
      \tikz[remember picture,baseline=(lower-hole1.base)]
      \node[draw, dashed, rectangle, inner sep=3pt, font=\normalsize] (lower-hole1)
      {$\color{white}Cons$};
    }
    \RightLabel{\TyLIEA{}}
    \BIC{$\ml{
      \Gamma'\{y[W_1, W_2, X, Z] \\
      \quad : \LI{\mathit{dbllist}(W_2,Y,W_1)}{\mathit{dbllist}(X,Y,Z)}{W_1,W_2,X,Z)}\\
      \vdash_{\Pdl} \nu W_1 W_2. (y[W_1,W_2,X,Z], \\
      \qquad\qquad \nu W_3W_4.(\Cons(W_3,W_2,W_4,W_1),z[W_3],
      \Nil(W_1,W_4), Y \bowtie W_4)) \\
      : \mathit{dbllist}(X,Y,Z)
      }$}

    \RightLabel{\TyCongA{}}
    \UIC{$\ml{\Gamma' \vdash_{\Pdl}
          \nu W_1 W_2 W_3. (
          y[W_1,W_2,X,Z],
          \Cons(W_3,W_2,Y,W_1),
          z[W_3],
          \Nil(W_1,Y)) \\
          : \mathit{dbllist}(X,Y,Z)}$}
  \end{prooftree}
  \begin{tikzpicture}[remember picture,overlay]
    \begin{scope}[even odd rule]
      \draw[->,thick] (upper-root1.south) to[looseness=2.5,out=-90,in=90] (lower-hole1.north);
    \end{scope}
  \end{tikzpicture}

  where
  \begin{align*}
    \Gamma' & =
    \begin{array}{lll}
      y[W_1,W_2,X,Z]
      : \LI{\mathit{dbllist}(W_2,Y,W_1)}{\mathit{dbllist}(X,Y,Z)}{W_1,W_2,X,Z}, \\
      z[W_3]: \mathit{nat}(W_3)
    \end{array}
  \end{align*}

  Here,
  \TyCong{},  \TyLIE{},  \TyAlpha{},  and \TyProd{}  andare abbreviated as
  \TyCongA{}, \TyLIEA{}, \TyAlphaA{}, and \TyProdA{} respectively.

  \hrulefill
  \caption{Typing derivation of the case template in \eqref{eq:dbllist-removal-type-judge-eg}.}
  \label{fig:dbllist-removal-typing-derivation-eg-1}
\end{figure*}

\begin{figure*}[tb]
  \hrulefill

  \begin{prooftree}
    \AXC{}
    \RightLabel{\TyVarA{}}
    \UIC{$\ml{
          \Gamma, \Gamma' \vdash_{\Pdl}
          y[W_1,W_2,X,Z]                      \\
          : \LI{\mathit{dbllist}(W_2,Y,W_1)}{ \\
            \quad \mathit{dbllist}(X,Y,Z)}{W_1,W_2,X,Z} \\[-5mm]
        }$}
    \AXC{}
    \RightLabel{\TyProdA{}}
    \UIC{$\ml{\Gamma, \Gamma' \vdash_{\Pdl} \Nil(X,Z), Y \bowtie Z \\
          : \mathit{dbllist}(X,Y,Z)}$}
    \RightLabel{\TyAlphaA{}}
    \UIC{$\begin{array}{lll}
          \Gamma, \Gamma' \vdash_{\Pdl}
          \Nil(W_2,W_1), Y \bowtie W_1 \\
          : \mathit{dbllist}(W_2,Y,W_1)
        \end{array}$}
    \RightLabel{\TyLIEA{}}
    \BIC{$\Gamma, \Gamma' \vdash_{\Pdl}
        \nu W_1 W_2.(
        y[W_1,W_2,X,Z],
        \Nil(W_2,W_1), Y \bowtie W_1)
        : \mathit{dbllist}(X,Y,Z)$}
    \RightLabel{\TyCongA{}}
    \UIC{$\Gamma, \Gamma' \vdash_{\Pdl}
        \nu W.(y[Y,W,X,Z], \Nil(W,Y)) : \mathit{dbllist}(X,Y,Z)$}
  \end{prooftree}

  where
  \begin{align*}
    \Gamma  & = x[X,Y,Z]: \mathit{dbllist}(X,Y,Z)
    \\
    \Gamma' & =
    \begin{array}{lll}
      y[W_1,W_2,X,Z]
      : \LI{\mathit{dbllist}(W_2,Y,W_1)}{\mathit{dbllist}(X,Y,Z)}{W_1,W_2,X,Z}, \\
      z[W_3]: \mathit{nat}(W_3)
    \end{array}
  \end{align*}

  Here,
  \TyCong{}, \TyLIE{}, \TyLIIntro{}, \TyAlpha{}, \TyProd{}, \TyLITrans{}, and \TyLIElimZ{} andare abbreviated as
  \TyCongA{}, \TyLIEA{}, \TyLIIntroA{}, \TyAlphaA{}, \TyProdA{}, \TyLITransA{}, and \TyLIElimZA{} respectively.

  \hrulefill
  \caption{Typing derivation of the composed graph obtained after matching in \eqref{eq:dbllist-removal-type-judge-eg}.}
  \label{fig:dbllist-removal-typing-derivation-eg-2}
\end{figure*}

\begin{figure*}[tb]
  \hrulefill

  \begin{prooftree}
    \AXC{}
    \RightLabel{\TyVar{}}
    \UIC{
      \tikz[remember picture,baseline=(upper-root1.base)]
      \node [font=\normalsize] (upper-root1)
      {$x[X,Y,Z]: \mathit{dbllist}(X,Y,Z) \vdash_{\Pdl}
          x[X,Y,Z]: \mathit{dbllist}(X,Y,Z)$};
    }
  \end{prooftree}

  \medskip

  \begin{prooftree}
    \AXC{
      \tikz[remember picture,baseline=(lower-hole1.base)]
      \node[draw, dashed, rectangle, inner sep=3pt, font=\normalsize] (lower-hole1)
      {$\color{white}Cons$};
    }
    \AXC{(\cref{fig:dbllist-removal-typing-derivation-eg-2})}
    \AXC{(\cref{fig:dbllist-removal-typing-derivation-eg-1})}
    \AXC{
      \tikz[remember picture,baseline=(lower-hole2.base)]
      \node[draw, dashed, rectangle, inner sep=3pt, font=\normalsize] (lower-hole2)
      {$\color{white}Cons$};
    }
    \RightLabel{\TyCase{}}
    \QuaternaryInfC{$\Gamma \vdash_{\Pdl} (\caseof{e_1}{T}{e_2}{e_1}) : \mathit{dbllist}(X,Y,Z)$}
    \RightLabel{\TyArrow{}}
    \UIC{$\emptyset \vdash_{\Pdl} e :
        (\mathit{dbllist}(X,Y,Z) \to \mathit{dbllist}(X,Y,Z))(W)$}
  \end{prooftree}
  \begin{tikzpicture}[remember picture,overlay]
    \begin{scope}[even odd rule]
      \draw[->] (upper-root1.south) to[looseness=0.8,out=-90,in=90] (lower-hole1.north);
      \draw[->] (upper-root1.south) to[looseness=0.8,out=-90,in=90] (lower-hole2.north);
    \end{scope}
  \end{tikzpicture}

  where
  \begin{align*}
    \Gamma & = x[X,Y,Z]: \mathit{dbllist}(X,Y,Z)
    \\
    e_1    & = x[X,Y,Z]
    \\
    e_2    & = \nu W.(y[Y,W,X,Z], \Nil(W,Y))
    \\
    T      & =
    \begin{array}{lll}
      \nu W_1 W_2 W_3. (                                                               \\
      \quad y[W_1,W_2,X,Z]                                                             \\
      \qquad : \LI{\mathit{dbllist}(W_2,Y,W_1)}{\mathit{dbllist}(X,Y,Z)}{W_1,W_2,X,Z}, \\
      \quad \Cons(W_3,W_2,Y,W_1),
      z[W_3]: \mathit{nat}(W_3),
      \Nil(W_1,Y))
    \end{array}
    \\
    e      & = (\lambda\,x[X,Y,Z]:\mathit{dbllist}(X,Y,Z).\,
    \caseof{e_1}{T}{e_2}{e_1})(W)
  \end{align*}

  \hrulefill
  \caption{Typing derivation of the typing judgement \eqref{eq:dbllist-removal-type-judge-eg}
    (removing the last element from a doubly linked list).}
  \label{fig:dbllist-removal-typing-derivation-eg}
\end{figure*}

The typing relation \eqref{eq:lltree-typing-relation-eg}
for the expression $\eLLT$ that manipulates leaf-linked trees
can also be derived straightforwardly,
in a manner similar to \cref{fig:dbllist-removal-typing-derivation-eg}.

\section{Type Soundness with Linear Implication Types}\label{sec:sec3-soundness}

This section proves type preservation and progress for the type system of \lgt
extended with linear implication types.
The overall proof structure is the same as that of the core type system.
The only part that requires additional argument is the substitution lemma,
where we need to consider the typing rules newly introduced for linear implication types.

We first recall the substitution lemma.

\begin{lemma*}[\textbf{Lemma \ref{lemma-substitution-lemma}}, Substitution Lemma]
  If \(\Gamma, x[\Xs]:\tau' \vdash_P e : \tau\) and
  \(\Gamma \vdash_P G : \tau'\), then
  \(\Gamma \vdash_P e[G / x[\Xs]] : \tau\).
\end{lemma*}

\begin{proof}
  By induction on the derivation tree of the typing relation
  \(\Gamma, x[\Xs]:\tau' \vdash_P e : \tau\).
  We split the cases by the rule used in the final step.

  The cases for the typing rules inherited from the core type system are the same as
  in the proof of Lemma~\ref{lemma-substitution-lemma}.
  Therefore, we only show the additional cases for linear implication types.

  \noindent
  \textbf{Case \TyLIIntro{}:} 
  By assumption, we have a proof ending with the form
  \begin{prooftree}
    \AXC{}
    \RightLabel{\TyLIIntro{}}
    \UIC{$\Gamma, x[\Xs]:\tau'  \vdash_P (C(\Ys), \seq{U \bowtie V}) : \LI{\taus}{\alpha(\Ws)}{\Ys}$}
  \end{prooftree}
  where $\alpha(\Ws) \lto \nu \Zs.(C(\Ys), \taus, \seq{U \bowtie V}) \in P$.
  From this side condition and \TyLIIntro{}, we also have the following.
  \begin{prooftree}
    \AXC{}
    \RightLabel{\TyLIIntro{}}
    \UIC{$\Gamma \vdash_P (C(\Ys), \seq{U \bowtie V}) : \LI{\taus}{\alpha(\Ws)}{\Ys}$}
  \end{prooftree}

  Since \((C(\Ys), \seq{U \bowtie V})[G / x[\Xs]] =
  (C(\Ys), \seq{U \bowtie V})\), the above can be written also as
  \begin{prooftree}
    \AXC{}
    \RightLabel{\TyLIIntro{}}
    \UIC{$\Gamma \vdash_P (C(\Ys), \seq{U \bowtie V})[G/x[\Xs]] : \LI{\taus}{\alpha(\Ws)}{\Ys}$}
  \end{prooftree}
  thus establishing the claim.

  \noindent
  \textbf{Case \TyLITrans{}:} 
  By assumption, we have a proof ending with the form
  \begin{prooftree}
    \AXC{$\begin{array}{r@{~}l}
          \Gamma, x[\Xs]:\tau'
           & \vdash_P T_1 : \LI{\seq{\tau_0}}{\tau_1}{\Zs}         \\
          \Gamma, x[\Xs]:\tau'
           & \vdash_P T_2 : \LI{\tau_1, \seq{\tau_2}}{\tau_3}{\Ys}
        \end{array}$}
    \RightLabel{\TyLITrans{}}
    \UIC{$\Gamma, x[\Xs]:\tau' \vdash_P \nu \Vs.(T_1, T_2)
        : \LI{\seq{\tau_0}, \seq{\tau_2}}{\tau_3}{\Ws}$}
  \end{prooftree}
  where \(\{\Ws\} = (\{\Zs\} \cup \{\Ys\}) \setminus \{\Xs\}\).
  By applying the induction hypothesis to the premises, we get the following:
  \[\begin{array}{rl}
      \Gamma & \vdash_P T_1[G/x[\Xs]] : \LI{\seq{\tau_0}}{\tau_1}{\Zs},         \\
      \Gamma & \vdash_P T_2[G/x[\Xs]] : \LI{\tau_1, \seq{\tau_2}}{\tau_3}{\Ys}.
    \end{array}\]
  Applying \TyLITrans{} to them gives us
  \[\mathmakebox[\dimexpr\width-2pt\relax][l]{%
    \Gamma \vdash_P
    \nu \Vs.(T_1[G/x[\Xs]], T_2[G/x[\Xs]]):
    \LI{\seq{\tau_0}, \seq{\tau_2}}{\tau_3}{\Ws}.}
  \]
  From the definition of graph substitution, the graph substitution
  can be factored out (without  \(\alpha\)-conversion) and we get
  \[
    \Gamma \vdash_P
    (\nu \Vs.(T_1, T_2))[G/x[\Xs]]:
    \LI{\seq{\tau_0}, \seq{\tau_2}}{\tau_3}{\Ws}.
  \]

  \noindent
  \textbf{Case \TyLIElimZ{}:} 
  By assumption, we have a proof ending with the following form.
  \begin{prooftree}
    \AXC{$\Gamma, x[\Xs]:\tau' \vdash_P T : \LIempty{\tau}{\Ys}$}
    \RightLabel{\TyLIElimZ{}}
    \UIC{$\Gamma, x[\Xs]:\tau' \vdash_P \nu\Zs.T : \tau$}
  \end{prooftree}
  By applying the induction hypothesis to the premise, we get
  \(\Gamma \vdash_P T[G/x[\Xs]] : \LIempty{\tau}{\Ys}\).
  Using this as the premise of \TyLIElimZ{}, we get
  \[
    \Gamma \vdash_P
    \nu\Zs.T[G/x[\Xs]]: \tau,
  \]
  where we note that graph substitution does not change the set of free links.

  \noindent
  \textbf{Case \TyLIIntroZ{}:} 
  By assumption, we have a proof ending with the following form.
  \begin{prooftree}
    \AXC{$\Gamma, x[\Xs]:\tau' \vdash_P T : \tau_1$}
    \RightLabel{\TyLIIntroZ{}}
    \UIC{$\Gamma, x[\Xs]:\tau' \vdash_P
        T
        : \LIempty{\tau_1}{\Ys}$}
  \end{prooftree}
  %
  By applying the induction hypothesis to the premise, we get
  \(\Gamma \vdash_P T[G/x[\Xs]] : \tau_1\).
  Using this as the premise of \TyLIIntroZ{}, we get
  \[
    \Gamma \vdash_P
    T[G/x[\Xs]]
    : \LIempty{\tau_1}{\Ys},
  \]
  where we note that graph substitution does not change the set of free links.
\end{proof}

\section{Conservativity for Complete Graph Types}\label{sec:elim-limprules}

Linear implication types are introduced to type incomplete graph structures.
They should not increase the class of complete graph values typable at ordinary graph types.
Otherwise, linear implication types would act as a hidden mechanism for typing ordinary graph values,
rather than as an explicit type former for incomplete graphs.

This section shows that this does not occur.
More precisely, we prove that the extended type system is conservative over the core type system
for judgements whose assumptions and conclusions contain no linear implication types.

\begin{theorem}[Conservativity for complete graph types]\label{thm:conservativity-complete-graph-types}
  Suppose that neither \(\Gamma\) nor \(\tau\) contains a linear implication type.
  If
  $\Gamma \vdash_P G : \tau$
  is derivable in the type system extended with linear implication types, then
  $\Gamma \vdash_P G : \tau$
  is derivable without using any of
  \(\TyLIIntro\), \(\TyLIIntroZ\), \(\TyLITrans\), or \(\TyLIElimZ\).
\end{theorem}

The proof is given at the end of this section, after establishing the
normalisation results used to eliminate linear-implication rules.

The rest of this section is organised as follows.
First, \cref{subsec:elim-limprules-eg} illustrates the basic idea of the
elimination by a representative example.
Next, \cref{subsec:graph-prefix-normalisation} isolates the graph-typing prefix
of a derivation and shows that administrative uses of \(\TyAlpha{}\) and
\(\TyCong{}\) can be removed or postponed.
Then, \cref{subsec:elim-limp} proves the main elimination result for
linear-implication rules in such graph-typing prefixes.
Finally, \cref{subsec:proof-of-conservativity} applies this elimination result
recursively to derivations with \(\TyArrow{}\) premises and proves
\cref{thm:conservativity-complete-graph-types}.

\textbf{Conventions:}
We use the following conventions.
Molecules, such as \(T_1, T_2, T_3, \dots\), are assumed to associate to the left.
By \cref{def:TypingRelation}, we also assume without further notice that the
free links on both sides of a typing judgement coincide; that is, for every
judgement $\Gamma \vdash_P T : \tau$, we have \(\fn(T) = \fn(\tau)\).

\subsection{Example of \(\TyLIs{}\) Elimination}\label{subsec:elim-limprules-eg}

We show how a graph (value) that is not of a linear implication type
can be typed without using the typing rules of \(\TyLIs{}\),
using the following production rules as an example:
\[\begin{array}{@{\kern-1pt}r@{\kern1pt}l@{}}
    \li(X)  & \lto \Nil(X),                                             \\
    \li(X)  & \lto \nu W_1 W_2.(\Cons(W_1,W_2,X), \nat(W_1), \li(W_2)), \\
    \nat(X) & \lto \one(X),                                             \\
    \nat(X) & \lto \two(X).
  \end{array}\]

\begin{sidewaysfigure}[tbp]
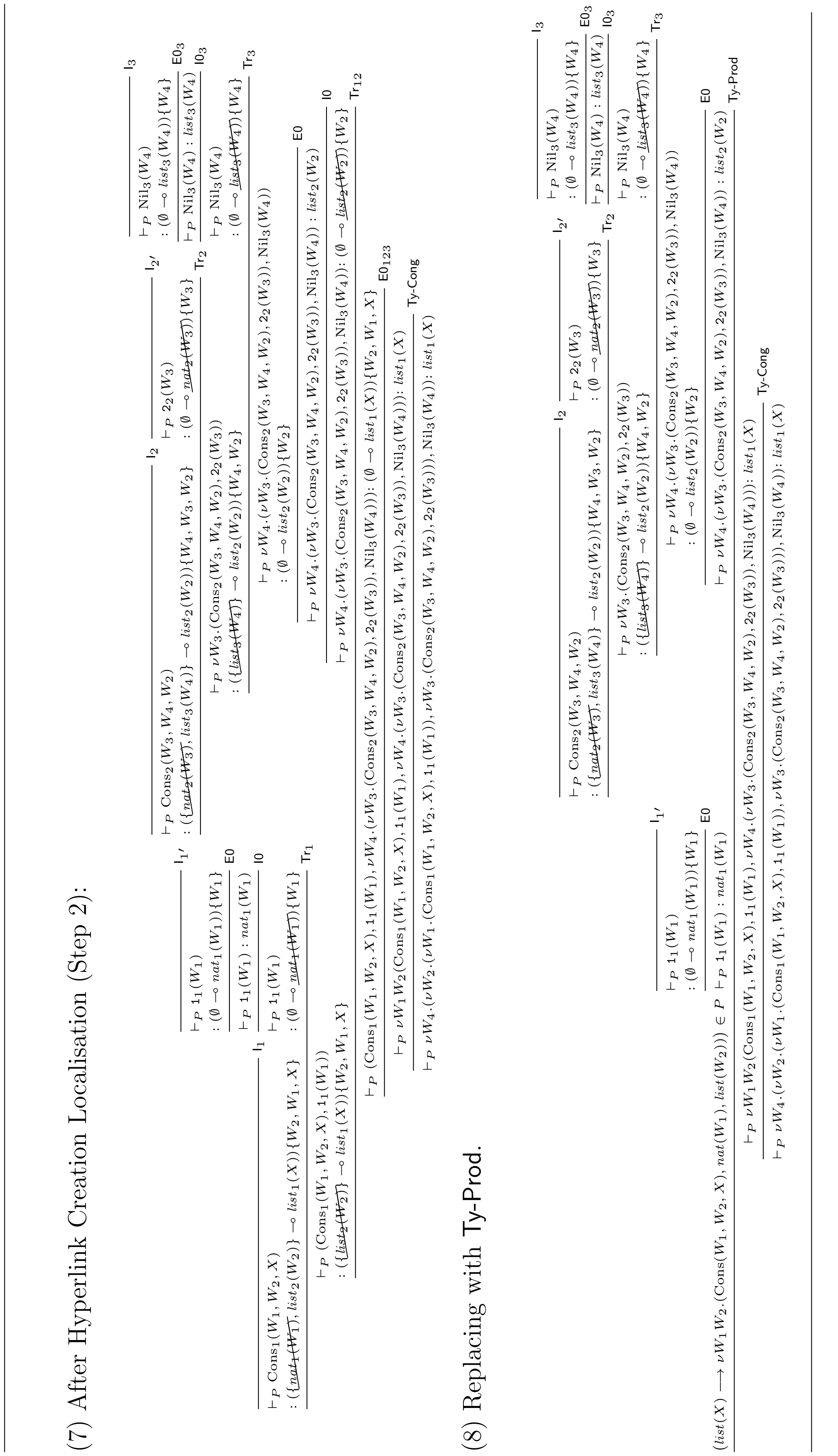

  \captionsetup{justification=centering,singlelinecheck=false}
  \hrulefill{}

  \def\ScoreOverhang{0pt}
  \def\defaultHypSeparation{\hskip .05in}
  \hspace*{-60pt}

  (7)
  After Hyperlink Creation Localisation (Step 2):
  \begin{prooftree}\tiny
    \AXC{}
    \RightLabel{$\TyLIIntroA{}_1$}
    \UIC{$\ml{\vdash_P \Cons_1(W_1,W_2,X)\\
          : \LI{\cancel{\nat_1(W_1)}, \li_2(W_2)}{\li_1(X)}{W_2,W_1,X}}$}
    \AXC{}
    \RightLabel{$\TyLIIntroA{}_{1'}$}
    \UIC{$\ml{\vdash_P \one_1(W_1) \\
          : \LIempty{\nat_1(W_1)}{W_1}}$}
    \RightLabel{$\TyLIElimZA{}$}
    \UIC{$\ml{\vdash_P \one_1(W_1)
          : \nat_1(W_1)}$}
    \RightLabel{$\TyLIIntroZA{}$}
    \UIC{$\ml{\vdash_P \one_1(W_1) \\
          : \LIempty{\cancel{\nat_1(W_1)}}{W_1}}$}
    \RightLabel{$\TyLITransA{}_1$}
    \BIC{$\ml{\vdash_P (\Cons_1(W_1,W_2,X), \one_1(W_1))\\
          : \LI{\cancel{\li_2(W_2)}}{\li_1(X)}{W_2,W_1,X}}$}
    \AXC{}
    \RightLabel{$\TyLIIntroA{}_2$}
    \UIC{$\ml{\vdash_P \Cons_2(W_3,W_4,W_2)\\
          : \LI{\cancel{\nat_2(W_3)},\li_3(W_4)}{\li_2(W_2)}{W_4,W_3,W_2}}$}
    \AXC{}
    \RightLabel{$\TyLIIntroA{}_{2'}$}
    \UIC{$\ml{\vdash_P \two_2(W_3)\\
          : \LIempty{\cancel{\nat_2(W_3)}}{W_3}}$}
    \RightLabel{$\TyLITransA{}_2$}
    \BIC{$\ml{\vdash_P \nu W_3.(\Cons_2(W_3,W_4,W_2), \two_2(W_3))\\
          : \LI{\cancel{\li_3(W_4)}}{\li_2(W_2)}{W_4,W_2}}$}
    \AXC{}
    \RightLabel{$\TyLIIntroA{}_3$}
    \UIC{$\ml{\vdash_P \Nil_3(W_4)\\
          : \LIempty{\li_3(W_4)}{W_4}}$}
    \RightLabel{$\TyLIElimZA{}_3$}
    \UIC{$\vdash_P \Nil_3(W_4) : \li_3(W_4)$}
    \RightLabel{$\TyLIIntroZA{}_3$}
    \UIC{$\ml{\vdash_P \Nil_3(W_4)\\
          : \LIempty{\cancel{\li_3(W_4)}}{W_4}}$}
    \RightLabel{$\TyLITransA{}_3$}
    \BIC{$\ml{\vdash_P
          \nu W_4.(\nu W_3.(\Cons_2(W_3,W_4,W_2), \two_2(W_3))
          ,\Nil_3(W_4))\\
          : \LIempty{\li_2(W_2)}{W_2}}$}
    \RightLabel{$\TyLIElimZA{}$}
    \UIC{$\ml{\vdash_P
          \nu W_4.(\nu W_3.(\Cons_2(W_3,W_4,W_2), \two_2(W_3))
          ,\Nil_3(W_4))
          : \li_2(W_2)}$}
    \RightLabel{$\TyLIIntroZA{}$}
    \UIC{$\ml{\vdash_P
          \nu W_4.(\nu W_3.(\Cons_2(W_3,W_4,W_2), \two_2(W_3))
          ,\Nil_3(W_4))}{
          : \LIempty{\cancel{\li_2(W_2)}}{W_2}}$}
    \RightLabel{$\TyLITransA{}_{12}$}
    \BIC{$\ml{\vdash_P
          (\Cons_1(W_1,W_2,X), \one_1(W_1)
          ,\nu W_4.(\nu W_3.(\Cons_2(W_3,W_4,W_2), \two_2(W_3))
          ,\Nil_3(W_4)))}{
          : \LIempty{\li_1(X)}{W_2,W_1,X}}$}
    \RightLabel{$\TyLIElimZA{}_{123}$}
    \UIC{$\ml{\vdash_P
          \nu W_1 W_2
          (\Cons_1(W_1,W_2,X), \one_1(W_1)
          ,\nu W_4.(\nu W_3.(\Cons_2(W_3,W_4,W_2), \two_2(W_3))
          ,\Nil_3(W_4)))}{
          : \li_1(X)}$}
    \RightLabel{$\TyCong{}$}
    \UIC{$\ml{\vdash_P
          \nu W_4.(\nu W_2.(\nu W_1.(\Cons_1(W_1,W_2,X), \one_1(W_1))
          ,\nu W_3.(\Cons_2(W_3,W_4,W_2), \two_2(W_3)))
          ,\Nil_3(W_4))}{
          : \li_1(X)}$}
  \end{prooftree}

  (8)
  Replacing with \TyProd{}.
  \begin{prooftree}\tiny
    \AXC{$\paren{\li(X) \lto \nu W_1 W_2.(\Cons(W_1,W_2,X), \nat(W_1), \li(W_2))} \in P$}
    \AXC{}
    \RightLabel{$\TyLIIntroA{}_{1'}$}
    \UIC{$\ml{\vdash_P \one_1(W_1) \\
          : \LIempty{\nat_1(W_1)}{W_1}}$}
    \RightLabel{$\TyLIElimZA{}$}
    \UIC{$\ml{\vdash_P \one_1(W_1)
          : \nat_1(W_1)}$}
    \AXC{}
    \RightLabel{$\TyLIIntroA{}_2$}
    \UIC{$\ml{\vdash_P \Cons_2(W_3,W_4,W_2)\\
          : \LI{\cancel{\nat_2(W_3)},\li_3(W_4)}{\li_2(W_2)}{W_4,W_3,W_2}}$}
    \AXC{}
    \RightLabel{$\TyLIIntroA{}_{2'}$}
    \UIC{$\ml{\vdash_P \two_2(W_3)\\
          : \LIempty{\cancel{\nat_2(W_3)}}{W_3}}$}
    \RightLabel{$\TyLITransA{}_2$}
    \BIC{$\ml{\vdash_P \nu W_3.(\Cons_2(W_3,W_4,W_2), \two_2(W_3))\\
          : \LI{\cancel{\li_3(W_4)}}{\li_2(W_2)}{W_4,W_2}}$}
    \AXC{}
    \RightLabel{$\TyLIIntroA{}_3$}
    \UIC{$\ml{\vdash_P \Nil_3(W_4)\\
          : \LIempty{\li_3(W_4)}{W_4}}$}
    \RightLabel{$\TyLIElimZA{}_3$}
    \UIC{$\vdash_P \Nil_3(W_4) : \li_3(W_4)$}
    \RightLabel{$\TyLIIntroZA{}_3$}
    \UIC{$\ml{\vdash_P \Nil_3(W_4)\\
          : \LIempty{\cancel{\li_3(W_4)}}{W_4}}$}
    \RightLabel{$\TyLITransA{}_3$}
    \BIC{$\ml{\vdash_P
          \nu W_4.(\nu W_3.(\Cons_2(W_3,W_4,W_2), \two_2(W_3))
          ,\Nil_3(W_4))\\
          : \LIempty{\li_2(W_2)}{W_2}}$}
    \RightLabel{$\TyLIElimZA{}$}
    \UIC{$\ml{\vdash_P
          \nu W_4.(\nu W_3.(\Cons_2(W_3,W_4,W_2), \two_2(W_3))
          ,\Nil_3(W_4))
          : \li_2(W_2)}$}
    \RightLabel{$\TyProd{}$}
    \TIC{$\ml{\vdash_P
          \nu W_1 W_2
          (\Cons_1(W_1,W_2,X), \one_1(W_1)
          ,\nu W_4.(\nu W_3.(\Cons_2(W_3,W_4,W_2), \two_2(W_3))
          ,\Nil_3(W_4)))}{
          : \li_1(X)}$}
    \RightLabel{$\TyCong{}$}
    \UIC{$\ml{\vdash_P
          \nu W_4.(\nu W_2.(\nu W_1.(\Cons_1(W_1,W_2,X), \one_1(W_1))
          ,\nu W_3.(\Cons_2(W_3,W_4,W_2), \two_2(W_3)))
          ,\Nil_3(W_4))}{
          : \li_1(X)}$}
  \end{prooftree}

  \hrulefill{}
  \medskip
  \caption{\hbox{Transformations of a Proof Tree (7)--(8).}}
  \label{fig:BeforeAfterTransformation-3}
\end{sidewaysfigure}

\Cref{fig:BeforeAfterTransformation-1,fig:BeforeAfterTransformation-2,fig:BeforeAfterTransformation-3}
shows
transformations of a derivation tree.
The subscript \(i\) attached to symbols such as \(\Cons_i\)
is merely for readability, to indicate the correspondence between constructor atoms
within the proof tree.
Similarly, the subscript \(i\) on type atoms such as \(\dbllist_i\)
marks those that will be cancelled,
so that the correspondence can be easily seen.

\begin{sidewaysfigure}[tbp]
  \captionsetup{justification=centering,singlelinecheck=false}
  \hrulefill{}

  \def\ScoreOverhang{0pt}
  \def\defaultHypSeparation{\hskip .05in}
  \hspace*{-60pt}

  (1)
  Before all the transformations:
  \begin{prooftree}\tiny
    \AXC{}
    \RightLabel{$\TyLIIntroA{}_1$}
    \UIC{$\ml{\vdash_P \Cons_1(W_1,W_2,X)\\
          : \LI{\cancel{\nat_1(W_1)}, \li_2(W_2)}{\li_1(X)}{W_2,W_1,X}}$}
    \AXC{}
    \RightLabel{$\TyLIIntroA{}_{1'}$}
    \UIC{$\ml{\vdash_P \one_1(W_1) \\
          : \LIempty{\cancel{\nat_1(W_1)}}{W_1}}$}
    \RightLabel{$\TyLITransA{}_1$}
    \BIC{$\ml{\vdash_P \nu W_1.(\Cons_1(W_1,W_2,X), \one_1(W_1))\\
          : \LI{\cancel{\li_2(W_2)}}{\li_1(X)}{W_2,X}}$}
    \AXC{}
    \RightLabel{$\TyLIIntroA{}_2$}
    \UIC{$\ml{\vdash_P \Cons_2(W_3,W_4,W_2)\\
          : \LI{\cancel{\nat_2(W_3)},\li_3(W_4)}{\li_2(W_2)}{W_4,W_3,W_2}}$}
    \AXC{}
    \RightLabel{$\TyLIIntroA{}_{2'}$}
    \UIC{$\ml{\vdash_P \two_2(W_3)\\
          : \LIempty{\cancel{\nat_2(W_3)}}{W_3}}$}
    \RightLabel{$\TyLITransA{}_2$}
    \BIC{$\ml{\vdash_P \nu W_3.(\Cons_2(W_3,W_4,W_2), \two_2(W_3))\\
          : \LI{\li_3(W_4)}{\cancel{\li_2(W_2)}}{W_4,W_2}}$}
    \RightLabel{$\TyLITransA{}_{12}$}
    \BIC{$\ml{\vdash_P
          \nu W_2.(\nu W_1.(\Cons_1(W_1,W_2,X), \one_1(W_1))
          ,(\nu W_3.\Cons_2(W_3,W_4,W_2), \two_2(W_3)))}{
          : \LI{\cancel{\li_3(W_4)}}{\li_1(X)}{W_4,X}}$}
    \AXC{}
    \RightLabel{$\TyLIIntroA{}_3$}
    \UIC{$\ml{\vdash_P \Nil_3(W_4)\\
          : \LIempty{\cancel{\li_3(W_4)}}{W_4}}$}
    \RightLabel{$\TyLITransA{}_3$}
    \BIC{$\ml{\vdash_P
          \nu W_4.(\nu W_2.(\nu W_1.(\Cons_1(W_1,W_2,X), \one_1(W_1))
          ,\nu W_3.(\Cons_2(W_3,W_4,W_2), \two_2(W_3)))
          ,\Nil_3(W_4))}{
          : \LIempty{\li_1(X)}{X}}$}
    \RightLabel{$\TyLIElimZA{}_{123}$}
    \UIC{$\ml{\vdash_P
          \nu W_4.(\nu W_2.(\nu W_1.(\Cons_1(W_1,W_2,X), \one_1(W_1))
          ,\nu W_3.(\Cons_2(W_3,W_4,W_2), \two_2(W_3)))
          ,\Nil_3(W_4))}{
          : \li_1(X)}$}
  \end{prooftree}

  \medskip

  (2)
  After Prenex Normal Form Transformation (Step 1):
  \begin{prooftree}\tiny
    \AXC{}
    \RightLabel{$\TyLIIntroA{}_1$}
    \UIC{$\ml{\vdash_P \Cons_1(W_1,W_2,X)\\
          : \LI{\cancel{\nat_1(W_1)}, \li_2(W_2)}{\li_1(X)}{W_2,W_1,X}}$}
    \AXC{}
    \RightLabel{$\TyLIIntroA{}_{1'}$}
    \UIC{$\ml{\vdash_P \one_1(W_1) \\
          : \LIempty{\cancel{\nat_1(W_1)}}{W_1}}$}
    \RightLabel{$\TyLITransA{}_1$}
    \BIC{$\ml{\vdash_P (\Cons_1(W_1,W_2,X), \one_1(W_1))\\
          : \LI{\cancel{\li_2(W_2)}}{\li_1(X)}{W_2,W_1,X}}$}
    \AXC{}
    \RightLabel{$\TyLIIntroA{}_2$}
    \UIC{$\ml{\vdash_P \Cons_2(W_3,W_4,W_2)\\
          : \LI{\cancel{\nat_2(W_3)},\li_3(W_4)}{\li_2(W_2)}{W_4,W_3,W_2}}$}
    \AXC{}
    \RightLabel{$\TyLIIntroA{}_{2'}$}
    \UIC{$\ml{\vdash_P \two_2(W_3)\\
          : \LIempty{\cancel{\nat_2(W_3)}}{W_3}}$}
    \RightLabel{$\TyLITransA{}_2$}
    \BIC{$\ml{\vdash_P \nu W_3.(\Cons_2(W_3,W_4,W_2), \two_2(W_3))\\
          : \LI{\li_3(W_4)}{\cancel{\li_2(W_2)}}{W_4,W_2}}$}
    \RightLabel{$\TyLITransA{}_{12}$}
    \BIC{$\ml{\vdash_P
          ((\Cons_1(W_1,W_2,X), \one_1(W_1))
          ,\nu W_3.(\Cons_2(W_3,W_4,W_2), \two_2(W_3)))}{
          : \LI{\cancel{\li_3(W_4)}}{\li_1(X)}{W_4,W_2,W_1,X}}$}
    \AXC{}
    \RightLabel{$\TyLIIntroA{}_3$}
    \UIC{$\ml{\vdash_P \Nil_3(W_4)\\
          : \LIempty{\cancel{\li_3(W_4)}}{W_4}}$}
    \RightLabel{$\TyLITransA{}_3$}
    \BIC{$\ml{\vdash_P
          (((\Cons_1(W_1,W_2,X), \one_1(W_1))
          ,(\nu W_3.\Cons_2(W_3,W_4,W_2), \two_2(W_3)))
          ,\Nil_3(W_4))}{
          : \LIempty{\li_1(X)}{W_4,W_2,W_1,X}}$}
    \RightLabel{$\TyLIElimZA{}_{123}$}
    \UIC{$\ml{\vdash_P
          \nu W_1 W_2 W_3 W_4.
          ((\Cons_1(W_1,W_2,X), \one_1(W_1)
          ,\nu W_3.(\Cons_2(W_3,W_4,W_2), \two_2(W_3)))
          ,\Nil_3(W_4))}{
          : \li_1(X)}$}
    \RightLabel{$\TyCong{}$}
    \UIC{$\ml{\vdash_P
          \nu W_4.(\nu W_2.(\nu W_1.(\Cons_1(W_1,W_2,X), \one_1(W_1))
          ,\nu W_3.(\Cons_2(W_3,W_4,W_2), \two_2(W_3)))
          ,\Nil_3(W_4))}{
          : \li_1(X)}$}
  \end{prooftree}

  (3)
  After Hyperlink Creation Localisation (Step 2):
  \begin{prooftree}\tiny
    \AXC{}
    \RightLabel{$\TyLIIntroA{}_1$}
    \UIC{$\ml{\vdash_P \Cons_1(W_1,W_2,X)\\
          : \LI{\cancel{\nat_1(W_1)}, \li_2(W_2)}{\li_1(X)}{W_2,W_1,X}}$}
    \AXC{}
    \RightLabel{$\TyLIIntroA{}_{1'}$}
    \UIC{$\ml{\vdash_P \one_1(W_1) \\
          : \LIempty{\cancel{\nat_1(W_1)}}{W_1}}$}
    \RightLabel{$\TyLITransA{}_1$}
    \BIC{$\ml{\vdash_P (\Cons_1(W_1,W_2,X), \one_1(W_1))\\
          : \LI{\cancel{\li_2(W_2)}}{\li_1(X)}{W_2,W_1,X}}$}
    \AXC{}
    \RightLabel{$\TyLIIntroA{}_2$}
    \UIC{$\ml{\vdash_P \Cons_2(W_3,W_4,W_2)\\
          : \LI{\cancel{\nat_2(W_3)},\li_3(W_4)}{\li_2(W_2)}{W_4,W_3,W_2}}$}
    \AXC{}
    \RightLabel{$\TyLIIntroA{}_{2'}$}
    \UIC{$\ml{\vdash_P \two_2(W_3)\\
          : \LIempty{\cancel{\nat_2(W_3)}}{W_3}}$}
    \RightLabel{$\TyLITransA{}_2$}
    \BIC{$\ml{\vdash_P \nu W_3.(\Cons_2(W_3,W_4,W_2), \two_2(W_3))\\
          : \LI{\li_3(W_4)}{\cancel{\li_2(W_2)}}{W_4,W_2}}$}
    \RightLabel{$\TyLITransA{}_{12}$}
    \BIC{$\ml{\vdash_P
          ((\Cons_1(W_1,W_2,X), \one_1(W_1))
          ,\nu W_3.(\Cons_2(W_3,W_4,W_2), \two_2(W_3)))}{
          : \LI{\cancel{\li_3(W_4)}}{\li_1(X)}{W_4,W_2,W_1,X}}$}
    \AXC{}
    \RightLabel{$\TyLIIntroA{}_3$}
    \UIC{$\ml{\vdash_P \Nil_3(W_4)\\
          : \LIempty{\cancel{\li_3(W_4)}}{W_4}}$}
    \RightLabel{$\TyLIElimZA{}_3$}
    \UIC{$\vdash_P \Nil_3(W_4) : \li_3(W_4)$}
    \RightLabel{$\TyLIIntroZA{}_3$}
    \UIC{$\ml{\vdash_P \Nil_3(W_4)\\
          : \LIempty{\cancel{\li_3(W_4)}}{W_4}}$}
    \RightLabel{$\TyLITransA{}_3$}
    \BIC{$\ml{\vdash_P
          (((\Cons_1(W_1,W_2,X), \one_1(W_1))
          ,(\nu W_3.\Cons_2(W_3,W_4,W_2), \two_2(W_3)))
          ,\Nil_3(W_4))}{
          : \LIempty{\li_1(X)}{W_4,W_2,W_1,X}}$}
    \RightLabel{$\TyLIElimZA{}_{123}$}
    \UIC{$\ml{\vdash_P
          \nu W_1 W_2 W_4.
          ((\Cons_1(W_1,W_2,X), \one_1(W_1)
          ,\nu W_3.(\Cons_2(W_3,W_4,W_2), \two_2(W_3)))
          ,\Nil_3(W_4))}{
          : \li_1(X)}$}
    \RightLabel{$\TyCong{}$}
    \UIC{$\ml{\vdash_P
          \nu W_4.(\nu W_2.(\nu W_1.(\Cons_1(W_1,W_2,X), \one_1(W_1))
          ,\nu W_3.(\Cons_2(W_3,W_4,W_2), \two_2(W_3)))
          ,\Nil_3(W_4))}{
          : \li_1(X)}$}
  \end{prooftree}

  \hrulefill{}
  \medskip
  \caption{Transformations of a Proof Tree (1)--(3).}
  \label{fig:BeforeAfterTransformation-1}
\end{sidewaysfigure}

\begin{sidewaysfigure}[tbp]
  \captionsetup{justification=centering,singlelinecheck=false}
  \hrulefill{}

  \def\ScoreOverhang{0pt}
  \def\defaultHypSeparation{\hskip .05in}
  \hspace*{-60pt}

  (4)
  After Proof Tree Normalisation (Step 3):
  \begin{prooftree}\tiny
    \AXC{}
    \RightLabel{$\TyLIIntroA{}_1$}
    \UIC{$\ml{\vdash_P \Cons_1(W_1,W_2,X)\\
          : \LI{\cancel{\nat_1(W_1)}, \li_2(W_2)}{\li_1(X)}{W_2,W_1,X}}$}
    \AXC{}
    \RightLabel{$\TyLIIntroA{}_{1'}$}
    \UIC{$\ml{\vdash_P \one_1(W_1) \\
          : \LIempty{\cancel{\nat_1(W_1)}}{W_1}}$}
    \RightLabel{$\TyLITransA{}_1$}
    \BIC{$\ml{\vdash_P (\Cons_1(W_1,W_2,X), \one_1(W_1))\\
          : \LI{\cancel{\li_2(W_2)}}{\li_1(X)}{W_2,W_1,X}}$}
    \AXC{}
    \RightLabel{$\TyLIIntroA{}_2$}
    \UIC{$\ml{\vdash_P \Cons_2(W_3,W_4,W_2)\\
          : \LI{\cancel{\nat_2(W_3)},\li_3(W_4)}{\li_2(W_2)}{W_4,W_3,W_2}}$}
    \AXC{}
    \RightLabel{$\TyLIIntroA{}_{2'}$}
    \UIC{$\ml{\vdash_P \two_2(W_3)\\
          : \LIempty{\cancel{\nat_2(W_3)}}{W_3}}$}
    \RightLabel{$\TyLITransA{}_2$}
    \BIC{$\ml{\vdash_P \nu W_3.(\Cons_2(W_3,W_4,W_2), \two_2(W_3))\\
          : \LI{\cancel{\li_3(W_4)}}{\li_2(W_2)}{W_4,W_2}}$}
    \AXC{}
    \RightLabel{$\TyLIIntroA{}_3$}
    \UIC{$\ml{\vdash_P \Nil_3(W_4)\\
          : \LIempty{\li_3(W_4)}{W_4}}$}
    \RightLabel{$\TyLIElimZA{}_3$}
    \UIC{$\vdash_P \Nil_3(W_4) : \li_3(W_4)$}
    \RightLabel{$\TyLIIntroZA{}_3$}
    \UIC{$\ml{\vdash_P \Nil_3(W_4)\\
          : \LIempty{\cancel{\li_3(W_4)}}{W_4}}$}
    \RightLabel{$\TyLITransA{}_3$}
    \BIC{$\ml{\vdash_P
          (\nu W_3.(\Cons_2(W_3,W_4,W_2), \two_2(W_3))
          ,\Nil_3(W_4))}{
          : \LIempty{\cancel{\li_2(W_2)}}{W_4,W_2}}$}
    \RightLabel{$\TyLITransA{}_{12}$}
    \BIC{$\ml{\vdash_P
          (\Cons_1(W_1,W_2,X), \one_1(W_1)
          ,(\nu W_3.(\Cons_2(W_3,W_4,W_2), \two_2(W_3))
          ,\Nil_3(W_4)))}{
          : \LIempty{\li_1(X)}{W_4,W_2,W_1,X}}$}
    \RightLabel{$\TyLIElimZA{}_{123}$}
    \UIC{$\ml{\vdash_P
          \nu W_1 W_2 W_4.
          (\Cons_1(W_1,W_2,X), \one_1(W_1)
          ,(\nu W_3.(\Cons_2(W_3,W_4,W_2), \two_2(W_3))
          ,\Nil_3(W_4)))}{
          : \li_1(X)}$}
    \RightLabel{$\TyCong{}$}
    \UIC{$\ml{\vdash_P
          \nu W_4.(\nu W_2.(\nu W_1.(\Cons_1(W_1,W_2,X), \one_1(W_1))
          ,\nu W_3.(\Cons_2(W_3,W_4,W_2), \two_2(W_3)))
          ,\Nil_3(W_4))}{
          : \li_1(X)}$}
  \end{prooftree}

  (5)
  After Hyperlink Creation Localisation (Step 2):
  \begin{prooftree}\tiny
    \AXC{}
    \RightLabel{$\TyLIIntroA{}_1$}
    \UIC{$\ml{\vdash_P \Cons_1(W_1,W_2,X)\\
          : \LI{\cancel{\nat_1(W_1)}, \li_2(W_2)}{\li_1(X)}{W_2,W_1,X}}$}
    \AXC{}
    \RightLabel{$\TyLIIntroA{}_{1'}$}
    \UIC{$\ml{\vdash_P \one_1(W_1) \\
          : \LIempty{\cancel{\nat_1(W_1)}}{W_1}}$}
    \RightLabel{$\TyLITransA{}_1$}
    \BIC{$\ml{\vdash_P (\Cons_1(W_1,W_2,X), \one_1(W_1))\\
          : \LI{\cancel{\li_2(W_2)}}{\li_1(X)}{W_2,W_1,X}}$}
    \AXC{}
    \RightLabel{$\TyLIIntroA{}_2$}
    \UIC{$\ml{\vdash_P \Cons_2(W_3,W_4,W_2)\\
          : \LI{\cancel{\nat_2(W_3)},\li_3(W_4)}{\li_2(W_2)}{W_4,W_3,W_2}}$}
    \AXC{}
    \RightLabel{$\TyLIIntroA{}_{2'}$}
    \UIC{$\ml{\vdash_P \two_2(W_3)\\
          : \LIempty{\cancel{\nat_2(W_3)}}{W_3}}$}
    \RightLabel{$\TyLITransA{}_2$}
    \BIC{$\ml{\vdash_P \nu W_3.(\Cons_2(W_3,W_4,W_2), \two_2(W_3))\\
          : \LI{\cancel{\li_3(W_4)}}{\li_2(W_2)}{W_4,W_2}}$}
    \AXC{}
    \RightLabel{$\TyLIIntroA{}_3$}
    \UIC{$\ml{\vdash_P \Nil_3(W_4)\\
          : \LIempty{\li_3(W_4)}{W_4}}$}
    \RightLabel{$\TyLIElimZA{}_3$}
    \UIC{$\vdash_P \Nil_3(W_4) : \li_3(W_4)$}
    \RightLabel{$\TyLIIntroZA{}_3$}
    \UIC{$\ml{\vdash_P \Nil_3(W_4)\\
          : \LIempty{\cancel{\li_3(W_4)}}{W_4}}$}
    \RightLabel{$\TyLITransA{}_3$}
    \BIC{$\ml{\vdash_P
          \nu W_4.(\nu W_3.(\Cons_2(W_3,W_4,W_2), \two_2(W_3))
          ,\Nil_3(W_4))\\
          : \LIempty{\li_2(W_2)}{W_2}}$}
    \RightLabel{$\TyLIElimZA{}$}
    \UIC{$\ml{\vdash_P
          \nu W_4.(\nu W_3.(\Cons_2(W_3,W_4,W_2), \two_2(W_3))
          ,\Nil_3(W_4))
          : \li_2(W_2)}$}
    \RightLabel{$\TyLIIntroZA{}$}
    \UIC{$\ml{\vdash_P
          \nu W_4.(\nu W_3.(\Cons_2(W_3,W_4,W_2), \two_2(W_3))
          ,\Nil_3(W_4))}{
          : \LIempty{\cancel{\li_2(W_2)}}{W_2}}$}
    \RightLabel{$\TyLITransA{}_{12}$}
    \BIC{$\ml{\vdash_P
          (\Cons_1(W_1,W_2,X), \one_1(W_1)
          ,\nu W_4.(\nu W_3.(\Cons_2(W_3,W_4,W_2), \two_2(W_3))
          ,\Nil_3(W_4)))}{
          : \LIempty{\li_1(X)}{W_2,W_1,X}}$}
    \RightLabel{$\TyLIElimZA{}_{123}$}
    \UIC{$\ml{\vdash_P
          \nu W_1 W_2
          (\Cons_1(W_1,W_2,X), \one_1(W_1)
          ,\nu W_4.(\nu W_3.(\Cons_2(W_3,W_4,W_2), \two_2(W_3))
          ,\Nil_3(W_4)))}{
          : \li_1(X)}$}
    \RightLabel{$\TyCong{}$}
    \UIC{$\ml{\vdash_P
          \nu W_4.(\nu W_2.(\nu W_1.(\Cons_1(W_1,W_2,X), \one_1(W_1))
          ,\nu W_3.(\Cons_2(W_3,W_4,W_2), \two_2(W_3)))
          ,\Nil_3(W_4))}{
          : \li_1(X)}$}
  \end{prooftree}

  (6)
  After Proof Tree Normalisation (Step 3):
  Same as (5).

  \hrulefill{}
  \medskip
  \caption{Transformations of a Proof Tree (4)--(6).}
  \label{fig:BeforeAfterTransformation-2}
\end{sidewaysfigure}

\subsection{Graph-Prefix Normalisation}\label{subsec:graph-prefix-normalisation}

We first isolate the part of a derivation in which graph-specific rules are applied.
This allows us to separate the treatment of graph construction from the treatment of
ordinary term-forming rules such as \(\TyArrow{}\).


\begin{lemma}[Rules in graph-typing prefixes]\label{lem:graph-prefix-rules}
  For any graph $G$,
  the typing is derived solely by using
  \(\TyAlpha\), \(\TyCong\), \(\TyProd\), and \(\TyLIs\)
  in the largest prefix of its derivation tree
  that does not contain premises of \(\TyArrow\)
\end{lemma}

\begin{proof}
  We proceed by induction on the structure of the typing derivation.
  Since \( G \) is a graph and not a case expression or an application,
  the rules \(\TyCase\) and \(\TyApp\) are not applicable.
  As \( G \) is a graph and not a graph variable,
  the rule \(\TyVar\) cannot be applied.
  Thus, only the rules
  \(\TyAlpha\), \(\TyCong\), \(\TyProd\), \(\TyLIs\), and \(\TyArrow\)
  are applicable.
  If one of
  \(\TyAlpha\), \(\TyCong\), \(\TyProd\), or \(\TyLIs\)
  is applied,
  then its premises must be typing judgements of graphs,
  by the definition of these rules.
  Therefore, the claim holds by the induction hypothesis.
\end{proof}

The previous lemma identifies the rules that may occur in the graph-typing prefix.
We next remove administrative uses of \(\TyAlpha{}\) from this prefix.

\begin{lemma}[\(\TyAlpha\) elimination in graph prefixes]\label{lem:elim-ty-alpha}
  Any typing derivation for a graph can be transformed
  so that \(\TyAlpha\) does not occur
  in the largest prefix of its derivation tree
  that contains no premises of \(\TyArrow\).
\end{lemma}

\begin{proof}
  Since link substitution is defined compositionally for graphs,
  the rule \(\TyAlpha\) can be propagated upward
  through the derivation tree.

  When the transformation reaches a rule
  \(\TyArrow\), \(\TyProd\), or \(\TyLIIntro\),
  the application of \(\TyAlpha\) can be eliminated at that point.
  \begin{itemize}
    \item
          In the case of \(\TyArrow\),
          no restriction is imposed on the free links
          of the $\lambda$-abstraction atom or the arrow type.
          Hence, the resulting typing judgement
          can be obtained directly without using \(\TyAlpha\).
    \item
          For \(\TyProd\) and \(\TyLIIntro\),
          the link names in the applying production rule
          can be freely renamed.
          Therefore, the typing judgement after applying \(\TyAlpha\)
          can be derived directly without using \(\TyAlpha\).
  \end{itemize}
\end{proof}

After eliminating \(\TyAlpha{}\), structural congruence is the remaining
administrative rule in the graph-typing prefix.
The next lemma shows that all such uses can be postponed to the end.

\begin{lemma}[\(\TyCong\) collection in graph prefixes]\label{lem:elim-ty-cong}
  Any typing derivation for a graph that uses only
  \(\TyCong\), \(\TyProd\), and \(\TyLIs\)
  in the largest prefix of its derivation tree
  that contains no premises of \(\TyArrow\)
  can be transformed
  so that \(\TyCong\) occurs only once as the final rule.
\end{lemma}

\begin{proof}
  We proceed by induction on the structure of the typing derivation.
  Note that the number of rule applications strictly decreases.

  \noindent \textbf{Base Case.}
  Consider the case in which the typing derivation uses only a single rule.

  In the typing derivation tree,
  the only rules with no premises
  are \(\TyProd\)
  whose production rule has no child types
  and \(\TyLIIntro\).
  If the typing judgement \( \Gamma \vdash_P G : \tau \) is derived using only
  \TyProd{} and \TyLIIntro{},
  then by the reflexivity of structural congruence \( G \equiv G \),
  we can add \TyCong{} to derive the same typing judgement.

  \noindent \textbf{Induction Step.}
  We perform case analysis on the last rule used in the derivation:
  \TyProd{}, \TyLIIntro{}, \TyLITrans{},
  \TyLIIntroZ{}, \TyLIElimZ{}, or \TyCong{}.

  \begin{description}
    \item[Case \TyProd{}.]
          Assume the last rule applied in the derivation is \TyProd{}.
          By induction hypotheses the premises
          can be derived using \TyCong{} only once at the end.
          Thus, the following derivation can be obtained:
          \begin{prooftree}
            \AXC{$\vdots$}
            \noLine
            \UIC{$\Gamma \vdash_P G_1': \tau_1'$}
            \RightLabel{\TyCong{}}
            \UIC{$\Gamma \vdash_P G_1: \tau_1$}
            \AXC{$\cdots$}
            \AXC{$\vdots$}
            \noLine
            \UIC{$\Gamma \vdash_P G_n': \tau_n'$}
            \RightLabel{\TyCong{}}
            \UIC{$\Gamma \vdash_P G_n': \tau_n'$}
            \RightLabel{\TyProd{}}
            \TIC{$\Gamma \vdash_P
                \Zs.(C(\Ys), G_1, \dots, G_n, \seq{U \bowtie V}): \tau$}
          \end{prooftree}
          Since \( G_i' \equiv G_i \),
          it follows from structural congruence rules (e.g., E6, E7) that
          \begin{align*}
                     & \Zs.(C(\Ys), G_1', \dots, G_n', \seq{U \bowtie V}) \\
            \equiv\  & \Zs.(C(\Ys), G_1, \dots, G_n, \seq{U \bowtie V}).
          \end{align*}
          Therefore, the following derivation is valid:
          \begin{prooftree}
            \AXC{$\vdots$}
            \noLine
            \UIC{$\ml{\Gamma \vdash_P G_1': \tau_1'}$}
            \AXC{$\cdots$}
            \AXC{$\vdots$}
            \noLine
            \UIC{$\ml{\Gamma \vdash_P G_n': \tau_n'}$}
            \RightLabel{\TyProd{}}
            \TIC{$\ml{\Gamma \vdash_P\tallstrut \Zs.(C(\Ys),G_1',\dots,G_n',\seq{U \bowtie V}): \tau}$}
            \RightLabel{\TyCong{}}
            \UIC{$\ml{\Gamma \vdash_P\tallstrut \Zs.(C(\Ys),G_1,\dots,G_n,\seq{U \bowtie V}): \tau}$}
          \end{prooftree}
          This derivation uses \TyCong{} only once at the final step,
          and thus satisfies the lemma.

    \item[Case \TyLITrans{}.]
          Assume the last rule applied in the derivation is \TyLITrans{}.
          By induction hypotheses the premises
          can be derived using \TyCong{} only once at the end.
          Thus, the following derivation can be obtained,
          where $\TyCongA$ stands for $\TyCong$ and
          $\TyLITransA$ stands for $\TyLITrans$ :
          \begin{prooftree}
            \AXC{$\vdots$}
            \noLine
            \UIC{$\Gamma \vdash_P G_1' : \LI{\seq{\tau_0}}{\tau_1}{\Zs}$}
            \RightLabel{\TyCongA{}}
            \UIC{$\Gamma \vdash_P G_1 : \LI{\seq{\tau_0}}{\tau_1}{\Zs}$}
            \AXC{$\vdots$}
            \noLine
            \UIC{$\Gamma \vdash_P G_2' : \LI{\tau_1, \seq{\tau_2}}{\tau_3}{\Ys}$}
            \RightLabel{\TyCongA{}}
            \UIC{$\Gamma \vdash_P G_2 : \LI{\tau_1, \seq{\tau_2}}{\tau_3}{\Ys}$}
            \RightLabel{\TyLITransA{}}
            \BIC{$\Gamma \vdash_P \nu \Xs.(G_1, G_2) : \LI{\seq{\tau_0}, \seq{\tau_2}}{\tau_3}{\Ws}$}
          \end{prooftree}
          Since \(G_1' \equiv G_1\) and \(G_2' \equiv G_2\),
          it follows from structural congruence rules (e.g., E6, E7) that
          \(\nu \Xs.(G_1, G_2) \equiv \nu \Xs.(G_1', G_2')\).
          Therefore, the following derivation is valid:
          \begin{prooftree}
            \AXC{$\vdots$}
            \noLine
            \UIC{$\Gamma \vdash_P G_1' : \LI{\seq{\tau_0}}{\tau_1}{\Zs}$}
            \AXC{$\vdots$}
            \noLine
            \UIC{$\Gamma \vdash_P G_2' : \LI{\tau_1, \seq{\tau_2}}{\tau_3}{\Ys}$}
            \RightLabel{\TyLITrans{}}
            \BIC{$\Gamma \vdash_P \nu \Xs.(G_1', G_2')
                : \LI{\seq{\tau_0}, \seq{\tau_2}}{\tau_3}{\Ws}$}
            \RightLabel{\TyCong{}}
            \UIC{$\Gamma \vdash_P \nu \Xs.(G_1, G_2)
                : \LI{\seq{\tau_0}, \seq{\tau_2}}{\tau_3}{\Ws}$}
          \end{prooftree}
          This derivation uses \TyCong{} only once at the final step,
          and thus satisfies the lemma.

    \item[Case \TyLIElimZ{}.]
          Assume the last rule applied in the derivation is \TyLIElimZ{}.
          By induction hypotheses the premises
          can be derived using \TyCong{} only once at the end.
          Thus, the following derivation can be obtained:
          \begin{prooftree}
            \AXC{$\vdots$}
            \noLine
            \UIC{$\Gamma \vdash_P\shortstrut
                T' : \LIempty{\tau}{\Ys}$}
            \RightLabel{\TyCong{}}
            \UIC{$\Gamma \vdash_P\shortstrut
                T : \LIempty{\tau}{\Ys}$}
            \RightLabel{\TyLIElimZ{}}
            \UIC{$\Gamma \vdash_P\shortstrut T : \tau$}
          \end{prooftree}
          The above can be transformed as follows.
          \begin{prooftree}
            \AXC{$\vdots$}
            \noLine
            \UIC{$\Gamma \vdash_P\shortstrut
                T' : \LIempty{\tau}{\Ys}$}
            \RightLabel{\TyLIElimZ{}}
            \UIC{$\Gamma \vdash_P\shortstrut T' : \tau$}
            \RightLabel{\TyCong{}}
            \UIC{$\Gamma \vdash_P\shortstrut T : \tau$}
          \end{prooftree}
          This derivation uses \TyCong{} only once at the final step,
          and thus satisfies the lemma.

    \item[Case \TyLIIntroZ{}.]
          Assume the last rule applied in the derivation is \TyLIIntroZ{}.
          By induction hypotheses the premises
          can be derived using \TyCong{} only once at the end.
          Thus, the following derivation can be obtained:
          \begin{prooftree}
            \AXC{$\vdots$}
            \noLine
            \UIC{$\Gamma \vdash_P\shortstrut T' : \tau$}
            \RightLabel{\TyCong{}}
            \UIC{$\Gamma \vdash_P\shortstrut T : \tau$}
            \RightLabel{\TyLIIntroZ{}}
            \UIC{$\Gamma \vdash_P
                T : \LIempty{\tau}{\Ys}$}
          \end{prooftree}
          The above can be transformed as follows.
          \begin{prooftree}
            \AXC{$\vdots$}
            \noLine
            \UIC{$\Gamma \vdash_P\shortstrut T' : \tau$}
            \RightLabel{\TyLIIntroZ{}}
            \UIC{$\Gamma \vdash_P\shortstrut T' : \LIempty{\tau}{\Ys}$}
            \RightLabel{\TyCong{}}
            \UIC{$\Gamma \vdash_P\shortstrut
                T : \LIempty{\tau}{\Ys}$}
          \end{prooftree}
          This derivation uses \TyCong{} only once at the final step,
          and thus satisfies the lemma.

    \item[Case \TyCong{}.]
          If multiple consecutive applications of \(\TyCong\) occur,
          they can be merged into a single application.
          \begin{prooftree}
            \AXC{$\vdots$}
            \noLine
            \UIC{$\Gamma \vdash_P\shortstrut G_1: \tau$}
            \RightLabel{\TyCong{}}
            \UIC{$\Gamma \vdash_P\shortstrut G_2: \tau$}
            \RightLabel{\TyCong{}}
            \UIC{$\Gamma \vdash_P\shortstrut G_3: \tau$}
          \end{prooftree}
          Since \(G_1 \equiv G_2\) and \(G_2 \equiv G_3\),
          \(G_1 \equiv G_3\).
          Therefore the above can be transformed as follows.
          \begin{prooftree}
            \AXC{$\vdots$}
            \noLine
            \UIC{$\Gamma \vdash_P\shortstrut G_1: \tau$}
            \RightLabel{\TyCong{}}
            \UIC{$\Gamma \vdash_P\shortstrut G_3: \tau$}
          \end{prooftree}
  \end{description}
\end{proof}

By the preceding lemmas, the essential part of the argument is reduced to
derivations whose graph-typing prefix consists only of \(\TyProd{}\) and
linear-implication rules, up to a final \(\TyCong{}\).
We now show that the linear-implication rules in such a prefix can be eliminated.

\subsection{Elimination of Linear Implication Rules}\label{subsec:elim-limp}

The key point is the form of the right premise of \(\TyLITrans{}\).
When this premise has an empty left-hand side and its graph exposes exactly the
free links required by the resulting linear implication type, the corresponding
use of \(\TyLITrans{}\) can be reorganised so that the incomplete graph structure
is generated directly by a production rule.

\begin{figure}[tbp]
  \scriptsize
  \hrulefill{}

  \def\ScoreOverhang{0pt}

  Before transformation:
  \begin{prooftree}
    \AXC{$\paren{\alpha (\Xs) \lto \nu
          \Zs.(C(\Ys), \seq{U \bowtie V}, \taus)} \in P$}
    \RightLabel{$\TyLIIntroA{}$}
    \UIC{$\vdash_P
        T_{n+1} : \LI{\tau_n,\dots}{\alpha(\Xs)}{\Us, \Vs, \Ys}$}
    \AXC{$\vinf{\vdash_P T_n : \LI{\dots}{\tau_n}{\dots}}$}
    \RightLabel{$\TyLITransA_n$}
    \BIC{$\begin{array}{c}
          \vdash_P
          \nu\seq{Z_n}.(T_{n+1},T_n) : \LI{\tau_{n-1},\dots}{\tau}{\dots} \\
          \quad \ddots                                                    \\[1pt]
          \begin{array}{r@{~}l}
            \vdash_P
            \nu\seq{Z_2}.(\dots(
            \nu\seq{Z_{i-1}}.(
            \nu\seq{Z_i}.(
            \nu\seq{Z_{i+1}}.(\dots(
            \nu\seq{Z_n}.(T_{n+1},T_n),\quad & \\
            \dots),
            T_{i+1}), T_i), T_{i-1}), \dots), T_2)
            : \LI{\tau_1}{\tau}{\dots}         \\
          \end{array}
        \end{array}$}
    \AXC{\raisebox{-21pt}{\hspace*{-55pt}$\vinf{\vdash_P T_1 : \LIempty{\tau_1}{\dots}}$}}
    \RightLabel{$\TyLITransA_1$}
    \BIC{$\vdash_P
        \nu\seq{Z_1}.(
        \nu\seq{Z_2}.(\dots(
        \nu\seq{Z_{i-1}}.(
        \nu\seq{Z_i}.(
        \nu\seq{Z_{i+1}}.(\dots(
        \nu\seq{Z_n}.(T_{n+1},T_n),
        \dots), T_{i+1}), T_i), T_{i-1}), \dots), T_2), T_1)
        : \LIempty{\tau}{\dots}$}
    \RightLabel{$\TyLIElimZA{}$}
    \UIC{$\vdash_P
        \nu\seq{X}
        \seq{Z_1}.(
        \nu\seq{Z_2}.(\dots(
        \nu\seq{Z_{i-1}}.(
        \nu\seq{Z_i}.(
        \nu\seq{Z_{i+1}}.(\dots(
        \nu\seq{Z_n}.(T_{n+1},T_n),
        \dots), T_{i+1}), T_i), T_{i-1}), \dots), T_2), T_1)
        : \tau$}
  \end{prooftree}

  where $T_{n+1} \triangleq (C(\Ys), \seq{U \bowtie V})$ and\\
  $\{\Xs\} = \fn(
  \nu\seq{Z_1}.(
  \nu\seq{Z_2}.(\dots(
  \nu\seq{Z_{i-1}}.(
  \nu\seq{Z_i}.(
  \nu\seq{Z_{i+1}}.(\dots(
  \nu\seq{Z_n}.(T_{n+1},T_n),
  \dots), T_{i+1}), T_i), T_{i-1}), \dots), T_2), T_1)) \setminus \fn(\tau)$

  \medskip
  After transformation:
  \begin{prooftree}
    \AXC{$\paren{\alpha (\Xs) \lto \nu
          \Zs.(C(\Ys), \seq{U \bowtie V}, \taus)} \in P$}
    \RightLabel{$\TyLIIntroA{}$}
    \UIC{$\vdash_P
        T_{n+1} : \LI{\tau_n,\dots}{\alpha(\Xs)}{\Us, \Vs, \Ys}$}
    \AXC{$\vinf{\vdash_P T_n : \LI{\dots}{\tau_n}{\dots}}$}
    \RightLabel{$\TyLITransA_n$}
    \BIC{$\begin{array}{c}
          \vdash_P
          (T_{n+1},T_n) : \LI{\tau_{n-1},\dots}{\tau}{\dots} \\
          \quad \ddots                                       \\
          \qquad \vdash_P
          (T_{n+1},T_n, \dots, T_{i+1}, T_i, T_{i-1}, \dots, T_2)
          : \LI{\tau_1}{\tau}{\dots}                         \\
        \end{array}$}
    \AXC{\raisebox{-13pt}{\hspace*{-20pt}$\vinf{\vdash_P T_1 : \LIempty{\tau_1}{\dots}}$}}
    \RightLabel{$\TyLITransA_1$}
    \BIC{$\vdash_P
        (T_{n+1},T_n, \dots, T_{i+1}, T_i, T_{i-1}, \dots, T_2, T_1)
        : \LIempty{\tau}{\dots}$}
    \RightLabel{$\TyLIElimZ{}$}
    \UIC{$\vdash_P \nu \seq{X}.(T_{n+1},T_n, \dots, T_{i+1}, T_i, T_{i-1}, \dots, T_2, T_1) : \tau$}
  \end{prooftree}

  where $T_{n+1} \triangleq (C(\Ys), \seq{U \bowtie V})$ and
  $\{\Xs\} = \fn((T_{n+1}, \dots, T_1)) \setminus \fn(\tau)$

  Here,
  \TyProd{},
  \TyLIIntro{},
  \TyLITrans{},
  \TyLIElimZ{}, and
  \TyLIIntroZ{},
  are abbreviated as
  \TyProdA{},
  \TyLIIntroA{},
  \TyLITransA{},
  \TyLIElimZA{}, and
  \TyLIIntroZA{}, respectively.

  \hrulefill{}
  \caption{Prenex Normal Form Transformation.}\label{fig:prenex-trans}
\end{figure}

\begin{figure}[tbp]
  \scriptsize
  \hrulefill{}

  \def\ScoreOverhang{0pt}
  \def\defaultHypSeparation{\hskip .05in}

  Before transformation:
  \begin{prooftree}
    \AXC{$\vinf{\vdash_P (T_{n+1}, \dots, T_{i+1})\\
          : \LI{\tau_i,\dots}{\tau}{\dots}}$}
    \AXC{$\vinf{\boxed{\vdash_P \boxed{T_i} : \LIempty{\tau_i}{\dots}}}$}
    \RightLabel{$\TyLITransA_i$}
    \BIC{$\mlc{\vdash_P
          (T_{n+1}, \dots, T_{i+1}, \boxed{T_i})\\
          : \LI{\tau_{i-1},\dots}{\tau}{\dots}}$}
    \AXC{$\vinf{\vdash_P T_{i-1} : \LIempty{\tau_{i-1}}{\dots}}$}
    \RightLabel{$\TyLIElimZA_{i-1}$}
    \UIC{$\vdash_P \nu\seq{Y_{i-1}}.T_{i-1} : \tau_{i-1}$}
    \RightLabel{$\TyLIIntroZA_{i-1}$}
    \UIC{$\vdash_P \nu\seq{Y_{i-1}}.T_{i-1} : \LIempty{\tau_{i-1}}{\dots}$}
    \RightLabel{$\TyLITransA_{i-1}$}
    \BIC{$\mlc{
          \vdash_P
          (T_{n+1}, \dots, T_{i+1}, \boxed{T_i}, \nu\seq{Y_{i-1}}.T_{i-1}) : \LI{\tau_{i-2},\dots}{\tau}{\dots} \\[6pt]
          \quad \ddots \\[6pt]
          \vdash_P
          (T_{n+1},T_n, \dots, T_{i+1}, \boxed{T_i}, \nu\seq{Y_{i-1}}.T_{i-1}, \dots, \nu\seq{Y_2}.T_2)
          : \LI{\tau_1}{\tau}{\dots}
        }$}
    \AXC{\hspace*{-40pt}$\vinf{\vdash_P T_1 : \LIempty{\tau_1}{\dots}}$}
    \RightLabel{$\TyLIElimZA_1$}
    \UIC{\hspace*{-40pt}$\vdash_P \nu\seq{Y_1}.T_1 : \tau_1$}
    \RightLabel{$\TyLIIntroZA_1$}
    \UIC{\hspace*{-40pt}$\vdash_P \nu\seq{Y_1}.T_1 : \LIempty{\tau_1}{\dots}$}
    %
    \RightLabel{$\TyLITransA_1$}
    \BIC{$\vdash_P
        (T_{n+1},T_n, \dots, T_{i+1}, \boxed{T_i},
        \nu\seq{Y_{i-1}}.T_{i-1}, \dots, \nu\seq{Y_2}.T_2, \nu\seq{Y_1}.T_1)
        : \LIempty{\tau}{\dots}$}
    \RightLabel{$\TyLIElimZA{}$}
    \UIC{$\vdash_P \nu \seq{X}.(T_{n+1},T_n, \dots, T_{i+1}, \boxed{T_i},
        \nu\seq{Y_{i-1}}.T_{i-1}, \dots, \nu\seq{Y_2}.T_2, \nu\seq{Y_1}.T_1) : \tau$}
  \end{prooftree}

  \medskip
  where
  $\{\Xs\} = \fn((T_{n+1},T_n, \dots, T_{i+1}, \boxed{T_i}, \nu\seq{Y_{i-1}}.T_{i-1}, \dots, \nu\seq{Y_1}.T_1)) \setminus \fn(\tau)$

  \medskip

  After transformation:
  \begin{prooftree}
    \AXC{$\vinf{\vdash_P (T_{n+1}, \dots, T_{i+1})\\
          : \LI{\tau_i,\dots}{\tau}{\dots}}$}
    \AXC{$\vinf{\boxed{\vdash_P T_i : \LIempty{\tau_i}{\dots}}}$}
    \RightLabel{$\TyLIElimZA_i$}
    \UIC{$\vdash_P \nu\seq{Y_i}.T_i : \tau_i$}
    \RightLabel{$\TyLIIntroZA_i$}
    \UIC{$\vdash_P \boxed{\nu\seq{Y_i}.T_i} : \LIempty{\tau_i}{\dots}$}
    \RightLabel{$\TyLITransA_i$}
    \BIC{$\mlc{\vdash_P
          (T_{n+1}, \dots, T_{i+1}, \boxed{\nu\seq{Y_i}.T_i})\\
          : \LI{\tau_{i-1},\dots}{\tau}{\dots}}$}
    \AXC{$\vinf{\vdash_P T_{i-1} : \LIempty{\tau_{i-1}}{\dots}}$}
    \RightLabel{$\TyLIElimZA_{i-1}$}
    \UIC{$\vdash_P \nu\seq{Y_{i-1}}.T_{i-1} : \tau_{i-1}$}
    \RightLabel{$\TyLIIntroZA_{i-1}$}
    \UIC{$\vdash_P \nu\seq{Y_{i-1}}.T_{i-1} : \LIempty{\tau_{i-1}}{\dots}$}
    \RightLabel{$\TyLITransA_{i-1}$}
    \BIC{$\mlc{
          \vdash_P
          (T_{n+1}, \dots, T_{i+1}, \boxed{\nu\seq{Y_i}.T_i}, \nu\seq{Y_{i-1}}.T_{i-1}) : \LI{\tau_{i-2},\dots}{\tau}{\dots} \\[3pt]
          \quad \ddots \\[3pt]
          \vdash_P
          (T_{n+1}, \dots, T_{i+1}, \boxed{\nu\seq{Y_i}.T_i}, \nu\seq{Y_{i-1}}.T_{i-1}, \dots, \nu\seq{Y_2}.T_2)
          : \LI{\tau_1}{\tau}{\dots}
        }$}
    \AXC{\hspace{-40pt}$\vinf{\vdash_P T_1 : \LIempty{\tau_1}{\dots}}$}
    \RightLabel{$\TyLIElimZA_1$}
    \UIC{\hspace*{-40pt}$\vdash_P \nu\seq{Y_1}.T_1 : \tau_1$}
    \RightLabel{$\TyLIIntroZA_1$}
    \UIC{\hspace*{-40pt}$\vdash_P \nu\seq{Y_1}.T_1 : \LIempty{\tau_1}{\dots}$}
    \RightLabel{$\TyLITransA_1$}
    \BIC{$\vdash_P
        (T_{n+1},T_n, \dots, T_{i+1}, \boxed{\nu\seq{Y_i}.T_i},
        \nu\seq{Y_{i-1}}.T_{i-1}, \dots, \nu\seq{Y_2}.T_2, \nu\seq{Y_1}.T_1)
        : \LIempty{\tau}{\dots}$}
    \RightLabel{$\TyLIElimZA{}$}
    \UIC{$\vdash_P \nu \seq{Z}.(T_{n+1},T_n, \dots, T_{i+1}, \boxed{\nu\seq{Y_i}.T_i},
        \nu\seq{Y_{i-1}}.T_{i-1}, \dots, \nu\seq{Y_2}.T_2, \nu\seq{Y_1}.T_1) : \tau$}
  \end{prooftree}

  \medskip
  where
  \begin{itemize}
    \item
          $\{\seq{Y_i}\} = \fn(T_i) \setminus \fn(\tau_i)$
    \item
          $\{\Zs\} = \fn((T_{n+1},\dots,T_{i+1}, \boxed{\nu\seq{Y_i}.T_i},\nu\seq{Y_{i-1}}.T_{i-1},\dots,\nu\seq{Y_1}.T_1)) \setminus \fn(\tau) = \{\Xs\} \setminus \{\seq{Y_i}\}$
  \end{itemize}

  Here,
  \TyProd{},
  \TyLIIntro{},
  \TyLITrans{},
  \TyLIElimZ{}, and
  \TyLIIntroZ{},
  are abbreviated as
  \TyProdA{},
  \TyLIIntroA{},
  \TyLITransA{},
  \TyLIElimZA{}, and
  \TyLIIntroZA{}, respectively.

  \hrulefill{}
  \caption{Hyperlink Creation Localisation.}
  \label{fig:hyperlink-hiding-localisations}
\end{figure}

\begin{figure}[tbp]
  \scriptsize \hrulefill{}

  Before transformation:

  \def\ScoreOverhang{0pt} \def\defaultHypSeparation{\hskip .05in}
  \begin{prooftree} \AXC{$\vinf{\vdash_P (T_{n+1}, \dots, T_{m+1}) \\ :
          \LI{\tau_m,\seq{\tau_{m+1}}}{\tau}{\dots}}$}
    \AXC{$\vinf{\vdash_P T_m : \LI{\tau_i,
            \seq{\tau_m}}{\tau_m}{\dots}}$} \RightLabel{$\TyLITrans_m$}
    \BIC{$\mlc{ \vdash_P (T_{n+1}, \dots, T_{m+1}, T_m) :
          \LI{\seq{\tau_{m+1}},\tau_i,\seq{\tau_m}}{\tau}{\dots}\\
          \ddots_2 \\ \vdash_P (T_{n+1},\dots,T_m,\dots,T_{i+1}) :
          \LI{\tau_i,\dots}{\tau}{\dots} }$} \AXC{$\vdash_P T_i :
        \tau_i$} \RightLabel{$\TyLIIntroZ_i$} \UIC{$\vdash_P T_i :
        \LIempty{\tau_i}{\dots}$} \RightLabel{$\TyLITrans_i$}
    \BIC{$\mlc{ \vdash_P (T_{n+1},\dots,T_m,\dots,T_{i+1},T_i) :
          \LI{\dots}{\tau}{\dots} \\ \ddots_1 \\ \vdash_P
          (T_{n+1}\dots,T_m,\dots, T_{i+1},T_i,T_{i-1},\dots,T_1) :
          \LIempty{\tau}{\dots} }$} \RightLabel{$\TyLIElimZ{}$}
    \UIC{$\vdash_P \nu
        \seq{X}.(T_{n+1},\dots,T_m,\dots,T_{i+1},T_i,T_{i-1},\dots,T_1)
        : \tau$} \end{prooftree}

  \medskip After transformation:

  \begin{prooftree} \AXC{$\vinf{\vdash_P (T_{n+1}, \dots,
          T_{m+1}) \\ :
          \LI{\tau_m,\seq{\tau_{m+1}}}{\tau}{\dots}}$}
    \AXC{$\vinf{\vdash_P T_m :
          \LI{\tau_i,\seq{\tau_m}}{\tau_m}{\dots}}$}
    \AXC{$\vdash_P T_i : \tau_i$}
    \RightLabel{$\TyLIIntroZ_i$} \UIC{$\vdash_P T_i :
        \LIempty{\tau_i}{\dots}$} \RightLabel{$\TyLITrans_i$}
    \BIC{$\vdash_P (T_m,T_i) :
        \LI{\seq{\tau_m}}{\tau}{\dots}$}
    \RightLabel{$\TyLITrans_m$} \BIC{$\mlc{ \vdash_P
          (T_{n+1}, \dots, T_{m+1}, (T_m, T_i)) :
          \LI{\seq{\tau_{m+1}},\seq{\tau_m}}{\tau}{\dots}\\
          \ddots_2 \qquad\\ \qquad \ddots_1 \\ \vdash_P
          (T_{n+1}\dots,(T_m,T_i),\dots,
          T_{i+1},T_{i-1},\dots,T_1) : \LIempty{\tau}{\dots} }$}
    \RightLabel{$\TyLIElimZ{}$} \UIC{$\vdash_P
        \nu\seq{X}.(T_{n+1},\dots,(T_m,T_i),\dots,T_{i+1},T_{i-1},\dots,T_1)
        : \tau$} \end{prooftree}

  \hrulefill{} \caption{Proof Tree Normalisation.}
  \label{fig:ProofTreeNormalization}
\end{figure}

Before proving the elimination theorem, we record the proof-tree transformations
used in the argument and justify that they preserve the side conditions of the
typing rules.

\begin{lemma}[Free Link Containment]\label{lem:free-link-containment}
  If \(\Gamma \vdash_P T : \LI{\seq{\tau}}{\tau}{\Xs}\) holds,
  then \(\fn(\tau) \subseteq \{\Xs\} \cup \fn(\seq{\tau})\).
\end{lemma}

\begin{proof}
  By induction on the structure
  of the typing derivation for linear implication.
  \begin{description}
    \item[Case \TyLIIntro{}]\mbox{}\\
          By assumption, we have a proof of the form.
          \begin{prooftree}
            \AXC{$\paren{\alpha (\Xs) \lto \nu
                  \Zs.(C(\Ys), \seq{U \bowtie V}, \taus)} \in P$}
            \RightLabel{$\TyLIIntro{}$}
            \UIC{$\Gamma \vdash_P (C(\Ys), \seq{U \bowtie V})
                : \LI{\taus}{\alpha(\Xs)}{\Us, \Vs, \Ys}$}
          \end{prooftree}
          Then, the lemma holds since
          \(\fn(\alpha(\Xs)) =
          \paren{\{\Ys\} \cup \fn(\seq{U \bowtie V}) \cup \fn(\taus)} \setminus \{\Zs\}
          \subseteq \paren{\{\Us, \Vs, \Ys\} \cup \fn(\taus))}\).

    \item[Case \TyLITrans{}]\mbox{}\\
          Since this property is precisely the link condition (v) of \TyLITrans{},
          it is trivially satisfied whenever that condition is explicitly enforced.

          Furthermore,
          when we perform the transformation in \cref{fig:prenex-trans},
          and the graph typed by \TyLITrans{} contains no leftmost hyperlink creation,
          the property holds even when the link condition (v) of \TyLITrans{}
          is not explicitly imposed,
          as shown below.

          By assumption, we have a proof of the form.
          \begin{prooftree}
            \AXC{$\ml{
                  \Gamma \vdash_P T_1 : \LI{\tau_2, \seq{\tau_1}}{\tau_1}{\Xs}
                  \\
                  \Gamma \vdash_P T_2 : \LI{\seq{\tau_2}}{\tau_2}{\Ys}
                }$}
            \RightLabel{$\TyLITrans{}$}
            \UIC{$\Gamma \vdash_P (T_1, T_2) :
                \LI{\seq{\tau_1},\seq{\tau_2}}{\tau_1}{\Xs, \Ys}$}
          \end{prooftree}

          By the induction hypothesis,
          we have
          \(\fn(\tau_1) \subseteq \{\Xs\} \cup \fn(\tau_2, \seq{\tau_1})\)
          and
          \(\fn(\tau_2) \subseteq \{\Ys\} \cup \fn(\seq{\tau_2})\).
          Thus, the following desired property holds.
          \begin{align*}
            \fn(\tau_1) & \subseteq \{\Xs\} \cup \fn(\tau_2) \cup \fn(\seq{\tau_1})                            \\
                        & \subseteq \{\Xs\} \cup \paren{\{\Ys\} \cup \fn(\seq{\tau_2})} \cup \fn(\seq{\tau_1}) \\
                        & =
            \{\Xs,\Ys\} \cup \fn(\seq{\tau_1},\seq{\tau_2})
          \end{align*}
    \item[Case \TyLIIntroZ{}]\mbox{}\\
          By assumption, we have a proof of the form.
          \begin{prooftree}
            \AXC{$\Gamma \vdash_P T : \tau$}
            \RightLabel{$\TyLIIntroZ{}$}
            \UIC{$\Gamma \vdash_P T : \LIempty{\tau}{\Ys}$}
          \end{prooftree}
          Since the free links of both sides of typing relation should be equal,
          \(\fn(T) = \fn(\tau) = \{\Ys\}\) must hold
          and thus
          \(\fn(\tau) \subseteq \{\Ys\} \cup \fn(\zero)\).
  \end{description}
\end{proof}

This containment property will be used to compare the links hidden by
\(\TyLIElimZ{}\) with the links that are already accounted for by the target type.
The following elementary set-theoretic equivalence is used for that comparison.

\begin{lemma}[Subset Duality]\label{lem:subset-duality}
  $(A \cap C \subseteq B) = (A \setminus B \subseteq A \setminus C)$.
\end{lemma}

\begin{proof}
  The LHS and the RHS are translated to
  $(x \in A \land x \in C \implies x \in B)$ and
  $((x \in A \land x \notin B) \implies (x \in A \land x \notin C))$,
  which are equivalent.
\end{proof}

We next justify the localisation step.
It moves hyperlink creation from the outermost graph to the subgraph where the
corresponding hidden links actually occur.

\begin{lemma}[Hyperlink Creation Localisation]\label{lem:nu-localisation}
  A graph with Hyperlink Creation Localisation as shown in
  \cref{fig:hyperlink-hiding-localisations}
  is congruent to the original graph.
\end{lemma}

\begin{proof}
  Consider the proof tree illustrated in the upper part of
  \cref{fig:hyperlink-hiding-localisations}.
  Its root rule is \(\TyLIElimZ\).
  The leftmost branch consists exactly of the rules \(\TyLITrans_1, \dots, \TyLITrans_i\).
  In every such derivation step,
  the left-hand side of the linear implication type in the second premise
  is empty.

  Here, the links $\Xs$ are defined as
  \[
    \{\Xs\} \triangleq
    \paren{\fn(T_{n+1}) \cup \dots \cup \fn(T_1)} \setminus \fn(\tau).
  \]
  and the rule \(\TyLIIntro\) is applied at the top left,
  \[
    T_{n+1} \triangleq (C(\Xs), \seq{U \bowtie V}).
  \]

  Then, the three side conditions on the links for \(\TyLITrans_i\)
  are given as follows.
  For all \(1 \le j, k \le n+1\),
  \begin{align}
    j > i & \implies \fn(T_j) \cap \fn(T_i)    \subseteq \fn(\tau_i) \label{C1}            \\
    j > i & \implies \fn(T_j) \cap \fn(\zero) = \emptyset \subseteq \fn(\tau_i) \label{C2} \\
    i > k & \implies \paren{\fn(T_i) \cup \fn(\zero)} \cap \fn(\tau_k)
    \notag                                                                                 \\
          & \qquad\qquad = \fn(T_i) \cap \fn(\tau_k) \subseteq \fn(\tau_i) \label{C3}
  \end{align}
  By replacing \(j, i\) in \eqref{C1} with \(i, k\),
  we obtain
  \[
    \fn(T_i) \cap \fn(T_k) \subseteq \fn(\tau_k).
  \]
  Substituting this into \eqref{C3} for \(\fn(\tau_k)\) yields
  \[
    \fn(T_i) \cap (\fn(T_i) \cap \fn(T_k)) = \fn(T_i) \cap \fn(T_k) \subseteq \fn(\tau_i).
  \]
  Hence, together with \eqref{C1}, we obtain
  \begin{align*}
    \forall j \neq i.\quad
    \fn(T_j) \cap \fn(T_i) \subseteq \fn(\tau_i).
    \label{C4}
  \end{align*}
  Therefore,
  \[
    \forall j \neq i.\quad
    \fn(T_j) \cap (\fn(T_i) \setminus \fn(\tau_i)) = \emptyset.
  \]
  Let
  \[
    \{\Ys\} \triangleq (\fn(T_i) \setminus \fn(\tau_i)).
  \]
  Then, the above can be rewritten as
  \begin{align}
    \forall j \neq i.\quad
    \fn(T_j) \cap \{\Ys\} = \emptyset.
    \label{C5}
  \end{align}
  By \cref{lem:free-link-containment}, we have
  \[
    \fn(\tau) \subseteq \fn(T_{n+1}) \cup \dots \cup \fn(T_{i+1}) \cup \fn(\tau_i) \cup \dots \cup \fn(\tau_1).
  \]
  Hence,
  \begin{align*}
     & \fn(T_i) \cap \fn(\tau)                                                                                                              \\
     & \qquad \subseteq
    \underbrace{(\fn(T_i) \cap \fn(T_{n+1})) \cup \dots \cup (\fn(T_i) \cap \fn(T_{i+1}))}_{\subseteq \fn(\tau_i) \; \because \eqref{C1}}   \\
     & \qquad\qquad \cup
    \underbrace{(\fn(T_i) \cap \fn(\tau_i))}_{\subseteq \fn(\tau_i)}                                                                        \\
     & \qquad\qquad \cup
    \underbrace{(\fn(T_i) \cap \fn(\tau_{i-1})) \cup \dots \cup (\fn(T_i) \cap \fn(\tau_1))}_{\subseteq \fn(\tau_i) \; \because \eqref{C3}} \\
     & \qquad \subseteq \fn(\tau_i)
  \end{align*}
  From this and \cref{lem:subset-duality}, it follows that
  \begin{align}
    \fn(T_i) \setminus \fn(\tau_i) \subseteq \fn(T_i) \setminus \fn(\tau).
    \label{C6}
  \end{align}
  Therefore, for links $\Xs$, we have
  \begin{align*}
    \{\Xs\}
     & \triangleq \bigl(\fn(T_n) \cup \dots \cup \fn(T_i)
    \cup \dots \cup \fn(T_1)\bigr) \setminus \fn(\tau)                                \\
     & =          \paren{\fn(T_n) \setminus \fn(\tau)} \cup \dots \cup
    {\paren{\fn(T_i) \setminus \fn(\tau)}}                                            \\
     & \qquad \cup \dots \cup \paren{\fn(T_1) \setminus \fn(\tau)}                    \\
     & \supseteq  {\paren{\fn(T_i) \setminus \fn(\tau)}}                              \\
     & \supseteq  {\paren{\fn(T_i) \setminus \fn(\tau_i)}} \qquad \because \eqref{C6} \\
     & =          {\{\Ys\}}.
  \end{align*}

  Let \(\{\Zs\} = \{\Xs\} \setminus \{\Ys\}\).
  From \eqref{C5} and \(\{\Ys\} \subseteq \{\Xs\}\), we obtain
  \begin{align*}
     & \nu \Zs.(T_{n+1}, T_n, \dots, T_{i+1}, \nu\Ys.T_i, T_{i-1}, \dots, T_2, T_1)
    \\
     & \qquad \equiv
    \nu \Xs.(T_{n+1}, T_n, \dots, T_{i+1}, T_i, T_{i-1}, \dots, T_2, T_1).
  \end{align*}
\end{proof}

The remaining transformation changes the order in which \(\TyLITrans{}\) is
applied.
The next lemma states that this reordering does not violate the link conditions
of \(\TyLITrans{}\).

\begin{lemma}[$\TyLITrans{}$ Link Condition Preservation]\label{lem:link-cond-preservation}
  Transforming proof trees as
  \cref{fig:ProofTreeNormalization}
  preserves the link conditions
  especially for \(\TyLITrans\).
\end{lemma}

\begin{proof}
  For the link condition (i) of \(\TyLITrans{}\), it holds trivially.
  The link condition (v) of \(\TyLITrans{}\) is established by \cref{lem:free-link-containment}.
  Therefore, we focus on conditions (ii)--(iv) in the latter discussion.
  For the link conditions of the rule \(\TyLITrans_i\)
  in the transformed typing derivation in
  \cref{fig:ProofTreeNormalization},
  the following holds.
  \begin{enumerate}
    \item[(ii)]
          \(\fn(T_m) \cap \fn(T_i) = \fn(T_m) \cap \fn(\tau_i) \subseteq \fn(\tau_i)\)\\
          \(\quad \because \fn(T_i) = \fn(\tau_i)\).
    \item[(iii)]
          \(\fn(T_m) \cap \fn(\zero) \subseteq \fn(\tau_i)\).
    \item[(iv)]
          \(\paren{\fn(T_i) \cup \fn(\zero)} \cap \fn(\zero) \subseteq \fn(\tau_i)\).
  \end{enumerate}

  For the link conditions of the rule \(\TyLITrans_m\),
  the following holds.
  \begin{enumerate}
    \item[(ii)]
          \(\fn(T_{m+1}) \cap \fn((T_m,T_i)) \subseteq \fn(\tau_m)\)
          because
          \begin{itemize}
            \item
                  \(\fn(T_{m+1}) \cap \fn(T_m) \subseteq \fn(\tau_m)\)
                  holds by the link condition (ii) of \(\TyLITrans_m\)
                  in the previous proof tree.
            \item
                  \(\fn(T_{m+1}) \cap \fn(T_i) =
                  \fn(T_{m+1}) \cap \fn(\tau_i) \subseteq \fn(\tau_m)\)
                  holds by the link condition (iii) of \(\TyLITrans_m\)
                  in the previous proof tree.
          \end{itemize}
    \item[(iii)]
          \(\fn(T_{m+1}) \cap \fn(\seq{\zero}) \subseteq \fn(\tau_m)\)
          trivially holds.
    \item[(iv)]
          $\fn((T_m,T_i)) \cap \fn(\tau_{m-1},\dots,\tau_{i+1},\tau_{i-1},\dots,\tau_1)$ \\
          $=
            \paren{\fn(T_m) \cup \fn(\tau_i)} \cap \fn(\tau_{m-1},\dots,\tau_{i+1},\tau_{i-1},\dots,\tau_1)$ \\
          $\qquad\qquad\quad\ \because \fn(T_i) = \fn(\tau_i)$ \\
          $\subseteq \fn(\tau_m)
            \quad\because
            \text{Condition (iv) of previous $\TyLITrans_i$.}$
  \end{enumerate}
  For the link conditions of the rule \(\TyLITrans_{i-1}\),
  the following holds.
  \begin{enumerate}
    \item[(ii)]
          \(\fn((T_{m+1},(T_m,T_i),\dots,T_{i+1})) \cap \fn(T_{i-1}) =
          \fn((T_{m+1},(T_m,T_i),\dots,T_{i+1})) \cap \fn(\tau_{i-1}) \subseteq \fn(\tau_{i-1})\)
          \(\quad \because \fn(T_{i-1}) = \fn(\tau_{i-1})\).
    \item[(iii)]
          \(\fn((T_{m+1},(T_m,T_i),\dots,T_{i+1}) \cap \fn(\zero) \subseteq \fn(\tau_{i-1})\)
          trivially holds.
    \item[(iv)]
          \(\paren{\fn(T_{i-1}) \cup \fn(\zero)} \cap \fn(\tau_{i-2}, \dots, \tau_1) =
          \fn(\tau_{i-1}) \cap \fn(\tau_{i-2}, \dots, \tau_1) =
          \fn(\tau_{i-1}) \subseteq \fn(\tau_{i-1})\).
  \end{enumerate}
  For the link conditions of the rule
  \(\TyLITrans_j \quad (n > j \neq i)\)
  the following holds.
  \begin{enumerate}
    \item[(ii)]
          \(\fn((T_{m+1},(T_m,T_i),\dots,T_{j+1})) \cap \fn(T_j) =
          \fn((T_{m+1},(T_m,T_i),\dots,T_{j+1})) \cap \fn(\tau_j) \subseteq \fn(\tau_j)\)
          \(\quad \because \fn(T_j) = \fn(\tau_j)\).
    \item[(iii)]
          \(\fn((T_{m+1},(T_m,T_i),\dots,T_{j+1}) \cap \fn(\zero) \subseteq \fn
          (\tau_j)\)
          trivially holds.
    \item[(iv)]
          \(\paren{\fn(T_j) \cup \fn(\zero)} \cap \fn(\paren{\tau_{j-1}, \dots, \tau_1} \setminus \tau_i) =
          \fn(\tau_j) \cap \fn(\paren{\tau_{j-1}, \dots, \tau_1} \setminus \tau_i) =
          \fn(\tau_j) \subseteq \fn(\tau_j)\).
  \end{enumerate}
\end{proof}

We can now combine the preceding transformations.
The following theorem is the main technical step:
it removes all uses of the
linear-implication rules from the graph-typing prefix.

\begin{theorem}[\(\TyLIs{}\) elimination]\label{thm:ty-lis-elim}
  Let \(\mathcal{D}\) be a typing derivation for a graph judgement.
  Suppose that, in the largest prefix of \(\mathcal{D}\) that contains no premises of \(\TyArrow\),
  every rule occurrence is an instance of either \(\TyProd\) or one of the linear-implication rules \(\TyLIs{}\).
  Then this prefix can be transformed into one that uses only \(\TyProd\) and \(\TyAlpha\),
  followed by at most one final application of \(\TyCong\).
\end{theorem}

\begin{proof}
  We follow a proof tree (without the final $\TyCong$ or $\TyAlpha$)
  upwards, repeating the operations described below, so that
  we can construct a proof tree of a graph template without using
  \(\TyLIs\).

  If the original proof tree does not contain
  \(\TyLIElimZ\), the claim holds vacuously because it does not
  involve linear implication types at all.

  If the original proof tree contains
  \(\TyLIElimZ\), we can split the cases into the following
  (i), (ii), (iii) based on the typing rule just above the \(\TyLIElimZ\).
  (Note that the conclusions of the other typing rules are not of linear
  implication rules.)

  \medskip\noindent
  \textbf{(i) Case \(\TyLIIntroZ\):}

  If \(\TyLIElimZ\) is preceded by \(\TyLIIntroZ\),
  since the free links of the both sides of a type relation is the same
  in \(\TyLIIntroZ\), no hyperlinks are hidden by \(\TyLIElimZ\).
  That is, the proof tree is of the following form.
  \begin{prooftree}
    \AXC{$\vinf{\Gamma \vdash_P T : \tau}$}
    \RightLabel{$\TyLIIntroZ{}$}
    \UIC{$\Gamma \vdash_P T : \LIempty{\tau}{\Ys}$}
    \RightLabel{$\TyLIElimZ{}$}
    \UIC{$\Gamma \vdash_P T : \tau$}
  \end{prooftree}

  Because the premise of
  \(\TyLIIntroZ\) and the conclusion of
  \(\TyLIElimZ\)
  are the same typing judgement,
  we can simply remove those \(\TyLIElimZ\) and \(\TyLIIntroZ\).
  \begin{prooftree}
    \AXC{$\vinf{\Gamma \vdash_P T : \tau}$}
  \end{prooftree}
  We repeat the same operation to the proof subtrees above this step.

  \medskip\noindent
  \textbf{(ii) Case \(\TyLIIntro\):}\label{ii-case-tyliintro}

  If \(\TyLIElimZ\) is preceded by \(\TyLIIntro\),
  since the LHS of the linear implication type in the premise of
  \(\TyLIIntroZ\) is empty, we have the following form:

  \begin{prooftree}
    \AXC{$\paren{\alpha (\Xs) \lto \nu
          \Zs.(C(\Ys), \seq{U \bowtie V})} \in P$}
    \RightLabel{$\TyLIIntro{}$}
    \UIC{$\Gamma \vdash_P
        \quad (C(\Ys), \seq{U \bowtie V})
        : \LIempty{\alpha(\Xs)}{\Us, \Vs, \Ys}$}
    \RightLabel{$\TyLIElimZ{}$}
    \UIC{$\Gamma \vdash_P \nu \Ws.(C(\Ys), \seq{U \bowtie V}) : \alpha(\Xs)$}
  \end{prooftree}

  Here, from the production rules and the link condition of
  \(\TyLIElimZ\), we have \(\{\Zs\} = \{\Ws\}\).
  Therefore, we can replace \(\TyLIElimZ, \TyLIIntro\) by \(\TyProd\)
  to obtain a typing to a graph structurally congruent to the original:

  \begin{prooftree}
    \AXC{$\paren{\alpha (\Xs) \lto \nu
          \Zs.(C(\Ys), \seq{U \bowtie V})} \in P$}
    \RightLabel{$\TyProd{}$}
    \UIC{$\Gamma \vdash_P
        \nu \Zs.(C(\Ys), \seq{U \bowtie V}) : \alpha (\Xs)$}
  \end{prooftree}

  where
  \(\Zs.(C(\Ys), \seq{U \bowtie V}) \equiv \Ws.(C(\Ys), \seq{U \bowtie V})\).

  \medskip\noindent
  \textbf{(iii) Case \(\TyLITrans\):}

  If \(\TyLIElimZ\) is preceded by \(\TyLITrans\),
  by following the proof subtree above
  \(\TyLIElimZ\) towards upper left, one finds a series of one ore more
  \(\TyLITrans\)'s followed by \(\TyLIIntro\) which terminates the
  subtree, as in the upper part (before transformation) in
  \cref{fig:prenex-trans}.
  Note that one cannot use \(\TyLIIntroZ\) here because
  the LHS of the linear implication type has one or more types.
  In this case,
  we first transform the derivation into its prenex normal form in Step~1.
  Then, by repeatedly applying Step~2 and Step~3,
  we obtain a proof tree in which, for each \(\TyLITrans\) step up to
  the left and uppermost \(\TyLIIntro\),
  the right premise has a linear implication type with an empty left-hand side,
  and the right subtree ends with \(\TyLIIntroZ\),
  as illustrated in \cref{fig:ty-prod-admissible-eg}.
  Hence, this fragment can be replaced by a single \(\TyProd\) rule.
  The same operation is applied recursively to the right proof subtree of
  \(\TyLITrans\) as well.

  \begin{description}
    \item[Step 1. Prenex Normal Form Transformation:]

          First, we move the link creations of the graph to the leftmost of the term.
          This transformation is carried out as shown in \cref{fig:prenex-trans}.
          If an unintended variable capture might occur,
          then such links should be $\alpha$-converted in advance
          by applying \(\TyAlpha{}\) above
          the right premise of \(\TyLITrans{}\).
          Since this yields a proof tree for a congruent graph,
          we obtain the original typing relation by applying
          $\TyCong$ at the very bottom.

          This transformation is valid, that is,
          it satisfies the side conditions of the typing rules.
          The link conditions (i)--(iv) of \(\TyLITrans{}\) are straightforward,
          and condition (v) is established in \cref{lem:free-link-containment}.


    \item[Step 2. Hyperlink Creation Localisation:]

          We follow a sequence of \(\TyLITrans\)'s from the bottom to upper
          left, and
          as long as linear implication type of the right premise has an
          empty LHS, we transform the right subtree to the one that uses
          \(\TyLIIntroZ\) and \(\TyLIElimZ\) as shown in
          \cref{fig:hyperlink-hiding-localisations}.
          The boxes in the figure are provided merely for visual clarity.
          In the transformation,
          the boxed term \( T_i \) is replaced by \(\nu \seq{Y_i}. T_i\),
          so that the rules \(\TyLIIntroZ\) and \(\TyLIElimZ\)
          can be introduced immediately after the boxed typing judgement
          \(\vdash_P T_i : \LIempty{\tau_i}{\dots}\).

          This localisation procedure stops when we reach the leftmost
          uppermost leaf or when the right premise has a linear
          implication type with a non-empty LHS.
          We will show in \cref{lem:nu-localisation} that the graph templates
          of the proof
          trees before and after the transformation are congruent.
          Hence, we obtain the original typing relation by applying
          $\TyCong$ at the very bottom.

    \item[Step 3. Proof Tree Normalisation:]


          We follow the sequence of the \(\TyLITrans\)'s towards upper left.
          Suppose that each of the first to the $(m-1)$-th \(\TyLITrans\)'s
          has the right premise with \(\TyLIIntroZ\).
          Suppose, for the $m$-th \(\TyLITrans_m\), the right premise has
          a linear implication type with a non-empty LHS.

          Then, by reordering the elements of the graph template, we can
          reduce the number of types of the LHS of the linear implication
          type of the right premise by the transformation shown in
          \cref{fig:ProofTreeNormalization}.
          The transformation is repeated until the LHS of the linear
          implication type of the right premise becomes empty.

          The transformation step is explained in \cref{fig:ProofTreeNormalization}.
          Before transformation,
          the first (\(m-1\)) \(\TyLITrans\)'s use \(\TyLIIntroZ\) as their
          right premises, and the \(m\)-th \(\TyLITrans_m\) has an linear
          implication type with a non-empty LHS.

          The graph templates of the two conclusions are obviously congruent
          and we obtain the original typing relation by applying
          $\TyCong$ at the very bottom.
          We will show in \cref{lem:link-cond-preservation} that the link condition
          of all \(\TyLITrans\)s are satisfied.
          The above transformation is repeated until the LHS of the linear
          implication type of the right premise of \(\TyLITrans_m\) becomes empty.
          When the LHS of the linear implication type of the right premise of
          \(\TyLITrans_m\) becomes empty,
          we apply Step 2 (Hyperlink Creation Localisation) to \(\TyLITrans_m\). \qedhere
  \end{description}
\end{proof}

\subsection{Proof of Conservativity}\label{subsec:proof-of-conservativity}

It remains to lift the graph-prefix elimination result to arbitrary derivations.
The only additional issue is that a derivation may contain sub-derivations above
premises of \(\TyArrow{}\).
The following simple observation ensures that the conservativity argument can be
applied recursively to those sub-derivations.

\begin{lemma}[Linear-implication freeness is inherited by \(\TyArrow{}\) premises]\label{lem:li-freeness-arrow-premises}
  Suppose that an instance of \(\TyArrow{}\) has conclusion
  \[
    \Gamma \vdash_P (\lambda\, x[\Xs] : \tau_1.e)(\Zs) : (\tau_1 \to \tau_2)(\Zs).
  \]
  If neither \(\Gamma\) nor \((\tau_1 \to \tau_2)(\Zs)\) contains a linear implication type,
  then neither \(\Gamma, x[\Xs] : \tau_1\) nor \(\tau_2\) contains a linear implication type.
\end{lemma}

\begin{proof}
  By the definition of occurrence of linear implication types in types,
  a linear implication type occurs in an arrow type \(\tau_1 \to \tau_2\)
  if and only if it occurs in either \(\tau_1\) or \(\tau_2\).
  Hence, since \((\tau_1 \to \tau_2)(\Zs)\) contains no linear implication type,
  both \(\tau_1\) and \(\tau_2\) contain no linear implication type.

  Moreover, by assumption, \(\Gamma\) contains no linear implication type.
  Since the only additional assumption in the premise of \(\TyArrow{}\) is
  \(x[\Xs] : \tau_1\), and \(\tau_1\) contains no linear implication type,
  the extended context \(\Gamma, x[\Xs] : \tau_1\) also contains no linear implication type.

  Therefore, the assumption context and the conclusion type of the premise
  \[
    \Gamma, x[\Xs] : \tau_1 \vdash_P e : \tau_2
  \]
  contain no linear implication type.
\end{proof}

We now prove the stated conservativity theorem.
The proof first normalises the graph-typing prefix and eliminates the
linear-implication rules there.
It then applies the same argument recursively to the sub-derivations above
\(\TyArrow{}\).

\begin{proof}[Proof of \cref{thm:conservativity-complete-graph-types}]
  Let \(\mathcal{D}\) be a derivation of
  \(\Gamma \vdash_P G : \tau\)
  in the extended type system.
  Consider the largest prefix of \(\mathcal{D}\) that contains no premises of
  \(\TyArrow\).

  By \cref{lem:graph-prefix-rules}, every rule occurrence in this prefix is an
  instance of one of
  \(\TyAlpha\), \(\TyCong\), \(\TyProd\), and the linear-implication rules
  \(\TyLIs\).
  By \cref{lem:elim-ty-alpha}, applications of \(\TyAlpha\) in this prefix
  can be propagated upward and eliminated, except for harmless renamings of
  bound links.
  By \cref{lem:elim-ty-cong}, applications of \(\TyCong\) in this prefix
  can be collected into at most one final application of \(\TyCong\).
  Hence the prefix can be transformed into a form in which, before the final
  \(\TyCong\), only \(\TyProd\) and the linear-implication rules occur.

  We can now apply \cref{thm:ty-lis-elim}.
  This eliminates all occurrences of
  \(\TyLIIntro\), \(\TyLIIntroZ\), \(\TyLITrans\), and \(\TyLIElimZ\)
  from the graph-typing prefix, replacing the corresponding uses of linear
  implication types by applications of \(\TyProd\), possibly together with
  \(\TyAlpha\) and a final \(\TyCong\).

  It remains to consider the sub-derivations above premises of \(\TyArrow{}\).
  By \cref{lem:li-freeness-arrow-premises}, linear-implication freeness of the
  conclusion and assumptions of an instance of \(\TyArrow{}\) is inherited by its
  premise.
  Hence every such premise has assumptions and conclusion type that contain no
  linear implication types.
  Applying the same argument recursively to these sub-derivations eliminates all
  remaining occurrences of the linear-implication rules.

  Therefore the original judgement
  \(\Gamma \vdash_P G : \tau\)
  has a derivation that does not use any of
  \(\TyLIIntro\), \(\TyLIIntroZ\), \(\TyLITrans\), or \(\TyLIElimZ\).
\end{proof}

\chapter[Towards Fully Static Type Safety]{Towards Static Type Safety for Pattern Matching in \lgt}\label{ch:fully-static}

In the previous type systems (Sections~\ref{ch:core-type} and~\ref{ch:limp-type}),
the typing rule \TyCase{} for \textbf{case} expressions
does not impose any special constraints on the type \(\tau_1\) of the expression \(e_1\).
Even when \(\tau_1\) does not match the type expected by the pattern,
the evaluation remains sound under dynamic checks:
if the match fails, the evaluation proceeds to the \textbf{otherwise} branch,
and if the match succeeds, the necessary checks are performed dynamically.

However, when dynamic checks at runtime are eliminated,
soundness can no longer be preserved by simply using the \TyCase{} rule as it is.
To address this, in this section we introduce additional constraints
that guarantee soundness of a fully static type system
without relying on the dynamic checks originally embedded in \textbf{case} expressions.

First, in \cref{sec:sec4-ty-case},
we present a counterexample demonstrating unsoundness,
and then discuss how to restore soundness.
In \cref{sec:sec4-constraints}, we give a formal definition of these constraints.
Finally, in \cref{sec:sec4-fully-static-soundness},
we provide a proof sketch of soundness for the extended type system
that incorporates these modifications and constraints.

\section{Characterising the Safe Fragment of Pattern Matching}\label{sec:sec4-ty-case}
In this section,
we first illustrate a case where soundness is violated
in \cref{sec:unsound-example}.
Subsequently,
we discuss how to prevent such unsoundness in \cref{sec:IntroductionOfRestrictions}.

\subsubsection{An Unsound Example}\label{sec:unsound-example}

Consider the following expression, 
which ideally should not be typable as it leads to type inconsistency.
\begin{align}\label{eq:unsound-dbllist-removal-eg-vis}
  \eDLun \triangleq
  \begin{array}{l}
    (\lambda\,\ctxx{x} : \dblltype{X}{Y}{Z}.       \\[1.5mm]
    \quad \mathbf{case}\; \ctxx{x}\; \mathbf{of}\; \\[1.5mm]
    \qquad
    \tikz[baseline=-0.5ex, scale=1]{%
      \mkcons{1}{W_1}{}{(0,0)}{1}
      \mkctxn{2}{y}{(1.5,0)}{W_2}{W_4}
      \begin{scope}[shift={(3,0)}]
        \node [ctxn] (C3)  at (0, 0) {$z$};
        \coordinate (p3AL) at ($(C3)+(-0.1,0.3)$) {};
        \coordinate (p3AR) at ($(C3)+(+0.1,0.3)$) {};
        \coordinate (p3R) at ($(p3AL)+( 0.5,0.5)$);
        \coordinate (p3L) at ($(p3AR)+(-0.5,0.5)$);
        \node [port] at (+0.250, +0.450) {2};
        \node [port] at (-0.250, +0.450) {3};
        \node [port] at (-0.400, -0.150) {1};
        \node [link,text=gray] at (-0.3, 1) {$(W_3)$};
      \end{scope}
      \node [link] (C0) at (-1, 0.8) {$Z$};
      \node [link] (LX) at (-1, 0.0) {$X$};
      \node [link] (C4) at (+4, 0.8) {$Y$};
      %
      \mkconslinkleft{0}{1}
      \draw (C1) to (LX);
      \mkconslink{1}{2}
      \mkconslink{2}{3}
      \mkconslinkright{3}{4}
    }
    \to \ctxx{z}                                   \\[0mm]
    \quad\: \texttt{|}\;\mathbf{otherwise}
    \to
    \mkleafatom{1}
    \\[-4mm]
    )(W)
  \end{array}
\end{align}

We demonstrate that the program \(\eDLun\)
leads to an inconsistency
when applied to doubly linked lists of length two and zero.

When applied to a doubly linked list of length two,
the expression evaluates to a doubly linked list of length zero.
\begin{align}\label{eq:unsound-reduction-1-eg-vis}
  \left(
  \eDLun
  \quad
  \tikz[baseline=-0.5ex]{%
    \mkcons{1}{W_1}{}{(0,0)}{1}
    \mkcons{2}{W_3}{W_2}{(1.5,0)}{2}
    \mknil{3}{(3,0)}
    \node [link] (C0) at (-1.0, 0.8) {$Z$};
    \node [link] (LX) at (-1.0, 0.0) {$X$};
    \node [link] (C4) at (+4.0, 0.8) {$Y$};
    \draw (C1) to (LX);
    \mkconslinkleft{0}{1}
    \mkconslink{1}{2}
    \mkconslink{2}{3}
    \mkconslinkright{3}{4}
  }
  \right)
  \
  \reducestwo
  \
  \tikz[baseline=-0.5ex]{%
    \mknil{1}{(0,0)}
    \node [link] (C0) at (-1.0, 0.8) {$Z$};
    \node [link] (LX) at (-1.0, 0.0) {$X$};
    \node [link] (C2) at (+1.0, 0.8) {$Y$};
    \draw (C1) to (LX);
    \mkconslinkleft{0}{1}
    \mkconslinkright{1}{2}
  }
\end{align}

On the other hand, when applied to a doubly linked list of length zero,
it does not return a doubly linked list but rather a leaf in a leaf-linked tree.
\begin{align}\label{eq:unsound-reduction-2-eg-vis}
  \left(
  \eDLun
  \quad
  \tikz[baseline=-0.5ex]{%
    \mknil{1}{(0,0)}
    \node [link] (C0) at (-1.0, 0.8) {$Z$};
    \node [link] (LX) at (-1.0, 0.0) {$X$};
    \node [link] (C2) at (+1.0, 0.8) {$Y$};
    \draw (C1) to (LX);
    \mkconslinkleft{0}{1}
    \mkconslinkright{1}{2}
  }
  \right)
  \quad
  \reducestwo
  \quad
  \mkleafatom{1}
\end{align}

This discrepancy in return types demonstrates a lack of type consistency.
Since the type system should reject programs exhibiting such inconsistencies,
this expression should not be typable.

However, contrary to this expectation,
the program is
typable under the production rules \(\Pdl\)
\eqref{eq:prod-rules-eg} (\cref{sec:core-type:examples})
within the type system introduced in \cref{ch:limp-type}.

To analyse the reason behind this,
we first present the
text version
of the program $\eDLun$ (previously introduced
visually) as follows:
\begin{align}\label{eq:unsound-dbllist-removal-eg}
  \eDLun \triangleq
  \begin{array}{l}
    (\lambda\,x[X,Y,Z]: \mathit{dbllist}(X,Y,Z).   \\
    \quad \mathbf{case}\; x[X,Y,Z]\; \mathbf{of}\; \\
    \qquad \nu W_1 W_2 W_3 W_4.(                   \\
    \qquad\quad \Cons(W_1,X,W_2,Z),                \\
    \qquad\quad w[W_1]: \mathit{nat}(W_1),         \\
    \qquad\quad y[Z,W_3,W_4,W_2]:                  \\
    \qquad\qquad \LI{\mathit{lltree}(Y,W_4,W_3)}{  \\
    \qquad\qquad\quad \mathit{dbllist}(Z, Y, W_2)  \\
    \qquad\qquad }{Z,W_3,W_4,W_2},                 \\
    \qquad\quad z[W_4,Y,W_3]:
    \mathit{lltree}(Y,W_4,W_3)                     \\
    \qquad ) \to z[X,Y,Z]                          \\
    \quad\: \texttt{|}\;\mathbf{otherwise} \to
    \Leaf(0,Y,Z), X \bowtie Z                      \\
    )(W).
  \end{array}
\end{align}

We demonstrate that the following type judgement
is successfully derived in the type derivation
presented in \cref{fig:unsound-dbllist-removal-typederiv-eg-1,fig:unsound-dbllist-removal-typederiv-eg}:

\begin{align}\label{eq:unsound-type-judge-eg}
  \emptyset \vdash_{\Pdl} \eDLun : (\mathit{dbllist}(X,Y,Z) \to \mathit{lltree}(X,Y,Z))(W).
\end{align}

In the type derivation in \cref{fig:unsound-dbllist-removal-typederiv-eg},
the derivation of \(\varphi\),
corresponding to the right-hand side of the first \(\to\) in the case expression,
leads to a type inconsistency in the return values.
While this step results in the type inconsistency,
it is not the fundamental source of the problem.

The primary issue arises in the type derivation of the pattern in the case expression,
as captured by
\cref{fig:unsound-dbllist-removal-typederiv-eg}.
Here, the type checking mechanism fails to
sufficiently enforce the expected types of subgraphs
that will be bound to
the graph variables \(y[Z,W_3,W_2,W_4]\) and \(z[W_4,Y,W_3]\).
In particular, the types of the subgraphs bound to these
graph variables will differ from
their expected annotated types,
leading to an unsound type derivation.

\begin{figure*}[tb]
  \hrulefill

  \begin{prooftree}
    \AXC{}
    \RightLabel{\TyVarA{}}
    \UIC{$\ml{
          \Gamma'\{ y[Z,W_3,W_2,W_4]               \\
          \quad : \LI{\mathit{lltree}(W_4,Y,W_3)}{ \\
            \qquad \mathit{dbllist}(Z,Y,W_2)}{Z,W_3,W_2,W_4}
          \} \vdash_P                              \\
          y[Z,W_3,W_2,W_4]                         \\
          : \LI{\mathit{lltree}(W_4,Y,W_3)}{       \\
            \quad \mathit{dbllist}(Z,Y,W_2))}{Z,W_3,W_2,W_4}
        }$}

    \AXC{}
    \RightLabel{\TyVarA{}}
    \UIC{$\ml{
          \Gamma'\{ z[W_4,Y,W_3]                         \\
          \quad : \mathit{lltree}(W_4,Y,W_3) \} \vdash_P \\
          z[W_4,Y,W_3]                                   \\
          : \mathit{lltree}(W_4,Y,W_3)
        }$}
    \RightLabel{\TyLIEA{}}
    \BIC{
      \tikz[remember picture,baseline=(upper-root1.base)]
      \node [font=\normalsize] (upper-root1)
      {$\begin{array}{lll}
            \Gamma'\{ y[Z,W_3,W_2,W_4]                                                             \\
            : \LI{\mathit{lltree}(W_4,Y,W_3)}{\mathit{dbllist}(Z,Y,W_2)}{Z,W_3,W_2,W_4}\} \vdash_P \\
            \nu W_3 W_4.(y[Z,W_3,W_2,W_4], z[W_4,Y,W_3])                                           \\
            : dbllist(Z,Y,W_2)
          \end{array}$};
    }
  \end{prooftree}

  \bigskip

  \begin{prooftree}
    \AXC{
      \tikz[remember picture,baseline=(lower-hole1.base)]
      \node[draw, dashed, rectangle, inner sep=3pt, font=\normalsize] (lower-hole1)
      {$\color{white}Cons$};
    }

    \AXC{}
    \RightLabel{\TyVarA{}}
    \UIC{$\Gamma' \vdash_P w[W_1]: \mathit{nat}(W_1)$}
    \RightLabel{\TyProdA{}}
    \BIC{$\ml{\Gamma' \vdash_P \\
          \nu W_1 W_2.(
          \Cons(W_1,X,W_2,Z), w[W_1],
          \nu W_3 W_4.(y[Z,W_3,W_2,W_4],
          z[W_4,Y,W_3])
          ) \\
          : \mathit{dbllist}(X,Y,Z)}$}
    \RightLabel{\TyCongA{}}
    \UIC{$\ml{\Gamma' \vdash_P \\
          \nu W_1 W_2 W_3 W_4.(
          \Cons(W_1,X,W_2,Z),
          w[W_1],
          y[Z,W_3,W_2,W_4],
          z[W_4,Y,W_3]
          ) \\
          : \mathit{dbllist}(X,Y,Z)}$}
  \end{prooftree}
  \begin{tikzpicture}[remember picture,overlay]
    \begin{scope}[even odd rule]
      \draw[->] (upper-root1.south) to[looseness=0.8,out=-90,in=90] (lower-hole1.north);
    \end{scope}
  \end{tikzpicture}

  where
  \begin{align*}
    \Gamma' & =
    \begin{array}{lll}
      y[Z,W_3,W_2,W_4]:
      \LI{\mathit{lltree}(Y,W_4,W_3)}{\mathit{dbllist}(Z,Y,W_2)}{Z,W_3,W_2,W_4}, \\
      z[W_4,Y,W_3]: \mathit{lltree}(W_4,Y,W_3),                                  \\
      w[W_1]: \mathit{nat}(W_1)
    \end{array}
  \end{align*}

  Here,
  \TyVar{},  \TyLIE{},  \TyProd{},  and \TyCong{} are abbreviated as
  \TyVarA{}, \TyLIEA{}, \TyProdA{}, and \TyCongA{} respectively.

  \hrulefill
  \caption{Typing derivation of unsound typing judgement \eqref{eq:unsound-type-judge-eg}.}
  \label{fig:unsound-dbllist-removal-typederiv-eg-1}
\end{figure*}

\begin{figure*}[tb]
  \hrulefill

  \begin{prooftree}
    \AXC{}
    \RightLabel{\TyVarA{}}
    \UIC{$\Gamma, \Gamma' \vdash_P z[W_4,Y,W_3]
        : \mathit{lltree}(W_4,Y,W_3)$}
    \RightLabel{\TyAlphaA{}}
    \UIC{
      \tikz[remember picture,baseline=(upper-root1.base)]
      \node [font=\normalsize] (upper-root1)
      {$\Gamma, \Gamma' \vdash_P z[X,Y,Z]: \mathit{lltree}(X,Y,Z)$};
    }
  \end{prooftree}

  \bigskip

  \begin{prooftree}
    \AXC{}
    \RightLabel{\TyVarA{}}
    \UIC{$\ml{\Gamma \vdash_P x[X,Y,Z] \\
          : \mathit{dbllist}(X,Y,Z)}$}

    \AXC{(\cref{fig:unsound-dbllist-removal-typederiv-eg-1})}
    \AXC{
      \tikz[remember picture,baseline=(lower-hole1.base)]
      \node[draw, dashed, rectangle, inner sep=5pt, font=\normalsize] (lower-hole1)
      {$\varphi$};
    }

    \AXC{}
    \RightLabel{\TyProdA{}}
    \UIC{$\ml{\Gamma \vdash_P \Leaf(0,R,X),L \bowtie X \\
          : \mathit{lltree}(L,R,X)}$}
    \RightLabel{\TyAlphaA{}}
    \UIC{$\ml{\Gamma \vdash_P \Leaf(0,Y,Z),X \bowtie Z \\
          : \mathit{lltree}(X,Y,Z)}$}

    \RightLabel{\TyCaseA{}}
    \QuaternaryInfC{$\ml{
          \Gamma \vdash_P\\
          (\caseof{x[X,Y,Z]}{T}{z[X,Y,Z]}{\Leaf(0,Y,Z),X \bowtie Z})\\
          : \mathit{lltree}(X,Y,Z)
        }$}

    \RightLabel{\TyArrowA{}}
    \UIC{$\emptyset \vdash_P e : (dbllist(X,Y,Z) \to \mathit{lltree}(X,Y,Z))(W)$}
  \end{prooftree}
  \begin{tikzpicture}[remember picture,overlay]
    \begin{scope}[even odd rule]
      \draw[->] (upper-root1.south) to[looseness=1.5,out=-90,in=90] (lower-hole1.north);
    \end{scope}
  \end{tikzpicture}

  where
  \begin{align*}
    T       & = \nu W_1 W_2 W_3 W_4.(
    \Cons(W_1,X,W_2,Z),
    w[W_1],
    y[Z,W_3,W_2,W_4],
    z[W_4,Y,W_3]
    )
    \\
    \Gamma  & = x[X,Y,Z]: \mathit{dbllist}(X,Y,Z)
    \\
    \Gamma' & =
    \begin{array}{lll}
      y[Z,W_3,W_2,W_4]:
      \LI{\mathit{lltree}(Y,W_4,W_3)}{\mathit{dbllist}(Z,Y,W_2)}{Z,W_3,W_2,W_4}, \\
      z[W_4,Y,W_3]: \mathit{lltree}(W_4,Y,W_3),                                  \\
      w[W_1]: \mathit{nat}(W_1)
    \end{array}
  \end{align*}

  Here,
  \TyVar{},  \TyAlpha{},  \TyProd{},  \TyCase{},  \TyCong{},  and \TyArrow{}  are abbreviated as
  \TyVarA{}, \TyAlphaA{}, \TyProdA{}, \TyCaseA{}, \TyCongA{}, and \TyArrowA{} respectively.

  \hrulefill
  \caption{Typing derivation of unsound typing judgement \eqref{eq:unsound-type-judge-eg}.}
  \label{fig:unsound-dbllist-removal-typederiv-eg}
\end{figure*}

This issue stems from a fundamental limitation in the type system,
which lacks a mechanism to ensure that a subgraph matched by a graph variable
is \emph{always} of the expected annotated type.
Previous approaches relied on dynamic type checking
to enforce the correctness of subgraph bindings,
ensuring that the subgraph bound to \(z[W_4,Y,W_3]\)
conformed to the expected annotation \(\mathit{lltree}(W_4,Y,W_3)\).
However, the objective of this study is to establish a statically verifiable guarantee
without resorting to runtime type enforcement.

To resolve this issue,
we introduce additional constraints to ensure
that any subgraph matched by a graph variable conforms to the expected type.

In the following section,
we discuss the constraints
on
the pattern of case expressions
and constraints on the type system
that prevent this form of unsoundness.

\subsubsection{Introduction of Restrictions}\label{sec:IntroductionOfRestrictions}



The key to ensuring the soundness of the type system
is the introduction of structural constraints
that enforce the types
of subgraphs matched in a case expression to be
uniquely determined.

The first essential constraint is to ensure that type production rules are disjoint:
each right-hand side of a production rule must contain exactly one constructor atom,
a node with a constructor name,
whose name is unique within the set of rules.
By enforcing disjointness,
every subgraph connected to a terminal atom is guaranteed
to have a uniquely determined type.

Furthermore,
in pattern matching,
we impose an additional constraint
that graph variables must not be directly connected to one another.
Instead, they must be connected via terminal atoms.
This constraint ensures that any subgraph matched by a graph variable
must necessarily pass through a terminal atom,
thereby guaranteeing the unique determination
of the type of the subgraph matching the graph variable.

We believe that the constraint on production rules
implies that the graphs must have bounded treewidth
and excludes certain classes of graphs.
However, we have found that even with this constraint,
useful types can still be expressed;
for example,
the production rules
$\Pdl$ \eqref{eq:prod-rules-eg}
all satisfy these constraints.
Thus, we consider this as a reasonable and practical design choice.
Examining the relationship between this constraint and treewidth,
as well as exploring the possibility of relaxing it,
remains an open challenge for future research.

We also believe that
the constraints on patterns in case expressions
do not substantially limit
the expressiveness of the language.
All examples presented in
\cref{ch:untyped}
and
\cref{ch:limp-type}
adhere to these constraints.
The program $\eDLun$ shown in
\eqref{eq:unsound-dbllist-removal-eg-vis}
intentionally violates these constraints;
however, it can be easily rewritten to comply with them
while preserving its semantics,
as demonstrated by $\eDLun'$
\eqref{eq:sound-dbllist-removal-eg-vis}.
\begin{align}\label{eq:sound-dbllist-removal-eg-vis}
  \eDLun' \triangleq
  \begin{array}{l}
    (\lambda\,\ctxx{x} : \dblltype{X}{Y}{Z}.       \\[1.5mm]
    \quad \mathbf{case}\; \ctxx{x}\; \mathbf{of}\; \\[1.5mm]
    \qquad
    \tikz[baseline=-0.5ex, scale=1]{%
      \mkctxn{1}{y}{(0,0)}{}{W_2}
      \mkcons{2}{W_3}{W_1}{(1.5,0)}{w}
      \mkctxn{3}{z}{(3,0)}{W_4}{}
      \node [link] (C0) at (-1, 0.8) {$Z$};
      \node [link] (LX) at (-1, 0.0) {$X$};
      \node [link] (C4) at (+4, 0.8) {$Y$};
      \draw (C1) to (LX);
      \mkconslinkleft{0}{1}
      \mkconslink{1}{2}
      \mkconslink{2}{3}
      \mkconslinkright{3}{4}
    }
    \to \ctxx{z}                                   \\[0mm]
    \quad\: \texttt{|}\;\mathbf{otherwise}
    \to
    \mkleafatom{1}
    \\[-4mm]
    )(W)
  \end{array}
\end{align}


In the expression \(\eDLun'\), 
since the tail of a \(\mathrm{Cons}\) atom,
a subgraph connected through its second argument,
is always of type \(\mathit{dbllist}\),
any subgraph that matches the graph variable \(z\) must necessarily have the type
\(\mathit{dbllist}\).
Consequently, in the corrected example,
type derivation fails because the types of the graph variable \(z\) and
the \(\Leaf\) atom
both appearing on the right-hand sides of \(\to\)s in the case expression
do not align.

\section{Formalisation of Static Constraints}\label{sec:sec4-constraints}

In this section, we formalise the constraints on the structure of production rules
and on the patterns occurring in \textbf{case} expressions.

We impose restrictions on graphs to enable unique traversal through specific links.
We introduce the notion of \emph{primary roots} for types, atoms, and graph variables.
We then impose additional restrictions so that, by traversing
these primary roots,
we obtain a unique spanning tree.
The introduction of primary roots effectively restricts the class of graphs
handled by \lgt to connected graphs.

We refer to the last arguments of types, atoms, and graph variables as their
\emph{primary roots}, and we use \(\Root\) to denote the function that extracts them.

\begin{mydef}[Primary Root]\label{def:primary-root}
  \begin{align*}
    \Root(\alpha(\dots, X))          & \triangleq X                                         \\
    \Root((\tau \to \tau)(\dots, X)) & \triangleq X                                         \\
    \Root(\LI{\dots}{\tau}{\dots})   & \triangleq \Root(\tau)                               \\
    \Root(p(\dots, X))               & \triangleq X                                         \\
    \Root(x[\dots, X]: \tau)         & \triangleq X \text{ where } X = \Root(\tau) \qedhere
  \end{align*}
\end{mydef}

The primary root of a linear implication type is
the primary root of the type on the right-hand side of \(\multimap\).
For a graph variable of the form \(x[\dots, X] : \tau\),
if the link \(X\)
is not the primary root of the annotated type $\tau$,
its primary root is undefined.

In what follows,
we require that the primary root always exists
for every type and variable.

We define a notation for expressing whether two links are
fused with given link fusions.
\begin{mydef}[Fusibility Relation]\label{def:fused-links}
  Let $F = (
    W_1 \bowtie U_1,
    \dots,
    W_n \bowtie U_n
    )$.
  $\areFused{X}{Y}{F}$
  denotes that
  $\angled{X, Y} \in R$,
  where $R$ is the smallest equivalence relation
  that contains
  $\{
    \angled{W_1, U_1},
    \dots,
    \angled{W_n, U_n}
    \}$.
\end{mydef}
Note that identical links are always considered fused.

In order to distinguish primary roots from other links,
we must ensure that certain links are not fused with primary roots.
To this end, we define the set of links of type atoms
that may potentially be fused with primary roots
by applying production rules as follows:
\begin{mydef}[Links Fusible with Primary Roots]\label{def:fused-links-ports}
  \[\begin{array}[b]{l}
      \FusedToRoot((\tau_1 \to \tau_2)(\Ys, X), P) \triangleq \{X\}
      \\
      \FusedToRoot(\alpha(X_1, \dots, X_n), P) \triangleq \\
      \quad \{X_i \mid                                    \\
      \qquad
      \paren{
        \alpha(Y_1, \dots, Y_n) \lto
        \nu \Zs. (C(\Vs), \seq{W \bowtie U}, \taus)
      }
      \in P                                               \\
      \qquad \land\, \areFused{Y_i}{Y_n}{\seq{W \bowtie U}}
      \} \hfill\qedhere\hspace*{-8pt}
    \end{array}\]
\end{mydef}

In the case of an arrow type \((\tau_1 \to \tau_2)(\Ys, X)\),
no further production rules are applied,
and thus the primary root will not be fused with any other links.
Therefore, the resulting set simply consists of the primary root \(X\).

For type variables, the situation is more complex.
The resulting set consists of a subset
containing the \(i\)-th link \(X_i\) of the type variable.
Here, the index \(i\) corresponds to the position of the link
for which there exists a production rule
\[
  \alpha(Y_1, \dots, Y_n) \lto
  \nu \Zs. (C(\Vs), \seq{W \bowtie U}, \taus)
\]
such that the \(i\)-th link \(Y_i\) and the primary root \(Y_n\)
of the type atom \(\alpha(Y_1, \dots, Y_n)\) on the left-hand side
LHS of the rule
are fused together on the right-hand side RHS of the rule
by \(\seq{W \bowtie U}\).

Note that the sets of link names obtained by \(\FusedToRoot(\tau, P)\)
naturally include the primary roots of \(\tau\).

Using the notations defined above,
\cref{def:disjoint-condition} states the constraints imposed on production rules.
\begin{mydef}[Restriction on Production Rules]\label{def:disjoint-condition}
  Each production rule in \(P\)
  must take the following form:
  \[
    \alpha(\Ws, X) \lto \nu \Ys.(C(\Zs, X), \seq{U \bowtie V}, \tau_1, \dots, \tau_n)
  \]
  and must satisfy the following conditions
  as well as the conditions given in \cref{ProductionRule}:
  \begin{enumerate}
    \item
          \(X \notin \{\Ys\}\);
    \item
          $\Root(\tau_i) \in \{\Zs\}$ for all $i$;
    \item
          for all $i,j,X_i,X_j$ such that\\
          $X_i \in \FusedToRoot(\tau_i, P)$ and
          $X_j \in \FusedToRoot(\tau_j, P)$,
          \begin{itemize}
            \item $\neg\bigl(\areFused{X_i}{X}{\seq{U \bowtie V}}\bigr)$
                  and
            \item $i \neq j \implies
                    \neg \bigl(\areFused{X_i}{X_j}{\seq{U \bowtie V}}\bigr)$;
          \end{itemize}
    \item
          for all
          $r_1, r_2 \in P$
          such that \\
          $r_1 = \bigl(\alpha_1 (\seq{X_1}) \lto
            \nu \seq{Z_1}.(C_1(\seq{Y_1}), \seq{W_1 \bowtie U_1}, \seq{\tau_1})\bigr)$
          and \\
          $r_2 = \bigl(\alpha_2 (\seq{X_2}) \lto
            \nu \seq{Z_2}.(C_2(\seq{Y_2}), \seq{W_2 \bowtie U_2}, \seq{\tau_2})\bigr)$,\\[0.5mm]
          $C_1 = C_2 \implies r_1 = r_2$. \qedhere
  \end{enumerate}
\end{mydef}

Condition~(1) ensures that \(X\) is a free link
so that the primary root \(X\) can be reached from outside.

Condition~(2) ensures that the subsequent types \(\taus\)
(i.e., the types of sub-components reachable from a constructor)
are connected to the constructor atom \(C(\Zs, X)\).

Condition~(3) ensures that each of its sub-components
can be traversed uniquely once the constructor atom \(C\) is reached.
One might think that, with respect to Condition~(3), the simple requirement
\(\Root(\tau_i) \neq \Root(\tau_j)\) for all \(i \neq j\) would suffice.
However, this condition is insufficient
because it still allows the possibility
that the primary roots \(\Root(\tau_i)\) and \(\Root(\tau_j)\)
could be fused at a later stage by the application of production rules.
Condition~(3) is therefore defined using \(\FusedToRoot\).

We show that the production rules $\Pdl$ \eqref{eq:prod-rules-eg}
satisfy Condition~(3) in \cref{eg:cond-dbllist} and \cref{eg:cond-lltree} below.
\begin{example}[Doubly Linked Lists]\label{eg:cond-dbllist}
  Recall that the production rules of \(\mathit{dbllist}\) in $\Pdl$
  \eqref{eq:prod-rules-eg}
  are as follows:
  \begin{align}
     & \mathit{dbllist}(X,Y,Z) \lto \notag{}       \\
     & \quad \;\; \Nil(X, Z), Y \bowtie Z
    \label{eg:prodrule-dbllist-1}                  \\
     & \quad \mid \nu W_1 W_2.(\Cons(W_1,X,W_2,Z),
    \mathit{nat}(W_1),\mathit{dbllist}(Z,Y,W_2)).
    \label{eg:prodrule-dbllist-2}
  \end{align}
  For the first production rule \eqref{eg:prodrule-dbllist-1},
  since no type atom occurs on the right-hand side,
  Condition~(3) is trivially satisfied.
  For the second production rule \eqref{eg:prodrule-dbllist-2}, we have
  \[\begin{array}{llll}
      \FusedToRoot(\mathit{nat}(W_1), P) =
      \{W_1\},
      \\
      \FusedToRoot(\mathit{dbllist}(Z,Y,W_2), P) =
      \{Y,W_2\}.
    \end{array}\]
  This production rule has no fusion,
  and the sets $\{W_1\}$, $\{Y, W_2\}$, and $\{X\}$ are pairwise disjoint.
  Therefore, this production rule satisfies Condition~(3).
\end{example}

\begin{example}[Leaf-Linked Trees]\label{eg:cond-lltree}
  Recall that the production rules of \(\mathit{lltree}\) in $\Pdl$
  \eqref{eq:prod-rules-eg}
  are as follows:
  \begin{align}
     & \mathit{lltree}(L,R,X) \lto \notag{}                                          \\
     & \quad \;\; \nu W_1. (\Leaf(W_1,R,X), L \bowtie X, \mathit{nat}(W_1))
    \label{eg:prodrule-lltree-1}                                                     \\
     & \quad \mid \nu W_1. (\mathrm{LeafM}(W_1,R,X), L \bowtie X, \mathit{nat}(W_1))
    \label{eg:prodrule-lltree-2}                                                     \\
     & \quad \mid \nu Y W_1 W_2.(\Node(W_1, W_2,X),
    \mathit{lltree}(L,Y, W_1), \mathit{lltree}(Y,R, W_1)).
    \label{eg:prodrule-lltree-3}
  \end{align}
  For the first and second production rules
  \eqref{eg:prodrule-lltree-1}\eqref{eg:prodrule-lltree-2},
  we have
  \[
    \FusedToRoot(\mathit{nat}(W_1), P) =
    \{W_1\},
  \]
  where \(W_1 \in \{W_1\}\) and \(X\) do
  not satisfy \(\areFused{W_1}{X}{L \bowtie X}\).
  Therefore, these production rules satisfy Condition~(3).
  For the third production rule \eqref{eg:prodrule-lltree-3}, we have
  \[\begin{array}{llll}
      \FusedToRoot(\mathit{lltree}(L, Y, W_1), P) =
      \{L, W_1\},
      \\
      \FusedToRoot(\mathit{lltree}(Y, R, W_2), P) =
      \{Y, W_2\}.
      \\
    \end{array}\]
  This production rule has no fusion,
  and the sets $\{L, W_1\}$, $\{Y, W_2\}$, and $\{X\}$ are pairwise disjoint.
  Therefore, this production rule satisfies Condition~(3).
\end{example}

Condition (4) requires
the constructor name \(C\) must be unique within the set $P$
of production rules.
Condition (4) is intended to uniquely determine both
the type of a graph whose root is a constructor atom
and the types of the subgraphs connected to that constructor atom.

We also impose a constraint on patterns in \textbf{case} expressions.
We first define the syntax of \textit{restricted templates},
which will be used to describe the patterns.
\begin{mydef}[Case Pattern Syntax]\label{def:pat-cond-syntax}
  The syntax of restricted templates $\Tx$ is
  defined recursively as follows:
  \[
    \begin{array}{l@{~}c@{~}l@{\quad}l@{}}
      \Tx & ::=
          & \nu \Ys.(x[\Zs, X]: \tau, \Tc_1, \dots, \Tc_n) \mid \Tc
          & \quad (n \geq 0),
      \\
      \Tc & ::=
          & \nu \Ys.(C(\Zs, X), \seq{U \bowtie V}, \Tx_1, \dots, \Tx_n)
          & \quad (n \geq 0).\qedhere
    \end{array}
  \]
\end{mydef}
The template \(\Tc\) denotes a template whose root contains a
constructor atom,
whereas the template \(\Tx\) allows a graph variable at the root.
When the root of a template is a graph variable,
its subtemplates must have constructor atoms at their root.
Note that the above definition allows two or more
constructor atoms to occur
consecutively.

To obtain the primary root of the leftmost graph variable or constructor atom
in restricted templates \(\Tx\),
we extend \(\Root\) as follows:
\begin{align*}
  \Root(\nu \Ys.(x[\Zs, X]: \tau, \dots)) & \triangleq X, \\
  \Root(\nu \Ys.(C(\Zs, X), \dots))       & \triangleq X.
\end{align*}

The constraint on links for case
patterns is formally defined as follows.
\begin{mydef}[Case Pattern Link Constraints]\label{def:pat-link}
  The pattern \(\Tx\) is required to satisfy \(\PatCond(\Tx)\)
  defined recursively as follows:
  \begin{align}
     & \PatCond(\nu \Ys.(x[\Zs, X]: \tau, \Tc_1, \dots, \Tc_n))
    \triangleq \notag{}                                                       \\
     & \quad X \notin \{\Zs\} \label{patc11}                                  \\
     & \quad \land\;
    \bigl(\forall i, j \in \{1, \dots, n\}.
    \; i \neq j \implies \notag{}                                             \\
     & \quad\qquad \land\; \VarLinks(\Tc_i) \cap \VarLinks(\Tc_j) = \emptyset
    \bigr)
    \label{patc31}                                                            \\
     & \quad \land\;
    \bigl(\forall i \in \{1, \dots, n\}. \notag{}                             \\
     & \quad\qquad \Root(\Tc_i) \in \{\Zs\} \label{patc21}                    \\
     & \quad\qquad \land\; \PatCond(\Tc_i)
    \bigr)
    \label{patc41}                                                            \\
     & \PatCond(
    \nu \Ys.(C(\Zs, X), \seq{U \bowtie V}, \Tx_1, \dots, \Tx_n)
    ) \triangleq \notag{}                                                     \\
     & \quad X \notin \{\Zs\} \label{patc12}                                  \\
     & \quad \land\;
    \bigl(\forall i, j \in \{1, \dots, n\}.
    \; i \neq j \implies \notag{}                                             \\
     & \quad\qquad \land\; \VarLinks(\Tx_i) \cap \VarLinks(\Tx_j) = \emptyset
    \bigr)
    \label{patc32}                                                            \\
     & \quad \land\;
    \bigl(\forall i \in \{1, \dots, n\}. \notag{}                             \\
     & \quad\qquad \Root(\Tx_i) \in \{\Zs\} \label{patc22}                    \\
     & \quad\qquad \land\; \PatCond(\Tx_i)
    \bigr)
    \label{patc42}
  \end{align}

  where $\VarLinks$ is defined recursively as follows:
  \begin{align}
     & \VarLinks(\nu \Ys.(x[\Zs]: \tau, \seq{\Tc})) \triangleq \notag{} \\
     & \quad
    \left(
    \bigcup\nolimits_{\Tc \in \seq{\Tc}} 
    \VarLinks(\Tc)
    \right) \setminus \{\Ys\},
    \label{patc:vl1}                                                    \\
     & \VarLinks(
    \nu \Ys.(C(\Zs, X), \seq{U \bowtie V}, \seq{\Tx})
    ) \triangleq \notag{}                                               \\
     & \quad
    \left\{ X \,\middle|
    \begin{array}{llll}
      \displaystyle
      X \bowtie_{\seq{U \bowtie V}} W \,\land \\[0.5mm]
      W \in \{X\} \cup \bigcup_{\Tx \in \seq{\Tx}} \VarLinks(\Tx)
    \end{array}
    \right\}
    \setminus \{\Ys\}. \label{patc:vl2}
  \end{align}
\end{mydef}

Given a graph template $\Tx$, $\VarLinks(\Tx)$
returns the set of links
that occur free and are either
the primary roots of constructor atoms or links fused with such links in the template.
The definition proceeds as follows:
\begin{enumerate}
  \item for a template whose root is a graph variable,
        it recurses into its subtemplates while eliminating hidden links \eqref{patc:vl1}.
  \item for a template whose root is a constructor atom,
        it collects the links that can be fused with
        the inductively obtained primary roots in its subtemplates
        or with the primary root of the constructor atom,
        while eliminating hidden links \eqref{patc:vl2}.
\end{enumerate}

Using \(\VarLinks\), \(\PatCond\) of \cref{def:pat-link}
checks that a template satisfies the following properties:
\begin{enumerate}
  \item the primary root of the leftmost graph variable
        or constructor atom is a free link \eqref{patc11}\eqref{patc12};
  \item each subtemplate is connected to the
        leftmost graph variable or constructor atom
        through its primary root \eqref{patc21}\eqref{patc22};
  \item \emph{separation condition:}
        the primary roots of constructor atoms in subtemplates
        are pairwise disjoint \eqref{patc31}\eqref{patc32}; and
  \item this property is checked recursively
        \eqref{patc41}\eqref{patc42}.
\end{enumerate}

The purpose of the case pattern syntax
(\cref{def:pat-cond-syntax})
and its link conditions
(\cref{def:pat-link})
is to guarantee the type safety of pattern matching in case expressions;
that is,
whenever pattern matching succeeds,
the type $\LI{\seq{\tau}}{\tau}{\Xs}$ assigned to
each graph variable is identical to
the type $\LI{\seq{\tau'}}{\tau'}{\seq{X'}}$ of the subgraph
that matches it.
Here, $\{\seq{\tau}\}$ and $\{\seq{\tau'}\}$ are identical as multisets and can be empty.
The consistency of free links, $\{\Xs\} = \{\seq{X'}\}$, is trivial,
since the free links of congruent graphs
and the left- and right-hand sides of the typing relation
are equivalent.
Below we explain how the case pattern syntax (\cref{def:pat-cond-syntax})
and its link conditions (\cref{def:pat-link})
ensure $\{\seq{\tau}\} = \{\seq{\tau'}\}$, and then how they ensure $\tau = \tau'$.

The case pattern syntax
(\cref{def:pat-cond-syntax})
and its link conditions
(\cref{def:pat-link})
ensure these properties as follows
(for more detail, see the proof sketch of \cref{prop:matching-consistency}):
\begin{description}
  \item[$\{\seq{\tau}\} = \{\seq{\tau'}\}$:]
        According to the syntactic conditions
        of
        \cref{def:pat-cond-syntax}
        and the link condition
        \eqref{patc11},
        every subtemplate connected to a graph variable via its primary root
        has a constructor atom at its root.
        Thus, by the constraints on production rules
        (\cref{def:disjoint-condition}),
        the types of these subtemplates
        and those of their corresponding matching subgraphs
        are uniquely determined.

        However, even if the types of these subtemplates
        and their matching subgraphs are uniquely determined,
        ambiguity may still arise when a subtemplate
        connected to the primary root of a graph variable
        is shared among other graph variables.
        To prevent this,
        the separation condition enforced by $\VarLinks$
        \eqref{patc31}\eqref{patc32}
        requires that
        the subtemplates of graph variables
        must not shared with other graph variables
        through their primary roots.

  \item[$\tau = \tau'$:]
        The syntactic conditions in
        \cref{def:pat-cond-syntax},
        together with the link conditions other than $\VarLinks$
        in \cref{def:pat-link}
        \eqref{patc11}\eqref{patc12}\eqref{patc21}\eqref{patc22}
        require that
        the primary root of each graph variable
        either (i) be connected to a constructor atom
        or (ii) serve as the primary root of the entire case pattern.
        For Case (i),
        by the constraints on production rules
        (\cref{def:disjoint-condition}),
        the type of the subtemplate and its matching subgraph
        are uniquely determined.
        For Case (ii),
        \TyCase{} in \cref{def:revised-ty-case} (and
        \cref{def:pattern-condition} coming later) ensures that
        the type of the case pattern coincides with
        the type of the scrutinee
        to be evaluated and
        its evaluation result, i.e.,
        the graph being matched.
\end{description}

We allow only patterns in \textbf{case} expressions
that are structurally congruent to a template that follows \cref{def:pat-cond-syntax}
and satisfy Case Pattern Link Conditions given in \cref{def:pat-link}.

For example,
$\eDLun'$ \eqref{eq:sound-dbllist-removal-eg-vis}
(\cref{sec:IntroductionOfRestrictions})
satisfies the constraints of
\cref{def:pat-cond-syntax} and \cref{def:pat-link}.
The case pattern of $\eDLun'$
is congruent to
\begin{align}\label{eq:sound-dbllist-removal-pat-eg}
  \patDLun' \triangleq
  \begin{array}{lll}
    \nu W_2 W_3.(y[X,W_3,W_2,Z]:                    \\
    \qquad \LI{\mathit{lltree}(W_3,Y,W_4)}{         \\
    \qquad\quad \mathit{dbllist}(X, Y, Z)           \\
    \qquad }{X,W_3,W_2,Z},                          \\
    \quad \nu W_1 W_4.(
    \Cons(W_1,W_2,W_4,W_3),                         \\
    \qquad w[W_1]: \mathit{nat}(W_1),               \\
    \qquad z[W_3,Y,W_4]: \mathit{lltree}(W_3,Y,W_4) \\
    \quad )                                         \\
    ).
  \end{array}
\end{align}
Here, the graph variables $y[X, W_3, W_2, Z]$ and $z[W_3, Y, W_4]$ do not
occur consecutively,
and the pattern $\patDLun'$ satisfies the syntactic conditions
specified in \cref{def:pat-cond-syntax}.
It also satisfies the link conditions defined in \cref{def:pat-link}.
In particular,
the primary root \(W_4\) of the graph variable \(z[W_3, Y, W_4]\)
is connected to the constructor atom \(\Cons(W_1, W_2, W_4, W_3)\)
as required by \eqref{patc22},
and it trivially satisfies the separation condition
since $\VarLinks$ always returns an empty set.

On the other hand,
\(\eDLun\) (defined in \eqref{eq:unsound-dbllist-removal-eg-vis})
does not satisfy the constraints of \cref{def:pat-cond-syntax} and \cref{def:pat-link}.
The case pattern of \(\eDLun\) is congruent to
\begin{align}\label{eq:unsound-dbllist-removal-pat-eg}
  \patDLun \triangleq
  \begin{array}{lll}
    \nu W_1 W_2.(
    \Cons(W_1,X,W_2,Z),                             \\
    \quad w[W_1]: \mathit{nat}(W_1),                \\
    \quad \nu W_3 W_4.(y[Z,W_3,W_4,W_2]:            \\
    \quad\qquad \LI{\mathit{lltree}(Y,W_4,W_3)}{    \\
    \quad\qquad\quad \mathit{dbllist}(Z, Y, W_2)    \\
    \quad\qquad }{Z,W_3,W_4,W_2},                   \\
    \qquad z[W_4,Y,W_3]: \mathit{lltree}(Y,W_4,W_3) \\
    \quad )                                         \\
    ).
  \end{array}
\end{align}
However, \(\patDLun\) does not satisfy the syntactic condition
in \cref{def:pat-cond-syntax},
because the graph variables $y[Z,W_3,W_4,W_2]$ and $z[W_4,Y,W_3]$
occur consecutively.
Since the graph variable $z[W_4,Y,W_3]$ has its primary root $W_3$
connected only to $y[Z,W_3,W_4,W_2]$ and not to any constructor atom,
even if we rearrange the order of the constructor atom \(\Cons(W_1,X,W_2,Z)\)
and the graph variables $y[Z,W_3,W_4,W_2]$ and $z[W_4,Y,W_3]$,
the pattern still fails to satisfy the link conditions
\eqref{patc22} of
\cref{def:pat-link}.

To represent typing derivations that involve only certain typing rules,
we define the following typing relation.
\begin{mydef}[Subset Type System]\label{def:subset-typesystem}
  Let $S$ be
  a subset of the set of the typing rules.
  $\Gamma \vdash_P^S e : \tau$
  denotes that
  $\Gamma \vdash_P e : \tau$
  can be derived only using typing rules in $S$.
\end{mydef}

The revised typing rule for \textbf{case} expressions is given as follows:
\begin{mydef}[Revised \TyCaseC{} with Pattern Constraints]\label{def:pattern-condition}
  \mbox{}
  \begin{prooftree}
    \def\extraVskip{1pt}
    \def\defaultHypSeparation{\hskip .1in}
    \AXC{$\begin{array}{r@{~}l}
          \Gamma                  & \vdash_P e_1 : \tau_1
          \\
          \Gamma, \CollectVars(T) & \vdash_P e_2 : \tau_2
        \end{array}$}
    \AXC{$\begin{array}{r@{~}l}
          \CollectVars(T) & \vdash_P^S \Tx : \tau_1
          \\
          \Gamma          & \vdash_P e_3 : \tau_2
        \end{array}$}
    \RightLabel{\TyCase{}}
    \BIC{$\ml{\Gamma \vdash_P\shortstrut (\caseof{e_1}{T}{e_2}{e_3}) : \tau_2}$}
  \end{prooftree}%
  where $\Tx \equiv T$,
  $\PatCond(\Tx)$, and
  $S = \{
    \TyProd{},
    \TyAlpha{},\allowbreak
    \TyVar{},
    \TyLIE{}
    \}$.
\end{mydef}
This revised \TyCase{} allows only patterns in \textbf{case} expressions
that are structurally congruent to the template \(\Tx\),
whose syntax is defined in \cref{def:pat-cond-syntax}
and which must satisfy the link constraints given in \cref{def:pat-link}.

Importantly,
these constraints
(\cref{def:disjoint-condition,def:pat-cond-syntax,def:pat-link})
do not significantly compromise the expressiveness of the language.
Most of the data types and operations illustrated in \Cite{sano2023}
remain expressible within this restricted framework.
Moreover,
the constrained language still supports algebraic data types,
as in conventional functional languages,
and additionally allows graph data structures with sharing and cycles.
Hence, it is more expressive than conventional functional languages with respect to data representation.

\section{Proof Sketch of Soundness}\label{sec:sec4-fully-static-soundness}

We conjecture that the type system described in \cref{sec:sec4-constraints},
obtained by imposing the constraints of
\cref{def:disjoint-condition} and \cref{def:pattern-condition}, is sound.
Here, we only discuss a proof sketch.

This statement is formalised as follows.
\begin{prop}[Preservation]\label{lem:preserve}
  If
  \(\Gamma \vdash_P e: \tau\)
  and
  \(e \reduces e'\),
  then
  \(\Gamma \vdash_P e' : \tau\)
\end{prop}
In what follows, we outline a proof sketch of \cref{lem:preserve}.

In the system described in \cref{sec:sec4-constraints},
dynamic checks in \textbf{case} expressions are eliminated.
It is therefore necessary to argue that the properties previously guaranteed
by dynamic checks are instead ensured statically.

More concretely, we consider the following proposition.
\begin{prop}[Matching Type Consistency]\label{prop:matching-consistency}
  Let $e$ be an expression $(\caseof{G}{T}{e_1}{e_2})$
  and $\thetas$ be graph substitutions.
  If $e$ is typable
  and
  $G \equiv T\thetas$,
  then
  $\emptyset \vdash_P x[\Xs]\thetas: \tau$
  for all $x[\Xs]: \tau \in \CollectVars(T)$.
\end{prop}
Intuitively, this proposition expresses that
if a subgraph matches a graph variable in a \textbf{case} expression,
the type of that subgraph must coincide with the type assigned to the graph variable.

To argue for this property, we proceed in two main steps.
First, we observe that any graph (value) matched in a \textbf{case} expression
can be transformed into a congruent \emph{Normal Form},
and that this normal form is unique.
Then, we analyse the matching process between templates and
matching graphs in their Normal Forms,
divide it into two exhaustive cases,
and argue that the property holds in each case.

We begin by defining the notion of a \emph{Normal Form} for graphs.
\begin{mydef}[Normal Form]\label{def:normal-form}
  Let \(G\) be a typable graph whose type is not a linear implication type,
  and whose typing derivation uses only \(\TyProd{}\)
  in derivations without premises of \(\TyArrow{}\).
  Then the graph \(G\) is said to be in \emph{Normal Form}.
  A graph is said to have a Normal Form
  if it is congruent to a graph that is in Normal Form.
\end{mydef}
Since the use of \TyCong{} is excluded from derivations,
the Normal Form of any set of congruent graphs is syntactically unique,
up to \(\alpha\)-renaming of local links.
This implies that the spanning tree of the graph whose
backbone is formed by primary roots
is uniquely determined.

We state that every typable graph (value) whose type is not a linear implication
has a unique Normal Form, up to \(\alpha\)-renaming of local link names.
We first state the existence of a Normal Form for such graphs.
\begin{prop}[Existence of a Normal Form]\label{lem:normal-form-exists}
  Let \(G\) be a typable graph whose type is not a linear implication type.
  Then, there exists a congruent graph $G'$
  whose
  typing derivation
  uses only \(\TyProd\) in
  the derivation
  without premises of \(\TyArrow\).
\end{prop}

Next, we state the uniqueness of the Normal Form.
\begin{prop}[Uniqueness of a Normal Form]\label{lem:normal-form-uniq}
  Let \(G\) and \(G'\) be graphs in Normal Form such that \(G \equiv G'\).
  Then \(G\) and \(G'\) share the same typing derivation
  among those that contain no premises of \(\TyArrow{}\).
\end{prop}


Detailed proofs of \cref{lem:normal-form-exists} and \cref{lem:normal-form-uniq}
are provided in Appendices \ref{app:normal-form-proof}.

Using \cref{lem:normal-form-exists} and \cref{lem:normal-form-uniq},
we give a proof sketch of \cref{prop:matching-consistency} below.
\begin{proof}[Proof sketch of \cref{prop:matching-consistency}]
  Consider the case expression
  $(\caseof{G}{T}{e_1}{e_2})$.
  Suppose that both \( G \) and \( T \) have the type \(\tau'\).
  Since the matching target graph in a \textbf{case} expression
  is typable,
  and because in \cref{sec:ext-type} we imposed the restriction that
  the type of the first argument of a \textbf{case} expression
  must not be a linear implication type,
  we can consider a Normal Form for the graph.

  We now analyse template-graph matching in the following
  two cases.
  \begin{enumerate}
    \item
          \textit{%
            A graph variable appears at the root of a graph,
            that is,
            the graph variable has a root link which is the root link of the type of a graph.
          }

          By the typing rule Ty-Case,
          the types of the expression $e_1$ and the template $T$ must be identical.
          Therefore, by the induction hypothesis,
          the graph reduced from $e_1$ has the same type as $T$.

          At this point, we consider two cases.
          \begin{enumerate}
            \item
                  If $T$ consists of only a graph variable,
                  then the type of the graph variable corresponds directly to the type of $T$,
                  which is also the type of the pattern being matched.

            \item
                  On the other hand, if $T$ is not solely a graph variable,
                  the typing must be derived using the rule Ty-LI-E.
                  According to \cref{def:pattern-condition} and \cref{def:disjoint-condition},
                  the root of each sub-spanning tree connected to the graph variable in $T$
                  and those matches in $G$
                  contains exactly the same constructor atom.
          \end{enumerate}

          As a result, the type of these sub-spanning trees remains consistent.
          Therefore,
          the type of the graph variable obtained by subtracting
          the types of sub-spanning tree from the overall types
          in both $T$ and the reduced graph $G$
          must be the same type.

    \item
          \textit{%
            The graph variable does not appear at the root of the graph.
          }

          We consider a sub-template containing a graph variable:
          \(
          T \triangleq \nu \Ys.(x[\Zs, X], \seq{\Tc}).
          \)
          According to \cref{def:pattern-condition},
          the graph variable must be located immediately under a constructor atom of the form
          $C(\dots, X, \dots)$, where $X$ is the root link of the graph variable.

          During matching, this constructor atom is expected to correspond to
          an identical constructor atom $C(\dots, X, \dots)$ in the target graph $G$.
          Since $G$ is assumed to admit a unique spanning tree,
          the sub-spanning tree of $G$ that matches the sub-template must be
          the one connected to $X$ via this constructor atom.

          At this point, we analyse the type consistency.
          Because of the disjointness constraint on the production rules,
          there exists a unique production rule that can be applied to any given constructor atom
          via the typing rule \TyProd{}.
          Therefore, the type of the sub-template $T$ and that of the corresponding subgraph in $G$
          must be identical.

          Finally
          we consider the cases whether the sub-template consists solely
          of a graph variable or not.
          In either case, the conclusion remains the same as in Case~(1).
          \qedhere
  \end{enumerate}
\end{proof}

\begin{proof}[Proof sketch of \cref{lem:preserve}]
  The preservation property stated in \cref{lem:preserve}
  can be established straightforwardly
  using the proof of \cref{sec:sec3-soundness},
  with \cref{prop:matching-consistency}.
\end{proof}

The soundness of the type system is established by \cref{lem:preserve},
in combination with the progress lemma which
can be established straightforwardly from
\cref{sec:sec3-soundness}.

\chapter{Formalising \lgt in Isabelle/HOL}\label{ch:formal}

This chapter presents a mechanised formalisation of \lgt in Isabelle/HOL.
The formalisation covers the core calculus and its extension with linear implication types.
The main result is a machine-checked proof of type soundness,
derived from preservation and progress.

The meta-theory of \lgt involves intricate interactions among link binders,
graph variables,
structural congruence,
and graph substitution.
These features go well beyond those found in the ordinary $\lambda$-calculus,
making it difficult to validate the meta-theory completely by paper-and-pencil reasoning alone.
For this reason,
mechanised proof is particularly important for \lgt.
It provides a rigorous means of checking the correctness of the proofs,
while also making the meta-theoretical results more transparent and trustworthy to third parties.

The mechanisation also provides a reusable formal foundation for future implementations of \lgt,
including interpreters and compilers.
Although the prototype interpreter described in \cref{ch:impl} has not yet been verified,
the Isabelle/HOL development is intended to support the verification of such implementations in future work.

The remainder of this chapter is organised as follows.
\Cref{sec:formal:structure} describes the structure of the Isabelle/HOL development,
from the STLC baseline to the core calculus and the linear-implication extension.
\Cref{sec:formal:semantics-types} presents the Isabelle/HOL mechanisation of the syntax,
operational semantics, and type system of \lgt.
\Cref{sec:formal:main-theorems} states the main mechanised results:
preservation,
progress,
and type soundness.
\Cref{sec:formal:proof-engineering} discusses the main proof-engineering challenges.
Finally,
\Cref{sec:formal:summary} summarises the chapter.

\section{Structure of the Isabelle/HOL Development}\label{sec:formal:structure}

The Isabelle/HOL development is organised into three separate theory files.
Each theory establishes type soundness by proving preservation and progress.

\begin{itemize}
  \item
        The theory \verb|STLC.thy| serves as a baseline formalisation.
        It formalises a simply typed $\lambda$-calculus with call-by-value evaluation,
        an explicit notion of values,
        list-based typing environments,
        and a progress theorem.

  \item
        The theory \verb|LambdaGT_Core.thy| formalises the core calculus of \lgt,
        as introduced in \cref{ch:core-type}.
        Compared with the STLC baseline,
        this theory introduces graph values,
        link binders,
        graph variables,
        structural congruence,
        and graph-level substitution.

  \item
        The theory \verb|LambdaGT_LI.thy| extends the core formalisation with linear implication types,
        as introduced in \cref{ch:limp-type}.
        This theory builds on the definitions and proof infrastructure developed for the core calculus,
        while adding the typing rules and meta-theoretic arguments required for linear implication.
\end{itemize}

In the remainder of this section,
we focus primarily on the two \lgt-specific theory files,
namely \verb|LambdaGT_Core.thy| and \verb|LambdaGT_LI.thy|.

The entire Isabelle/HOL development is available online\footnote{%
  \url{https://github.com/sano-jin/lambda-gt-formal}
}
and the entire development has been verified using Isabelle2025.
The session configuration explicitly specifies all dependencies via the \texttt{ROOT} file,
ensuring that the development can be reproduced in a uniform manner.

The development can be built using the standard Isabelle build tool as follows:
\begin{lstlisting}[style=shellplain,language=bash]
$ isabelle build -D . LambdaGT
\end{lstlisting}

In our environment,
the entire development is checked within approximately one minute.
A successful build produces output of the following form:

\begin{lstlisting}[style=shellplain,language={}]
Running LambdaGT ...
Finished LambdaGT (0:00:14 elapsed time, 0:01:06 cpu time, factor 4.62)
\end{lstlisting}

\section{Formalised Semantics and Type Systems}\label{sec:formal:semantics-types}

A central design issue in the Isabelle/HOL formalisation is the representation
of syntax and binding.
The paper presentation of \lgt uses named links and named graph variables,
which is convenient for readability.
In the Isabelle/HOL development,
however,
both links and graph variables are represented using de Bruijn indices.
This choice avoids a direct treatment of $\alpha$-conversion and turns
freshness and capture-avoidance conditions into explicit lemmas about
shifting,
renaming,
link substitution,
and graph substitution.

The syntax of graphs,
expressions,
and atom labels is defined by mutually recursive datatypes.
This reflects an essential feature of \lgt:
graphs may contain atom labels that bind graph variables,
and expressions may contain graph templates.
Consequently,
many definitions and lemmas in the Isabelle/HOL development are also mutually
recursive.

The representation used in the core formalisation is shown in
\cref{fig:formal:core-syntax}.
Links are represented as natural numbers,
and an ordered list of links represents the interface of an atom or graph
variable.

\begin{figure}[tbp]
  \par\noindent\hrule
  \vspace{0.6\baselineskip}
  {\small\ttfamily\setlength{\parindent}{0pt}
    type\_synonym\ link\ =\ nat\par
    type\_synonym\ links\ =\ "link\ list"\par
    \par
    datatype\ 'ty\ graph\ =\par
    \ \ \ \ Zero\par
    \ \ |\ Atom\ "'ty\ p"\ links\par
    \ \ |\ Fusion\ link\ link\par
    \ \ |\ Mol\ "'ty\ graph"\ "'ty\ graph"\par
    \ \ |\ Nu\ "'ty\ graph"\par
    and\ 'ty\ exp\ =\par
    \ \ \ \ Graph\ "'ty\ graph"\par
    \ \ |\ App\ "'ty\ exp"\ "'ty\ exp"\par
    \ \ |\ Case\ "'ty\ exp"\ "(links\ *\ 'ty)\ list"\
    "'ty\ graph"\ "'ty\ exp"\ "'ty\ exp"\par
    and\ 'ty\ p\ =\par
    \ \ \ \ GConstr\ string\par
    \ \ |\ GVar\ nat\par
    \ \ |\ GAbs\ links\ 'ty\ "'ty\ exp"\par
  }
  \vspace{0.6\baselineskip}
  \par\noindent\hrule
  \caption{Core Isabelle/HOL Representation of Links, Graphs, Expressions, and Atom Names.}
  \label{fig:formal:core-syntax}
\end{figure}

The definition of link substitution is given in \cref{fig:formal:lmap}.
Link substitution replaces link indices with other link indices.
When link substitution passes under a \texttt{Nu} binder,
indices must be shifted so that the locally bound link is not affected.

\begin{figure}[tbp]
  \par\noindent\hrule
  \vspace{0.6\baselineskip}
  {\small\ttfamily\setlength{\parindent}{0pt}
    primrec\
    lmap\ ::\ "(link\ \ensuremath{\Rightarrow}\ link)\ \ensuremath{\Rightarrow}\ 'ty\ graph\ \ensuremath{\Rightarrow}\ 'ty\ graph"\
    where\par
    \ \ "lmap\ f\ Zero\ =\ Zero"\par
    |\ "lmap\ f\ (Atom\ p\ xs)\ =\ Atom\ p\ (map\ f\ xs)"\par
    |\ "lmap\ f\ (Fusion\ x\ y)\ =\ Fusion\ (f\ x)\ (f\ y)"\par
    |\ "lmap\ f\ (Mol\ g1\ g2)\ =\ Mol\ (lmap\ f\ g1)\ (lmap\ f\ g2)"\par
    |\ "lmap\ f\ (Nu\ g)\ =\
    Nu\ (lmap\ (\ensuremath{\lambda}x.\ if\ x\ =\ 0\ then\ x\ else\ f\ (x\ -\ 1)\ +\ 1)\ g)"\par
  }
  \vspace{0.6\baselineskip}
  \par\noindent\hrule
  \caption{Definition of Link Substitution Over Graphs.}
  \label{fig:formal:lmap}
\end{figure}

Based on link substitution,
structural congruence is formalised as an inductively defined relation on graphs,
as shown in \cref{fig:formal:structural-congruence}.

\begin{figure}[tbp]
  \par\noindent\hrule
  \vspace{0.6\baselineskip}
  {\small\ttfamily\setlength{\parindent}{0pt}
  inductive\ gcong\ ::\ "'ty\ graph\ \ensuremath{\Rightarrow}\ 'ty\ graph\ \ensuremath{\Rightarrow}\ bool"\ (infix\ "\ensuremath{\simeq}"\ 50)\
  where\par
  \ \ refl\ [intro!]:\ "g\ \ensuremath{\simeq}\ g"\par
  |\ sym\ \ [intro]:\ \ "g1\ \ensuremath{\simeq}\ g2\ \ensuremath{\Longrightarrow}\ g2\ \ensuremath{\simeq}\ g1"\par
  |\ trans[trans]:\ \ "g1\ \ensuremath{\simeq}\ g2\ \ensuremath{\Longrightarrow}\ g2\ \ensuremath{\simeq}\ g3\ \ensuremath{\Longrightarrow}\ g1\ \ensuremath{\simeq}\ g3"\par
  \par
  |\ mol\_zero\ [simp]:\ "Mol\ g\ Zero\ \ensuremath{\simeq}\ g"\par
  |\ mol\_comm\ [simp]:\ "Mol\ g1\ g2\ \ensuremath{\simeq}\ Mol\ g2\ g1"\par
  |\ mol\_assoc[simp]:\ "Mol\ (Mol\ a\ b)\ c\ \ensuremath{\simeq}\ Mol\ a\ (Mol\ b\ c)"\par
  \par
  |\ mol\_cong[intro]:\ "g1\ \ensuremath{\simeq}\ g1'\ \ensuremath{\Longrightarrow}\ g2\ \ensuremath{\simeq}\ g2'\ \ensuremath{\Longrightarrow}\ Mol\ g1\ g2\ \ensuremath{\simeq}\ Mol\ g1'\ g2'"\par
  |\ nu\_cong\ [intro]:\ "g\ \ensuremath{\simeq}\ g'\ \ensuremath{\Longrightarrow}\ Nu\ g\ \ensuremath{\simeq}\ Nu\ g'"\par
  \par
  |\ nu\_subst\_fusion1:\par
  \ \ "y\ \ensuremath{\in}\ FL\ g\ \ensuremath{\Longrightarrow}\par
  \ \ \ Nu\ (Mol\ (Fusion\ 0\ y)\ g)\ \ensuremath{\simeq}\ Nu\ (lmap\ (\ensuremath{\lambda}x.\ if\ x\ =\ 0\ then\ y\ else\ x)\ g)"\par
  |\ nu\_subst\_fusion2:\par
  \ \ "0\ \ensuremath{\in}\ FL\ g\ \ensuremath{\Longrightarrow}\par
  \ \ \ Nu\ (Mol\ (Fusion\ 0\ y)\ g)\ \ensuremath{\simeq}\ Nu\ (lmap\ (\ensuremath{\lambda}x.\ if\ x\ =\ 0\ then\ y\ else\ x)\ g)"\par
  \par
  |\ nu\_nu\_fusion1\ [simp]:\ "Nu\ (Nu\ (Fusion\ 0\ 1))\ \ensuremath{\simeq}\ Zero"\par
  |\ nu\_nu\_fusion2\ [simp]:\ "Nu\ (Fusion\ 0\ 0)\ \ensuremath{\simeq}\ Zero"\par
  \par
  |\ nu\_zero\ [simp]:\ "Nu\ Zero\ \ensuremath{\simeq}\ Zero"\par
  \par
  |\ nu\_comm:\ "Nu\ (Nu\ g)\ \ensuremath{\simeq}\par
  \ \ \ \ \ \ \ \ \ \ \ \ Nu\ (Nu\ (lmap\ (\ensuremath{\lambda}x.\ if\ x\ =\ 0\ then\ 1\ else\ if\ x\ =\ 1\ then\ x\ else\ x)\ g))"\par
  \par
  |\ nu\_scope:\ "Mol\ (Nu\ g1)\ g2\ \ensuremath{\simeq}\ Nu\ (Mol\ g1\ (lmap\ Suc\ g2))"\par
  }
  \vspace{0.6\baselineskip}
  \par\noindent\hrule
  \caption{Inductive Definition of Structural Congruence.}
  \label{fig:formal:structural-congruence}
\end{figure}

\begin{figure}[tbp]
  \par\noindent\hrule
  \vspace{0.6\baselineskip}
  {\small\ttfamily\setlength{\parindent}{0pt}
    primrec\ subst\ ::\ "nat\ \ensuremath{\Rightarrow}\ links\ \ensuremath{\Rightarrow}\ 'ty\ graph\ \ensuremath{\Rightarrow}\ 'ty\ exp\ \ensuremath{\Rightarrow}\ 'ty\ exp"\par
    and\ subst\_graph\ ::\ "nat\ \ensuremath{\Rightarrow}\ links\ \ensuremath{\Rightarrow}\ 'ty\ graph\ \ensuremath{\Rightarrow}\ 'ty\ graph\ \ensuremath{\Rightarrow}\ 'ty\ graph"\par
    and\ subst\_p\ ::\ "nat\ \ensuremath{\Rightarrow}\ links\ \ensuremath{\Rightarrow}\ 'ty\ graph\ \ensuremath{\Rightarrow}\ 'ty\ p\ \ensuremath{\Rightarrow}\ links\ \ensuremath{\Rightarrow}\ 'ty\ graph"\par
    where\par
    \ \ "subst\ i\ xs\ G\ (Graph\ g)\ =\ Graph\ (subst\_graph\ i\ xs\ G\ g)"\par
    |\ "subst\ i\ xs\ G\ (App\ e1\ e2)\ =\par
    \ \ \ \ \ App\ (subst\ i\ xs\ G\ e1)\ (subst\ i\ xs\ G\ e2)"\par
    |\ "subst\ i\ xs\ G\ (Case\ e0\ \ensuremath{\Gamma}\ T\ e1\ e2)\ =\par
    \ \ \ \ \ Case\ (subst\ i\ xs\ G\ e0)\ \ensuremath{\Gamma}\ T\par
    \ \ \ \ \ \ (subst\ (i\ +\ length\ \ensuremath{\Gamma})\ xs\ ((lift\_graph\ 0\ \verb|^^| (length\ \ensuremath{\Gamma}))\ G)\ e1)\par
    \ \ \ \ \ \ (subst\ i\ xs\ G\ e2)"\par
    \mbox{}\\
    |\ "subst\_graph\ i\ xs\ G\ Zero\ =\ Zero"\par
    |\ "subst\_graph\ i\ xs\ G\ (Atom\ p\ ys)\ =\ subst\_p\ i\ xs\ G\ p\ ys"\par
    |\ "subst\_graph\ i\ xs\ G\ (Fusion\ x\ y)\ =\ Fusion\ x\ y"\par
    |\ "subst\_graph\ i\ xs\ G\ (Mol\ g1\ g2)\ =\par
    \ \ \ \ \ Mol\ (subst\_graph\ i\ xs\ G\ g1)\ (subst\_graph\ i\ xs\ G\ g2)"\par
    |\ "subst\_graph\ i\ xs\ G\ (Nu\ g)\ =\ Nu\ (subst\_graph\ i\ xs\ G\ g)"\par
    \mbox{}\\
    |\ "subst\_p\ i\ xs\ G\ (GConstr\ s)\ ys\ =\ Atom\ (GConstr\ s)\ ys"\par
    |\ "subst\_p\ i\ xs\ G\ (GAbs\ zs\ ty\ e0)\ ys\ =\par
    \ \ \ \ \ Atom\ (GAbs\ zs\ ty\ (subst\ (i\ +\ 1)\ xs\ (lift\_graph\ 0\ G)\ e0))\ ys"\par
    |\ "subst\_p\ i\ xs\ G\ (GVar\ j)\ ys\ =\par
    \ \ \ \ \ (if\ i\ <\ j\ then\ Atom\ (GVar\ (j\ -\ 1))\ ys\par
    \ \ \ \ \ \ else\ if\ j\ =\ i\ \ensuremath{\land}\ length\ xs\ =\ length\ ys\
    then\ lmap\ (lsubst\ (zip\ xs\ ys))\ G\par
    \ \ \ \ \ \ else\ Atom\ (GVar\ j)\ ys)"\par
  }
  \vspace{0.6\baselineskip}
  \par\noindent\hrule
  \caption{Mutually Recursive Definition of Graph Substitution.}
  \label{fig:formal:graph-substitution}
\end{figure}

Also based on link substitution,
graph substitution, which replaces graph variables with graphs,
is specified in \cref{fig:formal:graph-substitution}.

In this definition,
the operation \verb|lift_graph| is the graph-level counterpart
of the standard shifting operation used in de Bruijn representations.
of the standard shifting operation used in de Bruijn representations.
It adjusts the indices of free graph variables when graph substitution passes under a graph-variable binder.
This adjustment is necessary because the surrounding binding context changes beneath such a binder.
Without lifting, free graph variables in the substituted graph could be
inadvertently captured by the binder being entered.

The equality case for \texttt{GVar} is the essential part of graph substitution.
Substitution does not simply replace a variable with a graph,
but also adapts the free links of the substituted graph to the interface at the occurrence site.
This is where graph substitution differs from ordinary substitution in the $\lambda$-calculus.
When the variable index matches the substitution target and the two interfaces have the same length,
the substituted graph is relinked by applying the link substitution
generated from the correspondence between the original interface \texttt{xs}
and the occurrence interface \texttt{ys}.



The operational semantics is mechanised as an inductive relation,
as shown in \cref{fig:formal:cbv-semantics}.
The main rules are beta reduction,
successful case reduction,
failed case reduction,
and contextual evaluation rules for applications and case expressions.
Graphs are treated as values,
and graph matching is performed using the previously defined structural congruence relation and
graph substitution operation.

\begin{figure}[tbp]
  \par\noindent\hrule
  \vspace{0.6\baselineskip}
  {\small\ttfamily\setlength{\parindent}{0pt}
  inductive\ cbv\_ty\ ::\ "type\ exp\ \ensuremath{\Rightarrow}\ prodrule\ list\ \ensuremath{\Rightarrow}\ type\ exp\ \ensuremath{\Rightarrow}\ bool"\par
  \ \ ("\_\ ->\{\_\}\ \_"\ [60,0,61]\ 60)\par
  where\par
  \ \ beta\_v\_ty\ [simp,\ intro]:\par
  \ \ \ \ "is\_graph\_val\ v\ \ensuremath{\Longrightarrow}\ distinct\ xs\ \ensuremath{\Longrightarrow}\ FL\ v\ =\ set\ xs\ \ensuremath{\Longrightarrow}\par
  \ \ \ \ \ lam\ \ensuremath{\simeq}\ Atom\ (GAbs\ xs\ ty\ e)\ ys\ \ensuremath{\Longrightarrow}\par
  \ \ \ \ \ Graph\ lam\ \ensuremath{\cdot}\ Graph\ v\ ->\{P\}\ subst\ 0\ xs\ v\ e"\par
  \par
  |\ case1\_v\_ty\ [simp,\ intro]:\par
  \ \ \ \ "is\_graph\_val\ v\ \ensuremath{\Longrightarrow}\
  length\ \ensuremath{\Gamma}\ =\ length\ gs\ \ensuremath{\Longrightarrow}\par
  \ \ \ \ \ XsGs\ =\ zip\ \ensuremath{\Gamma}\ gs\ \ensuremath{\Longrightarrow}\
  v\ \ensuremath{\simeq}\ substs\_graph\ XsGs\ T\ \ensuremath{\Longrightarrow}\par
  \ \ \ \ \ (\ensuremath{\forall}((xs,\ ty),\ g)\ \ensuremath{\in}\ set\ XsGs.\par
  \ \ \ \ \ \ \ \ distinct\ xs\ \ensuremath{\land}\ FL\ g\ =\ set\ xs\ \ensuremath{\land}\
  []\ \ensuremath{\vdash}\{P\}\ Graph\ g\ :\ ty)\ \ensuremath{\Longrightarrow}\par
  \ \ \ \ \ Case\ (Graph\ v)\ \ensuremath{\Gamma}\ T\ e1\ e2\ ->\{P\}\ substs\ XsGs\ e1"\par
  \par
  |\ case2\_v\_ty\ [simp,\ intro]:\par
  \ \ \ \ "is\_graph\_val\ v\ \ensuremath{\Longrightarrow}\par
  \ \ \ \ \ Case\ (Graph\ v)\ \ensuremath{\Gamma}\ T\ e1\ e2\ ->\{P\}\ e2"\par
  \par
  |\ appL\_ty\ [simp,\ intro]:\par
  \ \ \ \ "e1\ ->\{P\}\ e1'\ \ensuremath{\Longrightarrow}\par
  \ \ \ \ \ e1\ \ensuremath{\cdot}\ e2\ ->\{P\}\ e1'\ \ensuremath{\cdot}\ e2"\par
  \par
  |\ appR\_ty\ [simp,\ intro]:\par
  \ \ \ \ "is\_val\ v\ \ensuremath{\Longrightarrow}\ e\ ->\{P\}\ e'\ \ensuremath{\Longrightarrow}\par
  \ \ \ \ \ v\ \ensuremath{\cdot}\ e\ ->\{P\}\ v\ \ensuremath{\cdot}\ e'"\par
  \par
  |\ caseL\_ty\ [simp,\ intro]:\par
  \ \ \ \ "e0\ ->\{P\}\ e0'\ \ensuremath{\Longrightarrow}\par
  \ \ \ \ \ Case\ e0\ \ensuremath{\Gamma}\ T\ e1\ e2\ ->\{P\}\ Case\ e0'\ \ensuremath{\Gamma}\ T\ e1\ e2"\par
  }
  \vspace{0.6\baselineskip}
  \par\noindent\hrule
  \caption{Call-By-Value Small-Step Operational Semantics.}
  \label{fig:formal:cbv-semantics}
\end{figure}

A noteworthy feature of this definition is that successful case reduction includes a dynamic type check.
This check appears in the premise of \texttt{case1\_v\_ty}:

\begin{quote}
  {\small\ttfamily\raggedright\setlength{\parindent}{0pt}
  (\ensuremath{\forall}((xs,\ ty),\ g)\ \ensuremath{\in}\ set\ XsGs.\\[0mm]
  \ \ distinct\ xs\ \ensuremath{\land}\ FL\ g\ =\ set\ xs\ \ensuremath{\land}\
  []\ \ensuremath{\vdash}\{P\}\ Graph\ g\ :\ ty)
  }
\end{quote}

This premise requires each graph \texttt{g} supplied at runtime to have the declared type \texttt{ty},
under the empty typing environment,
with an interface \texttt{xs} that is well formed and agrees exactly with the free links of \texttt{g}.
Thus,
the success branch of a case expression is taken only when the graph matching condition
holds and the graphs used for the match satisfy the required types.

The core type system is also formalised as an inductively defined judgement,
as defined in \cref{fig:formal:typing-core}.
The linear-implication extension, shown in \cref{fig:formal:typing-li},
adds typing rules for incomplete graph structures to the core type system.
The additional rules introduce linear implication types from production rules,
compose such implications,
eliminate a completed implication to obtain a base graph type,
and embed a base graph type as a trivial linear implication.

\begin{figure}[tbp]
  \par\noindent\hrule
  \vspace{0.6\baselineskip}
  {\small\ttfamily\setlength{\parindent}{0pt}
  inductive\ typing\ ::\
  "tyenv\ \ensuremath{\Rightarrow}\ prodrule\ list\ \ensuremath{\Rightarrow}\ type\ exp\ \ensuremath{\Rightarrow}\ type\ \ensuremath{\Rightarrow}\ bool"\par
  \ \ (\ensuremath{\langle}\_\ \ensuremath{\vdash}\{\_\}\ \_\ :\ \_\ensuremath{\rangle}\ [50,\ 50,\ 50,\ 50]\ 50)\par
  where\par
  \ \ TyVar:\par
  \ \ \ \ "i\ <\ length\ \ensuremath{\Gamma}\ \ensuremath{\Longrightarrow}\
  \ensuremath{\Gamma}\ !\ i\ =\ (xs,\ ty)\ \ensuremath{\Longrightarrow}\par
  \ \ \ \ \ length\ xs\ =\ length\ ys\ \ensuremath{\Longrightarrow}\
  distinct\ xs\ \ensuremath{\Longrightarrow}\ distinct\ ys\ \ensuremath{\Longrightarrow}\par
  \ \ \ \ \ set\ xs\ =\ FLty\ ty\ \ensuremath{\Longrightarrow}\par
  \ \ \ \ \ \ensuremath{\Gamma}\ \ensuremath{\vdash}\{P\}\ Graph\ (Atom\ (GVar\ i)\ ys)\ :\ lmap\_ty\ xs\ ys\ ty"\par
  \par
  |\ TyArrow:\par
  \ \ \ \ "(xs,\ ty1)\ \#\ \ensuremath{\Gamma}\ \ensuremath{\vdash}\{P\}\ e\ :\ ty2\ \ensuremath{\Longrightarrow}\par
  \ \ \ \ \ distinct\ xs\ \ensuremath{\Longrightarrow}\
  set\ xs\ =\ FLty\ ty1\ \ensuremath{\Longrightarrow}\par
  \ \ \ \ \ \ensuremath{\Gamma}\ \ensuremath{\vdash}\{P\}\ Graph\ (Atom\ (GAbs\ xs\ ty1\ e)\ ys)\
  :\ TArrow\ ty1\ ty2\ ys"\par
  \par
  |\ TyApp:\par
  \ \ \ \ "\ensuremath{\Gamma}\ \ensuremath{\vdash}\{P\}\ e1\ :\ TArrow\ ty1\ ty2\ xs\ \ensuremath{\Longrightarrow}\par
  \ \ \ \ \ \ensuremath{\Gamma}\ \ensuremath{\vdash}\{P\}\ e2\ :\ ty1\ \ensuremath{\Longrightarrow}\par
  \ \ \ \ \ \ensuremath{\Gamma}\ \ensuremath{\vdash}\{P\}\ e1\ \ensuremath{\cdot}\ e2\ :\ ty2"\par
  \par
  |\ TyCong:\par
  \ \ \ \ "\ensuremath{\Gamma}\ \ensuremath{\vdash}\{P\}\ Graph\ g1\ :\ ty\ \ensuremath{\Longrightarrow}\par
  \ \ \ \ \ g1\ \ensuremath{\simeq}\ g2\ \ensuremath{\Longrightarrow}\par
  \ \ \ \ \ \ensuremath{\Gamma}\ \ensuremath{\vdash}\{P\}\ Graph\ g2\ :\ ty"\par
  \par
  |\ TyAlpha:\par
  \ \ \ \ "\ensuremath{\Gamma}\ \ensuremath{\vdash}\{P\}\ Graph\ g\ :\ ty\ \ensuremath{\Longrightarrow}\par
  \ \ \ \ \ set\ xs\ =\ FL\ g\ \ensuremath{\Longrightarrow}\par
  \ \ \ \ \ length\ xs\ =\ length\ ys\ \ensuremath{\Longrightarrow}\
  distinct\ xs\ \ensuremath{\Longrightarrow}\ distinct\ ys\ \ensuremath{\Longrightarrow}\par
  \ \ \ \ \ \ensuremath{\Gamma}\ \ensuremath{\vdash}\{P\}\ Graph\ (lmap\ (lsubst\ (zip\ xs\ ys))\ g)\
  :\ lmap\_ty\ xs\ ys\ ty"\par
  \par
  |\ TyCase:\par
  \ \ \ \ "\ensuremath{\Gamma}\ \ensuremath{\vdash}\{P\}\ e0\ :\ ty1\ \ensuremath{\Longrightarrow}\par
  \ \ \ \ \ \ensuremath{\Gamma}2\ @\ \ensuremath{\Gamma}\ \ensuremath{\vdash}\{P\}\ e1\ :\ ty2\ \ensuremath{\Longrightarrow}\par
  \ \ \ \ \ \ensuremath{\Gamma}\ \ensuremath{\vdash}\{P\}\ e2\ :\ ty2\ \ensuremath{\Longrightarrow}\par
  \ \ \ \ \ \ensuremath{\Gamma}\ \ensuremath{\vdash}\{P\}\ Case\ e0\ \ensuremath{\Gamma}2\ T\ e1\ e2\ :\ ty2"\par
  }
  \vspace{0.6\baselineskip}
  \par\noindent\hrule
  \caption{Core Typing Rules Formalised as an Inductive Judgment.}
  \label{fig:formal:typing-core}
\end{figure}

\begin{figure}[tbp]
  \par\noindent\hrule
  \vspace{0.6\baselineskip}
  {\small\ttfamily\setlength{\parindent}{0pt}
    |\ TyLIIntro:\par
    \ \ \ \ "(lhs,\ rhs)\ \ensuremath{\in}\ set\ P\ \ensuremath{\Longrightarrow}\par
    \ \ \ \ \ \ensuremath{\Gamma}\ \ensuremath{\vdash}\{P\}\ Graph\ (make\_diffgraph\ rhs)\ :\ make\_LI\ (lhs,\ rhs)"\par
    \par
    |\ TyLITrans:\par
    \ \ \ \ "\ensuremath{\Gamma}\ \ensuremath{\vdash}\{P\}\ Graph\ T1\ :\ TLI\ (\{\#TBase\ a2\ xs2\#\}\ +\ tys1)\ a1\ xs1\ ys1\ \ensuremath{\Longrightarrow}\par
    \ \ \ \ \ \ensuremath{\Gamma}\ \ensuremath{\vdash}\{P\}\ Graph\ T2\ :\ TLI\ tys2\ a1\ xs2\ ys2\ \ensuremath{\Longrightarrow}\par
    \ \ \ \ \ ys\ =\ (ys1\ |\ensuremath{\cup}|\ ys2)\ \ensuremath{\Longrightarrow}\par
    \ \ \ \ \ \ensuremath{\Gamma}\ \ensuremath{\vdash}\{P\}\ Graph\ (Mol\ T1\ T2)\ :\ TLI\ (tys1\ +\ tys2)\ a1\ xs2\ ys"\par
    \par
    |\ TyLIElim0:\par
    \ \ \ \ "ys\ =\ fset\_of\_list\ ys'\ \ensuremath{\Longrightarrow}\
    set\ xs\ \ensuremath{\subseteq}\ FL\ T\ \ensuremath{\Longrightarrow}\par
    \ \ \ \ \ \ensuremath{\Gamma}\ \ensuremath{\vdash}\{P\}\ Graph\ T\ :\ TLI\ 0\ a\ xs\ ys\ \ensuremath{\Longrightarrow}\par
    \ \ \ \ \ \ensuremath{\Gamma}\ \ensuremath{\vdash}\{P\}\ Graph\ (lbinds\ (filter\ (\ensuremath{\lambda}y.\ y\ \ensuremath{\notin}\ set\ xs)\ ys')\ T)\ :\ TBase\ a\ xs"\par
    \par
    |\ TyLIIntro0:\par
    \ \ \ \ "\ensuremath{\Gamma}\ \ensuremath{\vdash}\{P\}\ Graph\ T\ :\ TBase\ a\ xs\ \ensuremath{\Longrightarrow}\par
    \ \ \ \ \ \ensuremath{\Gamma}\ \ensuremath{\vdash}\{P\}\ Graph\ T\ :\ TLI\ 0\ a\ xs\ (fset\_of\_list\ xs)"\par
  }
  \vspace{0.6\baselineskip}
  \par\noindent\hrule
  \caption{Typing Rules for the Linear-Implication Extension.}
  \label{fig:formal:typing-li}
\end{figure}

\section{Main Theorems}\label{sec:formal:main-theorems}

For both the core type system and its extension with linear implication types,
we prove preservation,
progress,
and type soundness.
The statements have the standard form used in type soundness proofs,
but their mechanised proofs depend on \lgt-specific infrastructure for
link substitution,
structural congruence, and
graph substitution.

The first main result is preservation,
also known as subject reduction.
It states that one step of call-by-value reduction preserves typing,
as shown in \cref{fig:formal:subject-reduction}.
From preservation,
we derive the corresponding multi-step preservation theorem\\
\texttt{subject\_reduction'}.
It states that typing is preserved by zero or more reduction steps.

\begin{figure}[tbp]
  \par\noindent\hrule
  \vspace{0.6\baselineskip}
  {\small\ttfamily\setlength{\parindent}{0pt}
    lemma\ subject\_reduction:\par
    \ \ assumes\ "\ensuremath{\Gamma}\ \ensuremath{\vdash}\{P\}\ e\ :\ \ensuremath{\tau}"\
    and\ "e\ ->\{P\}\ e'"\par
    \ \ \ \ shows\ "\ensuremath{\Gamma}\ \ensuremath{\vdash}\{P\}\ e'\ :\ \ensuremath{\tau}"\par
    using\ assms\par
    proof\ (induct\ arbitrary:\ e'\ set:\ typing)\par
    \ \ ...\par
  }
  \vspace{0.6\baselineskip}
  \par\noindent\hrule
  \caption{Mechanised Statement of Preservation.}
  \label{fig:formal:subject-reduction}
\end{figure}

The second main result is progress.
It states that a closed well-typed expression is either already a value or can
take a reduction step,
as shown in \cref{fig:formal:progress}.

\begin{figure}[tbp]
  \par\noindent\hrule
  \vspace{0.6\baselineskip}
  {\small\ttfamily\setlength{\parindent}{0pt}
  lemma\ progress:\par
  \ \ assumes\ "\ensuremath{\Gamma}\ \ensuremath{\vdash}\{P\}\ e\ :\ ty"\
  and\ "\ensuremath{\Gamma}\ =\ []"\par
  \ \ \ \ shows\ "is\_val\ e\ \ensuremath{\lor}\ (\ensuremath{\exists}e'.\ e\ ->\{P\}\ e')"\par
  using\ assms\par
  proof\ (induction\ arbitrary:\ ty\ e'\ set:\ typing)\par
  \ \ ...\par
  }
  \vspace{0.6\baselineskip}
  \par\noindent\hrule
  \caption{Mechanised Statement of Progress.}
  \label{fig:formal:progress}
\end{figure}

The final result is type soundness.
It combines multi-step preservation and progress.
If a closed well-typed expression reduces to another expression in zero or
more steps,
then the resulting expression is either a value or can take a further
reduction step,
as shown in \cref{fig:formal:type-soundness}.
The same pattern of results is established for the linear-implication
extension.

\begin{figure}[tbp]
  \par\noindent\hrule
  \vspace{0.6\baselineskip}
  {\small\ttfamily\setlength{\parindent}{0pt}
  theorem\ type\_soundness:\par
  \ \ assumes\ WT:\ "[]\ \ensuremath{\vdash}\{P\}\ e\ :\ \ensuremath{\tau}"\
  and\ STEPS:\ "e\ ->\{P\}*\ e'"\par
  \ \ \ \ shows\ "is\_val\ e'\ \ensuremath{\lor}\ (\ensuremath{\exists}u.\ e'\ ->\{P\}\ u)"\par
  proof\ -\par
  \ \ have\ "[]\ \ensuremath{\vdash}\{P\}\ e'\ :\ \ensuremath{\tau}"\par
  \ \ \ \ using\ STEPS\ WT\ subject\_reduction'\ by\ blast\par
  \ \ thus\ ?thesis\par
  \ \ \ \ using\ progress\ by\ blast\par
  qed\par
  }
  \vspace{0.6\baselineskip}
  \par\noindent\hrule
  \caption{Mechanised Statement of Type Soundness.}
  \label{fig:formal:type-soundness}
\end{figure}

\section{Proof Engineering Challenges}\label{sec:formal:proof-engineering}

The preservation,
progress,
and type-soundness theorems established in this chapter have the standard
shape expected from type-soundness proofs for typed $\lambda$ calculi.
However,
the proof infrastructure required to establish these theorems is not standard.
In particular,
the amount of auxiliary machinery grows rapidly as the formalisation moves from
ordinary simply typed $\lambda$ calculus to the core calculus of \lgt and then to
the linear-implication extension.

\Cref{tab:formal:size} summarises this growth in terms of source lines,
definitions and abbreviations,
and proved lemmas and theorems.
The simply typed $\lambda$ calculus formalisation in \verb|STLC.thy| consists of
696 lines,
6 definitions and abbreviations,
and 16 lemmas and theorems.
The core \lgt formalisation increases to 3,311 lines,
12 definitions and abbreviations,
and 97 lemmas and theorems.
The linear-implication extension in \verb|LambdaGT_LI.thy| further increases
the development to 4,789 lines,
21 definitions and abbreviations,
and 187 lemmas and theorems.

\begin{table}[tbp]
  \centering
  \caption{Size of the Isabelle/HOL formalisation.}
  \label{tab:formal:size}
  \begin{tabular}{lrrr}
    \toprule
    Theory file              & Lines & Definitions/abbreviations & Lemmas/theorems \\
    \midrule
    \verb|STLC.thy|          & 696   & 6                         & 16              \\
    \verb|LambdaGT_Core.thy| & 3,311 & 12                        & 97              \\
    \verb|LambdaGT_LI.thy|   & 4,789 & 21                        & 187             \\
    \bottomrule
  \end{tabular}
\end{table}

\begin{figure}[htb]
  \centering
  \includegraphics[width=0.12\linewidth]{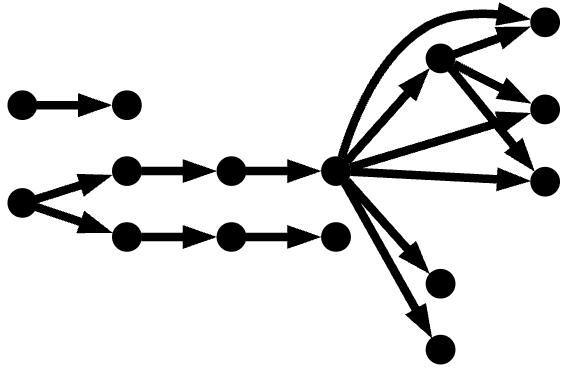}
  \quad
  \includegraphics[width=0.32\linewidth]{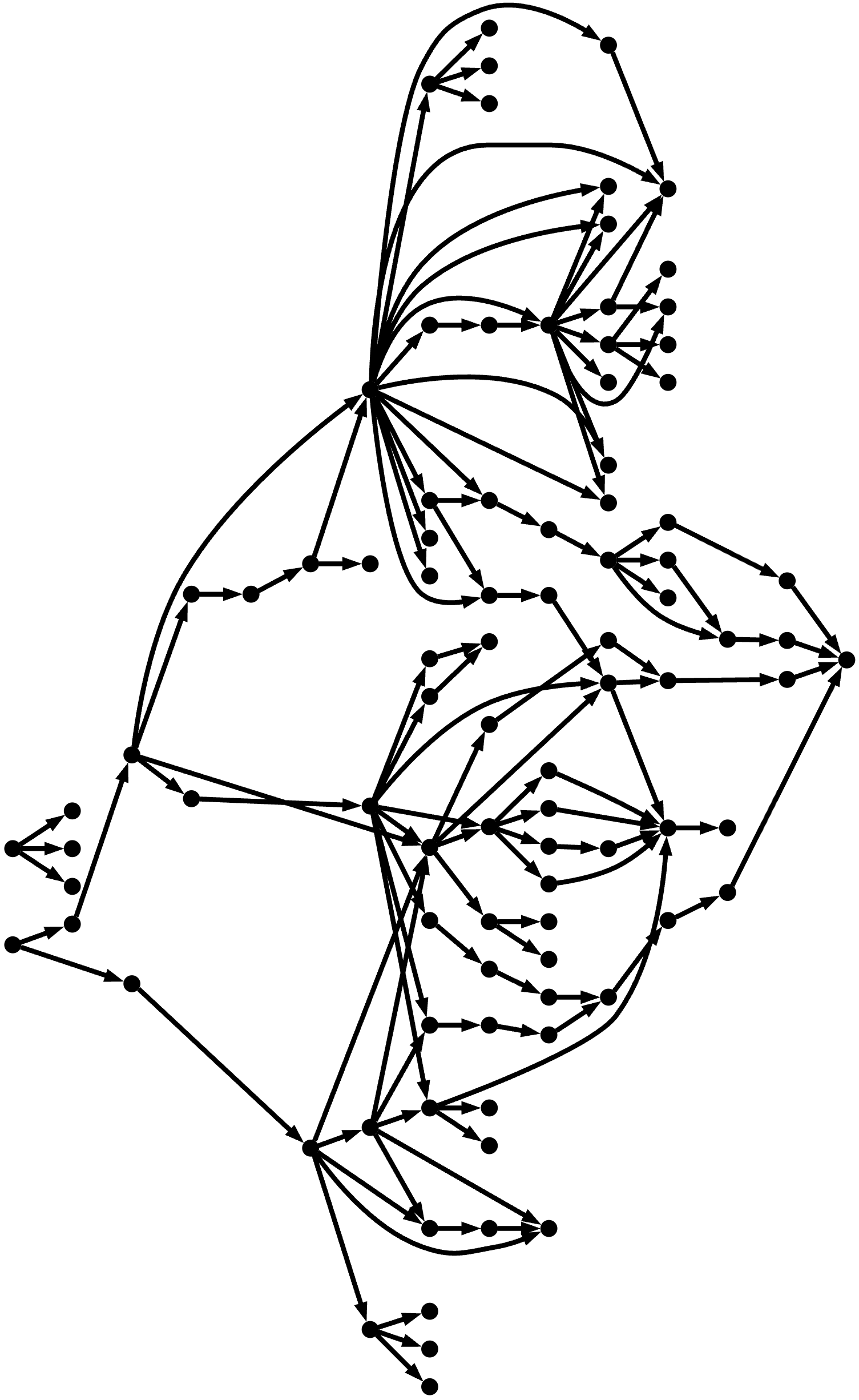}
  \quad
  \includegraphics[width=0.4\linewidth]{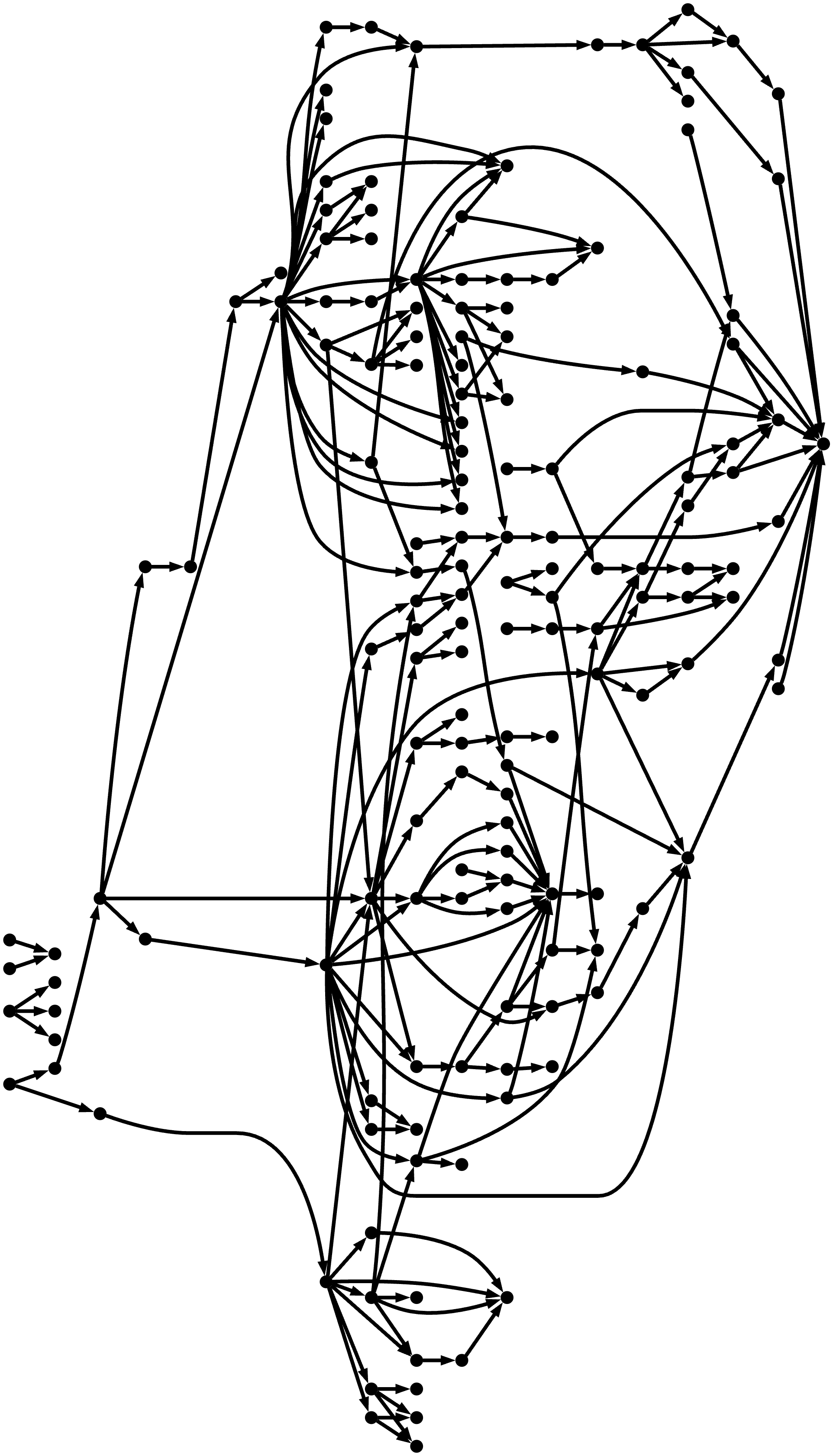}
  \caption{Dependency graphs for the formalisations of
    call-by-value simply typed $\lambda$-calculus,
    \lgt with the core type system,
    and \lgt with linear implication types, respectively.}
  \label{fig:depgraph-ESTLC}
\end{figure}

The increase in complexity is also visible from the perspective of
proof dependencies.
The dependency graphs in
\cref{fig:depgraph-ESTLC}
show where the additional proof infrastructure arises.
The main theorem statements remain compact,
but the supporting lemmas grow substantially as the formalisation moves from STLC
to the core calculus of \lgt and then to the linear-implication extension.

One major source of complexity is link binding.
In the paper presentation,
link binders are expressed using named links and informal freshness
conditions.
In the Isabelle/HOL formalisation,
these conditions are represented using de Bruijn indices,
free-link predicates,
renaming operations,
and link substitutions.
As a result,
paper-level arguments about $\alpha$-conversion and freshness become explicit
proof obligations about shifting,
substitution under binders,
and the preservation of free-link conditions.

A second source of complexity is structural congruence.
Unlike ordinary $\lambda$ terms,
graphs are considered modulo structural congruence.
Thus,
many arguments cannot proceed by direct syntactic inspection alone.
The proofs must also establish that typing is stable under graph equivalence
and that reduction rules behave correctly when graph components are rearranged
by associativity,
commutativity,
scope extrusion,
or fusion-related congruence rules.

A third source of complexity is graph substitution.
Substituting a graph for a graph variable is not the same as ordinary
$\lambda$-calculus substitution.
A graph-variable occurrence carries an ordered interface of links,
and the substituted graph must be reconnected according to that interface.
Therefore,
graph substitution depends on simultaneous link substitution.
Many preservation cases require lemmas showing that this operation preserves
typing,
free-link conditions,
and compatibility with structural congruence.

We also considered using Nominal Isabelle to handle binding.
Nominal Isabelle provides direct support for names,
freshness,
and $\alpha$-equivalence,
and is often effective for calculi with conventional binding structures.
However,
the binding structure of \lgt is less conventional.
The formalisation must handle ordered sequences of links,
simultaneous substitution,
multiple classes of variables, and
mutually recursive syntax.
In this setting,
a nominal representation made the required induction principles and
substitution lemmas difficult to manage.
We therefore adopted a de Bruijn representation for both links and graph
variables.

\section{Summary}\label{sec:formal:summary}

This chapter presented a mechanised formalisation of \lgt in Isabelle/HOL.
The formalisation covers the core calculus and its extension with linear implication types.
For both systems,
we formalised the syntax,
structural congruence,
call-by-value operational semantics,
typing judgements,
and the auxiliary operations required for substitution and binding.
On this basis,
we mechanically verified preservation,
progress,
and type soundness.

The main challenge was not the final form of the type-soundness theorem,
which follows the standard preservation-and-progress pattern,
but the proof infrastructure required for \lgt.
In particular,
the development must account for graph values,
link binders,
graph variables,
structural congruence,
link substitution,
and graph substitution.
These features make the formalisation substantially more involved than the ordinary $\lambda$-calculus.

The resulting Isabelle/HOL development provides machine-checked evidence
for the correctness of the type systems studied in this dissertation,
and serves as a reusable formal basis for future implementations and extensions of \lgt.

\chapter{Related Work}\label{ch:related}

This chapter positions \lgt in relation to existing work from four perspectives.
First, \Cref{sec:related:pfds} compares \lgt
with ordinary functional approaches to handling data structures.
Next, \Cref{sec:related:graph-comp} discusses approaches that treat graphs as objects of computation.
Then, \Cref{sec:related:pattern-matching} relates \lgt to mechanisms
for decomposing data beyond ordinary tree patterns.
Finally, \Cref{sec:related:static} compares the type system of \lgt
with static reasoning techniques for graph and pointer structures.

\section{Purely Functional Data Structures}\label{sec:related:pfds}

Functional programming is particularly effective for lists and trees.
Algebraic data types can describe their structure,
pattern matching decomposes that structure directly,
and type systems check programs against the declared structure.
These advantages apply naturally to tree-like data.
However, they do not directly extend to more complex data structures
that have sharing or cyclic references.
As a result, functional languages do not provide a direct way to program
with such data structures in the same natural manner as they do with lists and trees.

Purely functional data structures address this limitation
by representing complex data structures in terms of simpler functional structures.
Okasaki showed that many efficient data structures can be implemented using trees,
together with persistence, lazy evaluation, and amortised analysis~\cite{Okasaki1998}.
These data structures have been widely implemented as libraries and are highly effective,
especially because they provide persistence.
However, they should be understood as sophisticated encodings of complex data structures
rather than direct support for such structures.
Their implementations can therefore be complicated.

\lgt takes a different approach.
Rather than encoding complex data structures indirectly through trees and lists,
\lgt aims to extend the strengths of functional languages from trees to more complex data structures themselves.
We do not claim that this approach replaces purely functional data structures.
Indeed, efficiently supporting persistence may be difficult in this setting.
Nevertheless, \lgt may provide an alternative way:
a way to program directly with complex data structures in functional languages.

\section{Graphs as Objects of Computation}\label{sec:related:graph-comp}

We next consider approaches in which graphs are treated directly as objects of computation.

\subsection{Graph Transformation Languages}\label{sec:related-graph-trans}

Graph transformation languages provide a direct way to manipulate complex data structures, namely, graphs.
They describe computation by applying rewrite rules to graphs.
A rewrite rule typically consists of a pattern graph and a replacement graph:
computation proceeds by finding an occurrence of the pattern in a host graph and
replacing it with the corresponding instance of the replacement.

Many graph transformation systems have been developed in this tradition,
including AGG~\cite{agg2012}, GP/GP2~\cite{GP,GP2}, GROOVE~\cite{Groove},
GrGen.NET~\cite{grgen-net}, LMNtal~\cite{lmntal2009}, HyperLMNtal~\cite{hyperlmntal},
PORGY~\cite{PORGY2014}, and earlier graph-grammar-based systems~\cite{progress1999,structuredgamma}.

Graph transformation languages are well suited to state-based modelling,
because graphs can naturally represent the states of many systems.
Some graph transformation systems also provide verification mechanisms.
For example, SLIM, an implementation of LMNtal,
supports parallel state-space exploration and LTL model checking~\cite{slim-github},
which scales to state spaces with more than $10^9$ states.
However, these languages are not directly aligned with value-oriented functional languages,
where computation is usually organised around immutable values and function application.

\subsection{Functional Programming with Graph-Structured Data}
\label{sec:related-functional-graphs}

Functional programming languages can represent graphs in several ways.
Graphs can be implemented using references, mutable cells, or libraries such as FGL,
which provide data structures and operations for general graphs~\cite{fungraph-lib}.

Erwig's inductive graphs are one of the closest related approaches~\cite{erwig-active-patterns,erwig-functional-graphs,erwig-inductive-graphs}.
They provide a functional decomposition principle for graphs by separating a graph into a context and the remaining graph.
This supports recursive graph algorithms in a style analogous to programming with inductive data.

Structured graphs are also closely related~\cite{structured-graphs}.
Although functional languages appear, at the surface level, to support only trees as primitive data structures,
Oliveira and Cook showed that recursive binders can express sharing and cycles at the implementation level,
a mechanism commonly used for infinite lists.

Algebraic approaches to graphs in functional programming form another complementary line of work.
Mokhov's algebraic graphs provide a concise algebraic interface
for constructing and reasoning about graphs in Haskell~\cite{mokhov-algebraic-graphs}.
They focus on graph construction, equational reasoning, and transformation through algebraic operations.

Thus, the question addressed by \lgt is not whether graphs are representable.
Rather, \lgt asks how graph structure can be treated as primitive values in a core calculus,
thereby supporting both pattern matchings and typing with user-defined graph shapes.

\subsection{Graph Query and Transformation Languages}
\label{sec:related-graph-query}

The previous subsection discussed approaches that treat graphs
as data structures within functional programming languages.
A related but distinct line of work studies languages
for querying and transforming graphs stored outside the programming language,
such as graph databases and semi-structured data repositories.

Semi-structured data is data whose structure is not governed by a fixed schema.
Rather, structural information is often carried by the data itself,
for example through labels, nesting, and links.
Such data is therefore naturally represented by graph-based models.

UnQL was developed as a query language for semi-structured data,
and UnCAL provides the graph algebra underlying UnQL~\cite{uncal2000}.
UnCAL describes graph transformations by structural recursion over graph data.
FUnCAL later reformulated UnCAL-style graph transformations in a functional setting~\cite{funcal}.

These languages place graph transformation in the setting of database querying and semi-structured data.
The comparison with \lgt, however, is not only about this intended setting.
It also concerns the semantic basis on which graph structure is identified.
UnCAL graph semantics is designed to respect bisimulation.
In \lgt, graph pattern matching and typing are instead based on structural congruence of port hypergraphs.
Consequently, the concrete connection structure of a graph is part of the object being matched and typed.

\section{Pattern Matching beyond Ordinary Trees}\label{sec:related:pattern-matching}

Once graphs are treated as values,
the next question is how such values can be decomposed.

\subsection{Views, Contexts, and Graph Decomposition}
\label{sec:related-pattern-context}

Pattern matching is one of the central mechanisms of functional programming.
For algebraic data types, a pattern exposes the outer constructor of a value and binds its components to variables.
Several extensions of pattern matching have been proposed to apply this mechanism beyond ordinary constructor trees.
Wadler's views reconcile pattern matching with data abstraction
by allowing an abstract value to be matched through a suitable concrete representation~\cite{wadler-views}.
Erwig's active patterns pursue a related idea by applying preprocessing functions before matching~\cite{erwig-active-patterns}.

These mechanisms are related to \lgt because graph-specific views could in principle provide destructors for graph libraries.
In \lgt, however, graph decomposition is not delegated to an external view function.
Graphs, links, graph composition, and graph templates are part of the core syntax and semantics.
A pattern in \lgt is interpreted directly as a graph template modulo structural congruence,
and matching decomposes a graph value into subgraphs with explicit free-link interfaces.

Contextual patterns provide another related form of decomposition~\cite{ContextPatterns}.
They allow a sub-term to be matched even when it does not occur at a fixed position from the root,
thereby decomposing a tree-shaped value into a focused sub-term and its surrounding context.
This resembles graph pattern matching in \lgt,
but the corresponding context is no longer a one-hole tree context.
A separated subgraph may have multiple boundary links,
and its surrounding graph may contain sharing and cycles.
Thus, graph variables in \lgt represent subgraphs together with explicit interfaces.

This observation is also related to datatype derivatives and dissections.
McBride showed that the derivative of a regular type gives the type of one-hole contexts~\cite{mcbride-derivative},
and dissections further develop this view of data structures under traversal or transformation~\cite{mcbride-dissecting}.
The linear implication types of \lgt can be seen as a graph-oriented analogue of such context types:
they describe incomplete port hypergraphs that become complete when supplied with suitable missing parts.

\subsection{Functional Logic Programming}
\label{sec:related-functional-logic}

Functional logic programming provides another useful comparison point.
Logic programming generalises ordinary pattern matching by allowing variables to represent unknown parts of terms.
Using unification and search, a program can reason about incomplete trees as well as complete trees.

Functional logic languages combine this style of logic programming with features of functional programming.
For example, Curry combines higher-order functions, strong typing, and lazy evaluation
with free variables, non-determinism, narrowing, and built-in search~\cite{curry,flp}.
Curry also supports functional patterns,
which allow defined functions to appear in patterns and thereby express forms of deep pattern matching.

However, these mechanisms are still based mainly on terms and trees.
They do not directly provide pattern matching over general graph structures.
In this respect, \lgt is more expressive, since it treats general graphs as the target of pattern matching.

Conversely, ideas from logic programming also suggest possible extensions of \lgt.
For example, \lgt could be extended with search mechanisms analogous to those in logic programming.
Such an extension would make it possible to explore or enumerate all graph decompositions that satisfy a given pattern.
It would also be interesting to investigate whether Curry-style functional patterns can be adapted to graph pattern matching.

\section{Static Reasoning about Graph and Pointer Structures}\label{sec:related:static}

After representation and decomposition, the remaining question is how structural invariants can be checked or verified.

\subsection{Linearity, Affineness, and Ownership}
\label{sec:related-ownership-linearity}

Linear and affine types, as well as related type disciplines, provide techniques for controlling references~\cite{ownership2005,wadler1990linear}.
The basic idea is to restrict pointer sharing in order to prevent unintended propagation of the effects of destructive updates and to make memory management easier to handle mechanically.
A linear type requires a value to be used exactly once, whereas an affine type allows a value to be used at most once.

A prominent example of this idea is Rust~\cite{rust}.
Rust uses ownership, which can be seen as an affine type discipline, to provide memory-safety guarantees without relying on garbage collection.
In Rust, mutable references are restricted so that, at any point in time, there is at most one mutable reference to a value.
Although immutable references can be shared, many mutable data structures require destructive updates, and such updates must be performed through a unique mutable reference.

This discipline makes it difficult to directly express complex pointer structures with internal sharing, such as doubly linked lists or trees with parent pointers.
Therefore, linear or affine types are not well suited for complex data structures when they are applied naively to individual references.
Although fractional ownership and related approaches can ease this restriction, they do not by themselves ensure shape safety.

On the other hand, in \lgt, unrestricted copying of graphs may be inefficient.
One possible solution is to introduce linear or affine types for values in \lgt.
Since values in \lgt may themselves be complex graph structures,
such types would control copying at the level of values rather than forbidding sharing inside the graph.
Thus, linear or affine types could help avoid unnecessary copying without ruling out complex graph-structured values.

\subsection{Separation Logic and Shape Analysis}
\label{sec:related-separation-logic}

Separation logic and shape analysis provide powerful frameworks for reasoning about mutable heap structures.
Separation logic supports local reasoning about programs that manipulate pointers, and it can express heap invariants using separating conjunction, symbolic heaps, and inductive predicates~\cite{Reynolds02,separation-logic,beyond-symbolic-heaps,PiskacWZ13,EneaLSV14,IosifRS13}.
Shape analysis similarly aims to infer or verify structural properties of heap-allocated data, such as lists, trees, sharing patterns, and other pointer structures~\cite{shape-analysis}.

These approaches usually target programs that manipulate mutable heaps, where heap structures are described by logical assertions.
By contrast, \lgt treats complex data structures as immutable values and ensures structural safety through a lightweight type system.
Thus, \lgt differs from separation logic and shape analysis in that structural properties are expressed as types of immutable graph-structured values rather than as assertions over mutable heaps.

Nevertheless, the type system of \lgt is closely related to ideas from separation logic.
In particular, the type production rules of the type system of \lgt correspond to inductive predicates in separation logic~\cite{IosifRS13}.
Iosif et al.~\cite{IosifRS13} study restrictions on inductive definitions under which entailment becomes complete.
Such restrictions may provide useful guidance for identifying fragments of \lgt in which type checking can be made fully static, including pattern matching.

The linear implication types introduced in \cref{ch:limp-type} are also motivated by the magic wand of separation logic.
Studies of the magic wand may therefore provide useful insights for future extensions of \lgt,
such as type inference for type systems that include linear implication types.

\subsection{Graph Types and Reduction-Based Shape Typing}
\label{sec:related-graph-types}

Graph Types provide another related approach~\cite{graphtypes}.
They specify graph structures using spanning trees together with auxiliary edges described by routing expressions.
This representation differs from the graph formalisation used in \lgt.
In contrast, \lgt adopts a direct graph representation, which provides a simpler programming model and may offer a more straightforward basis for compilation.

Graph typing frameworks are the closest related work to the type system of \lgt.
Structured Gamma and ShapeType provide reduction-based approaches to describing and checking the shapes of graph structures~\cite{structuredgamma,shapetypes}.
In these systems, a graph type can be understood as a non-terminal symbol defined by production rules, and a graph has a type when it can be generated by reducing the corresponding type atom.
Yamamoto extended Structured Gamma to LMNtal and introduced difference types,
which correspond to linear implication types in the type system of \lgt~\cite{yamamoto2024}.

The core type system of \lgt follows a similar idea, using production rules based on hypergraph grammars to describe structural properties of graph values.
In \cref{sub:fgt-reduction}, we prove that, when graph variables and \(\lambda\)-abstraction atoms are ignored,
typing in \lgt corresponds to reduction from a type atom to a graph generated by production rules.
Thus, the type system of \lgt can be understood as an extension of reduction-based graph typing frameworks.

\chapter{Conclusions and Further Work}\label{ch:conclusion}

\section{Conclusions}

Enabling declarative and safe manipulation of complex data structures
is a fundamental challenge in programming language design.
Imperative languages can directly represent arbitrary structures using pointers
and destructive updates,
but it is difficult to ensure that the resulting structures satisfy their intended shapes.
Functional languages provide declarative manipulation, compositionality,
and static guarantees for algebraic data types,
but their standard mechanisms are most naturally suited to trees.
Graph transformation languages directly manipulate graphs
through pattern matching and rewriting,
but their state-transformation view does not fit neatly
with ordinary value- and function-oriented programming.

This dissertation has addressed this gap by developing \lgt{}.
The language treats complex data structures as port hypergraphs
and supports their declarative manipulation through graph pattern matching.
Its type system allows programmers to define shape invariants
and statically checks that graph values conform to them,
thereby ensuring that the intended structure of data is preserved during computation.
The main contributions of this dissertation are remarked as follows.

\begin{enumerate}
  \item
        \textbf{Formalisation and implementation of the core language of \lgt{}
          (\cref{ch:hyperlmntal,ch:untyped,ch:impl}).}
        To provide a foundation for the language,
        we first formalised HyperLMNtal as a term-based port hypergraph transformation formalism.
        On this basis,
        we defined the syntax and operational semantics of the untyped \lgt{} language,
        which extends a call-by-value \(\lambda\)-calculus with graph values modulo structural congruence
        and graph pattern matching.
        We also developed a compact prototype interpreter,
        showing that the core mechanisms of \lgt{},
        including graph decomposition through pattern matching
        and graph construction through composition,
        can be realised in a compact executable prototype.

  \item
        \textbf{Development of type systems for \lgt{}.
          (\cref{ch:core-type,ch:limp-type,ch:fully-static}).}
        The core type system allows programmers to define graph types
        that describe structural properties of port hypergraphs
        and checks that graph values conform to these specifications.
        This makes it possible to express global shape constraints of data structures,
        rather than only local node interfaces.
        We prove type soundness
        and also show that the type system can be viewed as a natural extension of
        existing graph-grammar-based typing frameworks
        for graph-transformation-based languages.
        We then extend the type system with \emph{linear implication types},
        which provide a way to type incomplete graph structures
        that naturally arise during pattern matching and graph transformation.
        For the extended system,
        we establish type soundness in the presence of dynamic checks during pattern matching
        and prove conservativity over complete graph types.
        We also identify a statically safe fragment of \lgt{}
        in which such dynamic checks can be safely eliminated.

  \item
        \textbf{Mechanisation of core parts of the meta-theory in Isabelle/HOL.
          (\cref{ch:formal}).}
        The meta-theory of \lgt is challenging because it combines
        several features that go beyond the ordinary \(\lambda\)-calculus,
        including graph values,
        multiple kinds of variables,
        structural congruence,
        and graph pattern matching.
        Mechanising the definitions and meta-theoretic results
        strengthens confidence in the soundness arguments
        and provides a reliable foundation for future extensions.
\end{enumerate}

The primary intended users of \lgt are library developers
and systems programmers who implement complex graph-shaped data structures.
Here,
systems programmers refer to programmers who develop foundational software such as database systems,
operating systems,
compilers,
runtime systems,
and other infrastructure-level software.
For such programmers,
it is desirable to describe the intended shape of internal data structures explicitly and to rule out structural errors before execution.
Since such software often requires efficient execution,
the performance aspects of \lgt are also important and are discussed later in this chapter.

The indirect beneficiaries of \lgt are programmers
who use libraries whose internal implementations rely on graph-shaped data structures.
In many cases,
library users interact only with the APIs
provided by libraries and do not need to know the internal representation of the data structures.
Nevertheless,
if these internal implementations are written in \lgt,
library users can benefit from improved reliability without directly programming in \lgt themselves.

\section{Future Work}

Several challenges remain for future work,
both in the runtime system and in the type system.

\subsubsection{Further development of the runtime system.}
Although we have provided a prototype implementation,
the correspondence between the formal semantics and the implementation
has not yet been proven.
The interpreter is designed to be as concise as possible
and consists of about 500 lines of code.
We therefore believe that it is feasible to prove this correspondence completely,
possibly with a mechanised proof in a proof assistant.

Efficient execution requires further investigation.
Although graph values are immutable in the formal semantics,
they may be represented using pointer-based data structures in an implementation.
This suggests a possible compilation strategy in which destructive updates
are used internally while the user-visible language remains purely functional.
Such an implementation would require additional static analyses,
such as ownership analysis or link-polarity analysis,
to ensure that these internal destructive operations are safe.

\subsubsection{Further investigation of the type system.}

The current prototype interpreter implements the full operational semantics of \lgt{},
but it does not yet include a type checker.
Ideally,
the type checker should also be mechanically verified against the theory developed
in this dissertation.

Polymorphism is important direction.
The current type system is essentially monomorphic.
For example,
difference lists are defined for specific element types,
rather than by a general parametric definition.
Introducing parametric polymorphism would improve usability
and bring \lgt closer to conventional functional languages.
However,
polymorphism in \lgt is nontrivial,
because graph operations may transform data structures
without explicitly mentioning constructor atoms,
and the result type of an operation may depend on the structural types
of its inputs.

Similar difficulties arise for type inference.
Since \lgt allows graph transformations to be expressed
without explicitly naming all constructor atoms,
the type of a program may depend on implicit structural information
about the input graphs.
Inferring such information automatically is likely to be more difficult
than type inference for ordinary algebraic data types.

\bigskip

The results of this dissertation constitute a foundational step towards
declarative and safe programming with complex data structures.
Further theoretical investigation and implementation work are still required,
but the framework developed here provides a basis for future work in this direction.

\printbibliography[title=References,heading=bibintoc]

\appendix

\chapter{Adding Recursive Let Bindings}\label{chapter:letrec}


As mentioned in \cref{sec:untyped:examples},
recursive definitions using \texttt{let rec} appear in the examples.
This appendix gives their formal treatment by extending the syntax,
type system, and operational semantics of \lgt
with recursive let bindings.

\begin{definition}[Syntax of Recursive Let Bindings]
  The syntax of expressions is extended as follows:
  \[
    e ::= \cdots
    \mid
    \letrec{f[\Ys]}{x[\Xs]:\tau_1}{e}{e}.
  \]
\end{definition}

Intuitively,
\[
  \letrec{f[\Ys]}{x[\Xs]:\tau_1}{e_1}{e_2}
\]
defines a recursive function \(f\) whose argument is \(x[\Xs]\), whose body is
\(e_1\), and whose scope is \(e_2\).
The sequence \(\Ys\) is the interface of the function value itself, whereas
\(\Xs\) is the interface of the argument graph.

\begin{definition}[Typing Rule for Recursive Let Bindings]
  Recursive let bindings are typed by the following rule:
  \begin{prooftree}
    \footnotesize
    \AXC{$
        \Gamma, f[\Ys] : (\tau_1 \to \tau_2)(\Ys), x[\Xs] : \tau_1
        \vdash_P
        e_1 : \tau_2
        \qquad
        \Gamma, f[\Ys] : (\tau_1 \to \tau_2)(\Ys)
        \vdash_P
        e_2 : \tau
      $}
    \RightLabel{\TyLetRec}
    \UIC{$
        \Gamma
        \vdash_P
        \letrec{f[\Ys]}{x[\Xs]:\tau_1}{e_1}{e_2} :
        \tau
      $}
  \end{prooftree}
  where
  \[
    \mathit{fn}(\tau_1) = \{\Xs\}
    \quad\mbox{and}\quad
    \mathit{fn}((\tau_1 \to \tau_2)(\Ys)) = \{\Ys\}.
  \]
\end{definition}

The rule \textsc{Ty-LetRec} follows the standard typing principle for recursive
function definitions.
To type the recursive function body \(e_1\), the function name \(f[\Ys]\) is made
available in the typing environment with function type
\((\tau_1 \to \tau_2)(\Ys)\).
The argument \(x[\Xs]\) is also added to the environment with type \(\tau_1\).
The body \(e_1\) must then have the result type \(\tau_2\).
The expression \(e_2\), which is the scope of the recursive binding, is typed under
the same recursive assumption for \(f[\Ys]\), but without the argument variable
\(x[\Xs]\), since \(x[\Xs]\) is bound only in the function body.
The whole recursive let expression has the type of \(e_2\).

To define the operational semantics of recursive let bindings, we need a form of
substitution that replaces a graph context \(f[\Ys]\) with an expression of the form
\(\mathbf{let\ rec}\ \cdots\).
However, the substitution defined so far replaces graph contexts with graph
templates, not with arbitrary expressions.
We therefore introduce a substitution operation that allows graph contexts to be
replaced by expressions.

There is a technical point specific to the syntax of \lgt.
In \lgt, graphs and expressions are syntactically distinct, and
expressions cannot occur directly inside graphs.
Therefore, when defining \(e_1[e_2/x[\Xs]]\), we cannot simply replace an
occurrence of \(x[\Xs]\) inside a graph template \(T\) by the expression \(e_2\).
Indeed, such an occurrence of \(x[\Xs]\) appears as part of a graph template, not
as an expression by itself.

To address this issue, when a graph template \(T\) contains the graph context
\(x[\Xs]\), we replace the whole template \(T\) by the application
\[
  ((\lambda\,x[\Xs].T)()\; e_2).
\]
The additional link sequence attached to the abstraction is immaterial, because it
disappears immediately after the subsequent \(\beta\)-reduction.
For this reason, we use the empty sequence \(()\) in the notation above.

We now define expression substitution as an extension of graph substitution.
Let \(\theta \triangleq [e'/x[\Xs]]\).
The substitution \(e\theta\) is defined as follows.

\begin{definition}[Expression Substitution]
  Expression substitution is defined by the following clauses:
  \[\begin{array}{llll}
      (e_1\; e_2)\theta
      \triangleq
      (e_1\theta\; e_2\theta).         \\[5mm]
      \begin{array}{llll}
        (\caseof{e_1}{T}{e_2}{e_3})\theta
        \triangleq \\
        \begin{cases}
          (\caseof{e_1\theta}{T}{e_2}{e_3\theta})
           & \mbox{if } \mathit{functor}(x[\Xs]) \in \mathit{ff}(T),
          \\
          (\caseof{e_1\theta}{T}{e_2\theta}{e_3\theta})
           & \mbox{otherwise}.
        \end{cases}
      \end{array} \\[15mm]
      T[e'/x[\Xs]]
      \triangleq
      ((\lambda x[\Xs].T)()\; e').
    \end{array}\]
\end{definition}

The first clause is standard for applications.
The second clause treats case expressions carefully.
If the pattern \(T\) binds the same graph context as the substitution target, then
the corresponding occurrence is bound in the successful branch, and the
substitution must not enter \(e_2\).
The otherwise branch \(e_3\), however, is not under this binding and is therefore
substituted as usual.
If the pattern does not bind the graph context, both branches are substituted.
The final clause handles graph templates by reifying the replacement of a graph
context by an expression as an application of a \(\lambda\)-abstraction.

Using this substitution, the operational semantics of recursive let bindings is
defined by the following reduction rule.

\begin{definition}[Operational Semantics of Recursive Let Bindings]
  We add the following reduction rule to the reduction rules in \cref{table:lgt-reduction}.

  \begin{prooftree}
    \small
    \AXC{}
    \RightLabel{\RdLetRec}
    \UIC{%
      \(\begin{array}{ll}
        \letrec{f[\Ys]}{x[\Xs]}{e_1}{e_2}
        \\
        \reduces
        e_2
        \rbracket{%
        (\lambda\,x[\Xs].\,e_1)(\Ys)
        \rbracket{%
          \letrec{f[\Ys]}{x[\Xs]}{e_1}{f[\Ys]} \Big/ f[\Ys]
        } \Big/ f[\Ys]
        }
      \end{array}\)}
  \end{prooftree}
\end{definition}

The rule unfolds a recursive binding by substituting the recursive function name
\(f[\Ys]\) with a function value whose body \(e_1\) may itself refer recursively to
\(f[\Ys]\).
The inner substitution replaces recursive calls inside the function body by the
recursive binding itself, while the outer substitution makes the resulting recursive
function available in the body \(e_2\).
Thus, recursive definitions are interpreted by one-step unfolding, in the same
spirit as standard operational accounts of \(\mathbf{let\ rec}\).

\chapter{Proofs of Propositions on Normal Forms}\label{app:normal-form-proof}

In this appendix, we provide detailed proofs of
\cref{lem:normal-form-exists} and \cref{lem:normal-form-uniq},
which were used in the proof sketch of soundness in
\cref{sec:sec4-fully-static-soundness}.
These propositions establish the existence and uniqueness of Normal Forms.


We prove \cref{lem:normal-form-exists} using the graph-prefix
normalisation and linear-implication elimination results established in
\cref{sec:elim-limprules}.
In particular, we use
\cref{lem:graph-prefix-rules},
\cref{lem:elim-ty-alpha},
\cref{lem:elim-ty-cong}, and
\cref{thm:ty-lis-elim}.

\noindent
\textbf{Restatement of \cref{lem:normal-form-exists}}
(Existence of Normal Form).
\textit{
  Let \(G\) be a typable graph whose type is not a linear implication type.
  Then, there exists a congruent graph $G'$
  whose
  typing derivation
  uses only \(\TyProd\) in
  the derivation
  without premises of \(\TyArrow\).
}

\begin{proof}
  By \cref{lem:graph-prefix-rules},
  for any graph,
  the typing derivation,
  that is the largest prefix of its derivation tree
  that does not contain premises of \(\TyArrow\),
  is derived solely by using
  \(\TyAlpha\), \(\TyCong\), \(\TyProd\), and \(\TyLIs\)

  Using \cref{lem:elim-ty-alpha} and \cref{lem:elim-ty-cong}),
  such typing derivation can be transformed
  so that \(\TyAlpha\) does not occur
  and \(\TyCong\) occurs only once as the final rule
  in the largest prefix of its derivation tree
  that contains no premises of \(\TyArrow\).

  Therefore,
  as shown in \cref{sec:elim-limprules},
  Any typing derivation for a graph that uses only
  \(\TyProd\) and \(\TyLIs\)
  in the largest prefix of its derivation tree
  that contains no premises of \(\TyArrow\)
  can be transformed,
  into one in which the typing uses only \(\TyProd\),
  \TyAlpha{}, and \TyCong{}.

  Applying \cref{lem:elim-ty-alpha} and \cref{lem:elim-ty-cong} again,
  we obtain the typing derivation
  only consists of \(\TyProd\)
  for a congruent graph.
\end{proof}


We next show that the normal form established above is unique.
That is, congruent graphs in normal form have the same typing derivation
when the subderivations above applications of \(\TyArrow{}\) are ignored.

\noindent
\textbf{Restatement of \cref{lem:normal-form-uniq}}
(Uniqueness of Normal Form).
\textit{
  Let \(G\) and \(G'\) be graphs in Normal Form such that \(G \equiv G'\).
  Then \(G\) and \(G'\) share the same typing derivation
  when the subderivations above applications of \(\TyArrow{}\) are ignored.
}

\begin{proof}
  As pointed out after \cref{def:normal-form},
  the syntactic structure of a graph in Normal Form can be viewed as a spanning tree,
  where the nodes correspond to atoms and the edges correspond to
  their primary root links.
  Accordingly, the statement
  is equivalent to claiming that the structure of this spanning tree is uniquely determined.

  To establish the uniqueness of the spanning tree,
  we traverse the primary root links of atoms in the graph,
  starting from the free root link \( X \) of the type of the graph,
  and demonstrate that the resulting traversal structure is uniquely determined.

  When a graph $G$ whose type has a link $X$ as a root atom
  is in a Normal Form,
  it can be written in the following form:
  \begin{align}\label{eq:normal-form-01}
    \nu \Ys.(C(\Zs, X), \overrightarrow{W_1 \bowtie W_2}, \Gs).
  \end{align}
  This representation arises due to two key factors:
  the constraints imposed on type production rules by \cref{def:disjoint-condition},
  and the fact that only the typing rules \TyProd{} and
  \TyAlpha{} are used in the derivation.
  The use of \TyProd{} ensures that the syntactic form of the graph conforms
  to the structural constraints specified by \cref{def:disjoint-condition}.
  Although \TyAlpha{} permits the renaming of free link names,
  it does not affect the overall syntactic structure of the graph.

  Due to the conditions on links in \cref{def:disjoint-condition},
  specifically Conditions (3),
  there is no atom in $\Gs$ whose final argument is $X$
  except for $C(\Zs, X)$
  in \cref{eq:normal-form-01}.
  Therefore,
  identifying the atom with final link $X$ uniquely determines the first atom
  $C(\Zs,X)$.

  By structural induction over the form of $G$,
  it is straightforward to see that
  every atom in $G_j$ in $\Gs$ is traversable via primary root
  links from $Z_i$.
  Also, any atom in $G_k$ where $k \neq j$ does not have a root link $Z_i$
  due to the link conditions in
  \cref{def:disjoint-condition} (Conditions (3)).
  Therefore,
  we can uniquely determine which atom belongs to
  which sub-spanning tree $G_j$ for every atom in $G$.
  Thus the partition of $G_j$'s is uniquely determined.

  Also, since $Z_i$s are distinct by Condition (3)
  in \cref{def:disjoint-condition},
  The order of $G_j$'s is also uniquely determined.

  Finally, the representation of $\overrightarrow{W_1 \bowtie W_2}$
  is uniquely determined by the form of the production rules.

  Therefore, Normal Form of any typable graph is uniquely determined
  up to $\alpha$-conversion of links.
  This implies that the spanning tree of the graph,
  whose backbone consists of root links, is unique.
\end{proof}

\end{document}